\documentclass[12pt]{article}

\usepackage{dsfont}
\usepackage{mathrsfs}
\usepackage{hvdashln}
\usepackage{multirow}

\usepackage{mathrsfs}   
\usepackage{eucal}      
\usepackage{amssymb}    
\usepackage{amsmath}    
\usepackage{amsbsy}

\usepackage{array}

\usepackage[english]{babel}
\usepackage[OT4]{polski}
\usepackage[cp1250]{inputenc}                       

\usepackage[pdftex,usenames,dvipsnames]{color}      
\usepackage[pdftex,pagebackref=false,draft=false,pdfpagelabels=false,colorlinks=true,urlcolor=blue,linkcolor=black,citecolor=black,pdfstartview=FitH,pdfstartpage=1,pdfpagemode=UseOutlines,bookmarks=true,bookmarksopen=true,bookmarksopenlevel=2,bookmarksnumbered=true,pdfauthor={Marcin Szewczyk},pdftitle={Doktorat},pdfsubject={Praca doktorska},pdfkeywords={transient recovery voltage trv},unicode=true]{hyperref}   

\usepackage[colorlinks=true]{hyperref}
\usepackage[numbers,sort&compress]{natbib}  
\usepackage{hypernat}                       

\usepackage{extsizes}           
\usepackage[pdftex]{graphics}
\usepackage[a4paper,inner=2.5cm,outer=2.5cm,top=3cm,bottom=2cm]{geometry}
\usepackage{indentfirst}        
\usepackage{epsfig}
\usepackage{strona_tytulowa}

\usepackage[hide]{todo}            
\usepackage[basic,physics]{circ}            
\usepackage[rm,bf,outermarks,]{titlesec}     
\usepackage{tocloft}                        
\usepackage{expdlist}    
\usepackage{flafter}     
\usepackage{array}       
\usepackage{listings}    
\usepackage[format=hang,labelsep=period,labelfont={bf,small},textfont=small]{caption}   
\usepackage{appendix}    
\usepackage{floatflt}    
\usepackage{here}        
\usepackage{float}
\usepackage{makeidx}     

\usepackage{fancyhdr}          

\makeatletter

\def\numberline#1{\hb@xt@\@tempdima{#1.\hfil}}                      
\renewcommand*\@seccntformat[1]{\csname the#1\endcsname.\enspace}   

\makeatother

\makeatletter

\@addtoreset{equation}{section} 
\renewcommand{\theequation}{{\thesection}.\@arabic\c@equation} 

\@addtoreset{figure}{section}
\renewcommand{\thefigure}{{\thesection}.\@arabic\c@figure}

\@addtoreset{table}{section}
\renewcommand{\thetable}{{\thesection}.\@arabic\c@table}

\makeatother

\renewcommand{\todoitem}[2]{%
\item \label{todo:\thetodo}%
\ifx#1\todomark%
\else\textbf{#1 }%
\fi%
(str.~\pageref{todopage:\thetodo})\ #2}

\renewcommand{\todomark}{~uzupe�ni�}

\reversemarginpar 

\def\mathcal#1{%
    \mathscr{#1}%
    }

\titleformat{\section}        
{\LARGE \rmfamily \bfseries}{\thesection}{1em}{}

{
	\fancyhead[LE,RO]{\slshape \leftmark}}%
\author{ALINA CZAJKA}
\title{SUPERSYMMETRIC PLASMA SYSTEMS AND THEIR NONSUPERSYMMETRIC COUNTERPARTS}

\uczelnia{JAN KOCHANOWSKI UNIVERSITY IN KIELCE}
\wydzial{FACULTY OF MATHEMATICS AND NATURAL SCIENCES}
\instytut{INSTITUTE OF PHYSICS}
\album{}
\promotor{prof. dr hab. Stanis\l aw Mr\'owczy\'nski}
\praca{Ph. D. thesis}
\rok{2015}

\begin{document}

\renewcommand{\refname}{Bibliography}

\renewcommand{\figurename}{Fig.}  
\renewcommand{\tablename}{Tab.}     
\pagenumbering{roman}               
\thispagestyle{empty}               
\stronatytulowa                     



\newpage
\thispagestyle{plain}
\textcolor[rgb]{1.00,1.00,1.00}{.}
\vspace{-1cm}

\begin{center}
\LARGE \textbf{Abstract}
\end{center}

{\small
In this thesis we systematically compare supersymmetric plasma systems to their nonsupersymmetric counterparts. The work is motivated by the AdS/CFT correspondence and our aim is to check how much the plasma governed by the $\mathcal{N} = 4$ super Yang-Mills theory resembles the quark-gluon plasma studied experimentally in relativistic heavy-ion collisions. The analysis is done in a~weak coupling regime where perturbative methods are applicable. Since the Keldysh-Schwinger approach is used, not only equilibrium but also nonequilibrium plasmas, which are assumed to be ultrarelativistic, are under consideration. 

Using the functional techniques we introduce Faddeev-Popov ghosts into the Keldysh-Schwin- ger formalism of nonAbelian gauge theories. A generating functional is constructed and the Slavnov-Taylor identities are derived. One of the identities expresses the ghost Green function through the gluon one. Having the ghost Green functions opens up a possibility of perturbative calculus in a covariant gauge. 

Next the collective excitations of the $\mathcal{N} = 1$ SUSY QED plasma are considered and compared to those of the usual QED system. The analysis is repeated to confront with each other the plasmas governed by the $\mathcal{N} = 4$ super Yang-Mills and QCD theories. Consequently, the dispersion equations of quasiparticles of all fields occurring in the plasmas are written down and respective self-energies, which enter the equations, are computed in the hard-loop approximation. To obtain a gluon polarisation tensor we use the ghost Green functions found before. Polarisation tensors of all gauge bosons of supersymmetric plasmas have the same structures as those of their nonsupersymmetric counterparts. The same holds for fermion self-energies. Self-energies of scalars, which occur in the supersymmetric systems, are found to be independent of the wavevector. It is also shown that the self-energies of gauge boson, fermion, and scalar fields for a whole class of gauge theories have unique and universal forms. Having the self-energies, we construct an effective action in the hard-loop limit, which appears to be universal as well. The universality of the action has far-reaching consequences, as it makes that the long-wavelength features of all considered plasma systems, in particular the spectrum of collective excitations, are almost identical. 

In the last part of the thesis, transport properties of the systems are studied. Because of dimensional constraints, only some transport coefficients of supersymmetric systems are likely to exhibit qualitative differences with respect to the usual ones. Accordingly, energy loss and momentum broadening caused by the Compton scattering on selectron are computed. The process is independent of momentum transfer and therefore is qualitatively different from processes in the usual QED or QCD systems. The formulas of the energy loss and momentum broadening in the high energy limit of the test particle are shown to be surprisingly similar to those of QED plasma. These considerations are generalised for the nonAbelian systems. Since both collective excitations and transport characteristics of supersymmetric and nonsupersymmetric plasmas are very similar to each other we conclude that supersymmetric plasma systems are qualitatively the same as their nonsupersymmetric partners in weak coupling regime. The quantitative differences mostly reflect different numbers of degrees of freedom.
}




\newpage
\thispagestyle{plain}
\textcolor[rgb]{1.00,1.00,1.00}{.}
\vspace{-1cm}

\begin{center}
\LARGE \textbf{Streszczenie}
\end{center}

{\small
W niniejszej rozprawie por\'ownujemy supersymetryczne uk\l{}ady plazmowe z ich niesupersymetrycznymi odpowiednikami. Praca jest motywowana dualno\'sci\k{a} AdS/CFT, a g\l{}\'ownym jej celem jest sprawdzenie, na ile plazma rz\k{a}dzona teori\k{a} $\mathcal{N} = 4$ super Yang-Mills przypomina plazm\k{e} kwarkowo-gluonow\k{a}, kt\'ora jest badana eksperymentalnie w zderzeniach relatywistycznych ci\k{e}\.zkich jon\'ow. Analiza uk\l{}ad\'ow supersymetrycznych i niesupersymetrycznych jest przeprowadzona w~re\.zimie s\l{}abego sprz\k{e}\.zenia, kt\'ory umo\.zliwia zastosowanie rachunku perturbacyjnego. Poniewa\.z u\.zywamy podej\'scia Keldysha-Schwingera, rozwa\.zane s\k{a} zar\'owno uk\l{}ady r\'ownowagowe jak i~nier\'ow- nowagowe, zak\l{}adamy przy tym, \.ze s\k{a} one ultrarelatywistyczne. 

Przy u\.zyciu metod funkcjonalnych pokazujemy jak wprowadzi\'c duchy Faddeeva-Popova do nieabelowych teorii z cechowaniem w podej\'sciu Keldysha-Schwingera. Konstruujemy zatem funkcjona\l{} generuj\k{a}cy, a nast\k{e}pnie wyprowadzamy to\.zsamo\'sci Slavnova-Taylora. Jedna z~nich wyra\.za zwi\k{a}zek funkcji Greena duch\'ow z funkcjami gluonowymi, co pozwala wyznaczy\'c dwu-punktow\k{a} funkcj\k{e} Greena duch\'ow i otwiera tym samym mo\.zliwo\'s\'c stosowania cechowania kowariantnego w rachunkach perturbacyjnych. 

Nast\k{e}pnie badamy wzbudzenia kolektywne plazmy opisywanej teori\k{a} $\mathcal{N} = 1$ SUSY QED i~por\'ownujemy je ze wzbudzeniami zwyk\l{}ej plazmy elektromagnetycznej. T\k{e} analiz\k{e} powtarzamy dla uk\l{}ad\'ow plazmowych rz\k{a}dzonych teoriami nieabelowymi, mianowicie teori\k{a} $\mathcal{N} = 4$ super Yang-Mills i QCD. Wypisujemy wi\k{e}c r\'ownania dyspersyjne kwazicz\k{a}stek poszczeg\'olnych p\'ol, a nast\k{e}pnie znajdujemy w przybli\.zeniu twardych p\k{e}tli ich energie w\l{}asne wchodz\k{a}ce do tych r\'owna\'n. Obliczaj\k{a}c tensor polaryzacji gluon\'ow wykorzystujemy funkcje Greena duch\'ow znalezione wcze\'sniej. Tensory polaryzacji bozon\'ow cechowania uk\l{}ad\'ow supersymetrycznych maj\k{a} tak\k{a} sam\k{a} struktur\k{e} jak tensory ich niesupersymetrycznych partner\'ow. T\k{a} sam\k{a} w\l{}asno\'s\'c wykazuj\k{a} fermionowe energie w\l{}asne. Energie w\l{}asne p\'ol skalarnych, kt\'ore wyst\k{e}puj\k{a} tylko w uk\l{}adach supersymetrycznych, okazuj\k{a} si\k{e} by\'c niezale\.zne od wektora falowego. Pokazujemy tak\.ze, \.ze energie w\l{}asne bozon\'ow cechowania, fermion\'ow i skalar\'ow maj\k{a} unikalne i uniwersalne formy dla ca\l{}ej klasy teorii z cechowaniem. Dysponuj\k{a}c energiami w\l{}asnymi konstruujemy dzia\l{}anie efektywne w przybli\.zeniu twardych p\k{e}tli, kt\'ore okazuje si\k{e} by\'c r\'ownie\.z uniwersalne. W\l{}asno\'s\'c ta ma daleko id\k{a}ce konsekwencje, a mianowicie powoduje, \.ze charakterystyki d\l{}ugofalowe rozwa\.zanych uk\l{}ad\'ow plazmowych, w szczeg\'olno\'sci widmo wzbudze\'n kolektywnych, s\k{a} we wszystkich uk\l{}adach niemal identyczne. 

W ostatniej cz\k{e}\'sci pracy rozwa\.zamy w\l{}asno\'sci transportowe uk\l{}ad\'ow. Ze wzgl\k{e}du na ograni- czenia wynikaj\k{a}ce z analizy wymiarowej, tylko niekt\'ore wsp\'o\l{}czynniki transportowe w uk\l{}adach supersymetrycznych mog\k{a} by\'c jako\'sciowo r\'o\.zne od ich odpowiednik\'ow w uk\l{}adach niesupersymetrycznych. Z tego powodu obliczone zosta\l{}y straty energii oraz poszerzenie p\k{e}dowe powodo- wane rozpraszaniem Comptona na selektronach. Taki proces jest niezale\.zny od przekazu p\k{e}du, przez co jest jako\'sciowo r\'o\.zny od proces\'ow zachodz\k{a}cych w plazmie elektromagnetycznej czy te\.z kwarkowo-gluonowej. Uzyskane wyra\.zenia na straty energii i poszerzenie p\k{e}dowe s\k{a} w granicy du\.zej energii cz\k{a}stki testowej zadziwiaj\k{a}co podobne do odpowiednich wielko\'sci charakteryzuj\k{a}cych zwyk\l{}\k{a} plazm\k{e} elektromagnetyczn\k{a}. Te rozwa\.zania s\k{a} uog\'olnione dla uk\l{}ad\'ow nieabelowych. Poniewa\.z zar\'owno wzbudzenia kolektywne jak i w\l{}asno\'sci transportowe rozwa\.zanych uk\l{}ad\'ow s\k{a} bardzo podobne do siebie stwierdzamy, \.ze uk\l{}ady supersymetryczne i niesupersymetryczne s\k{a} jako\'sciowo takie same. R\'o\.znice ilo\'sciowe odzwierciedlaj\k{a} g\l{}\'ownie r\'o\.zne liczby stopni swobody.
}




\newpage
\thispagestyle{plain}
\textcolor[rgb]{1.00,1.00,1.00}{.}
\vspace{3cm}

\textbf{\Large The thesis is based on the following original papers:}
\vspace{1cm}

\begin{enumerate}

\item A.~Czajka and St.~Mr\'owczy\'nski,
\emph{Collective Excitations of Supersymmetric Plasma},\\
Phys.\ Rev.\ D {\bf 83}, 045001 (2011).
\item A.~Czajka and St.~Mr\'owczy\'nski,
\emph{Collisional Processes in Supersymmetric Plasma},\\
Phys.\ Rev.\ D {\bf 84}, 105020 (2011).
\item A.~Czajka and St.~Mr\'owczy\'nski,
\emph{N=4 Super Yang-Mills Plasma},\\
Phys.\ Rev.\ D {\bf 86}, 025017 (2012).
\item A.~Czajka and St.~Mr\'owczy\'nski,
\emph{Ghosts in Keldysh-Schwinger Formalism},\\
Phys.\ Rev.\ D {\bf 89}, 085035 (2014).
\item A.~Czajka and St.~Mr\'owczy\'nski,
\emph{Universality of the Hard-Loop Action},\\
Phys.\ Rev.\ D {\bf 91}, 025013 (2015).

\end{enumerate}



\renewcommand{\cftbeforesecskip}{8pt}
\renewcommand{\cftsecafterpnum}{\vskip 12pt}
\renewcommand{\cftparskip}{4pt}
\renewcommand{\cfttoctitlefont}{\LARGE\bfseries\rmfamily}
\renewcommand{\cftsecfont}{\bfseries\rmfamily}
\renewcommand{\cftsubsecfont}{\rmfamily}
\renewcommand{\cftsubsubsecfont}{\rmfamily}
\renewcommand{\cftparafont}{\rmfamily}



\newpage
\thispagestyle{plain}
\tableofcontents
\newpage







\newpage
\textcolor[rgb]{1.00,1.00,1.00}{.}
\thispagestyle{plain}
    

\begin{center}
\LARGE \textbf{Acknowledgements}
\end{center}
\vspace{1cm}

I am deeply grateful to my supervisor, Stanis\l aw Mr\'owczy\'nski, for his devoted guidance
throughout my research, for sharing with me his excitement and passion for physics, and for constantly reminding me of its beauty. I greatly appreciate all the time he has spent with me, his constant involvement, support and inspiration. I thank him for his warm and encouraging attitude, for teaching me so many things in physics, and for showing me the importance of clarity and preciseness.

\thispagestyle{plain}



\newpage
\textcolor[rgb]{1.00,1.00,1.00}{.}
\thispagestyle{empty}

\titlespacing{\section}{0pt}{2cm}{2cm}

\pagenumbering{arabic}


\def\lg{{\mathchoice{~\raise.58ex\hbox{$<$}\mkern-14.8mu\lower.52ex\hbox{$>$}~}
                    {~\raise.58ex\hbox{$<$}\mkern-14.8mu\lower.52ex\hbox{$>$}~}
                    {\raise.59ex\hbox{{$\scriptscriptstyle <$}}\mkern-12.8mu%
                     \lower.01ex\hbox{{$\scriptscriptstyle >$}}}   {}   }}
\def\gl{{\mathchoice{~\raise.58ex\hbox{$>$}\mkern-12.8mu\lower.52ex\hbox{$<$}~}
                    {~\raise.58ex\hbox{$>$}\mkern-12.8mu\lower.52ex\hbox{$<$}~}
                    {\raise.62ex\hbox{{$\scriptscriptstyle >$}}\mkern-12.0mu%
                     \lower.05ex\hbox{{$\scriptscriptstyle <$}}}  {}    }}

\def\ca{{\mathchoice{~\raise.58ex\hbox{$c$}\mkern-9.0mu\lower.52ex\hbox{$a$}~}
                    {~\raise.58ex\hbox{$c$}\mkern-9.0mu\lower.52ex\hbox{$a$}~}
                    {\raise.59ex\hbox{{$\scriptscriptstyle c$}}\mkern-7.0mu%
		     \lower.01ex\hbox{{$\scriptscriptstyle a$}}}   {} 	}} 
\def\ac{{\mathchoice{~\raise.58ex\hbox{$a$}\mkern-10.0mu\lower.52ex\hbox{$c$}~}
                    {~\raise.58ex\hbox{$a$}\mkern-10.0mu\lower.52ex\hbox{$c$}~}
		    {\raise.62ex\hbox{{$\scriptscriptstyle a$}}\mkern-9.0mu%
		     \lower.05ex\hbox{{$\scriptscriptstyle c$}}}  {} 	}}

\newcommand{\be}{\begin{equation}}
\newcommand{\ee}{\end{equation}}
\newcommand{\ba}{\begin{eqnarray}}
\newcommand{\ea}{\end{eqnarray}}
\newcommand{\ban}{\begin{eqnarray*}}
\newcommand{\ean}{\end{eqnarray*}}
\newcommand{\sla}{\!\!\!/ \,}

\newcommand{\ol}{\overline}
\newcommand{\pab}{\bar{\partial}}
\newcommand{\thb}{\bar{\theta}}
\newcommand{\zeb}{\bar{\zeta}}
\newcommand{\xib}{\bar{\xi}}
\newcommand{\psib}{\bar{\psi}}
\newcommand{\chib}{\bar{\chi}}
\newcommand{\phib}{\bar{\phi}}
\newcommand{\lambdab}{\bar{\lambda}}
\newcommand{\sigmab}{\bar{\sigma}}
\newcommand{\ad}{{\dot\alpha}}
\newcommand{\bd}{{\dot\beta}}
\newcommand{\nn}{\nonumber}
\newcommand{\pd}{\hat{p}_1}
\newcommand{\pc}{\hat{p}_2}
\newcommand{\ej}{\hat{\varepsilon}_1}
\newcommand{\et}{\hat{\varepsilon}_2}
\newcommand{\kj}{\hat{k}_1}
\newcommand{\kt}{\hat{k}_2}
\newtheorem{tw}{Twierdzenie}

\vspace{-2.7cm}
\section{Introduction}
\label{intro}

\thispagestyle{plain}
\vspace{-1cm}

\begin{tabular}{ m{5cm} l}
& \emph{First of all Chawos came into being.}
\\
& \emph{But then Gaia broad-chested (\dots)} 
\\
& \emph{From Chawos were born Erebos and black Night.} 
\\
& \emph{From Night, again, were born Aether and Day (\dots)} 
\end{tabular}

\vspace{.3cm}

\begin{tabular}{ m{5cm} r}
& Theogony, Hesiod
\end{tabular}

\vspace{2cm}

Soon after the Big Bang the matter existed in the state of a quark-gluon plasma which then turned into hadrons, which next formed atomic nuclei, and these formed atoms and so forth, pending the present form of the Universe filled by numerous clusters of galaxies. Although a quark-gluon plasma is very `old' state of matter, albeit only a few dozen years old concept of human awareness, there are still no satisfactory methods which have enabled us to determine and understand its properties fully and unequivocally. Some approaches, such as a perturbative quantum field theory or lattice QCD, are helpful in describing this state of matter but limits of their application make the knowledge of the plasma rather fragmentary. Little is known, for example, about exact parameters of phase transition from a hadron to quark-gluon stage or about mechanisms which lead the plasma to rapid thermalisation. These and other puzzles galore are conducive to searching new tools in order to face up to numerous difficulties. 

One of these very recent ideas is Maldacena's discovery of the anti-de Sitter/conformal field theory correspondence (AdS/CFT) \cite{Maldacena:1997re}. The correspondence exhibits a relationship between the weakly coupled gravity in 5-dimensional anti-de Sitter space and a conformal field theory of strong coupling which is the $\mathcal{N}=4$ supersymmetric Yang-Mills theory (SYM). The Maldacena duality draws more and more attention among high-energy physics community as it offers a systematic method to study strongly coupled systems though indirectly, that is, via weakly coupled classical gravity whose toolkit is quite well known. Having said that, one can ask how much properties of this rather artificial, but theoretically interesting, system governed by the $\mathcal{N}=4$ super Yang-Mills theory resemble these of natural quark-gluon plasma described by quantum chromodynamics (QCD). And, generally speaking, to what extent AdS/CFT may be useful in exploring properties of matter.

The aim of this thesis is to compare plasma systems governed by the QCD and $\mathcal{N}=4$ super Yang-Mills theory not in a strong but in a weak coupling regime, where perturbative calculus is applied. The comparison is done systematically so that we start with an analysis of simpler theories, namely, we compare first the usual electromagnetic (QED) plasma with its supersymmetric counterpart, the $\mathcal{N}=1$ SUSY QED plasma. Then, we broaden the studies to the systems described by nonAbelian theories. In all these systems collective excitations and transport characteristics are investigated. 

However, prior to any quantitative considerations let us briefly present the main objects of our study to shed some light on the background of our research program. These include the quark-gluon plasma and AdS/CFT duality. Later on the outline of the thesis is adumbrated.

\subsection{Quark-gluon plasma}
\label{qgp}

A quark-gluon plasma is a state of matter constituted by quarks - the matter particles, and gluons which are massless messengers of interaction. All these particles carry additional charge, called colour. The plasma is a strongly interacting system with not only quarks but also gluons interacting among each other and therefore its properties are qualitatively different than those of known systems, such as an electromagnetic plasma. As discussed later on, the strong colour forces make the plasma behave as a near-ideal liquid. At normal terrestrial conditions, where low energy densities or low temperatures prevail, quarks and gluons are confined in the interiors of colour-neutral hadrons. Under extremely high temperature and/or density hadrons start to overlap releasing their constituents, which, in turn, can propagate within the whole volume that the system occupies. The plasma looms large as it is believed that just a split second after the Big Bang the matter existed in such a state. With the expansion of the Universe the temperature was decreasing and other forms of matter started to gradually emerge.

\begin{center}
{\bf Quantum chromodynamics}
\end{center}

The idea of quarks and gluons first appeared as a theoretical concept in the 1960s. Namely the idea of quarks as constituents of hadrons was delivered first in 1964 by Gell-Mann \cite{GellMann:1964nj} and Zweig \cite{Zweig:1981pd,Zweig:1964jf}. Subsequently, Greenberg \cite{Greenberg:1964pe} and Han and Nambu \cite{Han:1965pf} broke new ground on degrees of freedom carried by quarks so that a new charge, later called the colour, was introduced. Han and Nambu introduced the idea that quarks may interact by exchanging gluons. These revelations launched fast development of the theory of strong interactions, {\it quantum chromodynamics}\footnote{The basics of QCD are also briefly discussed in Sec. \ref{subsec-QCD}.}. There are six types of quarks, named flavours: $u$, $d$, $s$, $c$, $b$, and $t$, that are, {\it up}, {\it down}, {\it strange}, {\it charm}, {\it bottom}, and {\it top}, respectively, and gluons. Hadrons consist of different combinations of quarks, thus we differentiate mesons, the quantum numbers of which correspond to a pair of quark and antiquark and barions whose quantum numbers correspond to three quarks. As far as the colour charge is concerned, hadrons do not carry it, they are said to be white, and it is impossible to separate constituent quarks from each other so that to isolate and directly observe a colour particle. This phenomenon is known as the {\it colour confinement}. One says that the phenomenon reflects a strong interaction between the hadron constituents which were also called {\it partons} in a different context. Feynman \cite{Feynman:1969wa} and Bjorken \cite{Bjorken:1968dy} suggested a way on how experiments of high energies can observe them. The first experimental indications that nucleons may indeed contain smaller objects in their interiors appeared in 1969 when the experiments of the deep inelastic scattering of electrons on hadrons were conducted at the Stanford Linear Accelerator Center (SLAC). Then, deflected electrons revealed structures of hadrons. The partons were soon identified with quarks of QCD. It was also implied that quarks have fractional electric charges.

These achievements led Gross, Wilczek, and Politzer \cite{Gross:1973ju,Gross:1973id,Politzer:1973fx} in 1973 to the observation that QCD reveals a property called {\it asymptotic freedom}. The asymptotic freedom means that the coupling constant of the strong interactions $\alpha_s$ decreases with the transfer of four-momentum $Q$ as
\be
\label{coupling}
\alpha_s(Q)=\frac{12\pi}{(33-2N_f) \ln Q^2/\Lambda^2_{\rm QCD}},
\ee
where $N_f$ is the number of quark flavours and $\Lambda_{\rm QCD}$ is the scale parameter of QCD, which amounts approximately to 200 MeV. From Eq. (\ref{coupling}) one sees that the coupling constant is small as long as the momentum transfer is large, that is, $Q^2 \gg \Lambda^2_{\rm QCD}$. The discovery has allowed us to implement perturbative quantum field theory techniques to study the interactions with a large momentum transfer. 

\begin{figure}[!h]
\centering
\includegraphics*[width=0.6\textwidth]{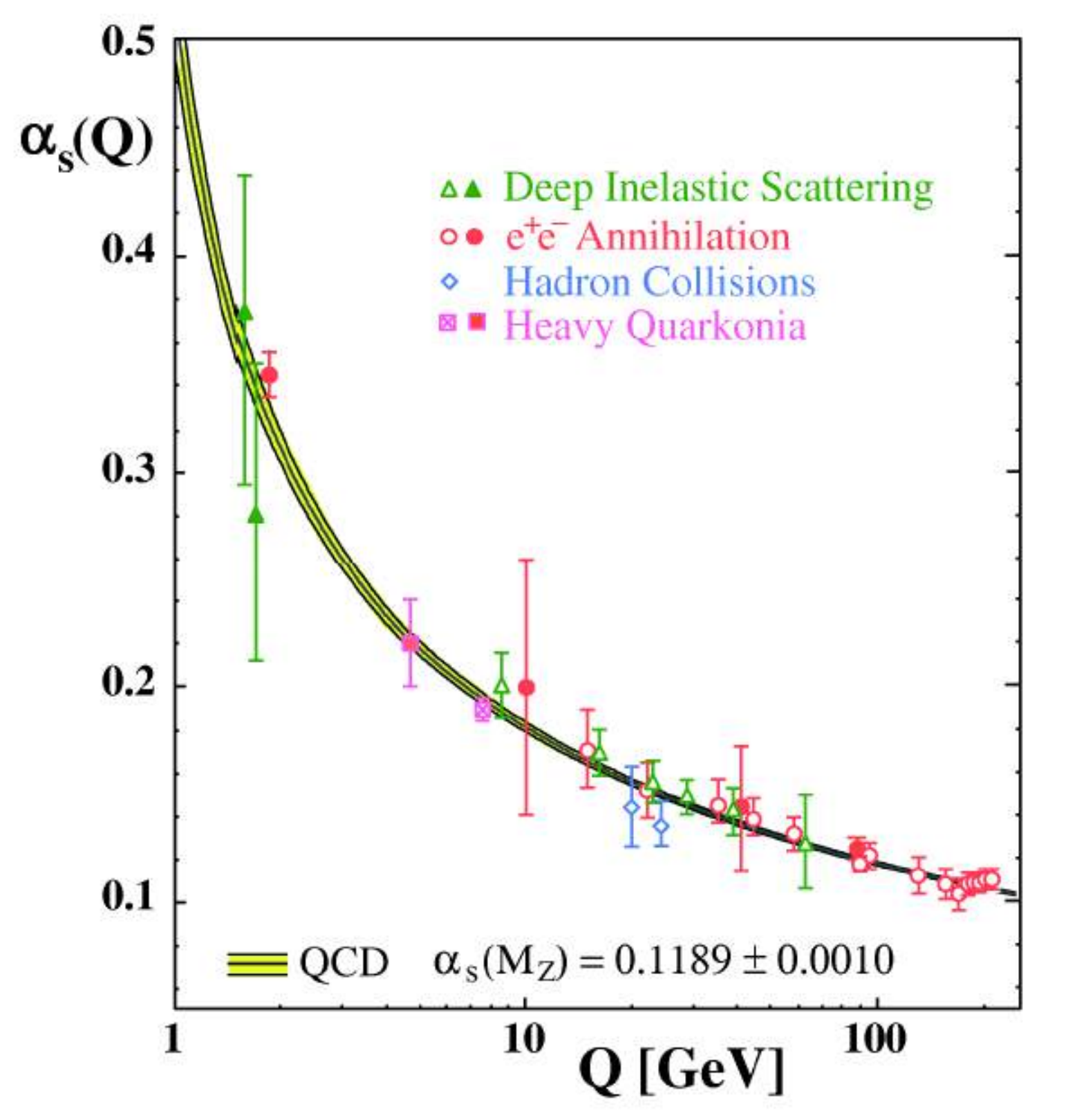}
\caption{Summary of measurements of $\alpha_s(Q)$ as a function of the respective energy scale. Figure taken from \cite{Bethke:2006ac}.}
\label{alpha}
\end{figure}

The coupling constant of strong interactions, $\alpha_s$, was measured experimentally as a function of the respective energy scale $Q$, for details see \cite{Bethke:2006ac}, and the outcome is presented in Fig. \ref{alpha}.
As seen, the experimental results are in a perfect agreement with QCD predictions, especially they confirm that $\alpha_s$ is a running coupling constant which makes that the strongly interacting nuclear matter asymptotically becomes an ideal gas of partons. Indeed, along with the increase in the temperature of a system, which is related to the average momentum transfer as $\langle Q^2\rangle \sim T^2$, the force between quarks becomes asymptotically weak. This, in turn, means that in a high-energy regime quarks can move freely and one says that the matter is in the phase of deconfinement. The confirmation of the asymptotic freedom delivered decent arguments to study quark-gluon plasma perturbatively, using the well known methods of many-body quantum field theory. The methods and some results are presented in the standard textbooks, such as \cite{Kapusta,Kapusta-Gale,Peskin,LeBellac,Pokorski,Weinberg:1995mt,Weinberg:1996kr,Weinberg:2000cr}.

Despite that matter does not exist in the state of a quark-gluon plasma in terrestrial conditions, there appeared in 1970s convincing arguments that the plasma is a natural object, see \cite{Collins:1974ky,Baym:1976yu,Chapline:1977rn,Morley:1978aq}. It is suggested there that the plasma may be concealed in dense centres of some compact astrophysical objects as neutron stars. Almost in the same period it was noted in \cite{Bohr:1977gj,Freedman:1976ub,Shuryak:1977ut,Shuryak:1978ij,Shuryak:1980tp,Kalashnikov:1979dp, Kapusta:1979fh} that the plasma may be produced in relativistic heavy-ion collisions and first attempts to study its properties at laboratories were undertaken. Accordingly, the heavy-ion collisions were established as a basic method of experimental studies of the quark-gluon plasma. Soon, experiments of colliding highly-energetic heavy ions started a new era of the quark-gluon plasma physics.

\begin{center}
{\bf Experimental programs}
\end{center}

At the beginning of this era, the particle physics community made use of existing accelerators which were adjusted to accelerate heavy ions to relativistic energies. Among them the Bevatron at the Lawrence Berkely National Laboratory (LBNL) was joined to the SuperHILAC to form the Bevalac where the energies of 1-2 GeV per nucleon were reached. Likewise the Dubna Syncrophasotron was modified. 

Broad research projects dedicated to truly ultrarelativistic heavy-ion collisions were established at the Brookhaven National Laboratory (BNL) and at the European Organisation for Nuclear Research (CERN) in 1986. The Alternating Gradients Synchrotron (AGS) at BNL started with boosting silicon ions at 14 Gev per nucleon whereas the Super Proton Synchrotron (SPS) at CERN accelerated oxygen ions at 60 and 200 GeV per nucleon in 1986 and sulphur ions at 200 GeV per nucleon in 1987. In 1992 BNL conducted experiments of acceleration of gold ions at 11 GeV per nucleon. In 1995, in turn, CERN initiated a program of boosting lead beams at 158 GeV per nucleon. 

In 2000 new machines were joined to AGS so that a new accelerator, the Relativistic Heavy Ion Collider, came into being. It was accustomed to accelerate fully stripped gold ions that collide with each other at the energy of 200 GeV per nucleon pair in the centre of mass frame. Hitherto RHIC has been exploring the following colliding systems: p+p, d+Au, Cu+Cu, Cu+Au, Au+Au, U+U at different energies. At RHIC there were conducted four experiments. Two of them, the smaller ones, PHOBOS and BRAHMS completed their programs in 2005 and 2006, respectively. The aim of PHOBOS was to measure total multiplicity of charged particles and particle correlations. BRAHMS was responsible for identification of particles over a wide range in rapidity and transverse momentum. Two other big experiments are PHENIX and STAR which have been working to date. PHENIX is aimed to detect rare and electromagnetic particles whereas STAR focuses on a detection of hadrons with its system of time projection chambers covering a large solid angle. At STAR there is other small experiment carried out, PP2PP, whose goal is to study spin dependence in proton-proton scattering. 

Certainly the biggest and the most powerful accelerator was completed in CERN in 2010. The initial energies of 3.5 TeV per beam that the Large Hadron Collider (LHC) started at were several times bigger than those achieved by RHIC and they have been gradually increased so that in 2015 the value of 6.5 TeV per beam was reached. At LHC proton and lead beams are produced so that proton-proton, lead-lead and proton-lead collisions take place. As RHIC was mainly concentrated on production and investigation of the quark-gluon matter, the prospects of LHC are much broader. Arguably, the discovery of Higgs boson in 2012 was a resounding success of LHC \cite{Aad:2012tfa,Chatrchyan:2012ufa}. Besides that, LHC is directed to  searching signals of physics beyond the Standard Model, such as supersymmetries or extra dimensions, explaining the nature of dark matter and, by and large, answering questions galore concerning properties of diverse particles and fundamental laws of physics. 

At LHC seven detectors are installed. The biggest two of them A Toroidal LHC Apparatus (ATLAS) and Compact Muon Solenoid (CMS) are detectors of general purposes. Then A Large Ion Collider Experiment (ALICE) is devoted to investigating properties of matter in the state of a quark-gluon plasma. The Large Hadron Collider beauty (LHCb) concentrates on testing CP-violation processes involving hadrons consisted of $b$ quark and therefore the excess of matter over antimatter might be explained. The smallest three detectors TOTEM, LHCf, and MoEDAL aim at very specialised topics such as the studies on diffractive processes or searching magnetic monopoles. While the chances that the so-called new physics will be discovered by experiments at LHC are very uncertain at reachable energies, the prospects to study the quark-gluon plasma experimentally are still very promising. More solid information and recent headway on the experimental outcomes from both LHC and RHIC colliders can be found in the series of the Quark Matter Proceedings, the last issues are listed here \cite{Braun-Munzinger:2014pya,Ullrich:2013qwa,Schutz:2011zz}.

\begin{center}
{\bf Main quark-gluon plasma features}
\end{center}

\begin{figure}[!h]
\centering
\includegraphics*[width=0.85\textwidth]{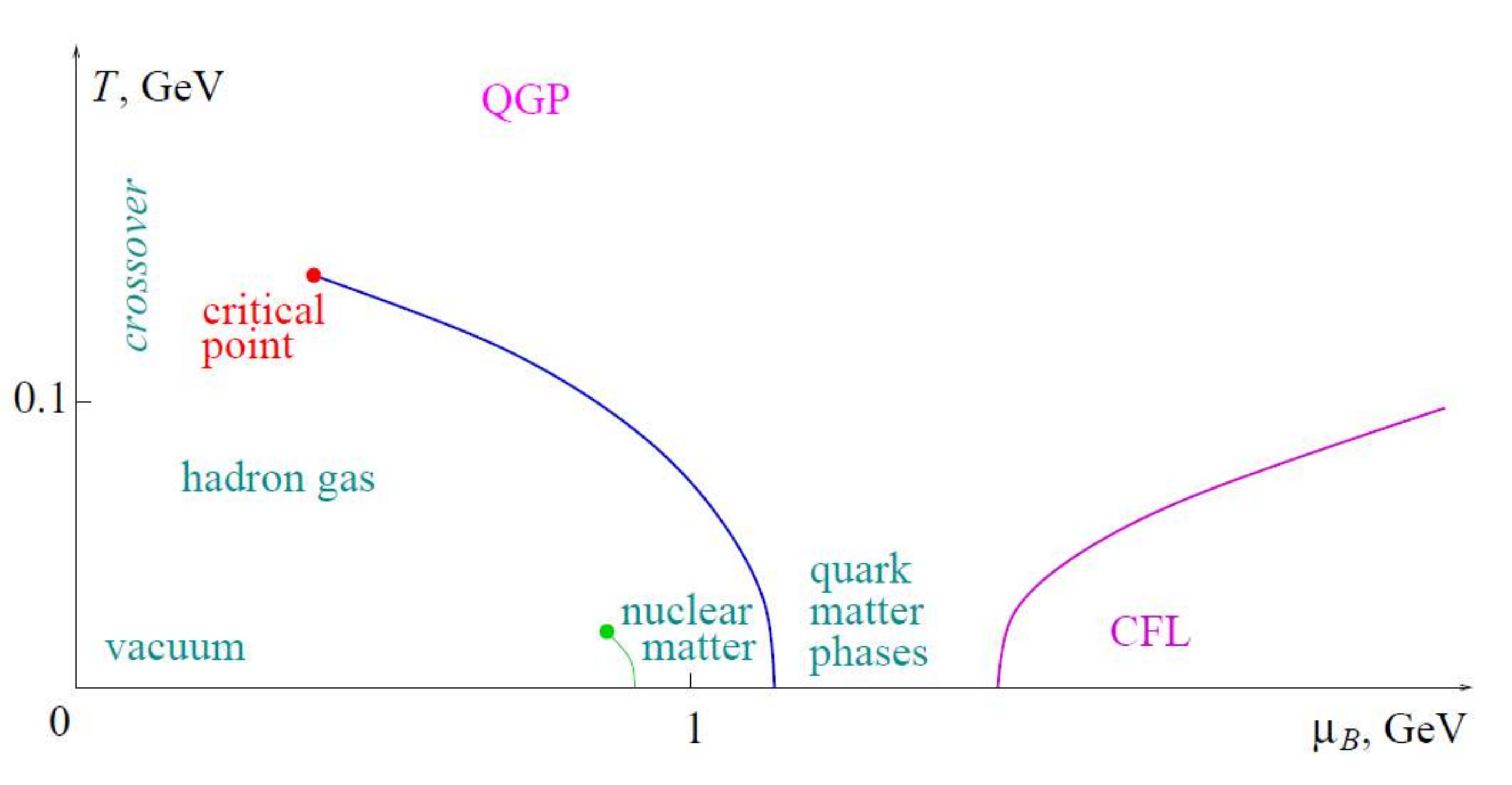}
\caption{Semi-quantitative phase diagram of QCD matter. Figure taken from \cite{Stephanov:2007fk}}
\label{phase-diagram}
\end{figure}

The long-term heavy ion programs at RHIC and LHC have delivered strong evidence for the creation of plasma droplets of quarks and gluons \cite{Heinz:2000bk}. Over the course of years experimental data in tandem with lattice QCD simulations and predictions of phenomenological models have been releasing different features of rather complex plasma dynamics. The findings, however, fail to build a coherent picture of the plasma system. Nevertheless, let us present them at a glance. First of all, a phase diagram of the nuclear matter has been sketched and it is shown in Fig. \ref{phase-diagram} in the plane of temperature and net baryon density. 
The diagram is widely discussed and is still getting improved, for some insight in it see \cite{Stephanov:2007fk,BraunMunzinger:2009zz}. A significant progress in this direction has been made within the framework of lattice QCD, see \cite{Bellwied:2015rza,Aoki:2006we,Aoki:2006br,Fodor:2004nz,Fodor:2001pe}, where it was established, for instance, that at zero net baryon density a hadronic matter changes into a quark-gluon plasma via a cross-over transition at temperatures around 150 - 200 MeV. One of the central issues is also to confirm the existence of a critical point and its location in the diagram. The problem, however, seems to remain unsolved.

Extensive analyses of heavy-ion collision data have shown also that the matter created during the collision exhibits a strongly collective hydrodynamic behaviour \cite{Heinz:2005ja}. Specifically, the azimuthal distribution of particles produced in an event is expressed by the Fourier decomposition as \cite{Poskanzer:1998yz}
\be
\label{mom-distr}
\frac{dN}{dp_T \, d\phi}=\frac{dN}{dp_T}(1+2 \upsilon_1 \cos\phi + 2\upsilon_2 \cos(2\phi) + 2 \upsilon_3 \cos (3\phi) + \dots),
\ee
where $\phi$ is the azimuthal angle of produced particles with respect to the reaction plane and the coefficients $\upsilon_i$ describe the momentum anisotropy so that $\upsilon_1$ corresponds to the direct flow and $\upsilon_2$ to the elliptic flow. The flow coefficients are deemed to represent the response of the system to spatial anisotropies in the initial state. The measurements of the elliptic flow receive a significant attention as they appear to be of quite large value, of the order of $0.1$, see \cite{Aamodt:2010pa}, which is in accordance with nearly ideal hydrodynamics. Since the hydrodynamics works when local thermal equilibrium is reached, large values of the elliptic flow are the hallmarks of very fast thermalisation of the matter. The equilibration time is estimated to be shorter than 1 fm/c. It also implies that dissipative effects are small, especially the ratio of shear viscosity to entropy ($\eta/s$) is not much bigger than its lower limit $1/4\pi$ \cite{Kovtun:2003wp,Kovtun:2004de}. Such a rapid equilibration and convincing arguments that the plasma is almost perfect fluid can be explained assuming that the plasma created in heavy-ion collisions is strongly coupled, as advocated, for example in \cite{Shuryak:2008eq}. These observations and, in general, the physics of the collisions of high-energy heavy ions are well covered by~\cite{Florkowski:2010zz}. 

Having said that, there are also other possible scenarios which reasonably describe these features of plasma behaviour and they base on the conviction that the coupling constant is not large. In particular, the colour glass condensate (CGC) approach \cite{McLerran:1993ni,McLerran:1993ka} provides the description of colliding nuclei and the state created immediately after the collision is characterised by small coupling constant, $\alpha_s \ll 1$, but the plasma is strongly coupled as the occupation numbers of gluonic fields are non-perturbatively large ($f \sim 1/\alpha_s$). And then again various explanations of how the system achieves equilibrium have been offered. One of them is that fast equilibration proceeds due to occurrence of unstable gluon modes at a very early non-equilibrium stage \cite{Kurkela:2011ti,Kurkela:2011ub,Arnold:2004ti,Mrowczynski:1994xv,Mrowczynski:1993qm}. The other possibility is the `bottom-up' thermalisation scenario \cite{Baier:2000sb}, which accentuates in turn the role of scattering processes. The ultimate answer on what leads the system to such a fast equilibration is not found but in \cite{Berges:2013fga,Berges:2013eia} there are some indications that the modified `bottom-up' scenario is more privileged than the others.  

Apart from the thermalisation and ideality of the plasma fluid there is also other curious phenomenon occurring within the nuclear matter which is worth underlying, that is, the jet quenching. Jets are the streams of very energetic groups of hadrons flying back-to-back which are created via hard scattering of incoming quarks and gluons. In proton-proton collisions these both streams of highly-energetic particles are seen whereas it is not the case in heavy-ion collisions. Then jets of a high transverse momentum can be quenched. The suggestion is that the jet quenching, which is an observed final state effect, reflects the energy loss of very energetic particles in the medium they traverse through. And again this observation may be naturally explained by a colour opacity of the strongly interacting system \cite{Adams:2004pa}.

Since the consistent and unanimous picture of what are the properties of the matter produced in heavy ion collisions at reachable energies is still missing there are strong indications that the matter is actually strongly interacting. In the light of this, new methods are welcome to study its properties systematically. And the appearance of Maldacena duality has offered new opportunities to broadly examine non-perturbative features of the quark-gluon plasma.

\subsection{AdS/CFT correspondence}
\label{adscft}

The AdS/CFT, or gauge/gravity, duality is a conjecture that constitutes a mapping between two very different and apparently unrelated theories: a conformal quantum field theory (CFT) that is strongly coupled and weakly coupled classical gravity. Since 1997 when Maldacena introduced the duality \cite{Maldacena:1997re}, it made a real revolution in approaching strongly interacting theories such as QCD.

\begin{center}
{\bf General remarks}
\end{center}

The AdS/CFT duality grew up on the fundamentals of the string theory which appeared at the end of 1960s as a theory of strong nuclear forces, for some insight into the string theory look at \cite{Green:1987sp}. Within the theory, point-like particles are replaced by strings of one dimension and interactions are represented by vibrations of the strings. At the beginning the string theory did not get broad interest among particle physics community as QCD offered a much more plausible way of studies of nuclear matter and, what is more, it was in agreement with experimental findings. However, over the course of years the string theory was getting more and more consistent and applicable rather not to strong forces but to multidimensional quantum gravity. Its fast development gave rise to many new concepts. Initial versions of the string theory were bosonic ones and the need of inclusion of fermions led to a superstring theory. The theory which works in 10 dimensions was soon extended to the so-called M-theory in 11 dimensions, which was considered as a candidate for the theory of grand unification. Since the world described by the superstring theory has to be 4-dimensional the extra-dimensions are subordinated to the procedure of compactification. The idea of D-branes, higher-than-one-dimensional objects, introduced in the mid 1990s contributed significantly to development of cosmological models and quantum gravity. These, in turn, allowed for better understanding, for example, the thermodynamic properties of black holes and enhanced some other developments. A further progress resulted in a discovery of the AdS/CFT duality by Maldacena which related gravity to quantum field theory. The duality which was soon improved and specified in \cite{Witten:1998qj,Gubser:1998bc}.

The AdS/CFT correspondence relates two different theories which means that all parameters obtained within one theory have their equivalents within the others. One of the key attributes of the correspondence is a strong/weak coupling duality, as the gauge theory is of strong coupling and the gravity is weakly coupled. The duality is also very successful realisation of the holographic correspondence which claims that the description of the surface of a space is a reflection of one higher dimensional volume of it, as first suggested by `t Hooft and Susskind \cite{'tHooft:1993gx,Susskind:1994vu}. Indeed, the gravity works in the anti-de Sitter space which is $n$-dimensional whereas the gauge theory is a reflection on the boundary of the space of $n-1$ dimensions.

The most famous example of the correspondance is the type IIB string theory on the product space $ \textrm{AdS}_5 \times S^5$ equivalence to the $\mathcal{N} = 4$ supersymmetric Yang-Mills theory in $4d$. The correspondence between these theories is given by the relations of their dimensionless parameters. On the AdS side there are the string coupling constant $g_s$ and the curvature scale of the space on which the theory works $l/l_s$, on the CFT side we have $N_c$ - the rank of the gauge group and the coupling constant $g$, which may be expressed via the 't Hooft coupling $\lambda=g^2 N_c$. Then the relations read
\be
\label{parameters}
4\pi g_s=g^2 = \frac{\lambda}{N_c}, \qquad \qquad \qquad 
\frac{l}{l_s} = (4\pi g_s N_c)^{1/4} = \lambda^{1/4}.
\ee
To make the gravity solution trustworthy (to suppress stringy corrections of the geometry) one has to keep $l$ large which next means $\lambda \gg 1$. Nevertheless, to suppress the quantum corrections $g_s$ has to be kept small. Thus, the correspondence is valid when the regime $N_c \gg \lambda \gg 1$ is held. In the limit $N_c \gg 1$ this is the 't Hooft constant (not $g$) that controls perturbative expansion. Therefore, the duality relates the strongly coupled CFT to weakly coupled gravity.

The fame of the duality engaging the $\mathcal{N} = 4$ super Yang-Mills theory bloomed when it turned out that the duality offers a promising perspective to study the quark-gluon plasma produced experimentally in heavy-ion collisions. The watershed took place when the gauge/gravity duality was applied to compute the shear viscosity of the super Yang-Mills theory in the limit of a large number of colours $N_c$ and strong `t Hooft coupling $\lambda$ \cite{Policastro:2001yc,Buchel:2004di}. The ratio of the shear viscosity density and entropy density $\eta/s$ equals
\be
\label{ration-eta-s}
\frac{\eta}{s}=\frac{1}{4\pi} \bigg(1 + \frac{135 \zeta(3)}{8 (2\lambda)^{3/2}}+ \cdots \bigg).
\ee
Then it was conjectured \cite{Kovtun:2003wp,Kovtun:2004de} that in all strongly-coupled theories there is a lower bound on $\eta/s$ that is $\eta/s=1/4\pi$. Then, an expectation arose that the $\mathcal{N} = 4$ super Yang-Mills theory resembles QCD near the phase transition \cite{Shuryak:2004tx} and, in general, that the supersymmetric theory may be useful in discovering different properties of QCD at strong coupling \cite{Herzog:2006gh,CasalderreySolana:2006rq,Liu:2006ug,CaronHuot:2006te}. 

\begin{center}
{\bf $\mathcal{N} = 4$ super Yang-Mills vs. QCD}
\end{center}

Although the Maldacena duality offers a unique tool to study strongly coupled systems there is a lot of criticism on possibilities of drawing some conclusions about the quark-gluon plasma from the $\mathcal{N} = 4$ super Yang-Mills plasma. In order to benchmark the AdS/CFT duality against the quark-gluon plasma let us briefly confront main features of both theories to each other.

In the vacuum both theories seem to be strikingly different. The super Yang-Mills theory includes supersymmetry which introduces new interactions to the system that the theory describes. The SYM theory includes a gauge field, six real scalars, and four Weyl spinors, all of them are in the adjoint representation\footnote{See Sec. \ref{subsec-super-YM}, where the fundamentals of the super Yang-Mills theory are presented.}. Unlike QCD, the super Yang-Mills is a finite theory as, due to supersymmetry, infinite expressions cancel out. The theory is conformal not only on the classical but also on the quantum level, so there is no mass scale. Likewise there is no running coupling. In effect, it has no confinement and the pure Coulombic potential is exhibited between colour sources. In $T=0$ limit quarks and gluons are confined in hadrons and then QCD is qualitatively different than SYM.

When temperature increases some qualitative distinctions between the theories disappear or are less and less important. Then, for example, supersymmetry is explicitly broken and the temperature introduces the only scale to the system. That having said, in case of QCD, when the temperature is bigger than the scale parameter $\Lambda_{\rm QCD}$, it gets the dominant scale in the system. The coupling constant $g$ in SYM can be fixed large and it remains large at any scale because of conformality. QCD, in turn, reaches asymptotic freedom for high enough energies and its properties are related to the energy scale. Nevertheless, at higher and higher energies QCD plasma becomes more and more scale independent. Besides that the equation of state of the super Yang-Mills plasma is exactly conformal which is not the case for QCD near the critical temperature $T_c$. The conformality causes that the bulk viscosity of SYM plasma is exactly zero. Moreover, the $\mathcal{N} = 4$ super Yang-Mills theory motivated by AdS/CFT should be considered when the limit $N_c \rightarrow \infty$ is taken whereas the quark-gluon plasma is described by QCD of $N_c=3$.

These and other more subtle problems in mapping QCD by the $\mathcal{N} = 4$ supersymmetric Yang-Mills theory are still investigated and it is hard to unanimously evaluate to what extent the $\mathcal{N} = 4$ super Yang-Mills theory mimics QCD. Usually, the possibility of an application of SYM to extract properties of real objects depends on the problem posed. Some illuminating reviews on applicability of AdS/CFT to model the properties of matter in the quark-gluon plasma state can be found here \cite{Janik:2010we,CasalderreySolana:2011us}. However, even if $\mathcal{N} = 4$ super Yang-Mills is in general different from QCD and AdS/CFT duality fails to give any reliable results about natural systems, it helps us constitute a context of studies on them and provides some reference points.

\subsection{Outline of the thesis}
\label{outline}

This thesis is organised as follows. In Sec. \ref{sec-gauge-theories} we shortly discuss all gauge theories which are taken into consideration. Not only are theories which are mainstays of the thesis described but also a few others are mentioned, as they are referred to at some points of this work. On description we try to underline differences between the theories. In Sec. \ref{basics} we show a comparison of basic characteristics of the quark-gluon plasma and the system governed by the $\mathcal{N}=4$ super Yang-Mills theory. Sec. \ref{sec-KS-form} is devoted to the Keldysh-Schwinger formalism, which, as appropriate to many-body systems, is the framework of our studies. Then, the real-time argument Green functions of all types of fields occurring in the theories discussed: gauge boson, fermion, and scalar ones, are derived. The Green functions are basic objects of perturbative computations and are used in order to extract physical properties of the plasma systems in next parts of the thesis. In Sec. \ref{sec-ghosts-KS} and the consecutive ones our original findings are presented. First, we show how to introduce the Faddeev-Popov ghosts into the Keldysh-Schwinger formulation of the Yang-Mills theory. Therefore, the Green functions of the ghost field are derived in terms of the path integral approach. In Sec. \ref{sec-collective} the self-energies of fermions, scalars, and gauge bosons of the $\mathcal{N}=1$ SUSY QED and $\mathcal{N}=4$ super Yang-Mills plasma systems are derived in the hard-loop approximation and compared to the respective ones of the QED and QCD plasmas. Since we work in the Feynman gauge the ghost Green functions obtained before are included in the calculations of the gluon polarisation tensor.  The self-energies are found to be of the universal forms and such is the effective hard-loop action constructed as well. We also investigate the question what are the consequences of this universality. It is discussed, in particular, what are spectra of collective excitations. We complete the discussion of the plasma systems' properties in Sec. \ref{sec-tr-charact}, where the transport characteristics are considered. There, we provide an explanation why only some of the transport coefficients are worth computing. Then, all cross sections of binary processes in the $\mathcal{N}=1$ SUSY QED plasma are calculated. Since there are processes whose cross section is qualitatively different from that caused by the Coulomb-like interaction, we calculate the collisional and radiative energy losses caused by this interaction. This analysis is generalised later on to the plasma governed by the $\mathcal{N}=4$ super Yang-Mills theory. The thesis is closed with summary and conclusions.

Throughout the thesis we use the natural system of units with $c= \hbar = k_B =1$; our choice of the signature of the metric tensor is $(+ - - -)$.

\newpage
\thispagestyle{plain}

\section{Gauge theories under consideration}
\label{sec-gauge-theories}

In this section we briefly present the gauge theories that govern dynamics of plasma systems studied and confronted to each other in the next paragraphs of the thesis. The main effort is put on stressing differences and similarities among the theories and also on fixing the notation. The content of this part is rather commonly known and is based on the classical books and reviews \cite{Peskin,Pokorski,Jackson}.

\subsection{Quantum electrodynamics}
\label{subsec-QED}

QED is the theory of electrons and positrons interacting with photons and its Lagrangian density reads
\ba
\label{L-QED}
{\cal L}_{\textrm{QED}} &=& -\frac{1}{4}F^{\mu \nu} F_{\mu \nu} 
+ \bar \Psi (iD\!\sla - m) \Psi,
\ea
where $m$ is a mass of an electron. $F^{\mu \nu}$, with the Lorentz indices $\mu,\nu=0,1,2,3$, is the electromagnetic field tensor that is expressed by the electromagnetic four-potential $A^\mu$ as
\ba
\label{str-tensor}
F^{\mu \nu} \equiv \partial^\mu A^\nu - \partial^\nu A^\mu.
\ea
$\Psi$ in the Lagrangian (\ref{L-QED}) means the Dirac spinor and the Dirac adjoint is defined as $\bar \Psi \equiv \Psi^\dagger \gamma^0$. We denote $ D\!\sla \equiv \gamma^\mu D_\mu$, where $\gamma^\mu$ are the Dirac matrices and the gauge covariant derivative equals
\ba
\label{cov-der-qed}
D^\mu \equiv \partial^\mu - ie A^\mu
\ea
with $e$ being a coupling constant which is the charge of an electron. The inclusion of an interaction of the fermion field with the electromagnetic one has been done via the operation of the minimal coupling $\partial_\mu \rightarrow D_\mu$. The Lagrangian (\ref{L-QED}) is invariant under the local gauge transformations
\ba
\label{gauge-qed-1}
\Psi(x) \rightarrow e^{i\alpha(x)}\Psi(x), \qquad\qquad\qquad A_\mu(x) \rightarrow A_\mu(x)-\frac{1}{e} \partial_\mu \alpha(x),
\ea 
which constitutes an Abelian $U(1)$ group of symmetries of the Lagrangian of QED. The covariant derivative (\ref{cov-der-qed}) of the Dirac field transforms in the same way as the field. The form of the Lagrangian (\ref{L-QED}) leads us to the Euler-Lagrange equation for $\Psi$
\ba
\label{Dirac-eq}
(iD\!\sla - m) \Psi=0,
\ea
which is the Dirac equation. The Euler-Lagrange equation for $A_\mu$ is given as
\ba
\label{A-eq}
\partial_\mu F^{\mu\nu}= j^\nu,
\ea
where $j^\mu$ is the electromagnetic current density
\ba
\label{current-em}
j^\nu=- e\bar\Psi \gamma^\nu \Psi,
\ea
which satisfies the continuity equation $\partial^\mu j_\mu=0$. This fact can be proven by acting a derivative on Eq. (\ref{A-eq})
\ba
\label{A-eq-cont}
\partial_\nu \partial_\mu F^{\mu\nu}=- \partial_\mu \partial_\nu F^{\nu\mu}=0=\partial_\nu j^\nu,
\ea
where we have used the fact that the strength tensor in antisymmetric.

\subsection{Scalar electrodynamics}
\label{subsec-scalar-QED}

A theory that governs the dynamics of a scalar complex field $\phi$ and a vector field $A^\mu$ is the scalar electrodynamics whose the Lagrangian reads
\ba
\label{L-ScQED}
{\cal L}_{\textrm{scalar\,QED}} &=& -\frac{1}{4}F^{\mu \nu} F_{\mu \nu} 
- (D^\mu \phi)^* D_\mu \phi - m^2 \phi^* \phi,
\ea
where $m$ is a mass of the scalar field and the covariant derivative is defined as in QED by the formula (\ref{cov-der-qed}). The equation of motion for the scalar field is
\ba
\label{eom-scalar}
\big(D^\mu D_\mu +m^2 \big) \phi=0,
\ea
and that of the electromagnetic one reads
\ba
\label{eom-emsc}
\partial_\mu F^{\mu\nu} = j^\nu,
\ea
where the current is defined as
\ba
\label{curr-sc-em}
j^\nu = -ie \big[ \phi^* D^\nu \phi - (D^\nu \phi)^* \phi \big].
\ea
Except for the interaction terms in the Lagrangian (\ref{L-ScQED}) which are $e(\partial^\mu \phi^*)  \phi  A_\mu$ and $ e\phi^* (\partial^\mu \phi)  A_\mu$, there is also a four-boson coupling $e^2\phi^* \phi A^\mu A_\mu$. Such a contact interaction is qualitatively different than that caused by a massless particle exchange. In absence of other interactions, it gives the scattering which is isotropic in the center-of-mass frame of colliding particles with characteristic energy and momentum transfers which are much bigger than those in one-photon exchange processes. Obviously, the contact interaction cannot be treated separately from the remaining ones as then the transition matrix element is gauge dependent.

\subsection{${\cal N}\!=\!1$ SUSY QED}
\label{subsec-super-QED}

A peculiar combination of QED and scalar QED is the ${\cal N}\!=\!1$ SUSY QED, see {\it e.g.} \cite{Hollik:1999xh} and \cite{Binoth:2002xg}. The theory consists of a vector multiplet $(A^\mu, \lambda_\alpha, \bar\lambda^{\dot \alpha})$ with the photon field $A^\mu$ and the Majorana photino $\Lambda$ expressed by the Weyl spinors $\lambda_\alpha, \bar\lambda^{\dot \alpha}$. It contains also two chiral multiplets $(\psi^\alpha_L,\phi_L)$ and $(\psi^\alpha_R,\phi_R)$ with the electron field represented by Weyl spinors $\psi^\alpha_L,\psi^\alpha_R$ and scalars which are superpartners of left- and righ-handed electrons. Let us add that the names of selectrons have nothing common with the chirality. The supersymmetric transformation here exchanges different members of a multiplet into each other. The Lagrangian of the ${\cal N}\!=\!1$ SUSY QED is of the form
\ba
\label{L-SUSY-QED}
{\cal L}_{\textrm{super\,QED}} &=&{\cal L}_{\textrm{QED}}
+\frac{i}{2} \bar \Lambda \partial \sla \Lambda
+(D_\mu \phi_L)^*(D^\mu \phi_L) + (D_\mu^* \phi_R)(D^\mu \phi_R^*)
\\ [2mm] \nn
&& +\sqrt{2} e \big( \bar \Psi P_R \Lambda \phi_L - \bar \Psi P_L \Lambda \phi_R^*
+ \phi_L^* \bar \Lambda P_L \Psi - \phi_R \bar \Lambda P_R \Psi \big)
\\ [2mm] \nn
&&
- \frac{e^2}{2} \big( \phi_L^* \phi_L - \phi_R^* \phi_R \big)^2  - m^2(\phi_L^* \phi_L + \phi_R^* \phi_R),
\ea
where ${\cal L}_{\textrm{QED}}$ is given by (\ref{L-QED}) and the projectors $P_L$ and $P_R$ are defined in a standard way
\be
\label{projection-op}
P_L \equiv \frac{1}{2}(1 - \gamma_5), \qquad\qquad \qquad P_R \equiv \frac{1}{2}(1 + \gamma_5). 
\ee
The Dirac and Majorana bispinors read as
\be
\label{M-D-bispinors}
\Psi  = \left(
   \begin{matrix}
   \psi_{L\alpha}  \cr
   \bar{\psi}_R^{\dot{\alpha}}  \cr
   \end{matrix}
   \right) ,
\;\;\;\;\;\;\;
\Lambda  = \left(
   \begin{matrix}
   -i\lambda_\alpha  \cr
   i\bar{\lambda}^{\dot{\alpha}}  \cr
   \end{matrix}
   \right).
\ee

The supersymmetric extension of QED describes a mixture of photons, Majorana and Dirac fermions, and scalars of two types with a variety of interactions. Except for the long-range one-photon exchanges, we have four-boson couplings and the Yukawa interactions of non-electromagnetic nature. The complete list of elementary processes, which is given in our paper \cite{Czajka:2011zn}, is thus very long and it makes the supersymmetric plasma very different at the microscopic level from the usual electromagnetic ones.

The nice feature of the Lagrangian (\ref{L-SUSY-QED}) is that it keeps left- and right-handed fermions separately which is important, as these fields transform differently under SU(2) gauge transformations. Therefore, the theory is a possible candidate for an extension of the electromagnetic sector of the Standard Model.  For the so-called extended supersymmetries $({\mathcal N}>1)$ this is not the case inasmuch as left- and right-handed fermions are mixed. However, supersymmetry, even if it is a symmetry of nature, must be broken as superparticles are supposed to have much larger masses than their nonsupersymmetric partners. Thus far there is no experimental confirmation of existence of superparticles. In spite of an ontological status of supersymmetry, the ${\cal N}\!=\!1$ SUSY QED, among other supersymmetric theories, raises a lot of interest because of its own attractive features. In particular, the presence of supersymmetry results in cancellations between the bosonic and fermionic degrees of freedom. Consequently, quadratic divergences in the ${\cal N}\!=\!1$ SUSY QED disappear.

\subsection{Yang-Mills theory}
\label{subsec-YM}

The pure Yang-Mills theory is a gauge theory of gluons with the ${\rm SU}(N_c)$ gauge group. The gauge field $A^\mu$, which describes gluons, is the four-vector as in the electrodynamics. 

The Lagrangian density of gluodynamics in the fundamental representation equals
\ba
\label{L-YM-fund}
{\cal L}_{\textrm{YM}} = -\frac{1}{4} {\rm Tr}[F^{\mu \nu} F_{\mu \nu}],
\ea
where  $F^{\mu \nu}$ is the chromodynamic strength tensor that equals
\ba
\label{F-fund}
F^{\mu \nu} = \partial^\mu A^\nu - \partial^\nu A^\mu - i g [A^\mu, A^\nu],
\ea
where $g$ is the coupling constant. If $D^\mu$ is the covariant derivative defined as 
\ba
\label{D-fund}
D^{\mu} \equiv \partial^\mu \, {\bf 1} - i g A^\mu
\ea
then the strength tensor can be expressed as
\ba
F^{\mu \nu} = \frac{i}{g} [D^\mu,D^\nu].
\ea
The transformation law of the gauge field is deduced from the requirement of gauge invariance of the Lagrangian (\ref{L-YM-fund}). Then, the chromodynamic field has to transform as
\ba
\label{gauge-transf-A-fund}
A^\mu (x) \rightarrow U(x) \, A^\mu (x) U^\dagger (x) 
+ \frac{i}{g} U(x) \partial^\mu U^\dagger (x) 
\ea
and the strength tensor as
\ba
\label{gauge-transf-F-fund}
F^{\mu \nu}(x) \rightarrow U(x) \, F^{\mu \nu} (x) U^\dagger (x).
\ea
The transformation matrix $U(x)$ belongs to the fundamental representation of ${\rm SU}(N_c)$ group and thus it is the $N_c \times N_c$ unitary matrix. The matrix $U(x)$ can be parametrized as
\ba
\label{U-gauge}
U(x) = e^{i \omega^a(x) \, \tau^a} ,
\ea
where $\omega^a(x) $, with $a = 1,\,2, \, \dots N_c^2-1$, are real functions and $\tau^a$ are the generators of the fundamental representation. The generators obey the commutation relations
\ba
\label{SUN-algebra}
[\tau^a, \tau^b] = i f^{abc} \tau^c,
\ea
where $f^{abc}$ are totally antisymmetric structure constants of the ${\rm SU}(N_c)$ group. The generators are Hermitian traceless matrices and then the matrix (\ref{U-gauge}) is automatically unitary and its determinant equals unity.  The generators are chosen to be normalized as
\ba
\label{tr-tau-tau}
{\rm Tr} [\tau^a \tau^b] =\frac{1}{2}\delta^{ab} .
\ea

The gauge field $A^\mu$ can be also expressed in the adjoint representation. The relation between the field in the fundamental representation and adjoint one is 
\ba
\label{ad-fun}
A^\mu = A^\mu_a \tau^a.
\ea
The field in the adjoint representation is obtained by means of the relation (\ref{tr-tau-tau}) from that one in the fundamental one that is $ A^\mu_a = 2 {\rm Tr}[A^\mu \tau_a$]. The trace is obviously taken over colour indices. In the adjoint representation there are $N_c^2 -1$ real functions $A^\mu_a$ with $a,b= 1,2, \dots N_c^2-1$. The Lagrangian of gluodynamics with the fields in the adjoint representation is given by
\ba
\label{L-YM-adj}
{\cal L}_{\textrm{YM}} =-\frac{1}{4}F^{\mu \nu}_a F_{\mu \nu}^a,
\ea
where the chromodynamic strength tensor $F_a^{\mu \nu}$ is expressed by the four-potential $A_a^\mu$ as 
\ba
\label{F-adj}
F^{\mu \nu}_a \equiv \partial^\mu A^\nu_a - \partial^\nu A^\mu_a + g f^{abc} A^\mu_b A^\nu_c.
\ea
The gauge transformation laws in the adjoint representation read
\ba
\label{gauge-transf-adjoint}
A^\mu_a (x) \rightarrow \mathcal{U}_{ab}(x) \, A^\mu_b (x) + \frac{i}{g} C^\mu_a(x),
\;\;\;\;\;\;\;\;
F^{\mu \nu}_a (x) \rightarrow \mathcal{U}_{ab}(x) \, F^{\mu \nu}_b (x) ,
\ea
where
\ba
\label{M-C-def}
\mathcal{U}_{ab}(x) \equiv 2{\rm Tr} [\tau^a U(x) \, \tau^b U^\dagger (x) ] ,
\;\;\;\;\;\;\;\;
 C^\mu_a(x) \equiv 2{\rm Tr}[ \tau^a  U(x) \partial^\mu U^\dagger (x) ].
\ea
The matrix $\mathcal{U}_{ab}$ and the vector $C^\mu_a$ acquire a simple form for infinitesimally small transformations. 
Substituting the parametrization (\ref{U-gauge}) into the definitions (\ref{M-C-def}) and keeping only the terms linear
in $\omega^a$, one gets
\ba
\label{gauge-transf-small}
\mathcal{U}_{ab}(x) \approx \delta^{ab} + f^{abc} \omega^c(x) ,
\;\;\;\;\;\;\;\;
 C^\mu_a(x) \approx -i \partial^\mu \omega^a(x) .
\ea

\subsection{Quantum chromodynamics}
\label{subsec-QCD}

Enriching the pure gluodynamics with quarks of $N_f$ flavors, which belong to the fundamental representation of the ${\rm SU}(N_c)$ gauge group, we get QCD. Save for gluon fields, there are the quark fields $\psi^i_q(x)$ that are Dirac bispinors which additionally carry a flavor index $q =1, \, 2, \, \ldots N_f$ and a colour one $i,j =1, \, 2, \, \ldots N_c$. The Lagrangian density of QCD is of the form
\ba
\label{L-QCD}
{\cal L}_{\textrm{QCD}} = {\cal L}_{\textrm{YM}} + \bar \psi^i_q \Big(i\gamma_\mu D^\mu_{ij} -m_q \, \delta_{ij}\Big) \psi^j_q,
\ea
where the summation convention is kept and the covariant derivative is given by (\ref{D-fund}).

The quark fields transform under the local gauge transformation as
\ba
\label{quark-gauge-transform}
\psi^i_q(x) \rightarrow U^{ij}(x) \psi^j_q(x) , 
\;\;\;\;\;\;\;
\bar{\psi}^i_q(x) \rightarrow \bar{\psi}^j_q(x) U_{ji}^\dagger(x) ,
\ea
where the tranformation matrix $U^{ij}(x)$ is defined by (\ref{U-gauge}).

The  Lagrangian (\ref{L-QCD}) leads to the equations of motion of the quark and gluon fields
\ba
\label{Dirac-eq}
&& i \gamma_\mu D_{ij}^\mu \psi^j_q  = 0,
\\ [2mm]
\label{YM-eq}
&&D^{ij}_\mu F_{jk}^{\mu \nu} =  j^\nu_{ik} ,
\ea
where the colour current is $ j^\mu_{ik} = j^\mu_a \tau_{ik}^a$ with 
\ba
j^\mu_a \equiv - g {\bar \psi}^i_q  \tau_{ij}^a \gamma^\mu \psi^j_q .
\ea
We note that the form of the covariant derivative depends whether it acts on colour vector as $\psi^j_q$ 
\ba
\label{cov-na-vec}
D_{ij}^\mu \psi^j_q \equiv (\partial^\mu \delta_{ij} - ig A^\mu_{ij}) \psi^j_q,
\ea
or colour tensor as $F_{ij}^{\mu \nu}$
\ba
\label{cov-na-tens}
D^{ij}_\mu F_{jk}^{\mu \nu}  \equiv \delta^{ij} \partial_\mu F_{jk}^{\mu \nu} 
- ig \big[A^{ij}_\mu, F_{jk}^{\mu \nu} \big].
\ea
We also observe that the current is not conserved but it is covariantly conserved that is $D_\mu j^\mu = 0$, which is seen from the operation
\ba
\label{ujk}
D_\nu D_\mu F^{\mu \nu} =- D_\mu D_\nu F^{\nu \mu} = 0.
\ea
The equation of motion of the chromodynamic field in the adjoint representation is
\ba
\label{YM-eq-adjoint}
D_\mu^{ab} F^{\mu \nu}_b  =  j^\nu_a
\ea
where the covariant derivative equals
\ba
\label{cov-D-adjoint}
D^\mu_{ab} \equiv \partial^\mu \delta^{ab} - g f^{abc} A^\mu_c.
\ea

\subsection{${\mathcal N} \! = \! 4$ super Yang-Mills}
\label{subsec-super-YM}

The last theory considered here is the ${\cal N}=4$ super Yang-Mills theory whose description is given in \cite{Brink:1976bc,Gliozzi:1976qd,Yamada:2006rx}. We follow here the presentation from \cite{Yamada:2006rx}.

The gauge group is assumed to be ${\rm SU}(N_c)$ and every field of the ${\cal N}=4$ super Yang-Mills theory belongs to its  adjoint representation.  The field content of the theory, which is summarized in Table~\ref{table-field-content}, is the following. There are gauge bosons (gluons) described by the vector field $A_\mu^a$ with $a, b, c, \dots = 1, 2, \dots N_c^2 -1$. There are four Majorana fermions represented by the Weyl spinors $\lambda^\alpha$ with $\alpha = 1,2$ which can be combined in the Dirac bispinors as
\be
\label{Majorana-bispinor}
\Psi  = \left(
   \begin{matrix}
   \lambda^\alpha  \cr
   \bar{\lambda}_{\dot{\alpha}}  \cr
   \end{matrix}
   \right) ,
\;\;\;\;\;\;\;
\bar{\Psi}  = (  \lambda_\alpha,  \bar{\lambda}^{\dot{\alpha}}  ) ,
\ee
where $\bar{\lambda}_{\dot{\alpha}} \equiv [ \lambda_\alpha]^\dagger$ with $\dagger$ denoting Hermitian conjugation. To numerate the Majorana fermions we use the indices $i, j = 1,2,3,4$ and the corresponding bispinor is denoted as $\Psi_i$. Finally, there are six real scalar fields which are assembled in the multiplet $\phi = (X_1, Y_1, X_2, Y_2, X_3, Y_3)$. The components of $\phi$ are either denoted as $X_p$ for scalars, and $Y_p$ for pseudoscalars, with $p,q =1,2,3$ or as $\phi_A$ with $A, B=1,2, \dots 6$.

{\small
\begin{table}[!h]
\caption{\label{table-field-content} Field content of the ${\cal N} =4$ super Yang-Mills theory.}
\centering
\begin{tabular}{cccc}
\hline \hline
Field& Range of the field's index & Spin & $N^{\rm dof}$
\\
\hline \hline
\vspace{-2mm}
\\
$A^\mu$ - vector &$\mu, \nu = 0,1,2,3$ & 1& $2 \times (N_c^2-1)$
\\[2mm]
$\phi_A$ - real (pseudo-)scalar & $A, B = 1,2, 3,4,5, 6$ & 0 & $6 \times (N_c^2-1)$
\\[2mm]
$\lambda_i$ - Majorana spinor & $i, j = 1,2,3,4$ & 1/2 & $8 \times (N_c^2-1)$
\\
\hline \hline
\end{tabular}
\end{table}
}

The Lagrangian density of the ${\cal N} =4$ super Yang-Mills theory can be written as
\ba
\label{Lagrangian-1}
{\cal L}
&=&
-\frac{1}{4}F^{\mu \nu}_a F_{\mu \nu}^a
+\frac{i}{2}\bar \Psi_i^a (D\!\sla \Psi_i)^a
+\frac{1}{2}(D_\mu \phi_A)_a (D^\mu \phi_A)_a
\\ [2mm] \nn
&&
-\frac{1}{4} g^2f^{abe} f^{cde} \phi_A^a \phi_B^b \phi_A^c \phi_B^d
-i\frac{g}{2} f^{abc} \Big( \bar \Psi_i^a  \alpha_{ij}^p  X_p^b \Psi_j^c  
+i\bar \Psi_i^a \beta_{ij}^p\gamma_5  Y_p^b \Psi_j^c \Big),
\ea
where $g$ is the coupling constant and $f^{abc}$ are the structure constants of the ${\rm SU}(N_c)$ group. The strength tensor is $F^{\mu \nu}_a = \partial^\mu A^\nu_a - \partial^\nu A^\mu_a + g f^{abc} A^\mu_b A^\nu_c$ and the action of the covariant derivatives is
\ba
\label{cov-1-sym}
(D\!\sla \Psi_i)^a &=& D^\mu_{ab} \Psi^i_b = \big(\partial \,\!\sla \delta_{ab} +g f^{abc} A_c \!\! \sla \big) \Psi_i^b,  
\\ [2mm]
(D^\mu \phi)_a &=& D^\mu_{ab} \phi_b = \big(\partial^\mu \delta_{ab} + gf^{abc}A^\mu_c \big)\phi_b.
\ea 
The $4 \times 4$ matrices $\alpha^p, \beta^p$ satisfy the relations
\be
\label{alpha-beta-relations}
\{\alpha^p, \alpha^q \} = - 2 \delta^{p q},
\;\;\;\;\;\;\;
\{\beta^p, \beta^q \} = - 2 \delta^{p q},
\;\;\;\;\;\;\;
[ \alpha^p, \beta^q] = 0 ,
\ee
and their explicit form can be chosen as
\ba
\label{alphas}
\alpha^1  &=& \left(\begin{matrix} 0 & \sigma_1   \cr  -\sigma_1 & 0  \cr \end{matrix} \right) ,
\;\;\;\;\;\;\;
\alpha^2  = \left( \begin{matrix} 0 & -\sigma_3   \cr  \sigma_3 & 0  \cr  \end{matrix} \right) ,
\;\;\;\;\;\;\;
\alpha^3  = \left( \begin{matrix} i \sigma_2 & 0  \cr 0 & i \sigma_2  \cr  \end{matrix} \right) ,
\\[2mm]
\label{betas}
\beta^1  &=& \left( \begin{matrix} 0 & i\sigma_2  \cr i \sigma_2 & 0  \cr \end{matrix} \right) ,
\;\;\;\;\;\;\;
\beta^2  = \left( \begin{matrix} 0 & \sigma_0  \cr -\sigma_0 & 0  \cr \end{matrix} \right) ,
\;\;\;\;\;\;\;
\beta^3  = \left( \begin{matrix} -i \sigma_2 & 0  \cr 0 & i \sigma_2  \cr \end{matrix} \right) ,
\ea
where the $2 \times 2$ Pauli matrices read
\ba
\label{Pauli}
\sigma^0  = \left( \begin{matrix} 1 & 0  \cr 0 & 1  \cr \end{matrix} \right) ,
\;\;\;\;\;\;\;
\sigma^1  = \left( \begin{matrix} 0 & 1  \cr 1 & 0  \cr \end{matrix} \right) ,
\;\;\;\;\;\;\;
\sigma^2  = \left( \begin{matrix} 0 & -i  \cr i & 0  \cr \end{matrix} \right) ,
\;\;\;\;\;\;\;
\sigma^3  = \left( \begin{matrix} 1 & 0  \cr 0 & -1  \cr \end{matrix} \right) .
\ea
As seen, the matrices $\alpha^p, \beta^p$  are antiHermitian: $(\alpha^p)^\dagger = -\alpha^p$ , $(\beta^p)^\dagger = -\beta^p$.

As in QCD, in the super Yang-Mills theory there are the three- and four-gluon couplings and the gluon interaction with the colour fermion current. Additionally there are the four-boson couplings $g^2\phi_A \phi_A A^\mu A_\mu$ and $g^2\phi_A \phi_B \phi_A \phi_B$. There is also the Yukawa interaction of fermions with scalars. The complete list of elementary interactions, which is given in \cite{Huot:2006ys}, is again rather long and it makes the super Yang-Mills plasma quite different at the microscopic level from the gluodynamic or QCD plasmas.

The $\mathcal{N}=4$ super Yang-Mills theory posses $\mathcal{N}=4$ pairs of generators of supersymmetry so it is a maximally supersymmetric field theory in a flat space. As already discussed, the theory is of massless particles as required by conformality. It is finite not only at classical but also at quantum level since supersymmetry gives rise to canceling out the fermionic and bosonic divergences. These properties render the theory especially exploitable in the context of computational purposes.

\newpage
\thispagestyle{plain}

\section{Basic characteristics of the $\mathcal{N}=4$ super Yang-Mills plasma}
\label{basics}

Prior to the step-by-step studies of the supersymmetric plasma systems let us give a brief account of basic $\mathcal{N}=4$ super Yang-Mills plasma characteristics which does not require us to involve any serious computational methods. These properties are considered at the perturbative level and such is the framework of further studies. All plasmas are considered as the ultrarelativistic ones and as far as supersymmetric systems are concerned the supersymmetry of the Lagrangians is exact. The discussion provided here is taken from our work \cite{Czajka:2012gq}.

Here we discuss the energy and particle densities, Debye mass and plasma parameter of the $\mathcal{N}=4$ super Yang-Mills plasma (SYMP) in equilibrium and thereafter compare the quantities to those of the quark-gluon plasma (QGP). For the beginning, however, a few comments are in order. 

In QGP there are several conserved charges: baryon number, electric and colour charges, strangeness. The net baryon number and electric charge are typically non-zero in QGP studied experimentally at RHIC and LHC while the total strangeness and colour charge vanish. Actually, the colour charge is usually assumed to vanish not only globally but locally as well. It certainly makes sense as the whitening of QGP appears to be the relaxation process of the shortest time scale \cite{Manuel:2004gk}. In SYMP, there are  conserved charges carried by fermions and scalars associated with the global ${\rm SU}(4)$ symmetry. One of these charges can be identified with the electric charge to couple ${\cal N} = 4$ super Yang-Mills to the electromagnetic field \cite{CaronHuot:2006te}. In the forthcoming the average ${\rm SU}(4)$ charges of SYMP are assumed to vanish and so are the associated chemical potentials. The constituents of SYMP carry colour charges but we further assume that the plasma is globally and locally colourless. 

Since there are conserved supercharges in supersymmetric theories, it seems reasonable to consider a statistical supersymmetric system with a non-zero expectation value of the supercharge. However, it is not obvious how to deal with a partition function customary defined as ${\rm Tr}e^{-\beta (H - \mu Q)}$ where $\beta \equiv T^{-1}$ is the inverse temperature, $H$ is the Hamiltonian, $Q$ is the supercharge operator and $\mu$ is the associated chemical potential. The problem is caused by a fermionic character of the supercharge $Q$. If $\mu$ is simply a number, as, say, the baryon chemical  potential, the partition function even of non-interacting system does not factorize into a product of partition functions of single momentum modes because the supercharges of different modes do not commute with each other. The supercharge is not an extensive quantity \cite{Kapusta:1984cp}. There were proposed two ways to resolve the problem. Either the chemical  potential remains a number but the supercharge is modified by multiplying it by an additional fermionic field $c$ \cite{Kapusta:1984cp,Mrowczynski:1986cu} or the chemical  potential by itself is a fermionic field \cite{Kovtun:2003vj}. Then, $\mu c Q$ and $\mu Q$ are both bosonic and the partition function can be computed in a standard way. The two formulations, however, are not equivalent to each other. According to the former one \cite{Kapusta:1984cp,Mrowczynski:1986cu}, properties of a supercharged system vary with an expectation value of the supercharge, within the latter one \cite{Kovtun:2003vj}, the partition function appears to be effectively independent of $Q$. Because of the ambiguity, we further consider SYMP where the expectation values of all supercharges vanish both globally and locally.

In view of the above discussion, SYMP is comparable to QGP where the conserved charges are all zero and so are the associated chemical potentials. We adopt the assumption whenever the two plasma systems are compared to each other. 

When the chemical potentials are absent, the temperature $T$ is the only dimensional parameter, which characterises the equilibrium plasma, and all plasma parameters are expressed through the appropriate powers of $T$. Taking into account the right numbers of bosonic and fermionic degrees of freedom in SYMP and QGP, the energy densities of equilibrium non-interacting plasmas equal
\ba
\varepsilon_{\rm SYMP} &=& \frac{\pi^2 T^4(N_c^2 -1)}{2}  , 
\\ [2mm]
\varepsilon_{\rm QGP} &=& \frac{\pi^2 T^4 \big(4(N_c^2 -1) + 7 N_f N_c\big)}{60}  
\ea
with $N_f$ light quark flavours. The quark is light when its mass is much smaller than the plasma temperature. For $N_c=N_f =3$, the energy density of SYMP is approximately 2.5 times bigger than that of QGP at the same temperature. The same holds for the pressure $p$ which, obviously, equals $\varepsilon/3$.

The particle densities in SYMP and QGP are found to be
\ba
n_{\rm SYMP} &=& \frac{14\zeta(3) T^3 (N_c^2 -1)}{\pi^2},
\\ [2mm]
n_{\rm QGP} &=& \frac{2\zeta(3)T^3 \big( 2(N_c^2 -1) + 3 N_f N_c\big)}{\pi^2} \,  ,
\ea
where $\zeta(3) \approx 1.202$ is the Riemann zeta function. For $N_c=N_f =3$ we have $n_{\rm SYMP}/n_{\rm QGP} \approx 1.3$ at the same temperature.

As we show in Sec.~\ref{sec-collective}, the gluon polarisation tensor has exactly the same structure in SYMP and QGP, and consequently the Debye mass in SYMP is defined in the same way as in QGP. The masses in both plasmas equal 
\ba
m_{D,{\rm SYMP}}^2 &=& 2g^2 T^2 N_c , 
\\ [2mm]
m_{D,{\rm QGP}}^2 &=& \frac{g^2 T^2 (2 N_c  + N_f) }{6}.
\ea
For $N_c=N_f =3$, the ratio of Debye masses squared is 2.4 at the same value of $gT$. The Debye mass determines not only the screening length $r_D = 1/m_D$ but it also gives the plasma frequency  $\omega_p = m_D/\sqrt{3}$ which is the minimal frequency of longitudinal and transverse plasma oscillations corresponding to the zero wavevector. The  plasma frequency  is also called the gluon thermal mass. 

Another important quantity characterising the equilibrium plasma is the so-called plasma parameter $\lambda$ which equals the inverse number of particles in the sphere of radius of the screening length, $\lambda \equiv (\frac{4}{3} \pi r_D^3 n)^{-1}$. When $\lambda$ is decreasing, the behaviour of plasma is more and more collective while inter-particle collisions are less and less important. For $N_c=N_f =3$, we have
\ba
\lambda_{\rm SYMP} & \approx & 0.257 g^3,
\\ [2mm]
\lambda_{\rm QGP} & \approx & 0.042 g^3,
\ea
As seen, the dynamics of QGP is more collective than that of SYMP at the same value of $g$.

The differences of $\varepsilon$ and $n$ for SYMP and QGP merely reflect the difference in numbers of degrees of freedom in the two plasma systems. In the case of $m_D$ and $\lambda$ it also matters that (anti-)quarks in QGP and fermions in SYMP  belong to different representations - fundamental and adjoint, respectively - of the ${\rm SU}(N_c)$ gauge group. In further parts of the thesis we provide deeper explanation of the plasma characteristics mentioned here.

\newpage
\thispagestyle{plain}

\section{Keldysh-Schwinger formalism}
\label{sec-KS-form}

The first attempts to combine relativistic quantum fields and many-body theories were undertaken at the end of fifties \cite{Matsubara:1955ws,Ezawa:1957rw,Martin:1959jp,Schwinger:1960qe}. However, it was not until the eighties that statistical quantum field theories were actively developed. 

Historically the oldest is the Matsubara or imaginary-time formalism \cite{Matsubara:1955ws} which was developed by a plethora of authors, see for example \cite{Kubo:1957mj,Thouless:1958zz,Abrikosov:1959,Fradkin:1959,Kadanoff:1962,Fradkin:1967,Fetter:1971,Mattuck:1976,
Lifshitz:1980}. The formalism is built on a formal analogy between inverse temperature and imaginary time which was first noticed by F.~Bloch \cite{Bloch:1932}. Consequently, the main objects of the approach are the temperature Green functions of imaginary time arguments which are used for a diagrammatic perturbative expansion of the partition function of grand canonical ensemble. Among the achievements of the approach there are the introduction of Fourier representation of the Matsubara Green functions, \cite{Ezawa:1957rw} and \cite{Abrikosov:1959,Fradkin:1959}, and the formulation of the theory in terms of functional integrals \cite{Feynman:1953zza,Feynman:1965,Bernard:1974bq,Popov:1983}. That said, the formalism has some serious limitations. The basic inconvenience is that it provides the unphysical representation of time, thus only static properties of a medium can be obtained. To study dynamical phenomena one needs to include a real time contour in the Matsubara formalism \cite{Abrikosov:1959,Fradkin:1959}. However, both approaches are valid only for systems in thermodynamical equilibrium. This is insufficient to investigate, in particular, a process of thermalisation of any physical system. 

A more relevant framework to study not only equilibrium but also non-equilibrium processes is the Keldysh-Schwinger or real-time formlism, which is used throughout this thesis. The method has its beginnings at sixties, when many efforts have been put to work out tools that combine quantum field theory with non-equilibrium statistical mechanics. The pioneering works were done by Schwinger \cite{Schwinger:1960qe} and others \cite{Mahanthappa:1962ex,Bakshi:1962dv,Craig:1968,Mills:1969}, and they have got commonly known thanks to Keldysh \cite{Keldysh:1964ud}. The main idea of the formalism is that time runs within a closed contour in the complex plane, see Fig. \ref{Keldysh-contour}. The understanding of this concept is as follows. In a vacuum field theory applied to a scattering problem a system evolves along real time and one can know its states in remote past and remote future. In case of a many-body system only the initial state can be known but not the final one. Thus, the time axis is turned in such a way that the evolution of the system goes through the closed path to end in the initial time state. That mentioned, the formalism is constructed in the language of the Green functions from which physical quantities may be extracted. Over the course of decades the approach has been developed to attack a variety of problems as reflected by a set of papers \cite{Niemi:1983nf,Niemi:1983ea,Chou:1984es,Rammer:1986zz,Landsman:1986uw,Niemi:1987pm,
Mrowczynski:1989bu,Calzetta:1989vs,Paz:1990jg,Botermans:1990qi,Mrowczynski:1992hq,Henning:1995sm,
Boyanovsky:1996xx,Boyanovsky:1999cy}.

In this chapter we present a comprehensive introduction to the Keldysh-Schwinger formalism. First, we provide a basic description of how it is formulated. Namely, we give the definitions of different, in general, non-equilibrium Green functions and next we write down some useful relations among them. For clarity of the presentation, we show the definitions of the Green functions for all types of fields studied here. Subsequently, we derive the Green functions of the scalar field starting with the corresponding equation of motion of the contour Green function. Next, we repeat the derivation of the functions for the electromagnetic field stressing some aspects characteristic of the gauge field. In Sec. \ref{ssec-gf-fermion} we write down the Green functions of the fermion field omitting, however, an extensive derivation.

\subsection{Basics of the Keldysh-Schwinger approach}
\label{ssec-basics-KS}

\subsubsection{Contour Green function}
\label{sssec-contour-GF}

\begin{figure}[h!]
\centering
\includegraphics[scale=0.4]{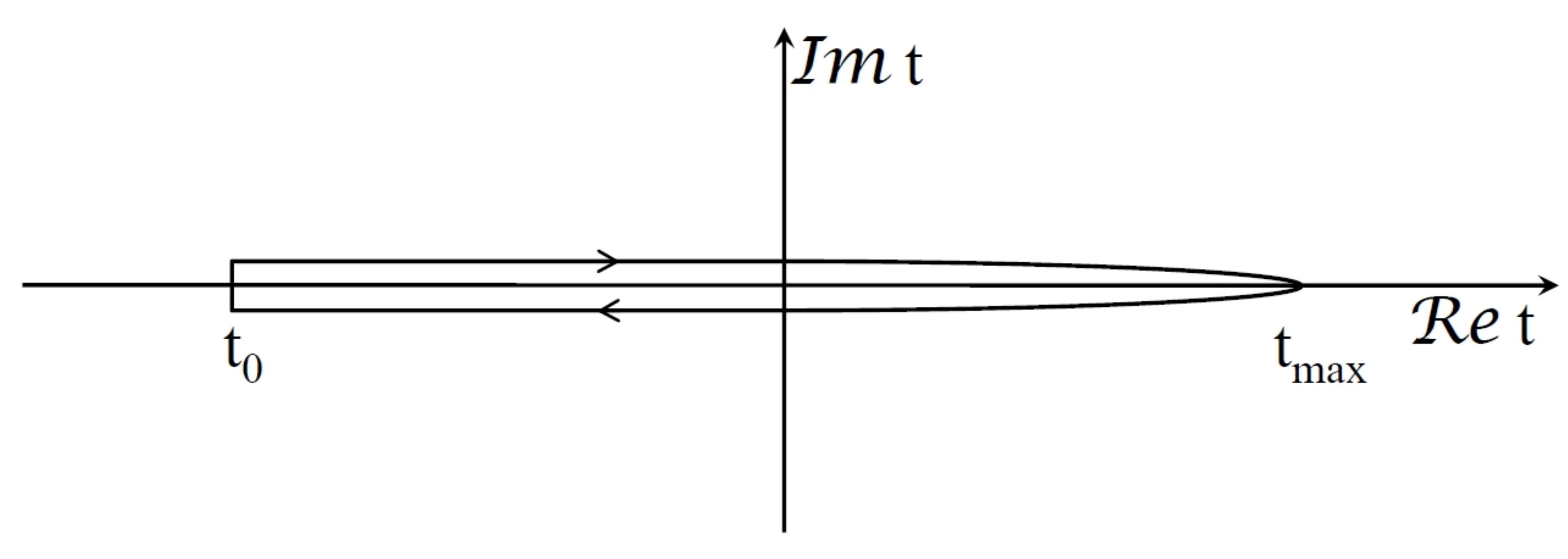}
\caption{The Keldysh contour.}
\label{Keldysh-contour}
\end{figure}

The main object in the Keldysh-Schwinger method is the contour-ordered Green function that is defined as
\ba
\label{contour-GF}
i \tilde \Delta(x,y) \equiv \langle \tilde T \phi(x) \phi^\dagger(y) \rangle
\ea
for a complex scalar field represented by the operator $\phi(x)$;
\ba
\label{contour-GF-em}
i \mathcal{D}_{\mu\nu}(x,y) \equiv \langle \tilde T A_\mu (x) A_\nu (y) \rangle
\ea
for a vector field represented by the operator $A_\mu(x)$ and
\ba
\label{contour-GF-fer}
i \mathcal{S}_{\alpha\beta}(x,y) \equiv \langle \tilde T \psi_\alpha(x) \bar\psi_\beta(y) \rangle
\ea
for a fermion field $\psi(x)$ where $\alpha,\beta=1,2,3,4$ are the spinor indices. The functions $\tilde \Delta$, $\mathcal{D}_{\mu\nu}$, and $\mathcal{S}_{\alpha\beta}$ describe interacting scalar, vector, and fermion fields, respectively. In the formulas (\ref{contour-GF})-(\ref{contour-GF-fer}) and further on we use the notation
\ba
\label{trace-not}
\langle \ldots \rangle \equiv \frac{{\rm Tr}[\hat \rho(t_0) \ldots]}{{\rm Tr}[\hat \rho(t_0)]}
\ea
where $\hat \rho (t_0)$ is a density operator, the trace is understood as a summation over all states of the system at a given initial time $t_0$ 
\ba
\label{tr}
{\rm Tr}[\ldots]=\sum_\alpha <\alpha|\ldots|\alpha>,
\ea
and $\tilde T$ is the operation of time ordering along the Keldysh contour shown in Fig. \ref{Keldysh-contour}.
The time arguments are complex with an infinitesimal positive or negative imaginary part which locates them on the upper or lower branch of the contour. The contour ordering operation of two arbitrary operators is defined as
\be
\label{contour-time-ordering}
\tilde T A(x)B(y)  \equiv \Theta(x_0,y_0) A(x)B(y) \pm \Theta(y_0,x_0)B(y)A(x),
\ee
where $\Theta(x_0,y_0)$ is the contour step function defined as
\ba
\label{theta-function}
\Theta(x_0,y_0)=\left\{ \begin{array}{lll} 1,&& \textrm {if $x_0$ succeeds $y_0$ along the contour}, \\ 0, && \textrm {if
$y_0$ succeeds $x_0$ along the contour}. \end{array} \right.
\ea
The plus sign in the formula (\ref{contour-time-ordering}) is relevant for bosonic operators of the scalar and vector field whereas the minus sign is proper for fermionic operators. The parameter $t_{\rm max}$ is shifted to $+ \infty$ and $t_0$ to $-\infty$ in calculations. 

\subsubsection{Real-time Green functions}

The contour function involves four Green functions of real time arguments. They can be thought of as corresponding to propagation along the top branch of the contour, from the top branch to the bottom one, along the bottom branch, and from the bottom branch to the top one. This can be expressed in the following way for the scalar field
\ba
\label{real-time-GF}
\tilde \Delta(x,y) &\rightarrow& \tilde \Delta^{>} (x,y) \;\;\;\;\; \textrm{for $x_0$ on the lower branch and $y_0$ on the upper one},
\nn \\ [2mm]
\tilde \Delta(x,y) &\rightarrow& \tilde \Delta^{<} (x,y) \;\;\;\;\; \textrm{for $x_0$ on the upper branch and $y_0$ on the lower one},
\nn \\ [2mm]
\tilde \Delta(x,y) &\rightarrow& \tilde \Delta^c (x,y) \;\;\;\;\;\; \textrm{for $x_0,y_0$ on the upper branch}, \nn \\ [2mm]
\tilde \Delta(x,y) &\rightarrow& \tilde \Delta^a (x,y) \;\;\;\;\;\; \textrm{for $x_0,y_0$ on the lower branch}.
\ea
The other types of fields comply with the same prescription.

Taking into account different combinations of time argument location on the contour, one defines the real-time Green functions of the scalar field as
\ba
\label{bigger-GF}
i \tilde \Delta^{>}(x,y) & \equiv & \langle \phi(x) \phi^\dagger(y) \rangle, 
\\ [2mm]
\label{smaller-GF}
i \tilde \Delta^{<}(x,y) & \equiv & \langle \phi^\dagger(y) \phi(x) \rangle, 
\\ [2mm]
\label{chronological-GF}
i \tilde \Delta^c(x,y) & \equiv & \langle T^c \phi(x) \phi^\dagger(y) \rangle, 
\\ [2mm]
\label{antichronological-GF}
i \tilde \Delta^a(x,y) & \equiv & \langle T^a \phi(x) \phi^\dagger(y) \rangle, 
\ea
these of the vector field are as follows
\ba
\label{bigger-GF-em}
i \mathcal{D}_{\mu\nu}^{>}(x,y) & \equiv & \langle A_\mu(x) A_\nu(y) \rangle, 
\\ [2mm]
\label{smaller-GF-em}
i \mathcal{D}_{\mu\nu}^{<}(x,y) & \equiv & \langle A_\nu(y) A_\mu(x) \rangle, 
\\ [2mm]
\label{chronological-GF-em}
i \mathcal{D}_{\mu\nu}^c(x,y) & \equiv & \langle T^c A_\mu(x) A_\nu(y) \rangle, 
\\ [2mm]
\label{antichronological-GF-em}
i \mathcal{D}_{\mu\nu}^a(x,y) & \equiv & \langle T^a A_\mu(x) A_\nu(y) \rangle, 
\ea
and these of the fermion field are
\ba
\label{bigger-GF-fer}
i \mathcal{S}_{\alpha\beta}^{>}(x,y) & \equiv & \langle \psi_\alpha(x) \bar \psi_\beta(y) \rangle, 
\\ [2mm]
\label{smaller-GF-fer}
i \mathcal{S}_{\alpha\beta}^{<}(x,y) & \equiv & -\langle \bar \psi_\beta(y) \psi_\alpha(x) \rangle, 
\\ [2mm]
\label{chronological-GF-fer}
i \mathcal{S}_{\alpha\beta}^c(x,y) & \equiv & \langle T^c \psi_\alpha(x) \bar\psi_\beta(y) \rangle, 
\\ [2mm]
\label{antichronological-GF-fer}
i \mathcal{S}_{\alpha\beta}^a(x,y) & \equiv & \langle T^a \psi_\alpha(x) \bar \psi_\beta(y) \rangle. 
\ea
In the definitions (\ref{bigger-GF})-(\ref{antichronological-GF-fer}) $T^c$ is a chronological time ordering
\be
\label{time-ordering}
T^c A(x) B(y)  \equiv \Theta(x_0-y_0) A(x) B(y) \pm \Theta(y_0-x_0) B(y) A(x)
\ee
and $T^a$ is an antichronological time ordering
\be
\label{anti-time-ordering}
T^a A(x) B(y)  \equiv \Theta(y_0-x_0) A(x) B(y) \pm \Theta(x_0-y_0) B(y) A(x),
\ee
where the plus sign is for bosonic operators and the minus for fermionic ones. All the functions of the real time arguments can be assembled in a $2 \times 2$ matrix which for the scalar field is written as
\ba
\label{matrix-GF}
\tilde \Delta(x,y)  = \left( \begin{matrix} \tilde\Delta_{11} &  \tilde\Delta_{12}  \cr  
\tilde\Delta_{21} &  \tilde\Delta_{22}  \cr \end{matrix} \right)
=\left( \begin{matrix} \tilde\Delta^c &  \tilde\Delta^>  \cr  \tilde\Delta^< &  \tilde\Delta^a  \cr \end{matrix} \right) .
\ea
As seen, the matrix elements with the index $i,j=1$ correspond to functions of time arguments located on the upper branch of the Keldysh contour and these indexed by $i,j=2$ refer to lower branch of the time contour. 

A physical meaning of the real-time functions is the following. The functions $\tilde\Delta^{>}$ and $\tilde\Delta^{<}$ play a role of the phase-space density of (quasi-)particles, so they can be treated as a quantum analog of classical distribution functions. These functions are discussed in detail in Sec.~\ref{sssec-meaning-scalars}. The function $\tilde\Delta^c$ describes a particle disturbance propagating forward in time, and an antiparticle disturbance propagating backward in time. The meaning of $\tilde\Delta^a$ is analogous but particles are propagated backward in time and antiparticles forward. In the zero density limit $\tilde\Delta^c$ coincides with the Feynman propagator~\cite{Bjorken}.

In some situations, it is useful to work with retarded $(+)$, advanced $(-)$ and symmetric Green functions. These propagators are defined as
\ba
\label{retarded-GF-sc}
i \tilde\Delta^{+}(x,y) & \equiv & \Theta(x_0-y_0) \langle [\phi(x), \phi^\dagger(y)] \rangle, 
\\ [2mm]
\label{advanced-GF-sc}
i \tilde\Delta^{-}(x,y) & \equiv & - \Theta(y_0-x_0) \langle [\phi(x), \phi^\dagger(y)] \rangle, 
\\ [2mm]
\label{symmetric-GF-sc}
i \tilde\Delta^{\rm sym}(x,y) & \equiv & \langle \{\phi(x), \phi^\dagger(y)\} \rangle,
\ea
\ba
\label{retarded-GF-em}
i \mathcal{D}_{\mu\nu}^{+}(x,y) & \equiv & \Theta(x_0-y_0) \langle [A_\mu(x), A_\nu(y)] \rangle, 
\\ [2mm]
\label{advanced-GF-em}
i \mathcal{D}_{\mu\nu}^{-}(x,y) & \equiv & - \Theta(y_0-x_0) \langle [A_\mu(x), A_\nu(y)] \rangle, 
\\ [2mm]
\label{symmetric-GF-em}
i \mathcal{D}_{\mu\nu}^{\rm sym}(x,y) & \equiv & \langle \{A_\mu(x), A_\nu(y)\} \rangle,
\ea
\ba
\label{retarded-GF-fer}
i \mathcal{S}_{\alpha\beta}^{+}(x,y) & \equiv & \Theta(x_0-y_0) \langle \{\psi_\alpha(x), \bar\psi_\beta(y)\} \rangle,
\\ [2mm]
\label{advanced-GF-fer}
i \mathcal{S}_{\alpha\beta}^{-}(x,y) & \equiv & - \Theta(y_0-x_0) \langle \{\psi_\alpha(x), \bar\psi_\beta(y)\} \rangle, 
\\ [2mm]
\label{symmetric-GF-fer}
i \mathcal{S}_{\alpha\beta}^{\rm sym}(x,y) & \equiv & \langle [\psi_\alpha(x), \bar\psi_\beta(y)] \rangle,
\ea
where $[\ldots,\ldots]$ indicates a commutator and $\{ \ldots,\ldots\}$ an anticommutator of operators. The retarded Green function describes the propagation of both particle and antiparticle disturbance forward in time, while the advanced one governs the evolution backward in time. 

There is also another common and useful Green function, the spectral function, which is defined as
\ba
\label{spectral-GF}
\mathcal{A}_s(x,y) 
& \equiv & 
\langle  [\phi(x), \phi^*(y)] \rangle = i \Big(\tilde \Delta^>(x,y) - \tilde\Delta^<(x,y)\Big),
\\ [2mm]
\label{spectral-GF-em}
\mathcal{A}_g(x,y) 
& \equiv & 
\langle  [A_\mu(x), A_\nu(y)] \rangle = i \Big(\mathcal{D}_{\mu\nu}^>(x,y) - \mathcal{D}_{\mu\nu}^<(x,y)\Big),
\\ [2mm]
\label{spectral-GF-fer}
\mathcal{A}_f^{\alpha\beta}(x,y) 
& \equiv & 
\langle  \{\psi_\alpha(x), \bar\psi_\beta(y)\} \rangle 
= i \Big(\mathcal{S}_{\alpha\beta}^>(x,y) - \mathcal{S}_{\alpha\beta}^<(x,y) \Big).
\ea
The spectral function gives us information about a spectrum of excitations of a system that is what types of (quasi-)particles we tackle with.

\subsubsection{Relations among different Green functions}

Here we show the relations among different Green functions of real time which can be obtained directly from the definitions (\ref{bigger-GF})-(\ref{antichronological-GF}). The relations are presented for the complex scalar field but the analogous relations hold for other fields as well. They read 
\be
\label{rel-2}
\tilde \Delta^{c}_a (x,y)=\Theta (x_0-y_0) \tilde \Delta^\gl (x,y)+\Theta (y_0-x_0) \tilde \Delta^\lg (x,y),
\ee
\ba
\label{rel-3}
\big(i \tilde \Delta^{c}_a (x,y)\big)^\dagger &=& i \tilde \Delta^{a}_c (x,y), 
\\ [2mm]
\label{rel-4}
\big(i \tilde \Delta^\gl(x,y)\big)^\dagger &=& i \tilde \Delta^\gl (x,y),
\ea
where $\dagger$ means the Hermitian conjugation which involves an interchange of the arguments of the functions. Using the relation (\ref{rel-2}) it can be easily shown that
\ba
\label{rel-1}
\tilde \Delta^{c}(x,y) + \tilde \Delta^{a}(x,y) = \tilde \Delta^{>}(x,y) + \tilde \Delta^{<}(x,y),
\ea
which reflects the fact that the four components of the contour Green function are not independent from each other, only three of them constitute a basis. 

Complying with the expressions (\ref{retarded-GF-sc})-(\ref{symmetric-GF-fer}), we immediately write down the relations between the retarded, advanced, and symmetric propagators ($\tilde \Delta^+$, $\tilde \Delta^-$, $\tilde \Delta^{\rm sym}$) and the original collection of the functions ($\tilde \Delta^>$, $\tilde \Delta^<$, $\tilde \Delta^a$, $\tilde \Delta^c$) which may be treated as a different basis. The relations read
\ba
\label{retarded-GF-id}
\tilde \Delta^{+}(x,y) & =& \Theta(x_0-y_0) \Big(\tilde \Delta^>(x,y) - \tilde \Delta^<(x,y)\Big), 
\\ [2mm]
\label{advanced-GF-id}
\tilde \Delta^{-}(x,y) &=& \Theta(y_0-x_0) \Big(\tilde \Delta^<(x,y) - \tilde \Delta^>(x,y)\Big), 
\\ [2mm]
\label{symmetric-GF-id}
\tilde \Delta^{\rm sym}(x,y) & = & \tilde \Delta^>(x,y) + \tilde \Delta^<(x,y).
\ea
Further manipulation on the functions leads us to the next identities
\be
\label{rel-5}
\tilde \Delta^\pm(x,y) = \tilde \Delta^c (x,y) - \tilde \Delta^\lg (x,y),
\ee
\be
\label{rel-6}
\tilde \Delta ^\gl (x,y) = \frac{1}{2} \big(\tilde \Delta^{\rm sym}(x,y) \pm \tilde \Delta^{+}(x,y) \mp \tilde \Delta ^{-}(x,y) \big),
\ee
\be
\label{rel-7}
\tilde \Delta^{+}(x,y) - \tilde \Delta ^{-}(x,y) = \tilde \Delta^>(x,y) - \tilde \Delta^<(x,y).
\ee

The relations (\ref{rel-2})-(\ref{rel-1}) and (\ref{rel-5})-(\ref{rel-7}) hold for both complex and real field, whereas there are an extra relations among the functions for pure real field and they are of the form
\ba
\label{rel-real1}
\tilde \Delta^\lg (x,y) &=& \tilde \Delta^\gl (y,x),
\\ [2mm]
\label{rel-real2}
\tilde \Delta^a_c (x,y) &=& \tilde \Delta^c_a (y,x).
\ea

\subsubsection{Meaning of $\tilde \Delta^>$ and $\tilde \Delta^<$}
\label{sssec-meaning-scalars}

In order to better understand what is the physical meaning of the functions $\tilde \Delta^>$ and $\tilde \Delta^<$ let us recall the Lagrangian density of the free massive charged (complex) scalar field, which is 
\be
\label{lagrangian-scalar}
\mathcal{L}(x)= \partial^\mu \phi(x) \partial_\mu \phi^*(x) -m^2\phi(x)\phi^*(x).
\ee 
It leads us to the following equations of motion
\ba
\label{eom-complex-field1}
\big[ \partial^\mu \partial_\mu +m^2 \big] \phi(x)=0, 
\\ [2mm]
\label{eom-complex-field2}
\big[ \partial^\mu \partial_\mu +m^2 \big] \phi^*(x)=0.
\ea
Due to the invariance of the Lagrangian (\ref{lagrangian-scalar}) under U(1) global transformations there is a conserved current
\ba
\label{current}
j^\mu(x)=i\phi(x) \stackrel{\leftrightarrow}{\partial^\mu}\phi^*(x),
\ea
where the action of the derivative $\stackrel{\leftrightarrow}{\partial^\mu}$ should be understood as 
\ba
\label{not-a}
A \stackrel{\leftrightarrow}{\partial_\mu} B =A \partial_\mu B - \big(\partial_\mu A \big)B.
\ea
One can check that the form of the four-current (\ref{current}) satisfies the continuity equation
\ba
\label{cont-eq}
\partial_\mu j^\mu (x) =0,
\ea
provided the fields obey the equations of motion (\ref{eom-complex-field1}) and (\ref{eom-complex-field2}). Let us also introduce the energy-momentum tensor which, as discussed in \cite{Forger:2003ut}, has the form
\ba
\label{e-m-tensor}
T^{\mu\nu}(x) = \partial^\mu \phi(x)\partial^\nu \phi^*(x) + \partial^\nu \phi(x) \partial^\mu \phi^*(x) -g^{\mu\nu} \mathcal{L}(x).
\ea
and, as the four-current, it is the conserved quantity
\ba
\label{tensor-cons-law}
\partial_\mu T^{\mu\nu}(x) =0,
\ea
which can be proven with the help of the Klein-Gordon equations (\ref{eom-complex-field1}) and (\ref{eom-complex-field2}). Having said that, the tensor (\ref{e-m-tensor}) can be modified in such a way that it remains the conserved quantity. Accordingly, we can subtract the total derivative terms from the expression (\ref{e-m-tensor})
\ba
\frac{1}{2}\partial^\mu \partial^\nu \big( \phi(x) \phi^*(x) \big) 
- \frac{1}{4}g^{\mu\nu} \partial^\sigma \partial_\sigma \big( \phi(x) \phi^*(x) \big)
\ea
to get the energy-momentum tensor in the convenient form
\ba
\label{e-m-tensor-1}
T^{\mu\nu}(x) = -\frac{1}{2} \phi(x) \stackrel{\leftrightarrow}{\partial^\mu} \stackrel{\leftrightarrow}{\partial^\nu} \phi^*(x),
\ea
which still satisfies the conservation law (\ref{tensor-cons-law}). Let us find a statistical average of the four-current and the energy-momentum tensor, that are
\ba
\label{current-average}
\big\langle j^\mu(x) \big\rangle 
&=& 
i \big\langle \phi(x) \stackrel{\leftrightarrow}{\partial^\mu}\phi^\dagger(x) \big\rangle, 
\\ [2mm]
\label{e-m-tensor-average}
\big\langle T^{\mu\nu}(x) \big\rangle 
&=& 
-\frac{1}{2} \big\langle \phi(x) \stackrel{\leftrightarrow}{\partial^\mu}
\stackrel{\leftrightarrow}{\partial^\nu} \phi^\dagger(x) \big\rangle.
\ea
where $\big\langle \dots \big\rangle$ is defined by (\ref{trace-not}). The expressions (\ref{current-average}) and (\ref{e-m-tensor-average}) can be expressed explicitly as
\ba
\label{current-average-1}
\big\langle j^\mu(x) \big\rangle 
&=& 
i \bigg(\big\langle \phi(x) \partial^\mu\phi^\dagger(x) \big\rangle
-\big\langle (\partial^\mu \phi(x)) \phi^\dagger(x) \big\rangle \bigg), 
\\ [2mm]
\label{e-m-tensor-average-1}
\big\langle T^{\mu\nu}(x) \big\rangle 
&=& 
- \frac{1}{2} \bigg( \big\langle \phi(x) \partial^\mu \partial^\nu \phi^\dagger(x) \big\rangle 
- \big\langle  \partial^\mu \phi(x) \partial^\nu \phi^\dagger(x) \big\rangle 
\\ [2mm] \nn
&& \qquad \qquad \qquad \qquad
- \big\langle  \partial^\nu \phi(x) \partial^\mu \phi^\dagger(x) \big\rangle 
+ \big\langle \partial^\mu \partial^\nu \phi(x)  \phi^\dagger(x) \big\rangle \bigg) 
\ea
and next as
\ba
\label{current-average-2}
\big\langle j^\mu(x) \big\rangle 
&=&
i \bigg(\partial^\mu_y \big\langle \phi(x) \phi^\dagger(y) \big\rangle
-\partial^\mu_x  \big\langle \phi(x) \phi^\dagger(y) \big\rangle \bigg) \bigg|_{x=y}, 
\\ [2mm]
\label{em-tensor-average-2}
\big\langle T^{\mu\nu}(x) \big\rangle 
&=& 
-\frac{1}{2} \bigg( \partial^\mu_y \partial^\nu_y \big\langle \phi(x) \phi^\dagger(y) \big\rangle 
- \partial^\mu_x \partial^\nu_y \big\langle \phi(x) \phi^\dagger(y) \big\rangle 
\\ [2mm] \nn
&& \qquad \qquad \qquad \qquad
- \partial^\nu_x \partial^\nu_y \big\langle \phi(x) \phi^\dagger(y) \big\rangle 
+ \partial^\mu_x \partial^\nu_y \big\langle \phi(x)  \phi^\dagger(y) \big\rangle \bigg) \bigg|_{x=y}.
\ea
In (\ref{current-average-2}) and (\ref{em-tensor-average-2}) one can recognize the unordered Green function
\ba
i \tilde\Delta^>(x,y) = \big\langle  \phi(x) \phi^\dagger(y) \big\rangle
\ea
which is defined by (\ref{bigger-GF}). Then the average of the four-current can be written as
\ba
\label{current-average-3}
\big\langle j^\mu(x) \big\rangle= \big(\partial^\mu_x - \partial^\mu_y  \big) \tilde \Delta^>(x,y) \Big|_{x=y},
\ea
and that of the energy-momentum tensor as
\ba
\label{e-m-tensor-average-3}
\big\langle T^{\mu\nu}(x) \big\rangle = 
-\frac{i}{2} \bigg( \partial^\mu_y \partial^\nu_y - \partial^\mu_x \partial^\nu_y - \partial^\nu_x \partial^\nu_y 
+ \partial^\mu_x \partial^\nu_y \bigg) \tilde \Delta^>(x,y) \Big|_{x=y}.
\ea

If the system under study is out of equilibrium, in particular it is not homogeneous, that is the translational invariance is broken, one usually introduces new variables $X$ and $u$ which are related to $x$ and $y$ in the following way
\ba
\label{Wigner-variables-0}
X=\frac{1}{2}(x+y), \qquad\qquad\qquad u=x-y
\ea
and then the old variables are given by
\ba
\label{old-variables}
x=X+\frac{1}{2}u, \qquad\qquad\qquad y=X-\frac{1}{2}u.
\ea
The derivatives are
\ba
\label{deriv}
\frac{\partial}{\partial x} = \frac{1}{2}\frac{\partial}{\partial X}+\frac{\partial}{\partial u}, 
\qquad\qquad\qquad
\frac{\partial}{\partial y} = \frac{1}{2}\frac{\partial}{\partial X}-\frac{\partial}{\partial u}.
\ea
In the new coordinates $\tilde \Delta^>(x,y)$ reads 
\ba
\label{FGG-new-coord}
\tilde \Delta^>(x,y) = \tilde \Delta^>\big(X+\frac{1}{2}u,X-\frac{1}{2}u\big).
\ea
The latter function in (\ref{FGG-new-coord}) is denoted as $\tilde \Delta^>(X,u)$ to simplify the notation. To go to the phase space we use the Wigner transform
\ba
\label{Wigner-transform-FGG}
\tilde \Delta^> (X,p) = \int d^4u \; e^{ipu} \; \tilde \Delta^> \big(X,u \big),
\ea
and the inverse Wigner transform
\ba
\label{Wigner-transform-inv}
\tilde \Delta^>(X,u) = \int \frac{d^4p}{(2\pi)^4} e^{-ipu} \tilde \Delta^> (X,p),
\ea
which hold for all real-time Green functions. Using the new variables $X$ and $u$ we immediately find the following relations
\ba
\partial^\mu_x - \partial^\mu_y 
&=& 
2 \partial^\mu_u, 
\\ [2mm]
\partial^\mu_y \partial^\nu_y - \partial^\mu_x \partial^\nu_y - \partial^\nu_x \partial^\nu_y 
+ \partial^\mu_x \partial^\nu_y 
& = & 
4 \partial^\mu_u \partial^\nu_u.
\ea
So, using the Wigner transform, we find $j^\mu$ and $T^{\mu\nu}$ as
\ba
\label{current-average-4}
\big\langle j^\mu(X) \big\rangle 
& = & 
2\partial^\mu_u  \int \frac{d^4p}{(2\pi)^4} e^{-ipu} \tilde \Delta^>(X,p) \Big|_{u=0}, 
\\ [2mm]
\label{e-m-tensor-average-4}
\big\langle T^{\mu\nu}(X) \big\rangle 
& = & 
-2i \; \partial^\mu_u \partial^\nu_u \int \frac{d^4p}{(2\pi)^4} e^{-ipu} \tilde \Delta^>(X,p) \Big|_{u=0}.
\ea
where we put $u=0$ after the differentiation over $u$. This leads us to 
\ba
\label{sc-current-average-5}
\big\langle j^\mu(X) \big\rangle 
& = & 
- 2 \int \frac{d^4p}{(2\pi)^4} p^\mu \;  i \tilde \Delta^>(X,p),
\\ [2mm]
\label{sc-e-m-tensor-average-5}
\big\langle T^{\mu\nu}(X) \big\rangle 
& = & 
2 \int \frac{d^4p}{(2\pi)^4} p^\mu p^\nu \; i \tilde \Delta^>(X,p).
\ea
The above derivation of the four-current $j^\mu(X)$ and the energy-momentum tensor $T^{\mu\nu}(X)$ can also be done with the help of $\tilde \Delta^<(x,y)$. From Eqs. (\ref{sc-current-average-5}) and (\ref{sc-e-m-tensor-average-5}) one sees that $i\tilde \Delta^>(X,p)$ (or $i\tilde \Delta^<(X,p)$) corresponds to the density of particles with four-momentum $p$ in a space-time point $X$, and consequently, as it has been already mentioned, it is a quantum analog of a classical distribution function. This interpretation is supported by the fact that both $i\tilde \Delta^>(X,p)$ and $i\tilde \Delta^<(X,p)$ are Hermitian, however, they are not positively defined and the probabilistic interpretation is only approximately valid. One should also observe that, in contrast to the classical distribution functions, $i\tilde \Delta^>(X,p)$ and $i\tilde \Delta^<(X,p)$ can be nonzero for the off-mass-shell four-momenta, when $p^2 \neq m^2$.

\subsection{Derivation of real-time Green functions of the scalar field}
\label{ssec-deriv-scalar}

This subsection is devoted to a derivation of explicit forms of the real-time Green functions. We consider here a real scalar field interacting with an external source. Starting with the equation of motion of the field, we find the equation of motion of the contour Green function. Next, we derive the Green functions of real time arguments. The functions are found for a non-equilibrium system, that is when it is, in general, inhomogeneous and a momentum distribution of plasma constituents is arbitrary. The derivation of the free equilibrium functions directly from the definitions (\ref{bigger-GF})-(\ref{antichronological-GF}) is given in Appendix~\ref{appendix-FG}. 

\subsubsection{Equation of motion of the contour Green function}
\label{sssec-eom-scalar}

In order to find an equation of motion of the contour Green function, let us start with some elementary remarks on the vacuum scalar field theory. Since the action is a fundamental quantity in the field theory, we start with the Lagrangian density of the field theory of real scalars interacting with an external current $j(x)$
\ba
\label{lagr-scal}
\mathcal{L}(x) = \frac{1}{2}\partial^\mu \phi(x) \partial_\mu \phi(x) - \frac{1}{2}m^2 \phi^2(x) +j(x)\phi(x).
\ea
The equation of motion of the field $\phi(x)$ is given by
\ba
\label{eom-field-inhomo}
\big(\square_x +m^2\big)\phi(x)=j(x),
\ea
and, as one sees, it is an inhomogeneous equation. Its solution can be written down in a general form
\ba
\label{eom-field-gen-sol}
\phi(x) =\phi_0(x) - \int d^4 x' \Delta(x-x') j(x'),
\ea
where $\phi_0(x)$ is a solution of the homogeneous equation which is the Klein-Gordon equation
\ba
\label{eom-field-homo}
\big(\square_x +m^2\big)\phi_0(x)=0,
\ea
and $\Delta(x-x')$ is the Green function. It is easy to guess that the Green function must satisfy the equation
\ba
\label{eom-field-green-f}
\big(\square_x +m^2\big)\Delta(x)=-\delta^{(4)}(x)
\ea
since then, the formula (\ref{eom-field-gen-sol}) solves Eq. (\ref{eom-field-inhomo}). 

The equation (\ref{eom-field-green-f}) is the equation of motion of the Green function of the scalar field. It appears that the Green function $\Delta(x)$, which is a solution to Eq. (\ref{eom-field-green-f}) with a properly chosen initial condition, coincides with the propagator of the scalar field, that is, the equation of motion of the propagator is given by (\ref{eom-field-green-f}). 

In the statistical theory the contour Green function of the scalar field is defined by (\ref{contour-GF}) and let us now find the equation of motion of the chronologically ordered Green function from the definition (\ref{chronological-GF}), that is
\ba
\label{eom-def-prop}
i\Delta^c(x,y) \equiv \langle T^c \phi(x) \phi(y) \rangle.
\ea
So, we have to find the result of the action of the Klein-Gordon operator on the Green function 
\ba
\label{eom-prop-0}
\big(\square_x +m^2\big)\Delta^c(x,y)
\ea
As the d'Alembert operator acts only on the field operators of the Green function (\ref{eom-def-prop}) and does not affect the density operator we can write the expression (\ref{eom-prop-0}) explicitly
\ba
\label{eom-prop-1}
\big(\square_x +m^2\big)\Delta^c(x,y) 
&=& 
-i\langle \partial^2_{x_0} \big(\theta(x_0-y_0) \phi(x) \phi(y)+ \theta(y_0-x_0) \phi(y) \phi(x)\big)
\\ [2mm] \nn
&&
+\big(-\nabla^2+m^2\big)\big(\theta(x_0-y_0) \phi(x) \phi(y)+ \theta(y_0-x_0) \phi(y) \phi(x)\big) \rangle,
\ea
Next, we perform the time differentiation
\ba
\label{eom-prop-2}
\big(\square_x +m^2\big)\Delta^c(x,y) 
&=& 
-i\langle \big(\partial^2_{x_0} \theta(x_0-y_0) \big) \phi(x) \phi(y)
+  \big(\partial^2_{x_0}\theta(y_0-x_0)\big) \phi(y) \phi(x) \quad 
\\ [2mm] \nn
&& 
+ 2\partial_{x_0} \theta(x_0-y_0) \partial_{x_0} \phi(x) \phi(y) 
+ 2\partial_{x_0} \theta(y_0-x_0) \phi(y) \partial_{x_0} \phi(x)
\\ [2mm] \nn
&&
+\theta(x_0-y_0) \big(\square_x+m^2\big) \phi(x) \phi(y)
+ \theta(y_0-x_0) \phi(y) \big(\square_x+m^2\big) \phi(x) \rangle.
\ea
As one can notice, we have grouped the terms in (\ref{eom-prop-2}) in such a way that the Klein-Gordon operator appears in the last line. We consider here the noninteracting scalar field, so there is no source, $j(x)=0$. Then, due to Eq.~(\ref{eom-field-homo}) the expression in the last line of Eq. (\ref{eom-prop-2}) vanishes. To compute the other terms we use the identity
\ba
\label{theta-delta}
\partial_{x_0} \theta(x_0-y_0) = \delta(x_0-y_0),
\ea
so we get
\ba
\label{eom-prop-3}
\big(\square_x +m^2\big)\Delta^c(x,y) 
&=& 
-i\langle \big(\partial_{x_0} \delta(x_0-y_0) \big) \phi(x) \phi(y)
+  \big(\partial_{x_0} \delta(y_0-x_0) \big) \phi(y) \phi(x) \quad\quad 
\\ [2mm] \nn
&& 
+ 2 \delta(x_0-y_0) \partial_{x_0} \phi(x) \phi(y) 
+ 2\delta(y_0-x_0) \phi(y) \partial_{x_0} \phi(x) \rangle.
\ea
In two first terms of the formula (\ref{eom-prop-3}) there are the expressions of the type $\partial_{x_0}
\delta(x_0-y_0)$ that should be understood as these suggested by the integral
\ba
\label{del-theta-int}
\int dx_0 \big(\partial_{x_0} \delta(x_0-y_0) \big) f(x) = - \int dx_0\; \delta(x_0-y_0) \partial_{x_0} f(x),
\ea
where the partial integration is performed. Applying the relation (\ref{del-theta-int}) to (\ref{eom-prop-3}), we obtain
\ba
\label{eom-prop-4}
\big(\square_x +m^2\big)\Delta^c(x,y) 
&=& 
-i\langle - \delta(x_0-y_0)  \big(\partial_{x_0}\phi(x)\big) \phi(y)
+  \delta(x_0-y_0)  \phi(y)  \partial_{x_0}\phi(x) \quad\quad 
\\ [2mm] \nn
&& 
+ 2 \delta(x_0-y_0) \big(\partial_{x_0} \phi(x)\big) \phi(y) 
- 2\delta(x_0-y_0) \phi(y) \partial_{x_0} \phi(x) \rangle,
\ea
which equals to
\ba
\label{eom-prop-4a}
\big(\square_x +m^2\big)\Delta^c(x,y) 
&=& 
- i \langle \delta(x_0-y_0)  \big(\partial_{x_0}\phi(x)\big) \phi(y)
- \delta(x_0-y_0)  \phi(y)  \partial_{x_0}\phi(x) \rangle.\quad\quad\quad
\ea
Remembering that in the canonical formalism the momentum conjugated to $\phi$ is
\be
\label{canonical-momentum}
\pi(x)=\dot\phi(x) =\partial_{x_0}\phi(x),
\ee
Eq. (\ref{eom-prop-4a}) gets the form
\ba
\label{eom-prop-5}
\big(\square_x +m^2\big)\Delta^c(x,y) &=& -i\langle  \delta(x_0-y_0)  \big(\pi(x) \phi(y) - \phi(y) \pi(x)\big) \rangle.
\ea
The equal-time commutation relation of the scalar field is
\be
\label{comm-rel}
\phi(t,{\bf y})\pi(t,{\bf x}) - \pi(t,{\bf x})\phi(t,{\bf y}) = i\delta^{(3)}({\bf y} - {\bf x}).
\ee
The delta function in front of the bracket in Eq. (\ref{eom-prop-5}) makes the whole expression non-zero only for $x_0=y_0$. Therefore, we can use the commutation relation (\ref{comm-rel}). Thus, the equation of motion of the chronologically ordered Green function is of the form
\ba
\label{eom-prop-final}
\big(\square_x +m^2\big)\Delta^c(x,y) &=& -\delta^{(4)}(x-y),
\ea
It is now clear that Eq.~(\ref{eom-prop-final}) is the same as Eq.~(\ref{eom-field-green-f}). The analogous derivation of the other Green functions of real-time arguments leads us to the final generalised formula which is the equation of motion of the contour Green function
\ba
\label{eom-prop-contour-final}
\big(\square_x +m^2\big)\Delta(x,y) &=& - \delta^{(4)}_C(x,y),
\ea
where the contour Dirac delta is defined as
\ba
\label{delta-contour}
\delta^{(4)}_C (x,y)=\left \{ \begin{array}{llll}
\delta^{(4)}(x-y) & \qquad \textrm{for} & x_0,y_0& \textrm{from the upper branch},\\
0 & \qquad \textrm{for} & x_0,y_0& \textrm{from the different branches},\\
-\delta^{(4)}(x-y) & \qquad \textrm{for} & x_0,y_0& \textrm{from the lower branch}.
\end{array} \right.
\ea

\subsubsection{The Green functions as solutions of equations of motion}
\label{sssec-GF-scalars}

Now we intend to derive the functions $\Delta^>,\Delta^<,\Delta^c,\Delta^a, \Delta^+$ and $\Delta^-$ of the system which is not homogeneous, that is, the translational invariance is not imposed. These functions are a basic tool to construct the perturbative calculus. In particular, we will use them further on to find some physical properties of plasma systems. 

\newpage
\begin{center}
{\bf Derivation of $\Delta^\gl$}
\end{center}

To find $\Delta^>$ we start with writing down the respective equations of motion, which, due to Eq. (\ref{eom-prop-contour-final}), read
\ba
\label{eom-FGG-x}
\big(\square_x +m^2\big)\Delta^>(x,y) &=& 0,
\\ [2mm]
\label{eom-FGG-y}
\big(\square_y +m^2\big)\Delta^>(x,y) &=& 0.
\ea
Using the variables $X$ and $u$, the equations of motion (\ref{eom-FGG-x}) and (\ref{eom-FGG-y}) have the following forms
\ba
\label{eom-FGG-u}
\bigg(\frac{1}{4}\partial^2_X+\partial_X \partial_u + \partial^2_u + m^2\bigg)\Delta^>(X,u) &=& 0,
\\ [2mm]
\label{eom-FGG-X}
\bigg(\frac{1}{4}\partial^2_X-\partial_X \partial_u + \partial^2_u + m^2\bigg)\Delta^>(X,u) &=& 0.
\ea
If we subtract Eq. (\ref{eom-FGG-X}) from (\ref{eom-FGG-u}), we obtain 
\ba
\label{eom-FGG-12}
\partial_X \partial_u  \Delta^>(X,u) &=& 0.
\ea
Further on, we use the Wigner transform given by (\ref{Wigner-transform-FGG}) to reach the result
\ba
\label{eom-FGG-kin-eq}
p_\mu \partial^\mu_X\Delta^>(X,p) &=& 0,
\ea
which is identified with the relativistic kinetic equation in absence of a collision term. After adding Eqs. (\ref{eom-FGG-u}) and (\ref{eom-FGG-X}) to each other, we are led to the relation
\ba
\label{eom-FGG-1}
\bigg(\frac{1}{4}\partial^2_X+ \partial^2_u + m^2\bigg)\Delta^>(X,u) &=& 0.
\ea
Applying the inverse Wigner transform (\ref{Wigner-transform-inv}) to Eq.~(\ref{eom-FGG-1}), we have
\ba
\label{eom-FGG-2}
\int \frac{d^4p}{(2\pi)^4} e^{-ipu}\bigg[\frac{1}{4}\partial^2_X-p^2 + m^2\bigg] \Delta^> (X,p) &=& 0,
\ea
so
\ba
\label{eom-FGG-3}
\bigg[\frac{1}{4}\partial^2_X-p^2 + m^2\bigg] \Delta^> (X,p) &=& 0.
\ea
Eq. (\ref{eom-FGG-3}), which is known as the mass-shell equation, shows that the Green function $\Delta^>(X,p)$ can be nonzero for the off-shell momenta, when $p^2 \neq m^2$. Nevertheless, the kinetic theory deals with the system's characteristics averaged over scales larger than the particle Compton wavelength of the order of $m^{-1}$. So, we impose the condition called the {\it quasi-particle approximation}
\ba
\label{quasi-particle}
\bigg|\frac{1}{m^2} \partial_X^2 \Delta^>(X,p)\bigg|\ll |\Delta^>(X,p)|,
\ea
which says that $\Delta^>(X,p)$ weakly depends on $X$ on the scale longer than the Compton wavelength. Then, we can neglect the first term of Eq. (\ref{eom-FGG-3}) and then it gets
\ba
\label{eom-FGG-4}
\big[p^2 - m^2\big] \Delta^> (X,p) &=& 0.
\ea
The solution is of the form
\ba
\label{eom-FGG-5}
\Delta^> (X,p) &\propto & \delta(p^2-m^2),
\ea
which says that the function $\Delta^>(X,p)$ is nonzero only for the on-shell four-momentum that is when $p^2=m^2$. Since the procedure of the derivation of $\Delta^<(X,p)$ is the same, the approximate form of $\Delta^<(X,p)$ is also given by
\ba
\label{eom-FGL-1}
\Delta^< (X,p) &\propto & \delta(p^2-m^2).
\ea

Since both the functions $\Delta^>(X,p)$ and $\Delta^<(X,p)$ are nonzero on-shell momenta, they correspond to real particles and thus the aim of the next part of the procedure is to express them in terms of a distribution function. Thereby, they can be written as
\ba
\label{eom-FGG-deriv-1}
i\Delta^> (X,p) &= & 2\pi \delta(p^2-m^2)h(X,p), 
\\ [2mm]
\label{eom-FGL-deriv-1}
i\Delta^< (X,p) &= & 2\pi \delta(p^2-m^2)g(X,p),
\ea
where $h(X,p)$ and $g(X,p)$ are unknown functions. To find them, let us first write $\Delta^>(X,p)$ and $\Delta^<(X,p)$ as combinations of positive and negative energy contributions, that are
\ba
\label{eom-FGG-deriv-2}
i\Delta^> (X,p) &= &\frac{\pi}{E_p} \Big[\delta(p_0-E_p)\theta(p_0) +\delta(p_0+E_p)\theta(-p_0) \Big]h(X,p), 
\\ [2mm]
\label{eom-FGL-deriv-2}
i\Delta^< (X,p) &= & \frac{\pi}{E_p} \Big[\delta(p_0-E_p)\theta(p_0) +\delta(p_0+E_p)\theta(-p_0) \Big]g(X,p),
\ea
Subsequently, one finds the positive energy contribution of the difference of $\Delta^>(X,p) - \Delta^<(X,p)$, that is 
\ba
\label{eom-positiv-part}
i\theta(p_0)\big[\Delta^>(X,p) - \Delta^<(X,p)\big] = \frac{\pi}{E_p} \delta(p_0-E_p)\theta(p_0) \big[h(X,p)-g(X,p)\big],
\ea
and that of the negative energy contribution
\ba
\label{eom-negativ-part}
i\theta(-p_0) \big[\Delta^>(X,p) - \Delta^<(X,p)\big] 
= \frac{\pi}{E_p} \delta(p_0+E_p) \theta(-p_0) \big[h(X,p)-g(X,p)\big].
\ea
The difference of $\Delta^>(x,y)$ and $\Delta^<(x,y)$, which is related to the spectral function, is known as the so-called Jordan function and is given by
\ba
\label{Green-Jordan}
i \big[\Delta^>(x,y) - \Delta^<(x,y)\big] = \int \frac{d^3p}{(2\pi)^3 2E_p}\Big[e^{-ip(x-y)} - e^{ip(x-y)}\Big].
\ea
Let us now manipulate the Jordan function to a form which reveals positive and negative parts of the difference of $\Delta^>(X,p)$ and $\Delta^<(X,p)$. Using the Wigner transform to the equality (\ref{Green-Jordan}), we produce the following identity
\ba
\label{Green-Jordan-wigner}
i\int \frac{d^4p}{(2\pi)^4} e^{-ipu} \big[\Delta^>(X,p) - \Delta^<(X,p)\big] 
= \int \frac{d^3p}{(2\pi)^3 2E_p}\Big[e^{-ipu} - e^{ipu}\Big]
\ea
and next
\ba
\label{Green-Jordan-wigner-1}
\int \frac{d^3p}{(2\pi)^3 2E_p}\Big[e^{-ipu} - e^{ipu}\Big] 
= \int \frac{d^4p}{(2\pi)^4} e^{-ipu}\bigg[\frac{\pi}{E_p}\delta(p_0-E_p)+\frac{\pi}{E_p}\delta(p_0+E_p)\bigg].
\ea
Thus, we find
\ba
\label{Green-Jordan-wigner5}
i\big[\Delta^>(X,p) - \Delta^<(X,p)\big] = \frac{\pi}{E_p} \Big[\delta(p_0-E_p) - \delta(p_0+E_p)\Big]
\ea
or, in terms of the positive and negative energy parts, as
\ba
\label{Green-Jordan-wigner5a}
i\theta(p_0) \big[\Delta^>(X,p) - \Delta^<(X,p)\big] &=& \frac{\pi}{E_p} \delta(p_0-E_p), 
\\ [2mm]
\label{Green-Jordan-wigner5b}
i\theta(-p_0) \big[\Delta^>(X,p) - \Delta^<(X,p)\big] &=&- \frac{\pi}{E_p} \delta(p_0+E_p).
\ea
Comparing (\ref{eom-positiv-part}) to (\ref{Green-Jordan-wigner5a}) and (\ref{eom-negativ-part}) to (\ref{Green-Jordan-wigner5b}), we get the equations
\ba
\label{comparison}
\delta(p_0-E_p) &=& \delta(p_0-E_p) \big[h(X,p)-g(X,p)\big], 
\\ [2mm]
-\delta(p_0+E_p) &=& \delta(p_0+E_p) \big[h(X,p)-g(X,p)\big],
\ea
which are solved by
\ba
\label{comparison-2}
h(X,E_p,{\bf p}) &=& 1+g(X,E_p,{\bf p}), 
\\ [2mm]
h(X,-E_p,{\bf p})+1 &=& g(X,-E_p,{\bf p}).
\ea
Since we would like $h(X,p)$ and $g(X,p)$ to be expressed by one function it is not difficult to guess that they are of
the forms
\ba
\label{hig}
h(X,p) &=& \theta(p_0) \big[f_s(X,{\bf p})+1\big] + \theta(-p_0) \bar f_s(X,-{\bf p}), 
\\ [2mm]
g(X,p) &=& \theta(p_0)f_s(X,{\bf p}) + \theta(-p_0) \big[\bar f_s(X,-{\bf p})+1\big],
\ea
where $f_s(X,{\bf p})$ and $\bar f_s(X,-{\bf p})$ are, as we show below, the distribution functions of particles and antiparticles. The distribution function is normalized in such a way that
\ba
\label{dist-fun-norm}
n_s(X)=\int \frac{d^3p}{(2\pi)^3} f_s(X,{\bf p}),
\ea
where $n_s(X)$ is the density of scalar particles. Finally, $\Delta^>(X,p)$ and $\Delta^<(X,p)$ have the following explicit forms
\ba
\label{Del->}
\Delta^>(X,p) 
&=&
- \frac{i\pi}{E_p} \Big[\delta(E_p-p_0)\big(f_s(X,{\bf p})+1\big) + \delta(E_p+p_0) \bar f_s(X,-{\bf p})\Big], 
\\
\label{Del-<}
\Delta^<(X,p) 
&=&
-\frac{i\pi}{E_p} \Big[\delta(E_p-p_0) f_s(X,{\bf p}) + \delta(E_p+p_0)\big( \bar f_s(X,-{\bf p})+1\big)\Big].
\ea
As we see, the Green functions (\ref{Del->}) and (\ref{Del-<}) have the same structures as these given by (\ref{bigger-GF-SF-FT-1}) and (\ref{smaller-GF-SF-FT}) which are found in the equilibrium limit. It is worth noting here that both $\Delta^>(X,p)$ and $\Delta^<(X,p)$ are very useful and convenient tool to find the other real-time argument Green functions.

To show the meaning of the functions (\ref{Del->}) and (\ref{Del-<}), it is illuminating to find the four-current and the energy-momentum tensor which have been discussed in Sec. \ref{sssec-meaning-scalars}. Inserting the function $i\Delta^>(X,p)$ given by (\ref{Del->}) to Eqs. (\ref{sc-current-average-5}) and (\ref{sc-e-m-tensor-average-5}), we have
\ba
\label{current-average-6}
\big\langle j^\mu(X) \big\rangle 
&\!=\!& 
- \int \frac{d^4p}{(2\pi)^3} \frac{p^\mu}{E_p}
\Big[\delta(E_p-p_0)\big(f_s(X,{\bf p})+1\big) + \delta(E_p+p_0) \bar f_s(X,-{\bf p})\Big],\quad\quad 
\\ [2mm]
\label{e-m-tensor-average-6}
\big\langle T^{\mu\nu}(X) \big\rangle 
&\! =\! & 
\int \frac{d^4p}{(2\pi)^3} \frac{p^\mu p^\nu}{E_p} 
\Big[\delta(E_p-p_0)\big(f_s(X,{\bf p})+1\big) + \delta(E_p+p_0) \bar f_s(X,-{\bf p})\Big]. \quad\quad
\ea
The integration over $p_0$ leads to
\ba
\label{sc-current-average-7}
\big\langle j^\mu(X) \big\rangle 
&=& 
- \int \frac{d^3p}{(2\pi)^3} \frac{p^\mu}{E_p} \Big(f_s(X,{\bf p}) - \bar f_s(X,{\bf p}) +1 \Big),
\\ [2mm]
\label{sc-e-m-tensor-average-7}
\big\langle T^{\mu\nu}(X) \big\rangle 
& = & 
\int \frac{d^3p}{(2\pi)^3} \frac{p^\mu p^\nu}{E_p} \Big(f_s(X,{\bf p}) + \bar f_s(X,{\bf p}) +1 \Big).
\ea
Let us add that we have changed the sign of the momentum ($-{\bf p} \rightarrow {\bf p}$) in case of the antiparticle distribution function to have the compact formulas of $j^\mu$ and $T^{\mu\nu}$. In the vacuum limit ($f_s(X,{\bf p}),\bar f_s(X,{\bf p}) \rightarrow 0$) both the four-current and the energy-momentum tensor should be zero for physical reasons. However, as one can see, in this limit the integrals in (\ref{sc-current-average-7}) and (\ref{sc-e-m-tensor-average-7}) are nonzero and they are of the forms, respectively,
\ba
\label{current-average-8}
- \int \frac{d^3p}{(2\pi)^3} \frac{p^\mu}{E_p}, 
\\ [2mm]
\label{e-m-tensor-average-8}
 \int \frac{d^3p}{(2\pi)^3} \frac{p^\mu p^\nu}{E_p}.
\ea
These types of divergences are well known in the field theory. In case of vacuum field theory they do not appear because of the normal-ordering of operators present in the definition of Green functions. Upon subtracting of the vacuum value from the right-hand sides of Eqs. (\ref{sc-current-average-7}) and (\ref{sc-e-m-tensor-average-7}), we get the finite expression of the current
\ba
\label{current-average-9}
\big\langle j^\mu(X) \big\rangle 
&=& 
- \int \frac{d^3p}{(2\pi)^3} \frac{p^\mu}{E_p} \Big(f_s(X,{\bf p}) - \bar f_s(X,{\bf p}) \Big),
\ea
and that of the energy-momentum tensor
\ba
\label{e-m-tensor-average-9}
\big\langle T^{\mu\nu}(X) \big\rangle 
& = & 
\int \frac{d^3p}{(2\pi)^3} \frac{p^\mu p^\nu}{E_p} \Big(f_s(X,{\bf p}) + \bar f_s(X,{\bf p}) \Big),
\ea
which coincide with standard expressions of relativistic kinetic theory, see \cite{DeGroot:1980dk}.

\newpage
\begin{center}
{\bf Derivation of $\Delta^{c,a}$}
\end{center}

To find $\Delta^c(X,p)$ we start with the relation 
\ba
\label{eom-FGC}
\Delta^c(x,y)=\theta(x_0-y_0)\Delta^>(x,y)+\theta(y_0-x_0)\Delta^<(x,y),
\ea
which expressed in the variables $X$ and $u$, defined by (\ref{Wigner-variables-0}), equals
\ba
\label{eom-FGC-1}
\Delta^c(X,u)=\theta(u_0)\Delta^>(X,u)+\theta(-u_0)\Delta^<(X,u).
\ea
The inverse Wigner transform of $\Delta^>(X,u)$ and $\Delta^<(X,u)$ gives us the relation (\ref{eom-FGC}) in the form
\ba
\label{eom-FGC-2}
\Delta^c(X,u)=\theta(u_0)\int \frac{d^4k}{(2\pi)^4}e^{-iku}\Delta^>(X,k)
+\theta(-u_0)\int \frac{d^4k}{(2\pi)^4}e^{-iku}\Delta^<(X,k).
\ea
In the next step we multiply both sides of Eq. (\ref{eom-FGC-2}) by the factor $e^{ipu}$ and integrate them over $u$ to have
\ba
\label{eom-FGC-3}
\int d^4u \; e^{ipu}\Delta^c(X,u) 
&=& 
\int d^4u \; e^{ipu}\theta(u_0)\int \frac{d^4k}{(2\pi)^4}e^{-iku}\Delta^>(X,k) 
\\ [2mm] \nn
&& \qquad \qquad \qquad
+\int d^4u \; e^{ipu}\theta(-u_0)\int \frac{d^4k}{(2\pi)^4}e^{-iku}\Delta^<(X,k).
\ea
It has been done to get $\Delta^c(X,u)$ in the phase space, so
\ba
\label{eom-FGC-3}
\Delta^c(X,p) 
&=& 
-i\pi \int_0^\infty du_0 \int d^3u \; e^{ipu} \int \frac{d^4k}{(2\pi)^4}e^{-iku} 
\\ [2mm] \nn
&& \qquad \qquad
\times \frac{1}{E_k} \Big(\delta(E_k-k_0)\big( f_s(X,{\bf k})+1\big)+\delta(E_k+k_0) \bar f_s(X,-{\bf k})\Big) 
\\ [2mm] \nn
&& 
-i\pi \int_{-\infty}^0 du_0 \int d^3u \; e^{ipu} \int \frac{d^4k}{(2\pi)^4} e^{-iku} 
\\ [2mm] \nn
&& \qquad \qquad
\times \frac{1}{E_k} \Big(\delta(E_k-k_0) f_s(X,{\bf k})+\delta(E_k+k_0)\big( \bar f_s(X,-{\bf k})+1\big)\Big),
\ea
where we have inserted the explicit forms of $\Delta^>(X,k)$ and $\Delta^<(X,k)$ given by (\ref{Del->}) and (\ref{Del-<}), respectively. Due to the integration over $u_0$, we reorganize the expression (\ref{eom-FGC-3}) and we write it down as the sum of three terms 
\ba
\label{eom-FGC-4}
\Delta^c(X,p) 
&=& 
-i\pi \bigg\{ \int d^4u \;e^{ipu} \int \frac{d^4k}{(2\pi)^4}e^{-iku} 
\\ [2mm] \nn
&& \qquad \qquad
\times \frac{1}{E_k} \Big(\delta(E_k-k_0) f_s(X,{\bf k}) +\delta(E_k+k_0) \bar f_s(X,-{\bf k})\Big) 
\\ [2mm] \nn
&& 
+ \int^{\infty}_0 du_0 \int d^3u \; e^{ipu} \int \frac{d^4k}{(2\pi)^4} e^{-iku}
\frac{\delta(E_k-k_0)}{E_k} 
\\ [2mm] \nn
&& 
+ \int_{-\infty}^0 du_0 \int d^3u \; e^{ipu} \int \frac{d^4k}{(2\pi)^4} e^{-iku}
\frac{\delta(E_k+k_0)}{E_k} \bigg\}.
\ea
For clarity of presentation, we denote them as
\ba
\label{eom-FGC-4a}
\Delta^c(X,p)= -i\pi (A+B+C).
\ea
Let us make the calculations term by term. The $A$ term corresponds to
\ba
\label{partA-0}
A = \int d^4u \int \frac{d^4k}{(2\pi)^4}e^{i(p-k)u}
\frac{1}{E_k} \Big(\delta(E_k-k_0) f_s(X,{\bf k}) +\delta(E_k+k_0) \bar f_s(X,-{\bf k})\Big).\qquad
\ea
Since 
\ba
\label{int-delta}
\int d^4u e^{i(p-k)u} = (2\pi)^4 \delta^{(4)}(p-k),
\ea
the $A$ term gets the form
\ba
\label{partA}
A = \int \frac{d^4k}{E_k} 
\Big(\delta(E_k-k_0) f_s(X,{\bf k}) +\delta(E_k+k_0) \bar f_s(X,-{\bf k})\Big) \delta^{(4)}(p-k).
\ea
Performing the integration over $k$ we obtain the result
\ba
\label{partA-2}
A = \frac{1}{E_p} \Big(\delta(p_0-E_p) f_s(X,{\bf p}) + \delta(p_0+E_p) \bar f_s(X,-{\bf p}) \Big).
\ea
The term $B$ is
\ba
\label{partB}
B = \int^{\infty}_0 du_0 \int d^3u \, e^{ipu} \int \frac{d^4k}{(2\pi)^4 E_k} e^{-iku} \delta(E_k-k_0)
\ea
and can be immediately changed into 
\ba
\label{partBa}
B = \int^{\infty}_0 du_0 \int d^3u \, e^{ipu} \int \frac{d^3k}{(2\pi)^4 E_k} e^{-i(E_ku_0-{\bf ku})},
\ea
where the integration over $k_0$ has been performed. The next integrations should be done in the same order as in the case of the $A$ term until we reach for the expression
\ba
\label{partB-1}
B = \frac{1}{2\pi E_p} \int^{\infty}_0 du_0 \; e^{i(p_0-E_p)u_0}.
\ea
The integral (\ref{partB-1}) is, however, ill defined and we have to change $E_p \rightarrow E_p-i0^+$ to make the limit $u_0 \rightarrow \pm\infty$ meaningful. Then, we have
\ba
\label{partB-2}
B = \frac{1}{2\pi E_p} \int^{\infty}_0 du_0 \; e^{i(p_0-E_p+i0^+)u_0} 
= \frac{1}{2\pi E_p} \frac{i}{p_0-E_p+i0^+}.
\ea
The part $C$ corresponds to
\ba
\label{partC}
C = \int_{-\infty}^0 du_0 \int d^3u \, e^{ipu} \int \frac{d^4k}{(2\pi)^4 E_k} e^{-iku} \delta(E_k+k_0).
\ea
Performing the same steps as in case of the derivation of the part $B$, we get the formula
\ba
\label{partC-3}
C = - \frac{1}{2\pi E_p} \frac{i}{p_0+E_p-i0^+}.
\ea
Adding all the terms $A$, $B$, and $C$ together, we find the chronologically ordered Green function
\ba
\label{eom-FGC-cont}
\Delta^c(X,p) 
&=& 
- \frac{i\pi}{E_p}\Big(\delta(p_0-E_p) f_s(X,{\bf p}) + \delta(p_0+E_p) \bar f_s(X,-{\bf p}) \Big) 
\\ [2mm] \nn
&& \qquad \qquad \qquad
+ \frac{1}{2E_p}\bigg(\frac{i}{p_0-E_p+i0^+} -\frac{i}{p_0+E_p-i0^+}\bigg),
\ea
which finally can be written as
\ba
\label{Del-c}
\Delta^c(X,p) = \frac{1}{p^2-m^2+i0^+} 
-\frac{i\pi}{E_p}\Big(\delta(p_0-E_p) f_s(X,{\bf p}) + \delta(p_0+E_p) \bar f_s(X,-{\bf p}) \Big). 
\ea

To find the anti-chronologically ordered Green function we need to perform the same computation starting with
\be
\label{eom-FGA}
\Delta^a(x,y)=\theta(x_0-y_0)\Delta^<(x,y)+\theta(y_0-x_0)\Delta^>(x,y).
\ee
Then, we obtain the final result
\ba
\label{Del-a}
\Delta^a(X,p) = -\frac{1}{p^2-m^2-i0^+} 
-\frac{i\pi}{E_p}\Big(\delta(p_0-E_p) f_s(X,{\bf p}) + \delta(p_0+E_p) \bar f_s(X,-{\bf p}) \Big), \qquad
\ea
where the replacement $E_p \rightarrow E_p+i0^+$ has been done to make the integrals well defined.   

\begin{center}
{\bf Derivation of $\Delta^\pm$ and $\Delta^{\rm sym}$}
\end{center}

In order to compute the retarded Green function we start with the following identity
\ba
\label{eom-FGP}
\Delta^+(X,p)=\Delta^c(X,p)-\Delta^<(X,p),
\ea
so we can use the known functions $\Delta^c(X,p)$ and $\Delta^<(X,p)$ given by Eqs. (\ref{Del-c}) and (\ref{Del-<}), respectively. Inserting the explicit functions, we have
\ba
\label{eom-FGP-1}
\Delta^+(X,p)=\frac{1}{p^2-m^2+i0^+}+\frac{i\pi}{E_p}\delta(E_p+p_0).
\ea
Due to the known mathematical identity
\ba
\label{eom-identity}
\frac{1}{x \pm i0^+}=\mathcal{P}\frac{1}{x} \mp i\pi \delta(x),
\ea
where $\mathcal{P}$ means the Cauchy principal value, the relation (\ref{eom-FGP-1}) can be rewritten as
\ba
\label{eom-FGP-1a}
\Delta^+(X,p)=\mathcal{P} \frac{1}{p^2-m^2} -  i\pi \delta(p^2-m^2) +\frac{i\pi}{E_p}\delta(E_p+p_0).
\ea
Applying the repeatedly used in this chapter property of the delta function
\ba
\label{eom-identity2}
\delta(p^2-m^2) = \frac{1}{2E_p} \Big( \delta(E_p-p_0) + \delta(E_p+p_0) \Big),
\ea
we obtain
\ba
\label{eom-FGP-2}
\Delta^+(X,p)=\mathcal{P} \frac{1}{p^2-m^2} - \frac{i\pi}{2E_p} \Big( \delta(E_p-p_0) - \delta(E_p+p_0) \Big),
\ea
which can be written as
\ba
\label{eom-FGP-3}
\Delta^+(X,p)=\mathcal{P} \frac{1}{p^2-m^2} - \frac{i\pi}{2E_p} 
\textrm{sgn}(p_0) \Big( \delta(E_p-p_0) + \delta(E_p+p_0) \Big).
\ea
Thus, recalling the identity (\ref{eom-identity}), the final form of the retarded Green function is 
\ba
\label{Del-ret}
\Delta^+(X,p)= \frac{1}{p^2-m^2+i\textrm{sgn}(p_0)0^+}.
\ea
The relation, which leads us to the advanced Green function, is as follows
\ba
\label{eom-FGM}
\Delta^-(X,p)=\Delta^c(X,p)-\Delta^>(X,p),
\ea
and repetition of the analogous procedure as for the retarded propagator gives the expression
\ba
\label{Del-adv}
\Delta^-(X,p)= -\frac{1}{p^2-m^2-i\textrm{sgn}(p_0)0^+}.
\ea
Finally, using the relation (\ref{symmetric-GF-id}), we find the symmetric Green function
\ba
\label{Del-sym}
\Delta^{\rm sym}(X,p) = -\frac{i\pi}{E_p}\Big[ \delta(E_p-p_0)\big(2f_s(X,{\bf p})+1 \big) 
+\delta(E_p+p_0)\big(2\bar f_s(X,-{\bf p})+1 \big) \Big].
\ea

Let us note that the functions $\Delta^\lg$ and $\Delta^{\rm sym}$ are nonzero only for on-mass-shell momenta and thus they describe real (quasi-)particles. The functions $\Delta^\lg$ include the distribution functions but the function $\Delta^{\rm sym}$ does not. The functions $\Delta^\pm$ describe virtual particles and they are independent of the distribution functions. The functions $\Delta^{c,a}$ mix up the virtual and real particles. All the real-time argument Green functions derived in this section will be used in Sec.~\ref{sec-collective} in which we calculate self-energies of fields of different plasma systems.

\subsection{Derivation of the real-time Green functions of the electromagnetic field}
\label{ssec-deriv-em}

\subsubsection{Equation of motion and canonical quantization}
\label{sssec-eom-can-quant-ym}

We start this part with writing down the Lagrangian density of the electromagnetic field interacting with an external source $j^\mu(x)$, which is of the form
\ba
\label{lagran-em}
\mathcal{L}=-\frac{1}{4}F^{\mu\nu}F_{\mu\nu} - j^\mu A_\mu,
\ea
where $F^{\mu\nu}=\partial^\mu A^\nu - \partial^\nu A^\mu$. The equation of motion is then
\ba
\label{eq-mot-em}
\partial_\nu F^{\nu\mu}=j^\mu,
\ea
or, expressed by $A^\mu$, it reads
\ba
\label{eq-mot-em-A}
\square A^\mu - \partial^\mu \partial_\nu A^\nu=j^\mu.
\ea
If we impose the Lorentz gauge condition on the vector potential $A^\nu$, which is
\ba
\label{lorentz-condition}
\partial_\nu A^\nu = 0,
\ea
Eq. (\ref{eq-mot-em-A}) takes the form
\ba
\label{eq-mot-em-A-1}
\square A^\mu=j^\mu.
\ea
The solution is given generally as
\ba
\label{sol-eom-em-A1}
A^\mu (x) = A^\mu_{0}(x) + \int d^4x' D^{\mu\nu}(x-x') j_\nu (x),
\ea
where $A^\mu_{0}(x)$ is the solution of the homogeneous equation
\ba
\label{eq-mot-em-A-2}
\square A_0^\mu=0,
\ea
and the Green function $D^{\mu\nu}(x)$ satisfies the equation
\ba
\label{sol-D}
\square D^{\mu\nu}(x) = g^{\mu\nu} \delta^{(4)}(x).
\ea

We intend to derive the Green function of the quantum theory obeying the equation of motion which appears to be the same as that given by (\ref{sol-D}). However, before we move on to it we would like to focus, for a while, on the quantization procedure of the electromagnetic field, as it is not so trivial to implement here the canonical formalism. When we try to apply the canonical quantization to the potential $A^\mu$ we immediately run into problems. If we define the conjugate momentum as
\ba
\label{can-mom}
\pi^\mu = \frac{\partial \mathcal{L}}{\partial \dot{A}_\mu},
\ea
we get
\ba
\label{can-mom-1}
\pi^k &=& \frac{\partial \mathcal{L}}{\partial \dot{A}_k}=-\partial A^k - \frac{\partial A^0}{\partial x^k} = E^k,
\\ [2mm]
\pi^0 &=& \frac{\partial \mathcal{L}}{\partial \dot{A}_0}=0.
\ea
Therefore, the conjugate momentum to the coordinate $A^0$ vanishes and does not allow us to use directly the canonical formalism. The problem has its origin in the fact that the field corresponding to massless spin-1 particle, $A^\mu $, has 4 components while photon has only two physical degrees of freedom. In the case of massive spin-1 field, it is sufficient to impose the Lorentz condition (\ref{lorentz-condition}) in order to remove one degree of freedom, leaving the theory still Lorentz-covariant. In the case of massless spin-1 field, we have another degree of freedom to be removed, and this cannot be done easily without an explicit breaking of the Lorentz-covariance. 

One approach is to impose the additional condition (e.g. $\nabla \cdot {\bf A} =0$) that explicitly restricts the number of degrees of freedom to two. Then, the procedure is no longer Lorentz-covariant. The Lorentz-covariance will be restored at the end when $S$ matrix is calculated. However, we adopt here an another way to quantize the electromagnetic field, keeping the Lorentz-covariant framework, which is known as the Gupta-Bleuler formalism \cite{Gupta:1949rh,Bleuler:1950cy}. In the method we always work with the 4-vector $A_\mu$ and consequently unphysical states enter the quantization procedure. The unphysical states are eliminated by imposing the Lorentz condition on states in such a way that the condition
\ba
\label{GB-cond}
\langle \chi | \partial^\mu A_\mu |\psi \rangle = 0
\ea
has to be fulfilled for any physical states $|\chi \rangle$ and $|\psi \rangle$ in the Fock space. But the price to pay is the appearance of states with negative norm. We have then to define the Hilbert space of the physical states as a sub-space where the norm is positive. Since we are solely interested in a quantization procedure, we skip many details of the Gupta-Bleuler formalism. 

To solve the difficulty of the vanishing $\pi^0$ we modify the Lagrangian (\ref{lagran-em}). We assume the theory is not interacting any more, so the source term in (\ref{lagran-em}) is omitted. However, a new extra term is added to the Lagrangian, so that, it is of the form
\ba
\label{lagr-term}
\mathcal{L}=-\frac{1}{4}F^{\mu\nu}F_{\mu\nu} + \frac{1}{2\alpha} \big(\partial_\mu A^\mu \big)^2,
\ea
where $\alpha$ is an arbitrary constant. The equation of motion is
\ba
\label{eom-em-gh}
\bigg[\square g^{\mu\nu} -\bigg(1-\frac{1}{\alpha}\bigg) \partial^\mu \partial^\nu\bigg] A_\nu=0.
\ea
The conjugate momenta are
\ba
\label{conj-mom}
\pi^\mu = \partial^\mu A^0 - \dot{A}^\mu - \frac{1}{\alpha} g^{\mu 0} \partial^\nu A_\nu,
\ea
that is
\ba
\label{conj-mom}
\pi^0 &=& - \frac{1}{\alpha} \partial^\nu A_\nu,
\\ [2mm]
\pi^k &=& E^k.
\ea
It is worth stressing that the Lagrangian (\ref{lagr-term}) and the equation of motion (\ref{eom-em-gh}) reduce to Maxwell theory under the Lorentz condition $\partial^\mu A_\mu=0$. This is why we say that the choice of the form of the Lagrangian (\ref{lagr-term}) corresponds to a class of Lorentz gauges with the parameter $\alpha$. And the value of $\alpha=1$ is known as the Feynman gauge and $\alpha=0$ as the Landau gauge. Although it is possible to continue with a general $\alpha$, from now on we choose the case $\alpha=1$. Then the equation of motion is of the form
\ba
\label{eom-A-A}
\square_x A^\mu (x)=0.
\ea
As $\pi^0=0$ is not any more vanishing, we can impose the canonical commutation relations at equal times
\ba
\label{com-rel-em-1}
\big[\pi^\mu(t, {\bf x}), A_\nu (t,{\bf y})  \big] &=& -ig^\mu_{\;\nu} \delta^{(3)}({\bf x}-{\bf y}),
\\ [2mm]
\label{com-rel-em-2}
\big[A_\mu(t, {\bf x}), A_\nu (t,{\bf y})  \big] &=& \big[\pi_\mu(t, {\bf x}), \pi_\nu (t,{\bf y})  \big]=0.
\ea
Knowing that $[A_\mu(t, {\bf x}), A_\mu (t,{\bf y})]=0$ we conclude that the space derivatives of $A^\mu$ also commute at equal times. Then, taking into account the formula (\ref{conj-mom}), we can find the relations (\ref{com-rel-em-1}) and (\ref{com-rel-em-2}) as
\ba
\label{com-rel-em-1a}
\big[\dot{A}_\mu(t, {\bf x}), A_\nu (t,{\bf y})  \big] 
&=& 
i g_{\mu\nu} \delta^{(3)}({\bf x}-{\bf y}),
\\ [2mm]
\label{com-rel-em-2a}
\big[A_\mu(t, {\bf x}), A_\nu (t,{\bf y})  \big] 
&=& 
\big[\dot{A}_\mu(t, {\bf x}), \dot{A}_\nu (t,{\bf y})  \big]=0.
\ea

After a short explanation how to quantize the electromagnetic field with the canonical commutation relations, we will derive the equation of motion of the Green functions of the free electromagnetic field. For this purpose we compute the action of the d'Alembert operator on the function $D^c_{\mu\nu}$
\ba
\square_x D^c_{\mu\nu}(x,y),
\ea
where the Green function $D^c_{\mu\nu}(x,y)$ is given by
\ba
\label{eom-def-A}
iD^c_{\mu\nu}(x,y) = \langle \theta(x_0-y_0) A_\mu(x) A_\nu(y) +\theta(y_0-x_0) A_\nu(y) A_\mu(x) \rangle.
\ea
To find the equation of motion of the Green function we follow the procedure already discussed for the scalar field. Throughout the computations we use the equation of motion of the field $A^\mu$ given by (\ref{eom-A-A}). Besides that the procedure is completely analogous to that presented for the scalar theory. Accordingly, we reach the formula
\ba
\label{eom-prop-A}
\square_x D^c_{\mu\nu}(x,y) = -i\langle \delta(x_0-y_0)  
\big(\dot{A}_\mu(x) A_\nu(y) - A_\nu(y) \dot{A}_\mu(x)\big) \rangle.
\ea
Then, we use the commutation relation (\ref{com-rel-em-1a}) justified before. Since the delta function in (\ref{eom-prop-A}) makes the operators $\dot{A}^\mu$ and $A^\nu$ act at the same time we can write Eq. (\ref{eom-prop-A}) as follows
\ba
\label{eom-prop-A-1}
\square_x D^c_{\mu\nu}(x-y) = g_{\mu\nu} \delta^{(4)}(x-y),
\ea
which is the equation of motion of the Green function of free electromagnetic field  in the Feynman gauge and it is of the same structure as Eq. (\ref{sol-D}). The equation of motion of the Green function can be generalised to an arbitrary gauge by recalling the $\alpha$ parameter. It can be simply done by replacement of the d'Alembert operator by the operator dictated by the equation of motion of the vector field (\ref{eom-em-gh}), that is
\ba
\label{dalembert-alpha}
\square \rightarrow \square g^{\mu\nu} -\bigg(1-\frac{1}{\alpha}\bigg) \partial^\mu \partial^\nu.
\ea
Then the equation of motion of the free Green function in a general covariant gauge gets the form
\ba
\label{eom-prop-A-1-arb}
\bigg[\square_x g^{\mu\nu} -\bigg(1-\frac{1}{\alpha}\bigg) \partial_{x}^\mu \partial_{x}^\nu \bigg] D^>_{\nu\rho}(x-y) 
= g^\mu_{\;\rho} \delta^{(4)}(x-y).
\ea

Repeating the same steps for the remaining Green functions of real-time arguments, we get the equation of motion of the contour Green function in the Feynman gauge
\ba
\label{eom-prop-A-2}
\square_x D_{\mu\nu}(x,y) = g^{\mu\nu} \delta_C^{(4)}(x,y),
\ea
where the contour delta function, which has been introduced in the context of the scalar field, is defined by the formula (\ref{delta-contour}). The equation of motion in an arbitrary covariant gauge is given by
\ba
\label{eom-prop-A-2-arb}
\bigg[\square_x g^{\mu\nu} -\bigg(1-\frac{1}{\alpha}\bigg) \partial_{x}^\mu \partial_{x}^\nu \bigg] D_{\nu\rho}(x,y) = g^\mu_{\;\rho} \delta_C^{(4)}(x,y).
\ea

\subsubsection{Green functions as solutions of equation of motion in arbitrary gauge}
\label{sssec-eom-ym}

Here we discuss some difficulties which arise when one would like to derive the Green function in an arbitrary covariant gauge. To sketch the problem we start this part with derivation of the vacuum Green functions where translational invariance is held. The equation of motion of the Green function in the phase space can be written as
\ba
\label{eom-prop-A-1-arb}
M^{\mu\nu} D_{\nu\rho}(k) = g^\mu_{\;\rho}
\ea
where the operator $M$ reads
\ba
\label{op-M}
M^{\mu\nu}=-k^2 g^{\mu\nu} +\bigg(1-\frac{1}{\alpha}\bigg) k^\mu k^\nu.
\ea
Eq. (\ref{eom-prop-A-1-arb}) can be written symbolically as
\be
\label{symb-eq}
M \cdot D = 1,
\ee
which means that the Green function is given as
\ba
\label{symb-eq1}
D = M^{-1}.
\ea
Therefore, we intend to find the Green function by inverting the operator $M$. To do so let us introduce the projection operators $T$ and $L$ which are defined by
\ba
\label{proj-oper-1}
T_{\nu\rho} &=& g_{\nu\rho} - \frac{k_\nu k_\rho}{k^2},
\\ [2mm]
\label{proj-oper-2}
L_{\nu\rho} &=& \frac{k_\nu k_\rho}{k^2},
\ea
and satisfy the following relations
\ba
\label{prop-proj-1}
L \cdot L &=& L,
\\ [2mm]
T \cdot T &=& T,
\\ [2mm]
\label{prop-proj-3}
L \cdot T &=& T \cdot L = 0.
\ea
Then, the operator $M$ can be written as
\ba
\label{M-proj-op}
M^{\mu\nu}=-k^2 (T^{\mu\nu}+ L^{\mu\nu}) + k^2\Big(1-\frac{1}{\alpha}\Big) L^{\mu\nu}
=k^2 T^{\mu\nu} + \frac{k^2}{\alpha} L^{\mu\nu}
\ea
and its inverse $M^{-1}$ as
\ba
\label{matrix-M}
M_{\mu\nu}^{-1}=a_T T_{\mu\nu} + a_L L_{\mu\nu}.
\ea
Keeping in mind that $MM^{-1}=1$ and $ L + T = 1$, the coefficients $a_T$ and $a_L$ are found as
\ba
a_T =- \frac{1}{k^2}, \qquad\qquad a_L = \frac{\alpha}{k^2},
\ea
and consequently
\ba
\label{inverse-matrix-M}
M_{\mu\nu}^{-1}=-\frac{1}{k^2} T_{\mu\nu} + \frac{\alpha}{k^2} L_{\mu\nu}.
\ea
Therefore, the Green function is of the form
\ba
\label{eom-prop-A-1-arb-sol}
D_{\mu\nu}(k) = -\frac{g_{\mu\nu}-(1-\alpha)\frac{k_\mu k_\nu}{k^2} }{k^2},
\ea
where the operators $ T$ and $ L$ given by (\ref{proj-oper-1}) and (\ref{proj-oper-2}) have been substituted. The form (\ref{eom-prop-A-1-arb-sol}) holds for $k^2 \neq 0$. When $k^2=0$, the function has to be redefined by modifying the denominator of (\ref{eom-prop-A-1-arb-sol}). The modification of the $k^2$ pole in the term $k_\mu k_\nu/k^2$ can be arbitrary, see \cite{Pokorski}. Thus, we obtain
\ba
\label{eom-prop-A-1-arb-sol-Fey}
D^c_{\mu\nu}(k) &=& -\frac{g_{\mu\nu}-(1-\alpha)\frac{k_\mu k_\nu}{k^2} }{k^2+i0^+},
\\ [2mm]
\label{eom-prop-A-1-arb--sol-antiFey}
D^a_{\mu\nu}(k) &=& \frac{g_{\mu\nu}-(1-\alpha)\frac{k_\mu k_\nu}{k^2} }{k^2-i0^+}
\ea
which obey the Feynman and anti-Feynman initial conditions, respectively. We can also produce the retarded and advanced Green functions, which are
\ba
\label{eom-prop-A-1-arb-sol-Fey}
D^+_{\mu\nu}(k) &=& -\frac{g_{\mu\nu}-(1-\alpha)\frac{k_\mu k_\nu}{k^2} }{k^2+i\textrm{sgn}(k_0)0^+},
\\ [2mm]
\label{eom-prop-A-1-arb--sol-antiFey}
D^-_{\mu\nu}(k) &=& -\frac{g_{\mu\nu}-(1-\alpha)\frac{k_\mu k_\nu}{k^2} }{k^2-i\textrm{sgn}(k_0)0^+}.
\ea
These computations show that in the vacuum an entire class of gauge choices is possible.

When we deal with a medium which, for simplicity, is translationally invariant, we need to start with Eq. (\ref{eom-prop-A-2-arb}). Let us consider the unordered Green function, which reads
\ba
\label{eom-prop-A-2-arb-KS-big-x}
\bigg[\square_x g^{\mu\nu} -\bigg(1-\frac{1}{\alpha}\bigg) \partial_{x}^\mu \partial_{x}^\nu \bigg] D^>_{\nu\rho}(x-y) 
= 0
\ea
and in the phase space it is found as
\ba
\label{eom-prop-A-2-arb-KS-big}
M^{\mu\nu} D^>_{\nu\rho}(k) = 0
\ea
with $M$ defined by (\ref{op-M}). Eq. (\ref{eom-prop-A-2-arb-KS-big}) is homogeneous so it has a solution if the determinant of the matrix $M$ vanishes
\ba
\label{eq-homo-FG}
\textrm{det}\; M^{\mu\nu}=0.
\ea
However, instead of computing the determinant we invert the matrix $M$, which is given by the formula (\ref{inverse-matrix-M}) and reads
\ba
M_{\mu\nu}^{-1}=-\frac{1}{k^2} T_{\mu\nu} + \frac{\alpha}{k^2} L_{\mu\nu},
\ea
where the operators $T$ and $L$ are given by (\ref{proj-oper-1}) and (\ref{proj-oper-2}). Since the zeros of ${\rm det}M$ correspond to the poles of $M^{-1}$, the solutions to Eq. (\ref{eom-prop-A-2-arb-KS-big}) exist under the condition that $k^2=0$. Thus, the Green function has to be of the form
\ba
\label{D>k20}
D^>_{\nu\rho}(k) = \delta(k^2) f_{\nu\rho}(k),
\ea
where $f_{\nu\rho}(k)$ is an arbitrary function which should be found. Substituting  the formula (\ref{D>k20}) into Eq.~(\ref{eom-prop-A-2-arb-KS-big}) we have
\ba
\label{eom-prop-A-2-arb-KS-big-22}
\bigg[k^2 g^{\mu\nu} -\bigg(1-\frac{1}{\alpha}\bigg) k^\mu k^\nu \bigg] \delta(k^2) f_{\nu\rho}(k) = 0.
\ea
Since $k^2 \delta(k^2) = 0$ and the first term drops, one finds
\ba
\label{eom-prop-A-2-arb-KS-big-23}
\delta(k^2) \bigg(1-\frac{1}{\alpha}\bigg) k^\mu k^\nu  f_{\nu\rho}(k) &=& 0.
\ea
Eq. (\ref{eom-prop-A-2-arb-KS-big-23}) is satisfied when $f_{\nu\rho}(k)$ is orthogonal to $k^\nu$ which demands
\ba
\label{f-ill-def}
f_{\nu\rho}(k)  \sim g_{\nu\rho} - \frac{k_\nu k_\rho}{k^2} .
\ea
This is, however, not possible because $k^2=0$ and the second term in the expression (\ref{f-ill-def}) would be ill defined.
To avoid the problem and fulfill Eq.~(\ref{eom-prop-A-2-arb-KS-big-23}), one is forced to choose $\alpha=1$, which corresponds to the Feynman gauge. We do so and further considerations are held in that gauge. Consequently, in the next subsection we present the comprehensive description of all Green functions of the electromagnetic field in the Feynman gauge when the system is, in general, not homogeneous.

\subsubsection{Green functions as solutions of equation of motion in the Feynman gauge}
\label{sssec-gf-fg}

Solving the equation of motion of the contour Green function in the Feynman gauge (\ref{eom-prop-A-2}) we intend to find $D_{\mu\nu}^>$, $D_{\mu\nu}^<$, $D_{\mu\nu}^c$, $D_{\mu\nu}^a$, $D_{\mu\nu}^+$, and $D_{\mu\nu}^-$ of the system which is not homogeneous. Since the derivation of the Green functions is analogous to that presented for the scalar field in Sec.~\ref{sssec-GF-scalars} we omit some steps which have been already discussed.

\begin{center}
{\bf Derivation of $D^\gl$}
\end{center}

In case of $D_{\mu\nu}^>$, the equations of motion read
\ba
\label{eom-FGG-x-em}
\big(\square_x +m^2\big)D_{\mu\nu}^>(x,y) &=& 0,
\\ [2mm]
\label{eom-FGG-y-em}
\big(\square_y +m^2\big)D_{\mu\nu}^>(x,y) &=& 0,
\ea
where we have introduced the artificial mass $m$ which is used later on as a scale parameter to apply the quasi-particle approximation. The mass can be identified with a plasmon mass but finally it is actually sent to zero. Next we go to the variables $X$ and $u$ to express Eqs. (\ref{eom-FGG-x-em}) and (\ref{eom-FGG-y-em}) in the following forms
\ba
\label{eom-FGG-u-em}
\bigg(\frac{1}{4}\partial^2_X+\partial_X \partial_u + \partial^2_u +m^2\bigg)D_{\mu\nu}^>(X,u) &=& 0,
\\ [2mm]
\label{eom-FGG-X-em}
\bigg(\frac{1}{4}\partial^2_X-\partial_X \partial_u + \partial^2_u +m^2\bigg)D_{\mu\nu}^>(X,u) &=& 0.
\ea
If we subtract the equation (\ref{eom-FGG-u-em}) from (\ref{eom-FGG-X-em}) we obtain 
\ba
\label{eom-FGG-12-em}
\partial_X \partial_u  D_{\mu\nu}^>(X,u) = 0,
\ea
which, after the Wigner transformation (\ref{Wigner-transform-inv}), is
\ba
\label{eom-FGG-kin-eq-em}
p_\lambda \partial^\lambda_X D_{\mu\nu}^>(X,p) = 0.
\ea
It is recognised as the relativistic kinetic equation. After adding the equations (\ref{eom-FGG-u-em}) and (\ref{eom-FGG-X-em}) we are led to 
\ba
\label{eom-FGG-1-em}
\bigg(\frac{1}{4}\partial^2_X+ \partial^2_u + m^2\bigg)D_{\mu\nu}^>(X,u) = 0.
\ea
which, after the Wigner transformation, reads
\ba
\label{eom-FGG-3-em}
\bigg(\frac{1}{4}\partial^2_X-p^2 + m^2\bigg) D_{\mu\nu}^> (X,p) = 0.
\ea
As argued in case of the scalar field, the Green function $D_{\mu\nu}^>(X,p)$ can be nonzero for the off-shell momenta, that is when $p^2 \neq m^2$. Nevertheless, the kinetic theory deals with the system characteristics averaged over scales larger than the particle Compton wavelength of the order of $m^{-1}$. So we impose the quasi-particle approximation
\ba
\label{quasi-particle-ap}
\bigg|\frac{1}{m^2} \partial_X^2 D_{\mu\nu}^>(X,p)\bigg|\ll |D_{\mu\nu}^>(X,p)|.
\ea
Then, we neglect the first term of Eq. (\ref{eom-FGG-3-em}) and get
\ba
\label{eom-FGG-4-em}
\big(p^2 - m^2\big) D_{\mu\nu}^> (X,p) = 0,
\ea
the solution to which is of the form
\ba
\label{eom-FGG-5-em}
D_{\mu\nu}^> (X,p) \propto  g_{\mu\nu} \delta(p^2-m^2),
\ea
where the metric tensor reflects the Lorentz structure in the Feynman gauge. Since the procedure of derivation of $D_{\mu\nu}^<(X,p)$ is the same, the approximate form of $D_{\mu\nu}^<(X,p)$ is given by
\ba
\label{eom-FGL-1-em}
D_{\mu\nu}^< (X,p) \propto  g_{\mu\nu} \delta(p^2-m^2).
\ea
Subsequent steps of derivation are the same as these of the scalar field so by repetition of the procedure beginning from the formula (\ref{eom-FGG-deriv-1}) we get final explicit forms of $D_{\mu\nu}^>(X,p)$ and $D_{\mu\nu}^<(X,p)$ 
\ba
\label{D->}
D_{\mu\nu}^>(X,p) 
&=& 
\frac{i\pi}{E_p} g_{\mu\nu} \Big[\delta(E_p-p_0)\big(f_\gamma(X,{\bf p})+1\big) 
+ \delta(E_p+p_0)f_\gamma(X,-{\bf p})\Big], 
\\ [2mm]
\label{D-<}
D_{\mu\nu}^<(X,p) 
&=& 
\frac{i\pi}{E_p} g_{\mu\nu} \Big[\delta(E_p-p_0)f_\gamma(X,{\bf p}) 
+ \delta(E_p+p_0)\big(f_\gamma(X,-{\bf p})+1\big)\Big],
\ea
where $f_\gamma(X,{\bf p})$ is the distribution function of photons. As one can recognize, the formulas (\ref{D->}) and (\ref{D-<}) can be expressed by $\Delta^>(X,p)$ and $\Delta^<(X,p)$, given by (\ref{Del->}) and (\ref{Del-<}), respectively. Then, they read
\ba
\label{D->-Del}
D_{\mu\nu}^>(X,p) &=& -g_{\mu\nu} \Delta^>(X,p), 
\\ [2mm]
\label{D-<-Del}
D_{\mu\nu}^<(X,p) &=& - g_{\mu\nu}\Delta^<(X,p).
\ea
Due to the connection between $D_{\mu\nu}^\lg(X,p)$ and $\Delta^\lg(X,p)$,  the consecutive steps of derivation of other Green functions are completely the same as these performed for the scalar field. A physical meaning of $D_{\mu\nu}^>(X,p)$ and $D_{\mu\nu}^<(X,p)$ is also analogous.

\begin{center}
{\bf Derivation of $D^{c,a}$}
\end{center}

To find $D_{\mu\nu}^c(X,p)$ we recall the relation 
\ba
\label{eom-FGC-em}
D_{\mu\nu}^c(x,y)=\theta(x_0-y_0)D_{\mu\nu}^>(x,y)+\theta(y_0-x_0)D_{\mu\nu}^<(x,y),
\ea
which due to (\ref{D->-Del}) and (\ref{D-<-Del}) can be expressed as
\ba
\label{eom-FGC-2-em}
D_{\mu\nu}^c(x,y)=-g_{\mu\nu} \Big(\theta(x_0-y_0)\Delta^>(x,y)+\theta(y_0-x_0)\Delta^<(x,y)\Big),
\ea
and next as
\ba
\label{eom-FGC-2-em-a}
D_{\mu\nu}^c(x,y)=-g_{\mu\nu} \Delta^c(x,y).
\ea
The relation (\ref{eom-FGC-2-em-a}) in the momentum space is
\ba
\label{D-c-Del}
D_{\mu\nu}^c(X,p)=-g_{\mu\nu} \Delta^c(X,p).
\ea

Since $\Delta^c(X,p)$ was derived and it is given by (\ref{Del-c}), the relation (\ref{D-c-Del}) gets the final form
\ba
\label{D-c}
D_{\mu\nu}^c(X,p)=-g_{\mu\nu}\bigg[ \frac{1}{p^2+i0^+} 
-\frac{i\pi}{E_p}\Big(\delta(p_0-E_p) f_\gamma(X,{\bf p}) + \delta(p_0+E_p) f_\gamma(X,-{\bf p}) \Big)\bigg]. \quad
\ea
In the formula (\ref{D-c}) we have neglected the mass term that, in case of equilibrium, is of the order of $gT$ which fixes the so-called {\it soft scale}. It is justified as long as we work in the limit of $p\gg m \sim gT$. 

The same steps can be repeated for $D_{\mu\nu}^a(X,p)$ which is 
\ba
\label{D-a-Del}
D_{\mu\nu}^a(X,p)=-g_{\mu\nu} \Delta^a(X,p)
\ea
and is found as
\ba
\label{D-a}
D_{\mu\nu}^a(X,p)= g_{\mu\nu}\bigg[ \frac{1}{p^2-i0^+} 
+\frac{i\pi}{E_p}\Big(\delta(p_0-E_p) f_\gamma(X,{\bf p}) + \delta(p_0+E_p) f_\gamma(X,-{\bf p}) \Big)\bigg].
\ea

\begin{center}
{\bf Derivation of $D^{\pm}$ and $D^{\rm sym}$}
\end{center}

In order to compute the retarded Green function of the electromagnetic field we use the identity
\ba
\label{eom-FGP-em}
D_{\mu\nu}^+(X,p)=D_{\mu\nu}^c(X,p)-D_{\mu\nu}^<(X,p),
\ea
which can also be presented as
\ba
\label{D-ret-Del}
D_{\mu\nu}^+(X,p)=-g_{\mu\nu} \big(\Delta^c(X,p)-\Delta^<(X,p)\big) = -g_{\mu\nu} \Delta^+(X,p).
\ea
The final formula of the retarded Green function can be immediately written down after inserting the scalar retarded Green function, given by (\ref{eom-FGP-3}), and omitting the mass. Then, we get
\ba
\label{D-ret}
D_{\mu\nu}^+(X,p)=- \frac{g_{\mu\nu}}{p^2+i\textrm{sgn}(p_0)0^+}.
\ea

The advanced Green function of the electromagnetic field is related to that of the scalar field through
\ba
\label{D-adv-Del}
D_{\mu\nu}^-(X,p) = -g_{\mu\nu} \Delta^-(X,p)
\ea
and is then of the form
\ba
\label{D-adv}
D_{\mu\nu}^-(X,p)=- \frac{g_{\mu\nu}}{p^2-i\textrm{sgn}(p_0)0^+}.
\ea

Finally, the symmetric Green function is found as
\ba
\label{D-sym-Del}
D_{\mu\nu}^{\rm sym}(X,p) = -g_{\mu\nu} \Delta^{\rm sym}(X,p)
\ea
and it is given explicitly as
\ba
\label{D-sym}
D_{\mu\nu}^{\rm sym}(X,p) 
= g_{\mu\nu} \frac{i\pi}{E_p}\Big[ \delta(E_p-p_0)\big(2f_\gamma(X,{\bf p})+1 \big) 
+\delta(E_p+p_0)\big(2 f_\gamma(X,-{\bf p})+1 \big) \Big]. \quad
\ea

All the functions computed here, $D^>_{\mu\nu}$, $ D^<_{\mu\nu}$, $D^c_{\mu\nu}$, $D^a_{\mu\nu}$, $D^+_{\mu\nu}$, $D^-_{\mu\nu}$, and $D^{\rm sym}_{\mu\nu}$  given by formulas (\ref{D->}), (\ref{D-<}), (\ref{D-c}), (\ref{D-a}), (\ref{D-ret}), (\ref{D-adv}), and (\ref{D-sym}), respectively, are exactly the same for the free gluon field yet the colour indices have to be taken into account. Therefore, the Kronecker delta $\delta^{ab}$ with the colour indices $a,b=1,2,...N_c^2-1$ must be included in all these formulas and the respective gluon distribution function $f_g$ instead of $f_\gamma$.

\subsection{Real-time argument Green functions of fermion field}
\label{ssec-gf-fermion}

Performing the analysis analogous to that of the scalar and electromagnetic field, we write down the equation of motion of the contour Green function of Dirac fermion field 
\be
\label{eom-fermion}
\Big(\big[ i\gamma^\mu \partial_\mu - m \big]S(x,y)\Big)_{\alpha\beta} = \delta_{\alpha\beta}\delta^{(4)}_C(x,y),
\ee
where $m$ is a mass and the Dirac delta function is defined by (\ref{delta-contour}). To solve it one needs to remember that the fermion field operators are anticommuting and they comply with the Grassmann algebra. Keeping it in mind we find the unordered Green functions 
\ba
\label{S->}
S^>_{\alpha\beta}(p) 
&=& 
\delta_{\alpha\beta}\frac{i\pi}{E_p} p\sla \Big[ \delta (E_p - p_0)  \big( f_f(X,{\bf p}) -1\big)
+ \delta (E_p + p_0) \bar f_f(X,-{\bf p}) \Big], 
\\ [2mm]
\label{S-<}
S^<_{\alpha\beta}(p) 
&=& 
\delta_{\alpha\beta}\frac{i\pi}{E_p} p\sla \Big[ \delta (E_p - p_0)  f_f(X,{\bf p})
+ \delta (E_p + p_0) \big(\bar f_f(X,-{\bf p})-1\big) \Big],
\ea
where $p\sla \equiv p^\mu \gamma_\mu$ and $f_f(X,{\bf p})$ and $\bar f_f(X,-{\bf p})$ are the distribution functions of fermions and of antifermions, respectively.  We assume here that both fermions and antifermions are unpolarized, that is, all spin states are equally probable. The distribution functions are normalized in such a way that the fermion density equals
\be
n_f = 2 \int \frac{d^3p}{(2\pi)^3}\, f_f({\bf p}) ,
\ee
where the factor of 2 takes into account two spin states of each fermion. The chronologically and antichronologically ordered Green functions are found as
\ba
\label{S-c}
S^c_{\alpha\beta}(p) 
&=& 
\frac{\delta_{\alpha\beta} p\sla}{p^2+ i 0^+} 
- \delta_{\alpha\beta}\frac{i\pi}{E_p} \Big[ \delta (E_p - p_0) f_f(X,{\bf p})
+ \delta (E_p + p_0) \bar f_f(X,-{\bf p}) \Big],\qquad 
\\ [2mm]
\label{S-a}
S^a_{\alpha\beta}(p) 
&=& 
- \frac{\delta_{\alpha\beta} p\sla}{p^2 - i 0^+} 
- \delta_{\alpha\beta}\frac{i\pi}{E_p} \Big[ \delta (E_p - p_0) f_f(X,{\bf p})
+ \delta (E_p + p_0) \bar f_f(X,-{\bf p}) \Big].
\ea
The retarded and advanced functions are in turn given by
\ba
\label{S-ret}
S^{+}_{\alpha\beta}(p) &=&  \frac{\delta_{\alpha\beta} p\sla}{p^2 + i\, {\rm sgn}(p_0)0^+}, 
\\ [2mm]
\label{S-adv}
S^{-}_{\alpha\beta}(p) &=&  \frac{\delta_{\alpha\beta} p\sla}{p^2 - i\, {\rm sgn}(p_0)0^+},
\ea
and the symmetric one 
\ba
\label{S-sym}
S^{\rm sym}_{\alpha\beta}(p)
=\delta_{\alpha\beta} \frac{i\pi}{E_p} p\sla \Big[ \delta(E_p-p_0)\big(2f_f(X,{\bf p})-1 \big) 
+\delta(E_p+p_0)\big(2\bar f_f(X,-{\bf p})-1 \big) \Big].
\ea
If a quark field is considered then the Kronecker delta $\delta^{ij}$, with the colour indices $i,j=1,2,..., N_c$, ought to be included in the formulas (\ref{eom-fermion})-(\ref{S-sym}).

\newpage
\thispagestyle{plain}

\section{Ghosts in Keldysh-Schwinger formalism}
\label{sec-ghosts-KS}

\vspace{-1cm}

\begin{tabular}{ m{5cm} l}
& \emph{What are these,}
\\
& \emph{So wither'd, and so wild in their attire;} 
\\
& \emph{That look not like the inhabitants o' th' earth,} 
\\
& \emph{And yet are on 't?} 
\end{tabular}
\vspace{.3cm}

\begin{tabular}{ m{5cm} r}
&  Macbeth, William Shakespeare
\end{tabular}

\vspace{2cm}

In the previous section we have shown how to obtain the Green functions of the Keldysh-Schwinger formalism of free fields that carry information about real objects of nature. We have derived thereby the Green functions of the scalar, gauge and fermion fields. Implementing some simple modifications, already discussed above, we are able to reconstruct the functions of fields carrying a colour charge as well. However, working with gauge theories one may also encounter the need for including artificial Faddeev-Popov ghost fields into calculations. 

In QED and its generalizations obeying a gauge symmetry there are four components of the vector field whereas a photon or other gauge boson physically exists only in two polarisation states. Furthermore, quantisation of the theory within the path integral formulation causes more serious difficulty since then an infinite number of equivalent configurations of a field is generated by a gauge symmetry of the theory. This excess of unphysical degrees of freedom may be reduced by properly chosen gauge condition. In a class of physical gauges the redundant degrees of freedom are eliminated completely but then the Lorentz invariance is lost and computations usually get complicated. To have manifestly Lorentz invariant formulation of a theory one needs to impose one of the covariant gauge fixing conditions. Such a gauge condition is sufficient within functional approach to obtain consistent QED with the proper number of degrees of freedom. The quantisation of a nonAbelian gauge theory in the Lorentz covariant way, however, demands an introduction of auxiliary fictitious fields which are called the Faddeev-Popov ghosts. Due to the coupling of the ghosts and gauge bosons a dynamics of the system becomes sensitive to ghosts. It is item possible to allow for the ghosts in U(1) gauge theory like QED or SU($N_c$) gauge theory with a physical gauge chosen yet then they are decoupled from the gauge boson field and do not influence a behaviour of the system.

It was Feynman who first noticed in 1963 the lack of unitarity of the Yang-Mills theory when unphysical degrees of freedom are not treated properly. In the work \cite{Feynman:1963ax} he attempted to use the commonly known form of the photon propagator to perform one-loop calculation of the gluon field. It turned out that one contribution from a scalar loop with the minus sign is missing in the computations to get a unitary scattering matrix. The problem was solved in 1967 almost simultaneously by DeWitt \cite{DeWitt:1967ub} and Faddeev and Popov \cite{Faddeev:1967fc}. That said, the approach of Faddeev and Popov was clearer and more intuitive and thus it has got widespread. A natural framework to formulate the procedure of quantization of gauge theories is provided by a path integral approach \cite{Feynman:1948ur,Feynman:1949zx,Feynman:1950ir} where the ghosts appear as a tricky representation of a Jacobian of the gauge transformation. Then, the generating functional of Green functions, which is found explicitly, determines the propagator of free ghost field. This is almost everything we need to include the ghosts in perturbative diagrammatic calculations, see {\it e.g.} \cite{Pokorski}. Further on in this section, we present the procedure in some detail.

In spite of the ghosts being well understood in the vacuum field theories, there is a question of how the ghosts should be included in statistical theories which are multifariously formulated. In the  Matsubara or imaginary time formalism, which applies to equilibrium systems, the ghosts are needed even in an Abelian theory \cite{Kapusta-Gale}. However, such non-interacting ghosts serve only to cancel unphysical degrees of freedom in the ideal gas contribution. In nonAbelian theories the ghosts are also included in the Feynman rules but the ghost propagator is obtained automatically when the explicit form of generating functional is computed \cite{Kapusta-Gale,LeBellac}. One should only remember that the fermionic ghost fields obey the bosonic periodic boundary conditions, as argued in \cite{Bernard:1974bq}, see also \cite{Hata:1980yr}. When a real time contour is included in the Matsubara approach one deals with the real time formalism of equilibrium systems which allows one to study time-dependent phenomena. The physical and unphysical degrees of freedom of gauge fields are usually treated on the same footing \cite{Landsman:1986uw,LeBellac}. The Faddeev-Popov ghosts are thermalised with the bosonic distribution function. Within the alternative `frozen ghosts' approach the ghosts are kept at zero temperature that is their free Green functions have no thermal contribution \cite{Landshoff:1992ne,Landshoff:1993ag}.

The problem of ghosts has been least understood in the Keldysh-Schwinger formalism which is applicable for non-equilibrium systems. One could expect to obtain the Green functions of a free ghost field by solving the respective equation of motion as in case of other fields which was presented in the previous section. Then, however, it is unclear what would be a distribution function of these artificial fields. Therefore, a more fundamental method to derive the ghosts in the Keldysh-Schwinger approach, which is the path integral formulation, is required. Yet the main difficulty is that the generating functional cannot be computed in an explicit form even in noninteracting theory because of, in general, an unknown density operator which enters the generating functional. Nevertheless the functional is still useful as it provides various relations among the Green functions. Deriving the Slavnov-Taylor identity which relates the gluon propagator to the ghost one we are able to obtain the Green function of ghosts. The analysis of the problem, which is based on our original results \cite{Czajka:2014eha}, is shown in this chapter, in Secs. \ref{ssec-gen-fun-mb}-\ref{ssec-ghosts}. We start, however, with the idea of ghosts in vacuum field theories.

\subsection{Introduction - ghosts in vacuum field theories}
\label{ssec-intro-ghosts}

In this part we discuss how ghosts emerge in vacuum field theories. For this purpose we formulate the path integral approach to gauge theories. We start with the electromagnetic field to sketch the procedure and to facilitate an introduction of ghosts to nonAbelian theories. Next we go on with the Yang-Mills theory. The generating functional, the Faddeev-Popov determinant and the way of generating Green functions are, in particular, discussed. We follow here the standard books, mostly \cite{Pokorski}.

\subsubsection{Generating functional of the electromagnetic field}
\label{sssec-gen-fun-em}

The starting point of our analysis is a generating functional of the electromagnetic field which by analogy with the scalar field can be written as
\ba
\label{W-electro-0}
W_0[J] = N \int \mathcal{D}A \exp \big( iS[A,J]  \big) 
= N \int \mathcal{D}A \exp \bigg( i\int d^4x \, \mathcal{L}(A,J) \bigg)
\ea
where $\mathcal{D}A \equiv \prod_{n=1}^N \prod_{\mu =0}^3 d A_\mu(x_n)$
The normalisation constant is chosen in such a way that $W_0[J=0]=1$ and the Lagrangian density reads
\ba
\mathcal{L}=-\frac{1}{4}F^{\mu\nu}F_{\mu\nu} + J_\mu A^\mu
\ea
with $J^\mu$ being the classical current and the strength tensor $F^{\mu \nu}$ is defined by (\ref{str-tensor}). Performing the integration by parts, the action is written as
\ba
\label{action-electro}
S[A_\mu,J] \equiv \int d^4x \mathcal{L}(x)
= -\frac{1}{2}\int d^4x \, A^\mu(x)K_{\mu\nu}A^\nu(x)-\int d^4x J^\mu(x)A_\mu(x),
\ea
where 
\ba
\label{K-operator}
K_{\mu\nu} \equiv -\partial^2g_{\mu\nu}+\partial_\mu \partial_\nu ,
\ea 
and the formula (\ref{W-electro-0}) gets the form
\ba
\label{W-electro-1}
W_0[J] = N \int \mathcal{D}A \exp \bigg(-\frac{i}{2}\int d^4x A^\mu K_{\mu\nu}A^\nu 
-i\int d^4x \, J_\mu A^\mu  \bigg) .
\ea
Since the integral in the functional (\ref{W-electro-1}) is of the Gaussian type, one may hope to compute it in the standard way as
\ba
\label{W-electro-2}
W_0[J] = N \exp \bigg(-\frac{i}{2}\int d^4x d^4y J^\mu(x) D_{\mu\nu}(x-y)J^\nu (y) \bigg),
\ea
where $D_{\mu\nu}(x)$ should be inverse to the operator $K_{\mu\nu}$, that is
\ba
\label{KG}
K_{\mu\nu}D^{\nu\lambda}(x) =  g_\mu^{\;\; \lambda} \delta^4(x).
\ea
However, the equation (\ref{KG}) does not have a solution since the operator $K$ defined by Eq.~(\ref{K-operator}) is proportional to a projection operator $P$. In the momentum space it reads
\ba
K_{\mu\nu}=p^2g_{\mu\nu}-p_\mu p_\nu = p^2 P_{\mu\nu} ,
\ea
with  
\ba
\label{pp}
P_{\mu\nu} \equiv g_{\mu\nu} - \frac{p_\mu p_\nu}{p^2} ,
\ea
which projects any tensor on the direction parallel to $p$.

The fact that the operator $K_{\mu\nu}$ defined by Eq.~(\ref{K-operator}) cannot be inverted has further consequences. Applying the local gauge transformation
\ba
\label{gauge-trans}
A^\mu(x) \rightarrow A^\mu(x) + \partial^\mu \Lambda (x) ,
\ea
the action (\ref{action-electro}) remains unchanged (provided $\partial_\mu J^\mu = 0$). This means that the field configurations that differ by the gauge transformations (\ref{gauge-trans}) have the same (infinite) weight which implies multiple counting of the physically equivalent configurations of $A_\mu$ in the integral (\ref{W-electro-1}). The functional (\ref{W-electro-1}) is consequently badly divergent which is particularly evident in the Euclidean formulation when the functional integral is truly Gaussian. 

The problem is solved by imposing a gauge condition on the field $A^\mu$ of the general form
\ba
\label{gauge-cond-em}
f[A] = 0,
\ea
where $f[A]$ is a local differentiable function of the gauge field or its derivatives. The potential $A$ is also assumed to be unique that is there is no gauge transformation which changes the potential and which leaves the condition unchanged. Then, the generating functional is written as 
\ba
\label{W-electro-gauge}
W_0[J] = N \int \mathcal{D}A \, \delta\big[ f[A] \big] \, \exp \big( iS[A,J]  \big),
\ea
where the delta Dirac functional should be understood as
\ba
\label{delta-function}
\delta \big[ f[A] \big] = \prod_x \delta \Big(f\big(A(x) \big) \Big) .
\ea

In practice one often uses the gauge condition (\ref{gauge-cond-em}) which does not fix the gauge completely. As an example let us consider the Lorentz condition
\ba
\label{Lorentz-gauge}
f[A] = \partial_\mu A^\mu (x) = 0.
\ea
One observes that the potential $A^\mu$, which obeys this condition, still satisfies Eq.~(\ref{Lorentz-gauge}) after the transformation (\ref{gauge-trans}) if $\Lambda$ solves the equation $\partial^2 \Lambda = 0$. So, the  Lorentz condition does not fix the gauge completely and the field configurations that differ by the gauge transformations (\ref{gauge-trans}) with $\partial^2 \Lambda = 0$ contribute repeatedly to the functional (\ref{W-electro-gauge}). Therefore, the contribution of a given field $A_\mu $ to the integrand in (\ref{W-electro-gauge}) must be divided by the volume of all configurations that obey the gauge condition (\ref{gauge-cond-em}) but differ by the gauge transformations (\ref{gauge-trans}).  The volume is typically denoted as $\Delta^{-1}[A_\mu]$ and the generating functional can be written as
\ba
\label{W-electro-gauge-1-1}
W_0[J] = N \int \mathcal{D}A \; \Delta_f[A] \; \delta \big[f[A] \big]\; \exp \big( iS[A,J]  \big),
\ea
where the index $f$ in $\Delta_f[A]$ states a link to the gauge condition (\ref{gauge-cond-em}). However, we still have to calculate the quantity $\Delta_f[A]$ which is known as the Faddeev-Popov determinant.

\begin{center}
{\bf The Faddeev-Popov determinant}
\end{center}

The volume of  the field configuration of interest equals a sum over all possible gauge transformations which preserve the gauge condition (\ref{gauge-cond-em}), that is
\ba
\label{F-P-det}
\Delta^{-1}_f[A] = \int \mathcal{D}\Lambda \; \delta \big[ f[A^{\Lambda}] \big],
\ea
where $A^{\Lambda}_\mu \equiv A_\mu + \partial_\mu \Lambda$. We now compute $\Delta^{-1}_f[A]$ treating $f[A^{\Lambda}]$ as a function of $\Lambda$ for $A_\mu$ fixed. Changing the variables from  $\Lambda$ to $f$ in the standard way, we get  the  Faddeev-Popov determinant
\ba
\label{F-P-det-11}
\Delta^{-1}_f[A] 
= \int \mathcal{D}\Lambda \; \delta \big[ f[A^{\Lambda}] \big]
= \int \mathcal{D}f \; {\textrm {det}} \left| \frac{\delta \Lambda}{\delta f} \right| \delta [f] 
= {\textrm {det}} \left| \frac{\delta \Lambda}{\delta f} \right|_{f=0}.
\ea
where the determinant is actually the Jacobian of the variable transformation. Let us stress out that the last equality in the formula (\ref{F-P-det-11}) is upheld when the gauge condition (\ref{Lorentz-gauge}) is fulfilled.

When the gauge condition (\ref{gauge-cond-em}) is linear in $A^\mu$, the Faddeev-Popov determinant is obviously independent of $A^\mu$. As an example we again consider the Lorentz condition (\ref{Lorentz-gauge}). Then,
\ba
\frac{\delta f[A^\Lambda(x)] }{\delta \Lambda (y) } 
= \frac{\delta \big( \partial^\mu A_\mu(x) + \partial^2 \Lambda(x) \big) }{\delta \Lambda (y) } 
= \partial^2 \delta^{(4)}(x-y) .
\ea
Keeping in mind that the delta is the unit operator in the functional calculus, the  Faddeev-Popov determinant is finally written as
\ba
\label{F-P-det-QED}
\Delta_f[A] 
= {\textrm {det}} \left| \frac{\delta f }{\delta \Lambda} \right|_{f=0} 
= {\textrm {det}} \left| \partial^2 \right| .
\ea
When the Faddeev-Popov determinant is independent of $A^\mu$, there is no need to include it in the generating functional (\ref{W-electro-gauge-1-1}) as it influences only the normalization constant which is anyway fixed by an additional condition. So, the Faddeev-Popov determinant is actually not needed at all.
\begin{center}
{\bf Final form of the generating functional}
\end{center}

To obtain the final form of the functional one manipulates on the delta function which fixes the gauge. It can be written as the Gaussian integral (in the Euclidean space)
\ba
\label{delta-alpha}
\delta \big[ f[A] \big] 
= \lim_{\alpha \rightarrow 0} \exp \Big( -\frac{i}{2\alpha} \int d^4 x \big( f[A]\big) ^2 \Big)
\ea
where $\alpha$ is a real constant. Then, the generating functional reads
\be
\label{W-electro-gauge-1}
W_0[J] = N \;\int \mathcal{D}A \; \exp \Big( -\frac{1}{2\alpha} \int d^4 x  \, \big( f[A]\big) ^2\Big) 
 \exp \big( iS[A,J]  \big) ,
\ee
where the limit $\alpha \rightarrow 0$ is implicitly assumed. Actually this limit does need to be taken as there is an alternative way to derive the generating functional. Then, one generalises the gauge condition (\ref{gauge-cond-em}) to a whole class of gauge conditions
\ba
\label{gauge-gen}
f[A_\mu]-C(x)=0,
\ea
where $C(x)$ is an arbitrary function. Since $C(x)$ does not depend on the gauge field $A^\mu$, the Faddeev-Popov determinant is still independent of $A^\mu$ and there is no need to include it in the generating functional which now is
\ba
\label{W-electro-gauge-gen}
W_0[J] = N \int \mathcal{D}A \; \delta \big[ f[A_\mu]-C(x) \big]\; \exp \big( iS[A_\mu,J]  \big).
\ea
We modify the expression (\ref{W-electro-gauge-gen}) by integrating it functionally over $C$ with an arbitrary weight functional $G[C]$. Such a trick will change only the constant $N$. Thus, we have
\ba
\label{W-electro-gauge-C}
W_0[J] = N \int \mathcal{D}A\; \int \mathcal{D}C \; \delta\big[ f[A_\mu]-C(x) \big] \; G[C] \;
\exp \big( iS[A_\mu,J]  \big) \;.
\ea
Choosing $G[C]$ as
\ba
\label{G}
G[C]=\exp \left( -\frac{i}{2\alpha} \int d^4 x [C(x)]^2 \right) 
\ea
and taking the trivial integral over $C$ in the formula (\ref{W-electro-gauge-C}) we reproduce the functional (\ref{W-electro-gauge-1}). The advantage of this method of derivation is that the limit $\alpha \rightarrow 0$ is not invoked. 

Both approaches lead us to the following form of the generating functional
\ba
\label{W-electro-gauge-eff}
W_0[J] = N \;\int \mathcal{D}A \; \exp \big( iS_{\rm eff}[A,J,\alpha]  \big) ,
\ea
where the effective Lagrangian density is 
\ba
\label{action-gauge}
\mathcal{L}_{\rm eff}= -\frac{1}{4}F^{\mu\nu}F_{\mu\nu} - \frac{(f[A])^2}{2\alpha} - J_\mu A^\mu , 
\ea
which is no longer gauge-invariant. Although the gauge parameter $\alpha$ is present in the Lagrangian it does not affect physical results. Choosing the Lorentz gauge (\ref{Lorentz-gauge}), the Lagrangian  equals
\ba
\label{lagr-gauge}
\mathcal{L}_{\rm eff} = \frac{1}{2} A_\mu K^{\mu\nu}_\alpha A_\nu - J_\mu A^\mu
\ea
with
\ba
\label{K-alpha-operator}
K^{\mu\nu}_\alpha \equiv g^{\mu\nu} \partial^2 - \frac{\alpha-1}{\alpha}\partial^\mu \partial^\nu ,
\ea 
which is no longer proportional to a projection operator and therefore can be inverted. Since the Lagrangian depends quadratically on fields, the functional integral in Eq.~(\ref{W-electro-gauge-eff}) is of the form of the Gaussian integral and it can be computed explicitly by using the integral formula
\ba
\label{gaussian}
\int d^n x \exp \bigg( -\frac{1}{2} {\bf x} \, O \, {\bf x} - {\bf b} \cdot {\bf x} \bigg)
=\sqrt{ \frac{(2 \pi )^n}{\det O} }\exp \bigg( \frac{1}{2} {\bf b} \, O^{-1} {\bf b} \bigg),
\ea
where $O$ is the invertible $m \times m$ matrix and ${\bf b}$ is the vector of $m$ components. With the substitutions $O \rightarrow K^{\mu\nu}_\alpha$ and ${\bf b} \rightarrow - iJ_\mu(x)$, we have
\be
\label{W-electro-gauge-final}
W_0[J] = \exp \left( -\frac{i}{2} \int d^4 x\, d^4y\; J_\mu (x) D^{\mu\nu}_\alpha(x-y) J_\nu (y)  \right)
\ee
where we have taken into account that the normalization constant, fixed by the condition $W_0[J=0]=1$, equals unity; and $D_\alpha^{\mu\nu}(x)$ is, as we show below, the photon propagator in the general covariant gauge.

\begin{center}
{\bf The photon propagator}
\end{center}
 
The photon propagator which enters the generating functional (\ref{W-electro-gauge-final}) satisfies the equation
\ba
\label{propag-gauge-eq}
K_{\mu\nu}^\alpha(x) D^{\nu \lambda}_\alpha(x) = -g_{\mu}^{\;\;\lambda} \delta^{(4)}(x) ,
\ea
where the Feynman boundary condition is assumed. The solution to Eq.~(\ref{propag-gauge-eq}) is of the form
\ba
\label{propag-gauge-sol}
 D^{\mu\nu}_\alpha(x) = \int \frac{d^4 k}{(2\pi)^4} \exp(-ikx) \,  D^{\mu\nu}_\alpha (k)
\ea
with 
\ba
\label{propag-gauge-k}
 D^{\mu\nu}_\alpha (k) = \frac{g^{\mu\nu} - (1 - \alpha) \frac{k^\mu k^\nu}{k^2}}{k^2 +i0^+} .
\ea
For $\alpha = 0$ we have the fully transverse Lorentz or Landau gauge. The case $\alpha = 1$ corresponds to the Feynman gauge. In the quatum field theory a choice of gauge does not mean constraining the potential $A^\mu$ but rather choosing a form of the longitudinal component of the photon propagator. 

The functional (\ref{W-electro-gauge-final}) is a basic object to generate free Green function of the electromagnetic field. Thus the 2-point Green function is obtained by the following operation
\ba
\label{propagator-gauge}
D_0^{\mu\nu} (x_1,x_2) 
\equiv  (-i)^2 \frac{\delta}{\delta J_\mu(x_1)} \frac{\delta}{\delta J_\nu(x_2)} W_0[J] \bigg|_{J\equiv 0}  
= i D^{\mu\nu}_\alpha (x_1 - x_2) .
\ea
The $n$-point Green function can be produced by acting of the derivative with respect to the source $J_\mu$  $n$ times on the functional (\ref{W-electro-gauge-final}).

\subsubsection{Generating functional of the Yang-Mills field}
\label{sssec-gen-fun-ym}

Here we derive the generating functional of gluodynamics. The Lagrangian density then reads
\ba
\label{lagrangian-YM}
{\cal L} = \mathcal{L}_{\rm YM} + J^a_\mu A^{\mu a},
\ea
where $\mathcal{L}_{\rm YM}$ is given by (\ref{L-YM-fund}). Since the Yang-Mills field shares some similarities with the electromagnetic field, the form of the corresponding generating functional is constructed analogously to that of QED which is already studied. However, we will also face some additional difficulties characteristic of nonAbelian theories. 

If we write the generating functional in the expected form
\ba
\label{W-gluon}
W[J] = N \int \mathcal{D}A \exp \bigg(i \int d^4 x \, \mathcal{L}_{\rm YM} + i\int d^4x \, J^a_\mu A^{\mu a}\bigg),
\ea
where $\mathcal{D}A \equiv \prod_{n=1}^N \prod_{\mu =0}^3 \prod_{a =1}^{N_c^2-1} d A_\mu^a(x_n)$,
we notice that not only physically different but also those configurations of gluon field that differ by gauge transformation 
\ba
\label{gauge-transf-adjoint}
A^\mu_a (x) \rightarrow A^\mu_a (x) + f^{abc} A^\mu_b (x) \omega^c(x)- \frac{1}{g}\partial^\mu \omega^a(x)
\ea
contribute to the integral (\ref{W-gluon}). In order to avoid multiple counting of the physically equivalent states, one divides the configuration space $\{A_\mu(x)\}$ into the equivalence classes $A_\mu^U(x)$ called the orbits of the gauge group. An orbit of the group includes all field configurations $A_\mu^U(x)$ related to $A_\mu (x)$ via all possible gauge transformations $U$ which belong to the gauge group ${\rm SU}(N_c)$. To get a correct expression of the generating functional, one sums not over all possible field configurations but over all possible orbits. Additionally we divide the integrand in (\ref{W-gluon}) by a volume of the corresponding orbit. The inverse orbit volume just equals the Faddeev-Popov determinant which plays here a much more important role than in QED.

After including the changes discussed above, the generating functional can be written as
\ba
\label{W-gluon-0}
W[J] = N \int \mathcal{D}A \; \Delta_f[A]\; \delta\big[f[A] \big] 
\exp \bigg( i \int d^4 x \, \mathcal{L}_{\rm YM} + i\int d^4x \, J^{a \mu} A_{\mu}^a  \bigg)
\ea
with $\Delta[A]$ being the Faddeev-Popov determinant. The gauge condition imposed on the field $A_\mu$ is of the known form
\ba
\label{gauge-cond}
f[A]=0,
\ea
where $f[A]=\prod_x f(A(x))$ and $f(A(x))$ is a local differentiable function of the gauge field or its derivatives. To have only one representative of the orbit, say $A_\mu$, there should be a unique solution $A_\mu^U$ to Eq.~(\ref{gauge-cond}). 

Using the explicit form of the strength tensor (\ref{F-adj}) we can write the Lagrangian density (\ref{lagrangian-YM}) as
\ba
\mathcal{L}_{\rm YM} 
&=& 
-\frac{1}{4}(\partial_\mu A_\nu^a-\partial_\nu A_\mu^a)(\partial^\mu A^{a\nu}
-\partial^\nu A^{a\mu})-gf^{abc} \partial_\mu A_\nu^a A^{b\mu} A^{c\nu}
\\ [2mm] \nn
&& \qquad \qquad \qquad \qquad \qquad \qquad
-\frac{1}{4}g^2 f^{abc} f^{ade} A_\mu^b A_\nu^c A^{d\mu} A^{e\nu}.
\ea
As we see, the Lagrange function is not only quadratic in the vector potential $A_\mu$, but it also consists of terms cubic and quartic in the fields which describe the three- and four-gluon interaction. Thus, the integral (\ref{W-gluon-0}) is not of the Gaussian type as in the case of Abelian field; only the free-field part is Gaussian. 

To find the explicit form of the generating functional of the chromodynamic field we first calculate the Faddeev-Popov determinant $\Delta_f[A_\mu]$.

\begin{center}
{\bf The Faddeev-Popov determinant}
\end{center}

The Faddeev-Popov determinant represents the inverse volume of an orbit of the group and it is defined as a sum over all possible gauge transformations inside one orbit bounded by the gauge condition (\ref{gauge-cond}), that is
\ba
\label{F-P-det}
\Delta^{-1}_f[A]=\int \mathcal{D}U \; \delta \big[f[A^U] \big],
\ea
where we have an integration measure of the group $\mathcal{D}U$ which is itself invariant under a gauge transformation
\ba
\label{measure-inv}
\mathcal{D}U=\mathcal{D}(UU\rq{}).
\ea
When the operator of gauge transformation is of the form (\ref{U-gauge}), the group measure is simply
\ba
\mathcal{D}U = \prod_{a=1}^{N_c^2-1} {\cal D} \omega_a(x) .
\ea
The invariance of the measure is evident in case of infinitesimal transformations 
$U(x) = 1 + i \omega_a(x) \, \tau^a$. Then, $U''(x) = U(x) U'(x) = 1 + i \big(\omega_a(x) + \omega'_a(x)\big) \, \tau^a$, and consequently ${\cal D} \omega''_a(x) = {\cal D} \big(\omega_a(x) + \omega'_a(x)\big) = {\cal D} \omega_a(x)$. Using the invariance of the measure (\ref{measure-inv}) we can prove the invariance of the functional $\Delta_f^{-1}[A_\mu]$
\ba
\label{inv-det}
\Delta_f^{-1}[A]=\int \mathcal{D}U\rq{}\rq{} \; \delta \big[f[A^{U\rq{}\rq{}}]\big]
=\int \mathcal{D}U \; \delta \big[f[A^{U\rq{} U}] \big]=\Delta_f^{-1}[A^{U\rq{}}]
\ea
where we have set $U\rq{}U=U\rq{}\rq{}$. 

Now we compute $\Delta_f[A]$ by treating $f[A^U]$ as a function of $U(x)$ for fixed $A_\mu$. So, we can change the integration variables in Eq.~(\ref{F-P-det}) from $U$ to $f$ in a standard way and
\ba
\label{F-P-det-1}
\Delta^{-1}_f[A]
=\int \mathcal{D} f \, {\textrm {det}}\left| \frac{\delta U}{\delta f} \right| \delta [f] 
= {\textrm {det}}\left| \frac{\delta U}{\delta f} \right|_{f=0},
\ea
where the determinant is the Jacobian of the change of variables. We denote the matrix in Eq.~(\ref{F-P-det-1}) as
\ba
\label{F-P-det-M}
M \equiv \frac{\delta f}{\delta U} \bigg|_{f=0},
\ea
and the Faddeev-Popov determinant then equals
\ba
\label{F-P-det-3}
\Delta_f[A]={\textrm {det}}\; M.
\ea

Further on, we look for the matrix $M$ for the nonAbelian gauge field under the Lorentz gauge condition which is
\ba
\label{landau-gauge}
f^a[A]=\partial^\mu A_\mu^a=0.
\ea
There are actually $N_c^2-1$ gauge conditions for every $a$. For this reason we have assigned the index `$a$' to $f$. With the infinitesimal gauge transformation (\ref{gauge-transf-adjoint}) the gauge condition takes the form
\ba
\label{lorentz-gauge-2}
f^a[A^U]=\partial^\mu \Big( A_\mu^a+f^{abc}\omega^b A_\mu^c -\frac{1}{g}\partial_\mu \omega^a \Big) = 0.
\ea
The differentiation of the functional $f^a [A^U]$  with respect to $U$ in Eq.~(\ref{F-P-det-M}) is performed as the differentiation of $f^a [A^U]$ over the parameters $\omega^b$ that is
\ba
\label{F-P-det-help}
M_f^{ab}(x,y)=\frac{\partial f^a \big[ A^U (x) \big]}{\partial \omega^b (y)}
&=& 
-\partial^\mu \Big( \partial_\mu \delta^{ab} -g f^{abc} A_\mu^c (x)   \Big) \delta^{(4)}(x-y) 
\\ [2mm] \nn
&=&
 -\partial^\mu D_\mu^{ab}\delta^{(4)}(x-y),
\ea
where $D_\mu^{ab}$ is the covariant derivative defined by Eq.~(\ref{cov-D-adjoint}). The Faddeev-Popov determinant thus equals
\ba
\label{F-P-det-final}
{\textrm {det}}\, M_f={\textrm {det}}\big(\partial^\mu D_\mu[A]\big).
\ea
We note that in contrast to QED the determinant (\ref{F-P-det-final}) explicitly depends on the gauge field, and consequently it influences not only the normalization constant of the generating functional but it produces a nontrivial contribution to the functional (\ref{W-gluon-0}). 

\begin{center}
{\bf Ghosts}
\end{center}

To proceed one needs to recall the identity which usually appears in the context of quantization of the fermion field. The identity relates the Gaussian integral over anti-commuting variables to a determinant and it reads
\ba
\label{integral-grassmann}
\int \mathcal{D}\chi \int \mathcal{D}\theta \exp \bigg[i\int d^4 x \Big(\chi A \theta + \chi\beta +\beta^* \theta \Big)\bigg] = \textrm{det}(iA) \exp \Big(- i \beta^* A^{-1}\beta \Big),
\ea
where $\chi$ and $\theta$ are anti-commuting numbers that belong to the Grassmann algebra. Some properties of Grassmann algebras are discussed in Appendix \ref{appendix-Grassmann-algebra}. As one sees, the Faddeev-Popov determinant  (\ref{F-P-det-final}) can be represented as a path integral over the auxiliary complex field $c(x)$ belonging to the Grassmann algebra. Then
\ba
\label{ghosts}
\textrm{Det}\;M_ f
= \int \mathcal{D}c \,\mathcal{D}c^* \exp \bigg( i \int d^4x \, d^4y \, c^*_a(x) M^{ab}_f(x,y) c_b(y) \bigg),
\ea
where the explicit form of the matrix $M$ (\ref{F-P-det-help}) has been plugged into. One observes that the Faddeev-Popov determinant in the form (\ref{ghosts}) equals the vacuum-vacuum transition matrix element of the complex fields  $c_a(x)$ coupled to  the gauge fields $\partial^\mu A^a_\mu(x)$. The action of the system is 
\ba
\label{action-ghosts}
 S_{\rm ghost} 
= \int d^4x \, d^4y \; c^*_a(x) M^{ab}_f(x,y) c_b(y) 
= \int d^4x \Big( -c^*_a \partial^\mu \partial_\mu c_a  
+ g  f^{abc} c^*_a \partial^\mu \big(A_\mu^c c_b \big) \Big).
\ea
One sees that in the last term of (\ref{action-ghosts}) there is the derivative which can be expressed as 
\ba
\label{action-ghosts-term}
g\int d^4x  f^{abc} c^*_a \partial^\mu \big(A_\mu^c c_b\big) 
=g \int d^4x  f^{abc} c^*_a \big(\partial^\mu A_\mu^c\big) c_b
+ g \int d^4x f^{abc} c^*_a A_\mu^c \partial^\mu c_b
\ea
and due to the Lorentz condition $\partial^\mu A_\mu^c=0$ only the last term in (\ref{action-ghosts-term}) survives and the action is of the form
\ba
\label{action-ghosts-1}
S_{\rm ghost} 
= \int d^4x \, d^4y \; c^*_a(x) M^{ab}_f(x,y) c_b(y) 
=  \int d^4x \Big( -c^*_a \partial^\mu \partial_\mu c_a + g f^{abc} c^*_a A_\mu^c (\partial^\mu c_b) \Big).
\ea
The Lagrangian inferred from the action (\ref{action-ghosts-1}) shows that the fields $c_a$ satisfy the Klein-Gordon equation as the spinless boson fields.  The fields, however, are quantized -- the path integral in Eq.~(\ref{ghosts}) represents the quantization -- as the fermion fields. Within the canonical quantization, the  fields $c(x)$ would obey anticommutation relations. The bosonic fields $c_a(x)$, which are quantized as fermions, are unphysical. The fields $c_a(x)$ called the Faddeev-Popov ghosts or simply ghosts show up only as intermediate states but are absent both in the asymptotic initial and final states. Therefore, they appear only as internal lines of Feynman diagrams.  We note that the Faddeev-Popov determinant of electromagnetic field (\ref{F-P-det-QED}) can be also represented by ghosts. In that case, however, the ghosts are free and consequently play no dynamical role in the system. 

\begin{center}
{\bf Explicit form of generating functional of gluodynamics}
\end{center}

When the Faddeev-Popov determinant  is expressed through the ghost fields as in Eq.~(\ref{ghosts}), 
the generating functional is written as
\ba
\label{W-gluon-1}
W[J,\chi,\chi^*] = N \int \mathcal{D}A \, \mathcal{D}c \, \mathcal{D}c^*\; \delta\big[\partial^\mu A_\mu\big] 
\exp \bigg( i \int d^4 x \mathcal{L}(x) \bigg) ,
\ea
where the Lagrangian is of the form
\ba
\label{lagr-gluons}
\mathcal{L} =\mathcal{L}_{\rm YM} - c^{*}_a \big(\partial^\mu \partial_\mu \delta^{ab}
-g\partial^\mu f^{abc} A_\mu^c\big) c_b + J^{\mu a} A^a_\mu +  \chi^*_a c_a  + \chi_a c^*_a.
\ea
where $\mathcal{L}_{\rm YM}$ is simply the Lagrangian of the Yang-Mills field, the second term of (\ref{lagr-gluons}) represents ghost contribution and the remaining parts are interaction terms of the fields with respective sources: $J$ being the source of the gluon field and $\chi$ and $\chi^*$, which are Grassmann fields, being the ghost sources. To deal with the delta function which fixes the gauge in Eq.~(\ref{W-gluon-1}), we can proceed as in Sec.~\ref{sssec-gen-fun-em}. As a result we get
\ba
\label{delta-gen}
\delta \big[\partial^\mu A_\mu \big] 
= \lim_{\alpha \rightarrow 0} \exp \bigg(-\frac{i}{2\alpha}\int d^4x \; \big(\partial^\mu A_\mu^a \big)^2 \bigg) 
\ea
and the generating functional is expressed as
\ba
\label{W-gluon-3}
W[J] 
= N \int \mathcal{D}A \, \mathcal{D}c \, \mathcal{D}c^* \exp \bigg( i \int d^4 x \, \mathcal{L}_{\rm eff}(x) \bigg) ,
\ea
where the effective Lagrangian density equals
\ba
\label{lagr-gluons-eff}
\mathcal{L}_{\rm eff} 
=\mathcal{L}_{\rm YM} - c^*_a \big(\partial^\mu \partial_\mu \delta^{ab} -g\partial^\mu f^{abc} A_\mu^c \big)c_b 
- \frac{1}{2\alpha} \big(\partial^\mu A_\mu^a \big)^2 + J^{\mu a} A^a_\mu 
+ \chi^*_a c_a + \chi_a c^*_a .
\ea
And as it was presented in the case of the electromagnetic field, the limit of $\alpha$ going to $0$ does not have to be taken into account.

As in case of QED, the Lagrangian (\ref{lagr-gluons-eff}) can be split into the free part $\mathcal{L}_0$, which is no more than quadratic in fields, and the interacting one $\mathcal{L}_I$, which includes higher order terms,  
\ba
\label{lagr-gluons-free+interact}
\mathcal{L}_{\rm eff} =\mathcal{L}_0+\mathcal{L}_{I}
\ea
with
\ba
\label{lagr-gluons-free}
\mathcal{L}_0 
&=&
-\frac{1}{4}\big(\partial_\mu A_\nu^a-\partial_\nu A_\mu^a\big)\big(\partial^\mu A^{\nu a}-\partial^\nu A^{\mu a}\big) 
+ \partial^\mu c^*_a \partial_\mu c_a - \frac{1}{2\alpha} \big(\partial^\mu A_\mu^a \big)^2 
\\ [2mm] \nn
&& \qquad \qquad \qquad \qquad \qquad \qquad \qquad \qquad
+ J^{\mu a} A^a_\mu + \chi^*_a c_a  + \chi_a c^*_a,
\\ [2mm]
\label{lagr-gluons-interact}
\mathcal{L}_I 
&=&
-\frac{1}{2}g f^{abc} \big(\partial_\mu A_\nu^a-\partial_\nu A_\mu^a\big) A^{\mu b} A^{\nu c} 
-\frac{1}{4}g^2 f^{abc} f^{ade}A_\mu^b A_\nu^c A^{\mu d} A^{\nu e} 
\\ [2mm] \nn
&& \qquad \qquad \qquad \qquad \qquad \qquad \qquad \qquad
+g f^{abc} c^*_a \big(\partial^\mu A_\mu^c\big) c_b .
\ea
Such a splitting enables us to express the generating functional in the following form
\ba
\label{W-gluon-free+interact}
W[J] = \exp \Big( i S_I [A, c, c^*] \Big) \, W^0_A[J] \, W^0_c[  \chi,   \chi^*] ,
\ea
where the free generating functionals of the gauge field $W^0_A[J]$ and of the ghosts $W^0_c[\chi, \chi^*]$ are
\ba
\label{W-free-A}
W^0_A [J] 
& \equiv& 
\int \mathcal{D} A \exp \bigg[ i \int d^4x 
\bigg(-\frac{1}{4} \big(\partial_\mu A_\nu^a-\partial_\nu A_\mu^a\big)^2 
- \frac{1}{2\alpha} \big( \partial^\mu A_\mu^a \big)^2 \bigg) \bigg] ,
\\ [2mm]
\label{W-free-c}
W^0_c [\chi, \chi^*] 
&\equiv& 
\int \mathcal{D} c \ \mathcal{D} c^* \exp \bigg[ i \int d^4x  \
\Big( \partial^\mu c^*_a \partial_\mu c_a + \chi^*_a c_a  + \chi_a c^*_a \Big) \bigg].
\ea

The free generating functional of the gluon field (\ref{W-free-A}) has the same structure as that of the electromagnetic one. Performing the functional integration and using the formula (\ref{gaussian}), we get
\ba
\label{W-free-A-1}
W^0_A[J] = \exp \bigg( \frac{i}{2} \int d^4x \, d^4 y \, J^{\mu a}(x) D_{\mu\nu}^{ab}(x-y) J^{\nu b}(y) \bigg) ,
\ea
where the Lorentz gauge condition was chosen. The free gluon propagator is 
\ba
\label{propag-gluon}
 D_{\mu\nu}^{ab}(x)
= \delta^{ab} \int \frac{d^4 k}{(2\pi)^4} \bigg[ \bigg( g_{\mu\nu} - \frac{k_\mu k_\nu}{k^2+i0^+} \bigg) 
+\alpha \frac{k_\mu k_\nu}{k^2+i0^+} \bigg] \frac{\exp(-ikx)}{k^2 +i0^+}.
\ea
Up to the delta $\delta^{ab}$ it coincides with the photon propagator discussed in Sec.~\ref{sssec-gen-fun-em}.

After writing down the free generating functional of ghost fields as 
\ba
\label{W-free-c-1}
W^0_c[\chi, \chi^*] = \int \mathcal{D} c \mathcal{D} c^* \exp \bigg[ -i \int d^4x 
\Big( c^*_a \partial^\mu \partial_\mu c_a  - \chi^*_a c_a - \chi_a c^*_a \Big) \bigg] ,
\ea
one easily performs the functional integrals and gets 
\be
\label{W-free-c-1}
W^0_c[\chi, \chi^*] = \exp \bigg( -i \int d^4x \, d^4y \, \chi^*_a (x) \Delta^{ab}(x-y) \chi_b (y) \bigg) ,
\ee
where the free propagator of ghost fields is
\be
\label{propag-ghost}
\Delta^{ab}(x)=- \delta^{ab} \int \frac{d^4 k}{(2\pi)^4} \frac{\exp(-ikx)}{k^2 +i0^+}.
\ee

The full generating functional of pure gluodynamics equals
\ba
\label{W-gluon-final}
W[J, \chi, \chi^*] 
&=& 
\exp \bigg( i S_I \bigg[\frac{1}{i}\frac{\delta}{\delta J^a_\mu}, \frac{1}{i}\frac{\delta}{\delta \chi_a}, \frac{1}{i}\frac{\delta}{\delta \chi^*_a} \bigg] \bigg)  
\\ [2mm] \nn
&& \qquad
\times \exp \bigg( \frac{i}{2} \int d^4x \, d^4 y \, J^{\mu a}(x) D_{\mu\nu}^{ab}(x-y) J^{\nu b}(y) \bigg)
\\ [2mm] \nn
&& \qquad \qquad
\times \exp \bigg( -i \int d^4x \, d^4y \, \chi^*_a(x) \Delta^{ab}(x-y) \chi_b (y) \bigg) ,
\ea
where we have replaced the fields by the corresponding derivatives over sources in the interaction contribution to the action 
\ba
\label{S_I}
S_I [A^a_\mu, c_a, c^*_a] 
&=& 
\int d^4x \, \bigg[-\frac{1}{2}g f^{abc} \big(\partial_\mu A_\nu^a-\partial_\nu A_\mu^a\big) A^{\mu b} A^{\nu c} 
\\ [2mm] \nn
&& \qquad \qquad 
-\frac{1}{4}g^2 f^{abc} f^{ade}A_\mu^b A_\nu^c A^{\mu d} A^{\nu e} 
+g f^{abc} c^*_a  \big(\partial^\mu A_\mu^c\big) c_b \bigg] ,
\ea
which is given by the Lagrangian ${\cal L}_I$ (\ref{lagr-gluons-interact}). Having the generating functional in he form (\ref{W-gluon-final}) we are able to perform a perturbative calculus on the gluon field. Moreover, the concept of ghosts was fundamental to prove renormalizability of the nonAbelian gauge theories by t'~Hooft \cite{'tHooft:1971rn}.

\subsubsection{Generating functional of full QCD}
\label{sssec-gen-fun-QCD}

To derive the generating functional of full QCD, the Lagrangian of pure gluodynamics has to be supplemented by the quark term
\be
{\cal L}^q \equiv {\bar \psi}_q  \left( i \gamma_\mu D^\mu - m_q \right) \psi_q  + \bar\eta_q \psi_q  +  \bar\psi_q \eta_q,
\ee
where $D^\mu \equiv \partial^\mu \, {\bf 1} - i g A^\mu$ and $\eta$ is a source of the fermion field. Quarks contribute to both ${\cal L}_0$ (\ref{lagr-gluons-free}) and ${\cal L}_I$  (\ref{lagr-gluons-interact}) as 
\ba
{\cal L}^q_0 &=& {\bar \psi}_q  \left( i \gamma_\mu \partial^\mu - m_q \right) \psi_q  + \bar\eta_q \psi_q  +  \bar\psi_q \eta_q,
\\[2mm] 
{\cal L}^q_I &=& g {\bar \psi}_q   \gamma^\mu  \tau^a \psi_q   A^a_\mu .
\ea
The free quark term ${\cal L}^q_0$ is responsible for the generating functional of free fermions while ${\cal L}^q_I$ produces an extra term in $S_I$. The generating functional of QCD with one quark flavor thus equals
\ba
\label{W-full-QCD}
W[J,   \chi,  \chi^*] &=& \exp 
\left\{ i S_I \left[\frac{1}{i}\frac{\delta}{\delta J^a_\mu}, \frac{1}{i}\frac{\delta}{\delta \chi_a}, \frac{1}{i}\frac{\delta}{\delta \chi^*_a} , \frac{1}{i}\frac{\delta}{\delta \eta_i} , \frac{1}{i}\frac{\delta}{\delta \bar\eta_i} \right] \right\}  
\\ [2mm] \nn
&& \;\;\;\;\; \times  \exp \left\{ \frac{i}{2} \int d^4x \, d^4 y \, J^{\mu a}(x) D_{\mu\nu}^{ab}(x-y) J^{\nu b}(y)    \right\}  
\\ [2mm] \nn
&& \;\;\;\;\;\;\;\;\;\; \times \exp \left\{ -i \int d^4x \, d^4y \,   \chi^*_a(x) \Delta^{ab}(x-y)   \chi_b (y) \right\}
\\ [2mm] \nn
&& \;\;\;\;\;\;\;\;\;\;\;\;\;\;\; \times \exp \left\{ i\int d^4 x \; d^4 y \; \bar\eta_i(x) \, S_F^{ij}(x-y) \, \eta_j (y) \right\},
\ea
where $i,j=1,2, ... \, N_c$ are colour indices in the fundamental representation, $S_F^{ij}(x)$ is the quark Feynman propagator
\be
S_F^{ij} (x) = \delta^{ij} \int \frac{d^4p}{(2\pi)^4} \frac{\gamma \cdot p+m}{m^2 - p^2 - i0^+} \, e^{-ipx}
\ee
and the action $S_I$ as a functional of fields equals
\ba
\label{S_I-QCD}
S_I [A^a_\mu, c_a, c^*_a,\psi_i, \bar\psi_i] &=& \int d^4x \, \Big[
-\frac{1}{2} g  f^{abc} (\partial_\mu A_\nu^a-\partial_\nu A_\mu^a)A^{\mu b} A^{\nu c} 
\\[2mm] \nn
&& 
-\frac{1}{4}g^2 f^{abc} f^{ade}A_\mu^b A_\nu^c A^{\mu d} A^{\nu e} 
+ \; g f^{abc} c^*_a  (\partial^\mu A_\mu^c) c_b 
+ g {\bar \psi}_i   \gamma^\mu  \tau^a_{ij} \psi_j   A^a_\mu  \Big].
\ea
It is straightforward to include in the generating functional (\ref{W-full-QCD}) several quark flavors. 

\subsection{Generating functional of many-body field theories}
\label{ssec-gen-fun-mb}

In this chapter we present a way on how to construct a generating functional of different statistical field theories\footnote{The subsections \ref{ssec-gen-fun-mb}-\ref{ssec-ghosts} are based on our work published as \cite{Czajka:2014eha}.}.

\subsubsection{Generating functional of the scalar field}
\label{sssec-gen-fun-scal}

The contour Green function of the scalar field is defined in Sec. \ref{sec-KS-form} and it equals
\ba
\label{FG-KS}
\tilde \Delta(x_1,x_2)=Z^{-1}\textrm{Tr} \big[ \hat\rho(t_0) \tilde T \phi(x_1) \phi(x_2) \big]
\ea
where $Z \equiv \textrm{Tr}[\hat \rho(t_0)]$. The density operator $\hat \rho(t_0)$ is given in terms of the eigenstates of the field operator $\phi(t_0=-\infty \pm i 0^+,{\bf x})$ as
\ba
\label{density-def}
\hat \rho = \int D\phi' ({\bf x}) \, D \phi'' ({\bf x})  \rho\big[\phi'({\bf x}) , \phi''({\bf x})\big] \, 
|\phi''({\bf x}) \rangle \langle \phi'({\bf x})|,
\ea
where 
\ba
\label{bound-cond} 
\phi'({\bf x})  = \phi(t =-\infty +i0^+,{\bf x}), 
\;\;\;\;\;\;\;\;\;\;\;
\phi''({\bf x})  = \phi(t =-\infty -i0^+,{\bf x})
\ea
and $-\infty +i0^+$ and $-\infty - i0^+$ correspond to the beginning of the upper branch and the end of the lower branch,
respectively, of the Keldysh contour shown in Fig. \ref{Keldysh-contour}. 

To determine the generating functional let us start with calculating $Z$ which is
\ba
\label{Omega}
Z= \textrm{Tr}[\hat \rho] 
\! &=& \!
\int D\phi_{-\infty}({\bf x}) \langle \phi_{-\infty} ({\bf x}) | \hat\rho | \phi_{-\infty} ({\bf x}) \rangle 
\\[2mm] \nn
\! &=& \!
\int D\phi_{-\infty} ({\bf x}) \; D\phi'({\bf x}) \; D\phi''({\bf x}) \; \rho\big[\phi'({\bf x})|\phi''({\bf x})\big] \, 
\langle \phi_{-\infty}({\bf x})|\phi''({\bf x}) \rangle \langle \phi'({\bf x})|\phi_{-\infty}({\bf x}) \rangle,
\ea
where
\ba
\phi_{-\infty} ({\bf x}) \equiv \phi(t =-\infty, {\bf x}) .
\ea
Since we have
\ba
\langle \phi_1 |\phi_2 \rangle = \delta[\phi_1-\phi_2]
\ea
the formula (\ref{Omega}) gets the form
\ba
Z = \int D\phi_{-\infty} \; D\phi' \; D\phi'' \, \rho[\phi'|\phi''] \,
\delta[\phi''-\phi_{-\infty}] \, \delta[\phi_{-\infty}-\phi'] =
\int D\phi \, \rho[\phi|\phi] .
\ea

Since the trace in the formula (\ref{FG-KS}) is taken over states in $t=-\infty$, so the contour Green function can be written
as
\ba
\label{FG-KS-1}
\tilde \Delta(x_1,x_2)=
Z^{-1}
\int D\phi''({\bf x}) \,  D \phi' ({\bf x}) \, \rho\big[\phi'({\bf x})\big|\phi''({\bf x}) \big] 
\big\langle \phi''({\bf x})\big| \tilde T  \phi(x_1) \phi(x_2) \big| \phi'({\bf x}) \big\rangle .
\ea
With the path-integral representation of the propagator
\ba
\label{path-rep}
\langle \phi''({\bf x}) | \tilde T \phi(x_1) \phi(x_2)  | \phi'({\bf x}) \rangle 
& = &
\int_{\begin{subarray}{l}\phi(-\infty +i0^+,{\bf x}) = \phi'({\bf x}) \\ 
\phi(-\infty -i0^+,{\bf x}) = \phi''({\bf x})\end{subarray}}
{\cal D}\phi(x) \, 
\\ [2mm] \nn
&& \qquad \qquad
\times \phi(x_1) \, \phi(x_2) \exp{\bigg[i\int_C d^4x \mathcal{L}(x)\bigg]},
\ea
where the functional integral is performed over the field configurations in the time and space with the boundary condition
given by Eq.~(\ref{bound-cond}). Since the integration over the Lagrangian density is now performed along the time
contour we have denoted
\be
\label{denote-measure}
\int_C d^4x \ldots \equiv \int_C dt \int d^3x \ldots
\ee
Inserting the path integral representation of the propagator (\ref{path-rep}) into (\ref{FG-KS-1}), the Green function gets the form
\ba
\label{FG-KS-2}
\tilde \Delta(x_1,x_2) 
& = & 
Z^{-1} \int D\phi'({\bf x}) \; D\phi''({\bf x}) \; \rho\big[\phi'({\bf x})\big|\phi''({\bf x}) \big] 
\\ [2mm] \nn
&& 
\times \int_{\begin{subarray}{l}\phi(-\infty +i0^+,{\bf x}) = \phi'({\bf x}) \\ 
\phi(-\infty -i0^+,{\bf x}) = \phi''({\bf x})\end{subarray}}
{\cal D}\phi(x) \, \phi(x_1) \, \phi(x_2) \exp{\bigg[i\int_C d^4x \mathcal{L}(x)\bigg]}.
\ea
The generating functional may be then written as 
\ba
\label{W-KS}
W[J] 
& = & 
N \int D\phi'({\bf x}) \; D\phi''({\bf x}) \; \rho\big[\phi'({\bf x})\big|\phi''({\bf x}) \big]
\\ [2mm] \nn
&& 
\times \int_{\begin{subarray}{l}\phi(-\infty +i0^+,{\bf x}) = \phi'({\bf x}) \\ 
\phi(-\infty -i0^+,{\bf x}) = \phi''({\bf x})\end{subarray}}
{\cal D}\phi(x) \, \exp{\bigg[i\int_C d^4x \Big( \mathcal{L}(x) + \phi(x) \, J(x) \Big) \bigg]} .
\ea
Then, the contour Green function can be generated through
\be
\label{CGF-gen}
\tilde \Delta(x_1,x_2)=\frac{1}{W[J=0]} \frac{1}{i^2} \frac{\delta}{\delta J(x_1)}\frac{\delta}{\delta J(x_2)} W[J] \bigg|_{J=0}.
\ee
Looking for any time-dependent Green function, we should use the following functional contour differentiation rule
\be
\label{contour-different}
\frac{\delta J(x)}{\delta J(y)} = \delta^{(4)}_C(x,y),
\ee
which takes into account the different positions of the source $J$ on the Keldysh contour. The contour Dirac delta is defined by
(\ref{delta-contour}). 

\begin{center}
{\bf Generation of $\Delta^>$}
\end{center}

To check whether the formula (\ref{CGF-gen}) works properly we are going to find $\Delta^>(x_1,x_2)$. Such a function should be generated through
\be
\label{CGF-gen-check}
\Delta^>(x_1,x_2)=\frac{1}{W[J=0]} \frac{1}{i^2} \frac{\delta}{\delta J_+(x_1)}\frac{\delta}{\delta J_-(x_2)} W[J]
\bigg|_{J=0},
\ee
where the indices `+' and `-' denote the positions of the source $J$ on the upper and lower branch, respectively. Since the
derivatives act only on the exponent of the source term of the functional (\ref{W-KS}), we calculate only the following expression
\ba
\frac{\delta}{\delta J_+(x_1)} \frac{\delta}{\delta J_-(x_2)} \exp\bigg[i\int_C d^4x \, \phi(x)J(x)\bigg].
\ea
We should remember that the integration over the contour can be split into two integrals, that is
\be
\label{express-contour}
\int_C d^4x \, \phi(x)J(x) = \int d^4x \, \phi_+(x)J_+(x) - \int d^4x \, \phi_-(x)J_-(x).
\ee
Making allowance for the expression (\ref{express-contour}) and for the following equalities
\ba
\frac{\delta J_+(x)}{\delta J_-(x_2)} = 0, \qquad\qquad\qquad 
\frac{\delta J_-(x)}{\delta J_-(x_2)} =- \delta^{(4)}(x-x_2),
\ea
one gets
\ba
&&
\frac{\delta}{\delta J_+(x_1)} \frac{\delta}{\delta J_-(x_2)} \exp \bigg[i\int_C d^4x \, \phi(x)J(x)\bigg]
\\ [2mm] \nn
&& \qquad \qquad \qquad \qquad \qquad \qquad
=  i^2 \, \phi_-(x_2) \, \phi_+(x_1) \, \exp \bigg[i\int_C d^4x \, \phi(x)J(x)\bigg]. 
\ea
Finally, after putting $J=0$ the full formula of $\Delta^>(x_1,x_2)$ is given by
\ba
\label{FG-check}
\Delta^>(x_1,x_2)
& = & 
Z^{-1}\int D\phi'({\bf x}) \; D\phi''({\bf x}) \; \rho\big[\phi'({\bf x})\big|\phi''({\bf x}) \big] 
\\ [2mm] \nn
&& \qquad 
\times \int_{\begin{subarray}{l}\phi(-\infty +i0^+,{\bf x}) = \phi'({\bf x}) \\ 
\phi(-\infty -i0^+,{\bf x}) = \phi''({\bf x})\end{subarray}}
{\cal D}\phi(x) \, \phi_-(x_2) \, \phi_+(x_1) \exp{\bigg[i\int_C d^4x \, \mathcal{L}(x)\bigg]}, \nn
\ea
which is in agreement with the formula (\ref{FG-KS-2}).

\subsubsection{Generating functional of the quantum electrodynamics}
\label{sssec-gen-fun-qed}

Following the analogous procedure as in case of the scalar field, we find the generating functional of the electrodynamics, 
which is given by
\ba
\label{W-KS-QED}
W[J, \bar\eta, \eta] &=& N \int DA'({\bf x}) \; DA''({\bf x}) \; D\Psi'({\bf x}) \; D\Psi''({\bf x}) \; 
D\bar\Psi'({\bf x}) \; D\bar\Psi''({\bf x})  
\\ [2mm] \nn
&& \qquad
\times \rho\Big[A'({\bf x}),\Psi'({\bf x}),\bar\Psi'({\bf x}) \big|A''({\bf x}),\Psi''({\bf x}),\bar\Psi''({\bf x})\Big] 
W_0[J, \bar\eta, \eta],
\ea
where $W_0[J, \bar\eta, \eta]$ is defined as follows
\ba
\label{W0-KS-QED}
W_0[J, \bar\eta, \eta]
&=& 
N_0 \int_{\begin{subarray}{l} A(-\infty +i0^+,{\bf x})=A'({\bf x}) \\ 
A(-\infty -i0^+,{\bf x})=A''({\bf x})\end{subarray} } \mathcal{D}A(x)
\int_{\begin{subarray}{l} \Psi(-\infty +i0^+,{\bf x}) =\Psi'({\bf x}) \\ 
\Psi(-\infty -i0^+,{\bf x}) =\Psi''({\bf x}) \end{subarray}}\mathcal{D}\Psi(x)
\\ [2mm] \nn
&& \qquad
\times \int_{\begin{subarray}{l} \bar\Psi(-\infty +i0^+,{\bf x}) =\bar \Psi'({\bf x}) \\ 
\bar\Psi(-\infty -i0^+,{\bf x}) = \bar\Psi''({\bf x}) \end{subarray}}\mathcal{D}\bar\Psi(x) 
\exp{\bigg[i\int_C d^4x \mathcal{L}_{\rm eff-QED}(x) \bigg]}
\ea
and it strongly resembles the functional of a vacuum field theory. The effective Lagrangian is 
\be
\label{lagr-QED-eff}
\mathcal{L}_{\rm eff-QED}=\mathcal{L}_{\rm QED} - \frac{1}{2\alpha}\big(\partial^\mu A_\mu \big)^2 
- \bar\eta \Psi - \bar\Psi \eta - J^\mu A_\mu, 
\ee
where $\mathcal{L}_{\rm QED}$ is given by (\ref{L-QED}).

\subsubsection{Generating functional of the Yang-Mills field}
\label{sssec-gen-fun-ym}

The generating functional of pure gluodynamics is found as
\ba
\label{W-KS-YM}
W[J, \chi, \chi^*] 
&=& 
N \int  DA'({\bf x}) \; DA''({\bf x}) \; Dc'({\bf x}) \; Dc''({\bf x}) \; 
D{c^{*}}'({\bf x}) \; D{c^{*}}''({\bf x}) 
\\ [2mm] \nn
&& 
\times \rho\Big[A'({\bf x}),c'({\bf x}),{c^{*}}'({\bf x})\Big|A''({\bf x}),c''({\bf x}),{c^{*}}''({\bf x})\Big] 
W_0[J, \chi, \chi^*], 
\ea
where
\ba
\label{W0-KS-YM}
W_0[J, \chi, \chi^*] 
&=& 
N_0 \int_{\begin{subarray}{l} A(-\infty +i0^+,{\bf x})=A'({\bf x}) \\ 
A(-\infty -i0^+,{\bf x})=A''({\bf x})\end{subarray} } \mathcal{D}A(x) 
\int_{\begin{subarray}{l} c(-\infty +i0^+,{\bf x})=c'({\bf x}) \\ 
c(-\infty -i0^+,{\bf x})=c''({\bf x})\end{subarray} } \mathcal{D}c(x) 
\\ [2mm] \nn
&& \qquad
\times
\int_{\begin{subarray}{l} c^{*}(-\infty +i0^+,{\bf x})={c^{*}}'({\bf x}) \\ 
c^{*}(-\infty -i0^+,{\bf x})={c^{*}}''({\bf x})\end{subarray} } \mathcal{D}c^{*}(x) 
\exp{\bigg[i\int_C d^4x \mathcal{L}_{\rm eff-YM}(x) \bigg]}.
\ea
The effective Lagrangian equals
\ba
\label{lagr-YM-eff}
\mathcal{L}_{\rm eff-YM} &=&\mathcal{L}_{\rm YM}
- c^*_a \big(\partial^\mu \partial_\mu \delta^{ab} -g\partial^\mu f^{abc} A_\mu^c \big) c_b 
- \frac{1}{2\alpha} \big(\partial^\mu A_\mu^a \big)^2 
\\ [2mm] \nn
&& + J^{\mu a} A^a_\mu + \chi^*_a c_a  +   \chi_a c^*_a,
\ea
where $\mathcal{L}_{\rm YM}$ is given by (\ref{L-YM-adj}).

\subsection{Derivation of the general Slavnov-Taylor identity}
\label{ssec-general-identity}

In this part we derive the general Slavnov-Taylor identity. To do so we use the generating functional of pure gluodynamics which is given by (\ref{W-KS-YM}).

The functional may be also written in a more compact form
\be
\label{W-gluon-sti-sec}
W[J,\chi,\chi^*] = N \int_{BC} \mathcal{D}A(x) \Delta[A]
\exp \bigg[ i \int_C d^4 x \, \mathcal{L}(x) \bigg] ,
\ee
with the Lagrangian being given as
\be
\label{lagr-YM}
\mathcal{L} = \mathcal{L}_{\rm YM} - \frac{1}{2\alpha} \big(\partial^\mu A_\mu^a \big)^2 
+ J^{\mu a} A^a_\mu.
\ee
In (\ref{W-gluon-sti-sec}) we have used a compressed notation of the functional $W[J,\chi,\chi^*]$, which reads
\ba
\label{sti-notation}
\int_{BC} \mathcal{D}A(x)  \dots 
&\equiv &  
\int  DA'({\bf x}) \,DA''({\bf x})\, 
Dc'({\bf x}) \,Dc''({\bf x}) \,D{c^{*}}'({\bf x}) \,D{c^{*}}''({\bf x}) 
\\ [2mm] \nn
&& \qquad
\times \rho\Big[A'({\bf x}),c'({\bf x}),{c^{*}}'({\bf x})\Big|A''({\bf x}),c''({\bf x}),{c^{*}}''({\bf x})\Big]
\\ [2mm] \nn
&& \qquad \qquad
\times \int_{\begin{subarray}{l} A(-\infty +i0^+,{\bf x})=A'({\bf x}) \\ 
A(-\infty -i0^+,{\bf x})=A''({\bf x})\end{subarray} } \mathcal{D}A(x) \dots.
\ea
and the functional $\Delta[A]$ is
\ba
\label{sti-notation-delta}
\Delta(A) 
&\equiv &  
\int_{\begin{subarray}{l} c(-\infty +i0^+,{\bf x}) =c'({\bf x}) \\ 
c(-\infty -i0^+,{\bf x}) =c''({\bf x}) \end{subarray}}\mathcal{D}c(x) \;\;
\int_{\begin{subarray}{l} c^*(-\infty +i0^+,{\bf x}) ={c^{*}}'({\bf x}) \\ 
c^*(-\infty -i0^+,{\bf x}) ={c^{*}}''({\bf x}) \end{subarray}}\mathcal{D}c^*(x) 
\\ [2mm] \nn
&& \qquad
\times \exp \bigg[ -i \int_C d^4 x \Big(c^*_a(\partial^\mu \partial_\mu \delta^{ab} -g f^{abc} A_\mu^c \partial^\mu)c_b -   \chi^*_a c_a  -   \chi_a c^*_a\Big) \bigg], 
\ea
which is an analog of the Fadeev-Popov determinant in the vacuum theory. One obtains the standard form of the determinant choosing the boundary conditions of the ghosts fields as $c'({\bf x})={c^{*}}'({\bf x})= c''({\bf x}) ={c^{*}}''({\bf x})$ and the density operator which acts on the ghost field as $|0 \rangle \langle 0|$. 

The infinitesimal gauge transformations of the gluon field, under which the Lagrangian of the Yang-Mills is invariant, is
\ba
\label{gauge-trans-sti}
A_\mu^a \rightarrow (A_\mu^a)^U
= A_\mu^a+f^{abc}\omega^b A_\mu^c - \frac{1}{g}\partial_\mu \omega^a +{\cal O}(\omega^2)
\ea
where $|\omega|\ll 1$. Since we assume that the gauge transformations (\ref{gauge-trans-sti}) do not work at
$t=-\infty$, that is $\omega(t=-\infty, {\bf x})=0$, the density matrix in the expression (\ref{sti-notation}) remains unchanged.

Expressing the generating functional of gluodynamics (\ref{W-gluon-sti-sec}) through the transformed fields and observing that the integration measure $\Delta(A) \mathcal{D}A$ is, as one can show, invariant under the transformation (\ref{gauge-trans-sti}), one finds it in the following form
\ba
\label{W-gauge-sti-1}
W'[J,\chi^*,\chi] 
&=& N \int_{BC} \mathcal{D}A(x) \Delta[A] \exp \bigg[ i\int_C d^4x \mathcal{L}'(x) \bigg],
\ea
where the changed Lagrangian is 
\ba
\label{changed-L}
\mathcal{L}' = \mathcal{L} - \frac{1}{g \alpha}\Big( -\partial^\mu\partial_\mu \delta^{ab}
+ g f^{abc}\partial^\mu A_\mu^c \Big)\omega^b \partial^\nu A_\nu^a 
-\frac{1}{g} J^{\mu a} \partial_\mu \omega^a
+J^{\mu a} f^{abc} A_\mu^c \omega^b.
\ea
The integration by parts of the term 
\ba
\label{int-omega}
\frac{1}{g}\int_C d^4x  J^{\mu a} \partial_\mu \omega^a 
= - \frac{1}{g} \int_C d^4x  \big(\partial_\mu J^{\mu a}\big) \omega^a
\ea
leads us to the following form of the Lagrangian (\ref{changed-L}) 
\ba
\label{changed-L-a}
\mathcal{L}' 
= \mathcal{L} - \frac{1}{g \alpha}M^{ab}\omega^b \partial^\nu A_\nu^a
+\frac{1}{g} \big(\partial_\mu J^{\mu a} \big) \omega^a
+J^{\mu a} f^{abc} A_\mu^c \omega^b,
\ea
where the matrix $M$ is
\ba
\label{M-sti}
M^{ab}[A|x] \equiv -\partial^\mu\partial_\mu \delta^{ab} + g f^{abc}\partial^\mu A_\mu^c(x).
\ea

The generating functional is expected to be invariant under the gauge transformations (\ref{gauge-trans-sti}) and therefore it should be independent of the gauge parameter $\omega$. Consequently, it should be independent of any function of $\omega$. We will exploit the fact that the functional is independent of the function $\xi_a(x)$ which is defined as
\ba
\label{f-ksi}
\xi_a(x)=M_{ab}[A|x]\omega_b(x) .
\ea
For this purpose we are going to differentiate the functional (\ref{W-gauge-sti-1}) with the Lagrangian (\ref{changed-L-a}) with respect to the function $\xi_a(x)$. Therefore, we have to express the functional through the function $\xi$. Using the Green function of the $M$ operator, which satisfies the equation
\ba
\label{eq-FG-M}
\int_C d^4y \; M_{ab}[A|x,y] \; M^{-1}_{bc}[A|y,z]= \delta_{ac}\delta_C^{(4)}(x,z)
\ea
the gauge parameter $\omega$ is expressed as
\ba
\label{omega-f}
\omega_a(x)=\int_C d^4y \; M^{-1}_{ab}[A|x,y] \; \xi_b(y).
\ea

Inserting the omega function given by (\ref{omega-f}) into the Lagrangian (\ref{changed-L-a}), we get
\ba
\mathcal{L}'(x) 
& =& 
\mathcal{L}(x) - \frac{1}{g\alpha} \partial^\nu A_\nu^a(x) M_{ab}[A|x] 
\int_C d^4y \; M^{-1}_{bd}[A|x,y] \, \xi_d(y) 
\\ [2mm] \nn
&& \qquad
+\frac{1}{g} \big(\partial_\mu J^{\mu a}(x)\big)  \int_C d^4y \; M^{-1}_{ad}[A|x,y] \, \xi_d(y)
\\ [2mm] \nn
&& \qquad \qquad
+J^{\mu a}(x) f^{abc} A_\mu^c(x) \int_C d^4y \; M^{-1}_{bd}[A|x,y] \, \xi_d(y). 
\ea
Now we are ready to differentiate the functional (\ref{W-gauge-sti-1}) over the function $\xi$. Performing the operation we remember that
\ba
\label{xi-diff}
\frac{\delta \xi_a(x)}{\delta \xi_b(y)} = \delta_{ab}\delta_C^{(4)}(x,y),
\ea
where the contour delta function is defined by (\ref{delta-contour}). In this way we obtain
\ba
\label{W-gauge-sti-3}
\frac{\delta W'[J,\chi^*,\chi]}{\delta \xi_g(z)} 
&=& 
N \int_{BC} \mathcal{D}A \, \Delta[A] 
\exp \bigg[ i\int_C d^4x \; \mathcal{L}'(x) \bigg] 
\\ [2mm] \nn
&& \qquad
\times \; i \int_C d^4 x \; \bigg[- \frac{1}{g \alpha} \partial^\nu A_\nu^a(x) \; M_{ab}[A|x] M^{-1}_{bg}[A|x,y] 
\\ [2mm] \nn
&& \qquad\qquad
+\frac{1}{g} \big(\partial_\mu J^{\mu a}(x)\big) M^{-1}_{ag}[A|x,z]
+J^{\mu a}(x) f^{abc} A_\mu^c(x)  M^{-1}_{bg}[A|x,z] \bigg], 
\ea
where we have used the identity
\ba
\int_C d^4y \; M^{-1}_{bg}[A|x,y] \delta_C^{(4)}(y,z) = M^{-1}_{bg}[A|x,z].
\ea
Taking into account the relation (\ref{eq-FG-M}), we get 
\ba
\label{W-gauge-sti-4}
\frac{\delta W'[J,\chi^*,\chi]}{\delta \xi_g(z)} 
&=& 
N \int_{BC} \mathcal{D}A \, \Delta[A] \exp \bigg[ i\int_C d^4x \; \mathcal{L}'(x) \bigg]
\\ [2mm] \nn
&&  \qquad
\times \; i\int_C d^4 x \; \bigg[- \frac{1}{g \alpha} \delta_C^{(4)}(x,z) \partial^\nu A_\nu^g(x)  
+\frac{1}{g} \big(\partial_\mu J^{\mu a}(x)\big) M^{-1}_{ag}[A|x,z]
\\ [2mm] \nn
&& \qquad \qquad \qquad \qquad \qquad \qquad \qquad \qquad 
+J^{\mu a}(x) f^{abc} A_\mu^c(x)  M^{-1}_{bg}[A|x,z]  \bigg].
\ea
Subsequently, we integrate the term containing the delta function
\ba
\label{W-gauge-sti-5}
\frac{\delta W'[J,\chi^*,\chi]}{\delta \xi_g(z)} 
&=& 
i N \int_{BC} \mathcal{D}A \, \Delta[A] \exp \bigg[ i\int_C d^4x \; \mathcal{L}'(x) \bigg]
\\ [2mm] \nn
&& \qquad
\times \bigg\{ -\frac{1}{g \alpha} \partial_{(z)}^\nu A_\nu^g(z)
+ \int_C d^4x \; \bigg[\frac{1}{g} \big(\partial_\mu J^{\mu a}(x)\big) M^{-1}_{ag}[A|x,z] 
\\ [2mm] \nn
&& \qquad \qquad \qquad \qquad \qquad \qquad \qquad \qquad 
+J^{\mu a}(x) f^{abc} A_\mu^c(x)  M^{-1}_{bg}[A|x,z]\bigg] \bigg\}. 
\ea
Putting the function $\xi =0$, we observe that
\be
\label{lagr-ch}
\mathcal{L}'(x) \rightarrow \mathcal{L}(x)
\ee
and the generating functional gets the final form
\ba
\label{W-gauge-sti-6}
\frac{\delta W'[J,\chi^*,\chi]}{\delta \xi_g(z)}\bigg|_{\xi=0} 
&=& 
i N \int_{BC} \mathcal{D}A \, \Delta[A] \exp \bigg[i\int_C d^4x \; \mathcal{L}(x) \bigg] 
\\ [2mm] \nn
&& \qquad 
\times \bigg\{ \int_C d^4 x \; \bigg[\frac{1}{g} \big(\partial_\mu J^{\mu a}(x)\big) \delta^{ab}
+J^{\mu a}(x) f^{abc} A_\mu^c(x) \bigg] M^{-1}_{bg}[A|x,z]  
\\ [2mm] \nn
&& \qquad \qquad \qquad \qquad \qquad \qquad \qquad \qquad \qquad \qquad 
- \frac{1}{g \alpha} \partial_{(z)}^\nu A_\nu^g(z) \bigg\}. \nn
\ea

The generating functional $W'[J,\chi^*,\chi]$ is just the functional $W[J,\chi^*,\chi]$ expressed through the fields transformed according to Eq.~(\ref{gauge-trans-sti}).  A change of the integration variables cannot change the value of the integral, and thus there appears the condition
\ba
\label{condition-gauge-sti}
\frac{\delta W'[J,\chi^*,\chi]}{\delta \xi_g(z)}\bigg|_{\xi=0}=0,
\ea
which gives 
\ba
\label{W-gauge-sti-7}
&& 
N \int_{BC} \mathcal{D}A \, \Delta[A] \exp \bigg[i\int_C d^4x\;  \mathcal{L}(x) \bigg]
\\ [2mm] \nn
&& \qquad
\times \bigg\{- \alpha \int_C d^4 x \bigg[J^{\mu a}(x) \partial_\mu \delta_{ab}
- gJ^{\mu a}(x) f^{abc} A_\mu^c(x) \bigg]  M^{-1}_{bg}[A|x,z] 
- \partial_{(z)}^\nu A_\nu^g(z) \bigg\}=0 .
\ea
Replacing the field $A_\mu$ by the corresponding derivative
\ba
\label{replacement-sti}
A_\mu^a(y) \rightarrow \frac{1}{i}\frac{\delta}{\delta J^{\mu a}(y)},
\ea
we get 
\ba
\label{WTI-general-W-sti}
&&
\Bigg\{i\partial_{(z)}^\mu \frac{\delta}{\delta J^{\mu g}(z)} 
\\ [2mm] \nn
&&
- \alpha \int_C d^4 x \bigg(J^{\mu a}(x) \partial_\mu \delta_{ab}
+ ig J^{\mu a}(x) f^{abc} \frac{\delta}{\delta J^{\mu c}(x)} \bigg)
M^{-1}_{bg}\bigg[\frac{1}{i}\frac{\delta}{\delta J}\bigg|x,z\bigg]\Bigg\} W[J,\chi^*,\chi]=0, 
\ea
where $W[J,\chi^*,\chi]$ is given by (\ref{W-gluon-sti-sec}). The relation (\ref{WTI-general-W-sti}) is the general Slavnov-Taylor identity and it is a starting point to derive some specific identities. One of them, the identity for the gluon propagator, is derived in the next subsection.

\subsection{The Slavnov-Taylor identity for the gluon propagator}
\label{ssec-identity-gluon}

Following Slavnov, we derive here the identity for the gluon propagator. Differentiating the general relation (\ref{WTI-general-W-sti}) over $J^{\nu d}(y)$, we get
\ba
\label{WTI-long-G-sti}
&& 
\Bigg\{i\partial_{(z)}^\mu \frac{\delta^2}{\delta J^{\mu g}(z) \delta J^{\nu d}(y)} 
- \alpha \int_C d^4 x \frac{\delta J_{\mu a}(x)}{ \delta J^{\nu d}(y)} \partial_{(x)}^\mu 
M^{-1}_{bg}\bigg[\frac{1}{i}\frac{\delta}{\delta J}\bigg|x,z\bigg]
\\ [2mm] \nn
&& \qquad\qquad
-i\alpha g \int_C d^4 x \bigg(\frac{\delta J_{\mu a}(x)}{\delta J^{\nu d}(y)} f^{abc} \frac{\delta }{\delta J_{\mu c}(x)}
+ J^{\mu a}(x) f^{abc} \frac{\delta^2}{\delta J^{\mu c}(x) \delta J^{\nu d}(y)} \bigg) 
\\ [2mm] \nn
&& \qquad\qquad\qquad\qquad \qquad \qquad  \qquad \qquad 
\times \; M^{-1}_{bg}\bigg[\frac{1}{i}\frac{\delta}{\delta J}\bigg|x,z\bigg] \Bigg\} W[J,\chi^*,\chi] =0. 
\ea
Using the formula of the functional differentiation $\frac{\delta J_{\mu a} (x)}{\delta J^{\nu b}(y)}=g_{\mu\nu} \delta^{ab} \delta_C^{(4)}(x,y)$, we find
\ba
\label{WTI-long-G-1-sti}
&& 
\Bigg\{i\partial_{(z)}^\mu \frac{\delta^2}{\delta J^{\mu g}(z) \delta J^{\nu d}(y)} 
-\alpha \int_C d^4 x g_{\mu\nu} \delta^{ad} \delta_C^{(4)}(x,y) \partial_{(x)}^\mu 
M^{-1}_{ag}\bigg[\frac{1}{i}\frac{\delta}{\delta J}\bigg|x,z\bigg] \qquad
\\ [2mm] \nn
&& \qquad \qquad
- i\alpha g \int_C d^4 x \bigg( g_{\mu\nu} \delta^{ad} \delta_C^{(4)}(x,y) f^{abc} 
\frac{\delta}{\delta J_{\mu c}(x)} 
+J^{\mu a}(x) f^{abc} \frac{\delta^2}{\delta J^{\mu c}(x) \delta J^{\nu d}(y)} \bigg)
\\ [2mm] \nn
&& \qquad \qquad \qquad \qquad \qquad \qquad \qquad \qquad 
\times \; M^{-1}_{bg}\bigg[\frac{1}{i}\frac{\delta}{\delta J}\bigg|x,z\bigg] \Bigg\} W[J,\chi^*,\chi] =0
\ea
and next
\ba
\label{WTI-long-G-1-sti-1}
&&
\!\!\Bigg\{ i\partial_{(z)}^\mu \frac{\delta^2}{\delta J^{\mu g}(z) \delta J^{\nu d}(y)} 
- \alpha \partial^{(y)}_\nu M^{-1}_{dg}\bigg[\frac{1}{i}\frac{\delta}{\delta J}\bigg|y,z\bigg] 
-i\alpha g f^{dbc} \frac{\delta}{\delta J_{\nu c}(y)} 
M^{-1}_{bg}\bigg[\frac{1}{i}\frac{\delta}{\delta J}\bigg|y,z\bigg]\qquad 
\\ [2mm] \nn
&& \qquad \qquad 
- i\alpha g \int_C d^4 x J^{\mu a}(x) f^{abc} \frac{\delta^2}{\delta J^{\mu c}(x) \delta J^{\nu d}(y)} 
M^{-1}_{bg}\bigg[\frac{1}{i}\frac{\delta}{\delta J}\bigg|x,z\bigg] \Bigg\} W[J,\chi^*,\chi] =0.\qquad
\ea
Putting $\chi=\chi^*=J=0$ the last term of (\ref{WTI-long-G-1-sti-1}) vanishes and then we obtain
\ba
\label{WTI-long-G-1-sti-2}
&& 
i\partial_{(z)}^\mu \frac{\delta^2 W[J,\chi^*,\chi]}
{\delta J^{\mu g}(z) \delta J^{\nu d}(y)}\bigg|_{\chi=\chi^*=J\equiv 0} 
\\ [2mm] \nn
&& \qquad\qquad\qquad
= \alpha \bigg(\partial^{(y)}_\nu \delta_{db}
-g f^{dbc} \frac{1}{i}\frac{\delta}{\delta J_{\nu c}(y)} \bigg) 
M^{-1}_{bg}\bigg[\frac{1}{i}\frac{\delta}{\delta J}\bigg|y,z\bigg] W[J,\chi^*,\chi] \bigg|_{\chi=\chi^*=J\equiv 0}.
\ea

As one remembers, the definition of the full (interacting) gluon propagator $\mathcal{D}_{\mu\nu}^{ab} (x,y)$ is 
\ba
\label{propagator-gauge-id-W-sti}
(-i)^2\frac{\delta^2 W[J,\chi^*,\chi]}{\delta J^{\mu a}(x) \delta J^{\nu b}(y)}\bigg|_{\chi=\chi^*=J\equiv 0} 
= i \mathcal{D}_{\mu\nu}^{ab} (x,y).
\ea
On the other hand, according to the definition of the matrix $M$ given by (\ref{M-sti}) and the equality (\ref{eq-FG-M}), we have
\ba
\label{MMinv}
-\partial_{(y)}^\nu \Big(\partial^{(y)}_\nu \delta_{ab} - g f^{abc} A^c_\nu(y)\Big) M^{-1}_{bd}[A|y,z]
= \delta_{ad}\delta^{(4)}_C(y,z).
\ea
Moreover, the Green function of free ghost field obeys the equation of motion
\ba
\label{eos-ghost-f}
-\partial_{(y)}^\nu \partial^{(y)}_\nu \Delta_{ad}(y,z)
= \delta_{ad}\delta^{(4)}_C(y,z).
\ea
Combining the equations (\ref{eos-ghost-f}) and (\ref{MMinv}) one observes that
\ba
\label{MM-eos}
\Big(\partial^{(y)}_\nu \delta_{ab} - g f^{abc} A^c_\nu(y)\Big) M^{-1}_{bd}[A|y,z]
= \partial^{(y)}_\nu \Delta_{ad}(y,z),
\ea
which holds up to the function independent of $y$ which is eliminated due to the boundary conditions obeyed by $M^{-1}[A|y,z]$ and $\Delta(y,z)$.

Using the results (\ref{propagator-gauge-id-W-sti}) and (\ref{MM-eos}), we rewrite Eq. (\ref{WTI-long-G-1-sti-2}) as 
\ba
\label{WTI-long-G-2-sti}
\partial_{(z)}^\mu \mathcal{D}_{\mu\nu}^{dg}(z,y) = 
\alpha \; \partial^{(y)}_\nu \Delta_{dg}(y,z),
\ea
which is the Slavnov-Taylor identity for the full interacting gluon propagator. The identity constrains a possible form of the gluon propagator relating it to the free ghost propagator. Locating the time arguments $y_0$ and $z_0$ on the upper or lower branch of the contour shown in Fig.~\ref{Keldysh-contour}, we get the relations for the Green functions of real-time arguments
\ba
\label{STI-long-G-1}
\frac{1}{\alpha} \partial_{(z)}^\mu  \big(\mathcal{D}_{\mu\nu}^{ab}\big)^\lg (z,y)
&=& 
\partial^{(y)}_\nu \big(\Delta_{ab}\big)^\gl (y,z)  ,
\\ [2mm]
\label{STI-long-G-2}
\frac{1}{\alpha} \partial_{(z)}^\mu  \big(\mathcal{D}_{\mu\nu}^{ab}\big)^\ca (z,y)
&=& 
\partial^{(y)}_\nu \big(\Delta_{ab}\big)^\ca (y,z)  .
\ea

In case of translationally invariant system, studied further on, we have
\be
\label{tr-inv-prop}
\mathcal{D}_{\mu\nu}^{dg}(z,y) 
=\mathcal{D}_{\mu\nu}^{dg}(z-y),
~~~~~~~~~~~~~
\Delta_{dg}(y,z)=\Delta_{dg}(y-z),
\ee
and the relation (\ref{WTI-long-G-2-sti}) can be written as
\ba
\label{WTI-long-G-2-sti-inv-0}
\frac{1}{\alpha} \partial_{(z)}^\mu 
\mathcal{D}_{\mu\nu}^{dg}(z-y) = \partial^{(y)}_\nu \Delta_{dg}(y-z).
\ea
Changing the variables in the right side of Eq. (\ref{WTI-long-G-2-sti-inv-0}), we get the equality
\ba
\label{WTI-long-G-2-sti-inv}
-\frac{1}{\alpha} \partial_{(z)}^\mu 
\mathcal{D}_{\mu\nu}^{dg}(z-y) = \partial^{(z)}_\nu \Delta_{dg}(z-y),
\ea
which after the Fourier transformation gets the desired form
\ba
\label{WTI-long-G-2-sti-mom}
-\frac{1}{\alpha} k^\mu 
\mathcal{D}_{\mu\nu}^{dg}(k) = k_\nu \Delta_{dg}(-k),
\ea
 which relates the longitudinal part of the gluon Green function to the free ghost function. Eq.~(\ref{WTI-long-G-2-sti-mom}) also expresses the well-known fact that the longitudinal part of the gluon Green function is not modified by interaction.
 
An attempt to derive the  Slavnov-Taylor identities within the Keldysh-Schwinger formalism was undertaken in \cite{Okano:2001id}. However, there were serious flaws in the derivation. The fields present in the generating functional (\ref{W-KS-YM}) were stated to obey periodic boundary conditions which effectively meant that the density matrix was diagonal. There was no justification for such an assumption. Since the global BRST transformation was used, the density matrix was assumed invariant under the transformation to guarantee the invariance of the generating functional. Again there was no justification for this assumption. It was also overlooked that the ghost contour Green function includes the medium contribution, which is shown explicitly in the subsequent subsection, and consequently the relations, which were obtained, were simply incorrect. 

\subsection{Green functions of the free ghost field}
\label{ssec-ghosts}

Here we write down the Green function of free ghost field which come from the identity (\ref{WTI-long-G-2-sti-mom}). It holds for every component of the contour function $D$ and $\Delta$.  With the explicit expressions of the gluon functions given by Eqs.~(\ref{D->}), (\ref{D-<}), (\ref{D-c}), and (\ref{D-a}), the relation (\ref{WTI-long-G-2-sti-mom}) together with (\ref{STI-long-G-1}), and (\ref{STI-long-G-2}) provide
\ba
\label{Del-g->}
\Delta_{ab}^>(p)
&=& 
- \delta^{ab}  \frac{i\pi}{E_p}\Big[ \delta(E_p-p_0)\big(f_g({\bf p})+1\big) + \delta(E_p+p_0)f_g(-{\bf p}) \Big], 
\\[2mm]
\label{Del-g-<}
\Delta_{ab}^<(p)
&=& 
- \delta^{ab}  \frac{i\pi}{E_p}\Big[ \delta(E_p-p_0)f_g({\bf p}) + \delta(E_p+p_0)\big(f_g(-{\bf p})+1\big) \Big], 
\\[2mm]
\label{Del-g-c}
\Delta_{ab}^c(p)
&=& 
\delta^{ab} \bigg[\frac{1}{p^2+i0^+} - \frac{i\pi}{E_p}
\Big(\delta(p_0-E_p) f_g({\bf p}) + \delta(p_0+E_p) f_g(-{\bf p}) \Big) \bigg], 
\\[2mm]
\label{Del-g-a}
\Delta_{ab}^a(p)
&=& 
- \delta^{ab} \bigg[\frac{1}{p^2-i0^+} 
+ \frac{i\pi}{E_p} \Big(\delta(p_0-E_p) f_g({\bf p}) + \delta(p_0+E_p) f_g(-{\bf p}) \Big)\bigg] .
\ea
As seen, the bosonic gluon distribution function $f_g({\bf p})$, which describes the physical transverse gluons, enters the ghost Green functions.

The relation (\ref{WTI-long-G-2-sti-mom}) provides also the retarded $(+)$, advanced $(-)$, and symmetric $({\rm sym})$ ghost Green functions
\ba
\label{Del-g-ret}
\Delta_{ab}^+ (p) &=& \frac{\delta_{ab}}{p^2 + i \textrm{sgn}(p_0)0^+}, 
\\[2mm]
\label{Del-g-adv}
\Delta_{ab}^- (p) &=& \frac{\delta_{ab}}{p^2 - i \textrm{sgn}(p_0)0^+}, 
\\[2mm]
\label{Del-g-sym}
\Delta_{ab}^{\textrm{sym}}(p)&=& - \delta^{ab} \frac{i\pi}{E_p}
\Big[ \delta(E_p-p_0)\big(2f_g({\bf p})+1\big) + \delta(E_p+p_0)\big(2f_g(-{\bf p})-1\big) \Big], 
\ea
which are used in the subsequent section.

\newpage
\thispagestyle{plain}

\section{Collective excitations}
\label{sec-collective}

Collective behaviour is a fundamental feature of plasma systems which leads to some specific phenomena such as screening, plasma oscillations, instabilities, etc. The mechanism governing these phenomena is based on the fact that a range of the interaction, the screening notwithstanding, is typically longer than average distance between plasma constituents. Accordingly, there are many particles that occur at the range of this effective interaction which consequently makes a motion of the plasma constituents highly correlated. Therefore, while studying properties of a medium it is more relevant to consider a propagation of collective modes, or quasi-particles, than behaviour of elementary particles propagating in the medium. Consequently, it is the spectrum of collective excitations that is a crucial attribute of any statistical system as it provides us with a multitude of information about thermodynamic and transport properties of a system in equilibrium and beyond. Additionally, it meaningfully affects a temporal evolution of nonequilibrium plasmas.

To be more specific, any plasma, which at the beginning is assumed to be a homogeneous and stationary medium where local charges and currents are not present, may be at some point perturbed by, say, a random fluctuation. As a consequence, local charges and currents appear and generate respective fields. While in electromagnetic plasmas electromagnetic fields are excited, in the plasmas with colour charges chromoelectric and chromomagnetic fields appear instead. These fields, in turn, make a feedback reaction to the medium as they interact with constituents of it contributing to their dynamics. Providing the wavelength of the perturbation is bigger than typical distance between plasma constituents, the collective motion of plasma particles is launched engaging all particles occurring in the range of interaction. These changes of charges and currents of high frequency are classically termed as {\it plasma oscillations} or {\it plasma (Langmuir) waves}. In the quantum-mechanical parlance we cope with the collective excitations or quasi-particles, which throughout this paper are used alternately.

Collective excitations are of different nature. Sometimes the interaction with a medium gives rise only to change a mass of a particle which is rather a small modification and then this quasi-particle can be easily identified with a corresponding elementary particle. In other cases such a connection is not so obvious. Therefore, we can speak of different categories of collective modes so that we may specify the excitations which are associated with elementary particles and such excitations which are a pure medium effect. Such a division is, however, not precise as one can study electron or quark quasiparticles and transverse photon or gluon ones whose resemblance to their elementary counterparts is obvious, but there are also longitudinal excitations of a photon or gluon, called plasmons, and fermionic modes, called plasminos, which exist only in a medium. In this thesis the elaboration on all the excitations corresponding to plasma constituents is provided. Having said that, we neglect phonons, which as density fluctuations are present in almost all systems, but are genuine phenomena of statistical systems and thus have not much to do with elementary particles.

Dealing with any equilibrium ultrarelativistic (isotropic) plasma characterized by a temperature $T$ and a coupling constant $g$, which is assumed to be small, one immediately observes some basic properties connected with collective motion of plasma particles. In particular, it can be noticed that there exists an energy scale, called the electric or soft scale, $gT$, which plays a crucial role since it emerges as the Debye mass and typical energy of collective modes. The average energy of an individual particle is then of the order $T$ and the average inter-particle distance in the plasma is in the natural units of the order $T^{-1}$. Then, the particle density is of the order $T^3$. The range of interaction or the screening length is given by the inverse of the Debye mass $m_D^{-1}=r_D \sim (gT)^{-1}$, which defines the Debye sphere wherein the average number of particles is $1/g^3$, which is a big number in case of weakly-coupled systems. Additionally, if $gT$ is much larger than a rest mass of a given particle, this mass can be neglected, which is a case in point within our considerations. 

A collective mode is represented by a dispersion relation $\omega({\bf k})$ which gives mode energy $\omega$ as a function of its momentum ${\bf k}$. In mathematical language the dispersion law appears as a real part of the pole of a respective propagator. When the imaginary part of the frequency of a given mode is equal to zero, $\Im\omega=0$ then its amplitude is constant as a function of time and such a mode is stable. Supposing the imaginary part of the frequency is negative the mode is damped, that is, its amplitude exponentially decreases in time as $e^{\Im \omega t}$. Thus, the next characteristic of a collective mode is the decay (damping) rate $\gamma({\bf k})$ which corresponds to an imaginary part of the pole of a propagator. The stable or damped modes are typical of equilibrium plasmas and then the decay rate is of the order $g^2T$, which is called the magnetic or ultrasoft scale. Incidentally, at such a scale the non-perturbative phenomena  (confinement) in the transverse gluon propagator emerge which makes a qualitative difference in properties of the quark-gluon medium against electromagnetic one. As a result, these systems are arguably rather different at this scale. Returning to quasi-particles, one can also encounter unstable modes of the amplitude growing in time and such modes appear when $\Im \omega>0$. The unstable modes are present in a plasma out of equilibrium and then they strongly influence a dynamics of the system, especially, they can lead the plasma to a faster thermalisation. 

As already mentioned, $gT$ is the energy scale at which collective modes exist and of such order is the wavelength of the mode. The wavelenght is, in turn, much bigger than a typical inter-particle spacing in the plasma. In the momentum space it means that the wavevector of the mode is much smaller than the momentum of a plasma constituent. Therefore one deals with soft collective excitations of hard plasma particles. It is the hard-loop approach that combines such the physics with diagrammatic methods of field theories. 

The hard-loop approach is a practical tool to describe plasma systems in question in a gauge invariant way which is free of infrared divergences, see the reviews \cite{Thoma:1995ju,Blaizot:2001nr,Litim:2001db,Kraemmer:2003gd}. Initially the approach was developed within the thermal field theory \cite{Braaten:1989mz,Taylor:1990ia,Braaten:1990az} but it was soon realized that it can be formulated in terms of quasiclassical kinetic theory \cite{Blaizot:1993be,Kelly:1994dh}. The plasma systems under consideration were assumed to be in thermodynamical equilibrium but the methods could be naturally generalised to plasmas out of equilibrium \cite{Carrington:1997sq,Mrowczynski:2000ed,Mrowczynski:2004kv}. The hard-loop approach is used in this chapter to obtain collective modes in the plasma systems under study.

As one may wonder how much a given plasma characteristic is different for different plasma systems we study in this chapter collective excitations of different plasma systems and compare them to each other. It has been known for a long time that the self-energies of gauge bosons, which dicate the form of collective modes, in the long-wavelength limit are of the same structure for QED and QCD plasmas \cite{Weldon:1982aq}. Consequently, the collective excitations and many other characteristics are the same, or almost the same, in the two plasma systems \cite{Mrowczynski:2007hb}. However, the situation seems likely to change when a supersymmetry is implemented. Accordingly, the aim of this section is to check how much collective modes of supersymmetric plasmas are different from their counterparts of non-supersymmetric systems.

We start our consideration with the ${\mathcal N}=1$ SUSY QED plasma. The dispersion equations are provided and then the self-energies of all fields occurring in the plasma are computed in the long-wavelength limit. Special attention is paid to the self-energy of a photino which, as a fermionic superpartner of a photon, is also studied in the context of instabilities. While elaborating on the supersymmetric electromagnetic plasma we also refer to self-energies of other plasmas which are governed by the electromagnetic interaction. Next we perform the same analysis for the ${\mathcal N}=4$ super Yang-Mills plasma. In Sec. \ref{ssec-eff-action} the effective action is constructed and emergence of the universality of self-energies and of the hard-loop action is discussed. Some physical aspects of the universality are also considered. The findings presented here are published in our original papers \cite{Czajka:2010zh,Czajka:2012gq,Czajka:2014gaa}. Finally, we describe in short the collective excitations which result from the respective self-energies.

\subsection{${\cal N} =1$ SUSY QED plasma}
\label{ssec-susy-qed}

Here we study the self-energies of all fields occuring in the system described by $\mathcal{N}=1$ SUSY QED\footnote{This subsection is based on our work published as \cite{Czajka:2010zh}.}. The plasma is assumed to be homogeneous but the momentum distribution is, in general, different from equilibrium one. Therefore, we use the the Keldysh-Schwinger formalism and the free Green functions, which have been derived in Sec. \ref{sec-KS-form}. Since the plasma is assumed to be homogeneous, all the Green functions, which are going to be used here, are independent of the position $X$. The computation is performed within the hard-loop approach. The plasma is assumed to be ultrarelativistic and thus masses of electrons and selectrons are neglected. We also assume that the system is electrically neutral and unpolarised and that the distribution function of electrons $f_e({\bf p})$ equals the distribution function of positrons $\bar f_e({\bf p})$. Analogous equality is assumed for selectrons: $f_s({\bf p}) = \bar f_s({\bf p})$. The additional assumption is that both left and right selectrons are described by the same function $f_s({\bf p})$.

\subsubsection{Dispersion equations}
\label{sssec-dis-eqs-sqed}

We start our consideration with writing down the dispersion equations of quasi-photons, quasi-electrons, quasi-photinos, and quasi-selectrons which determine dispersion relations of respective quasi-particle excitations.

\begin{center}
{\bf Photons}
\end{center}

The equation of motion of the mean electromagnetic field $A^{\mu}(k)$ in momentum space is of the form
\ba
\label{eom-A}
\Big[ k^2 g^{\mu \nu} -k^{\mu} k^{\nu} - \Pi^{\mu \nu}(k) \Big]
A_{\nu}(k) = 0 ,
\ea
where $\Pi^{\mu \nu}(k)$ is the retarded polarisation tensor which carries the information on the interaction of an electromagnetic field with the plasma. The field $A^{\mu}(k)$ should be understood as an expectation value of the gauge field operator or as a classical field.  This is justified as we consider long-wavelength collective modes. Since Eq. (\ref{eom-A}) is a homogeneous equation it has a solution under the condition that 
\ba
\label{dis-eq-A}
{\rm det}\Big[ k^2 g^{\mu \nu} -k^{\mu} k^{\nu} - \Pi^{\mu \nu}(k) \Big]
 = 0,
\ea
which constitutes the general photon dispersion equation. The solutions to Eq. (\ref{dis-eq-A}) are given as functions of energy of wavevector, $\omega({\bf k})$, which are just dispersion relations of collective modes. 

The dispersion equation (\ref{dis-eq-A}) may be obtained equivalently starting from the equation of motion of a propagator of the electromagnetic field. Then, the dispersion relations are given by positions of poles of the propagator. 

Due to the transversality of the polarisation tensor, $k_\mu \Pi^{\mu \nu}(k) =0$, which is required by gauge covariance, not all components of $\Pi^{\mu \nu}$ are independent from each other and consequently the dispersion equation (\ref{dis-eq-A}) can be much simplified by expressing the polarisation tensor through the dielectric tensor $\varepsilon^{ij}(k)$.

\begin{center}
{\bf Electrons}
\end{center}

The electron field $\psi (k)$ obeys the following equation
\ba
\label{eom-psi}
\Big[ k\sla  - \Sigma (k)  \Big] \psi (k) =0 , 
\ea
where $\Sigma (k)$ is the retarded electron self-energy and $k\sla \equiv k^\mu \gamma_\mu$. The dispersion equation is then given as
\ba
\label{dis-el}
{\rm det}\Big[ k\sla  - \Sigma (k) \Big]  = 0 . 
\ea
The same dispersion equation is found starting with the equation of motion of the electron propagator. Further on we assume that the spinor structure of $\Sigma(k)$ is 
\ba
\label{structure-electron1} 
\Sigma (k) = \gamma^{\mu} \Sigma_{\mu}(k) .
\ea 
Then, plugging the expression (\ref{structure-electron1}) in Eq.~(\ref{dis-el}) and computing the determinant as explained in Appendix 1 of \cite{Mrowczynski:1992hq}, we get 
\ba
\label{dis-el-2} 
\Big[\big( k^{\mu} - \Sigma^{\mu}(k) \big) \big(k_{\mu} - \Sigma_{\mu}(k) \big)\Big]^2  = 0 . 
\ea

\begin{center}
{\bf Photinos}
\end{center}

The photino equation of motion is
\ba
\Big[ k\sla  - \tilde \Pi (k)  \Big] \Lambda (k) =0, 
\ea
where $\Lambda$ is the photino Majorana bispinor and $\tilde \Pi$ is the respective retarded self-energy. The dispersion equation is then
\ba
\label{dis-photino-1} 
{\rm det}\Big[ k\sla  - \tilde \Pi (k) \Big] = 0 . 
\ea 
As previously this equation can be obtained from the equation of motion of the photino propagator. Since the expected spinor structure of $\tilde \Pi(k)$ is analogous to that given by Eq.~(\ref{structure-electron1}), the dispersion equation coincides with Eq.~(\ref{dis-el-2}).

\begin{center}
{\bf Selectrons}
\end{center}

The selectron fields $\phi_L (k)$ and $\phi_R (k)$ obey the Klein-Gordon equation
\ba
\label{dis-eq-selectron}
\Big[ k^2  + \tilde\Sigma_{L,R} (k) \Big] \phi_{L,R} (k) =0 ,
\ea
where $\tilde\Sigma_{L,R} (k)$ is the retarded self-energy of left or right selectrons. The equation of motion written for the selectron propagator is analogous and will lead us to the same dispersion equation, which is
\ba 
\label{dis-selectron1}
k^2 + \tilde\Sigma_{L,R} (k) = 0 .
\ea

\subsubsection{Self-energies}
\label{sssec-sqed-se}

In this section we compute the self-energies which enter the dispersion equations (\ref{dis-eq-A}), (\ref{dis-el}), (\ref{dis-photino-1}), and (\ref{dis-selectron1}). The self-energies can be computed, in general, in several ways but here we use a diagramatic technique. Therefore, the self-energies are calculated using the free Green functions found in the Keldysh-Schwinger representation in Sec.~\ref{sec-KS-form}. As already mentioned, the hard-loop approximation is applied to the self-energies, as we are interested in collective excitations occurring in the plasma. Within the hard-loop approach the external momentum $k$, which corresponds to the wavevector of a given excitation, is much smaller than the internal momentum $p$ that flows along the loop and is carried by a plasma constituent.

\newpage
\begin{center}
{\bf Polarisation tensor}
\end{center}
\label{photon-se-sqed}

\begin{figure}[!h]
\centering
\includegraphics*[width=0.7\textwidth]{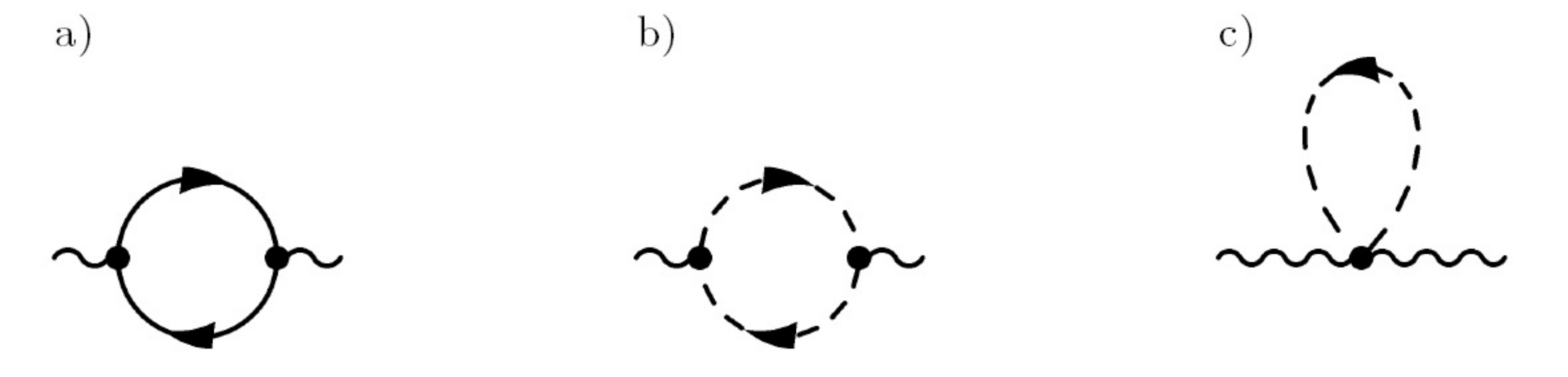}
\caption{Contributions to the photon self-energy. }
\label{fig-photon}
\end{figure}

The polarisation tensor $\Pi^{\mu \nu}$ can be defined by means of the Dyson-Schwinger equation
\ba
\label{DSE-photon}
i{\cal D}^{\mu \nu} (k) = i D^{\mu \nu} (k)
+ i D^{\mu \rho}(k) \, i\Pi_{\rho \sigma}(k)  \, i{\cal D}^{\sigma \nu}(k) ,
\ea
where ${\cal D}^{\mu \nu}$ and $D^{\mu \nu}$ is the interacting and free photon propagator, respectively. We limit the calculation of $\Pi$ only to the one-loop level. The lowest order contributions to $\Pi^{\mu \nu}$ are given by three diagrams shown in Fig.~\ref{fig-photon}. The solid, wavy and dashed lines denote, respectively, the electron, photon and selectron fields.

\begin{center}
{\it  Electron loop}
\end{center}
\label{electron-loop-sqed}

Applying the Feynman rules of the Keldysh-Schwinger formalism, which are discussed in, {\it e.g.}, Sec.~8 of \cite{Mrowczynski:1992hq}, the contribution to the contour polarisation tensor from the electron loop corresponding to the graph from Fig.~\ref{fig-photon}a is immediately written down in the coordinate space as
\ba
\label{pi-1a}
i {_{(a)}\Pi_{\mu \nu}} (x,y) 
= (-1)(-ie)^2 {\rm Tr} \big[\gamma _\mu iS (x,y) \gamma _\nu iS(y,x)\big] ,
\ea
where the factor $(-1)$ occurs due to the fermion loop and $S(x,y)$ is the contour free electron Green function. The trace in the formula (\ref{pi-1a}) is taken over the spinor indices. We are interested in the retarded polarisation tensor which is found due to the relation
\ba
\label{pi-1a-1}
\Pi_{\mu \nu}^{\pm} (x,y) = \Pi_\delta(x) \delta^{(4)}(x-y) 
\pm \Theta(\pm x_0) \Big( \Pi_{\mu \nu}^> (x,y) - \Pi_{\mu \nu}^< (x,y) \Big),
\ea
where the first term corresponds to the contribution coming from an one-point tadpole diagram. The polarisation tensors $\Pi^>$ and $\Pi^<$ are extracted from the contour tensor (\ref{pi-1a}) by locating the time arguments on the respective branches of the contour. Then, we get
\ba
\label{pi-1aa}
i {_{(a)}\Pi_{\mu \nu}^{\lg}} (x,y) 
= (-1)(-ie)^2 {\rm Tr} \Big[\gamma _\mu iS^{\lg} (x,y) \gamma _\nu iS^{\gl}(y,x) \Big] .
\ea
Since we consider here the translationally invariant system, the two-point functions depend on $x$ and $y$ only through their difference $x-y$. Accordingly, we can put $y=0$ and replace $S(x,y)$ by $S(x)$ and $S(y,x)$ by $S(-x)$. It gives
\ba
 {_{(a)}\Pi_{\mu \nu}^{\lg}} (x) 
= -i e^2 {\rm Tr} \Big[\gamma _\mu S^{\lg} (x) \gamma _\nu S^{\gl}(-x)\Big] .
\ea
Due to the relations
\ba
S^{\pm} (x) & = & \pm \Theta(\pm x_0) \Big(S^> (x) - S^< (x) \Big), 
\\ [2mm]
S^{\rm sym} (x) & = & S^> (x) + S^< (x)
\ea
the retarded polarisation tensor $_{(a)}\Pi_{\mu \nu}^+ (x)$ is found as
\be
\label{Pi-x-1a}
_{(a)}\Pi_{\mu \nu}^+ (x) 
= i \frac{e^2}{2} {\rm Tr} \Big[\gamma _\mu S^+(x) \gamma _\nu S^{\rm sym}(-x)
+ \gamma _\mu S^{\rm sym}(x) \gamma _\nu S^-(-x) \Big].
\ee
In the momentum space it reads
\be
\label{Pi-k-e-1a}
_{(a)}\Pi^{\mu \nu}(k) 
= i\frac{e^2}{2} \int \frac{d^4p}{(2\pi )^4}
{\rm Tr} \Big[\gamma^\mu S^+(p+k)\gamma ^\nu S^{\rm sym}(p)
+ \gamma ^\mu S^{\rm sym}(p) \gamma ^\nu S^-(p-k) \Big].
\ee
The index `$+$' of the polarisation tensor from Eq.~(\ref{Pi-k-e-1a}) has been dropped. Further on, we will consider only the retarded self-energies and thus the index `$+$' will not be used.

Plugging the functions $S^{+}$ (\ref{S-ret}), $S^{-}$ (\ref{S-adv}) and $S^{\rm sym}$ (\ref{S-sym}) in Eq.~(\ref{Pi-k-e-1a}) and performing the integration over $p_0$ one finds
\ba
\label{Pi-k-e-3a}
_{(a)}\Pi^{\mu \nu}(k) 
&=&
-\frac{e^2}{4} \int \frac{d^3p}{(2\pi )^3} \, \frac{2f_e({\bf p}) -1}{E_p}
\\ [2mm] \nn
&&
\times {\rm Tr} \bigg[ \frac{ \gamma^\mu (p\sla + k\sla)\gamma ^\nu  p\sla
+ \gamma ^\mu p\sla \gamma ^\nu(p\sla + k\sla)}{(p+k)^2 + i\, {\rm sgn}\big((p+k)_0\big)0^+}
+ \frac{\gamma ^\mu p\sla \gamma ^\nu(p\sla -k\sla) + \gamma^\mu (p\sla - k\sla)\gamma^\nu  p\sla}
{(p-k)^2 - i\, {\rm sgn}\big((p-k)_0\big)0^+} \bigg],
\ea
where $p^\mu \equiv (E_p, {\bf p})$ with $E_p \equiv |{\bf p}|$. In the formula (\ref{Pi-k-e-3a}) the momentum ${\bf p}$ has been changed into  $-{\bf p}$ in the positron contribution. It was also used that $f_e({\bf p}) = \bar f_e({\bf p}) $. 

Computing the traces of gamma matrices and taking into account that $p^2 =0$, one finds
\ba
\label{Pi-k-e-4a}
_{(a)}\Pi^{\mu \nu}(k) 
&=&
- 2e^2  \int \frac{d^3p}{(2\pi )^3} \, \frac{2f_e({\bf p}) -1}{E_p}
\\ [2mm] \nn
&& 
\times \bigg[\frac{2p^\mu p^\nu  + k^\mu p^\nu + p^\mu k^\nu - g^{\mu \nu} (k \cdot p)  }
{(p+k)^2 + i\, {\rm sgn}\big((p+k)_0\big)0^+}
+ \frac{2p^\mu p^\nu  -  k^\mu p^\nu - p^\mu k^\nu + g^{\mu \nu} (k \cdot p) }
{(p-k)^2 - i\, {\rm sgn}\big((p-k)_0\big)0^+} \bigg] .
\ea

We are interested in collective modes which occur when wavelength of a quasi-particle is much bigger than a characteristic interparticle distance in the plasma. Thus, we look for the polarisation tensor at $k^\mu \ll p^\mu$ which is the condition of the hard-loop approximation for anisotropic systems \cite{Mrowczynski:2000ed,Mrowczynski:2004kv}. The approximation is implemented by observing that 
\ba
\label{HLA-plus}
&&
\frac{1}{(p+k)^2 + i0^+} + \frac{1} {(p-k)^2 - i0^+}
\\ [2mm] \nn
&& \qquad\qquad\qquad
=\frac{2k^2}{(k^2)^2 - 4 (k\cdot p)^2 - i {\rm sgn}(k\cdot p) 0^+}
\approx -\frac{1}{2}  \frac{k^2}{(k\cdot p + i 0^+)^2} ,\qquad \;
\\ [2mm]
\label{HLA-minus}
&&
\frac{1}{(p+k)^2 + i0^+} - \frac{1} {(p-k)^2 - i0^+}
\\ [2mm] \nn
&& \qquad\qquad\qquad
=\frac{-4(k \cdot p)}{(k^2)^2 - 4 (k\cdot p)^2 - i {\rm sgn}(k\cdot p) 0^+}
\approx \frac{k\cdot p}{(k\cdot p + i 0^+)^2}. 
\ea
We note that $(p+k)_0 > 0$ and $(p-k)_0 > 0$ for $p^\mu \gg k^\mu$. With the above formulas, Eq.~(\ref{Pi-k-e-4a}) gives
\ba
\label{Pi-k-e-finala}
_{(a)}\Pi^{\mu \nu}(k) 
= 2e^2  \int \frac{d^3p}{(2\pi )^3} \, \frac{2f_e({\bf p})-1}{E_p} \,
\frac{k^2 p^\mu p^\nu  -  \big(k^\mu p^\nu + p^\mu k^\nu 
- g^{\mu \nu} (k \cdot p) \big) (k \cdot p)}{(k\cdot p + i 0^+)^2} ,
\ea
which is the well-known form of the polarisation tensor of photons and of gluons in ultrarelativistic plasmas, see {\it e.g.} the reviews \cite{Mrowczynski:2007hb,Blaizot:2001nr}. As seen from the expression (\ref{Pi-k-e-finala}), $_{(a)}\Pi^{\mu \nu}(k)$ is symmetric with respect to the Lorentz indices 
\ba
_{(a)}\Pi^{\mu \nu}(k) = {_{(a)}}\Pi^{\nu \mu}(k)
\ea
and transverse 
\ba
k_\mu {_{(a)}}\Pi^{\mu \nu}(k) = 0
\ea 
as required by the gauge invariance.

When $f_e({\bf p})$ vanishes the polarisation tensor (\ref{Pi-k-e-finala}) is still nonzero. It is infinite and represents vacuum contribution, which may be subtracted from the formula (\ref{Pi-k-e-finala}) as we are interested in medium effects.

\begin{center}
{\it  Selectron loop}
\end{center}
\label{selectron-loop-sqed}

The contribution to the polarisation tensor coming from the selectron loop depicted in Fig.~\ref{fig-photon}b is given by an appropriately modified Eq.~(\ref{Pi-k-e-1a}), that is
\ba
\label{Pi-k-s-1a}
_{(b)}\Pi^{\mu \nu}(k) 
&=& 
- i\frac{e^2}{2}
 \int \frac{d^4p}{(2\pi )^4}
\Big[(2p+k)^\mu (2p+k)^\nu \Delta^+(p+k) \Delta^{\rm sym}(p) 
\\ [2mm] \nn
&& \qquad\qquad\qquad\qquad
+ (2p-k)^\mu (2p-k)^\nu \Delta^{\rm sym}(p) \Delta^-(p-k) \Big],
\ea
where $\Delta^\pm$ and $\Delta^{\rm sym}$ are free Green functions of the scalar field. The sign is different than in Eq.~(\ref{Pi-k-e-1a}) as we deal here with the boson not the fermion loop. Substituting the functions $\Delta^{+}$, $\Delta^{-}$, and $\Delta^{\rm sym}$ given by Eqs.~(\ref{Del-ret}), (\ref{Del-adv}), and (\ref{Del-sym}), respectively, into Eq.~(\ref{Pi-k-s-1a}), one finds
\ba
\label{Pi-k-s-3a}
_{(b)}\Pi^{\mu \nu}(k) 
&=&
-\frac{e^2}{2}  \int \frac{d^3p}{(2\pi )^3} \,\frac{2f_s({\bf p})+1}{E_p} \,
\bigg[  \frac{(2p+k)^\mu (2p+k)^\nu}{(p+k)^2 + i\, {\rm sgn}\big((p+k)_0\big)0^+} 
\\ [2mm] \nn
&& \qquad\qquad\qquad\qquad\qquad\qquad
+ \frac{(2p-k)^\mu  (2p-k)^\nu}{(p-k)^2 - i\, {\rm sgn}\big((p-k)_0\big)0^+}\bigg],
\ea
where the change ${\bf p} \rightarrow -{\bf p}$ was made in the antiselectron part and we assumed that $\bar f_s({\bf p}) = f_s({\bf p})$. After adopting the hard-loop approximation, Eq.~(\ref{Pi-k-s-3a}) gives
\ba
\label{Pi-k-s-4a}
_{(b)}\Pi^{\mu \nu}(k) =  e^2  \int \frac{d^3p}{(2\pi )^3} \, \frac{2f_s({\bf p})+1}{E_p} \,
 \frac{k^2 p^\mu p^\nu - (p^\mu k^\nu + k^\mu p^\nu)(k \cdot p)}{(k \cdot p + i0^+)^2}.
\ea

\newpage
\begin{center}
{\it  Selectron tadpole}
\end{center}
\label{selectron-tadpole-sqed}

The contribution to the polarisation tensor coming from the selectron tadpole depicted in Fig.~\ref{fig-photon}c is
\ba
\label{Pi-k-s-t-1a}
i {_{(c)}}\Pi^{\mu \nu}(k) 
= 2i e^2 g^{\mu \nu} \int \frac{d^4p}{(2\pi )^4}  i\Delta^<(p) .
\ea 
Substituting the function $\Delta^<$ given by Eq.~(\ref{Del-<}) into Eq.~(\ref{Pi-k-s-t-1a}), one finds 
\ba
\label{Pi-k-s-t-2a} 
_{(c)}\Pi^{\mu \nu}(k) 
=  e^2 g^{\mu \nu} \int \frac{d^3p}{(2\pi )^3} \,\frac{2f_s({\bf p})+1}{E_p}  ,
\ea
where the equality $\bar f_s({\bf p}) = f_s({\bf p})$ was assumed.

We get the complete contribution from a single selectron field to the polarisation tensor by summing the contributions from the selectron loop and the selectron tadpole. Thus, one finds
\ba
\label{Pi-k-s-totala}
_{(b+c)}\Pi^{\mu \nu}(k) 
=  e^2  \int \frac{d^3p}{(2\pi )^3} \, \frac{2f_s({\bf p})+1}{E_p} \,
\frac{k^2 p^\mu p^\nu - \big(p^\mu k^\nu + k^\mu p^\nu - g^{\mu \nu} (k \cdot p)\big)(k \cdot p)}
{(k \cdot p + i0^+)^2}.
\ea
As seen, it is of exactly the same form as the electron contribution given by Eq.~(\ref{Pi-k-e-finala}) -- it is symmetric and transversal. Actually, the expression (\ref{Pi-k-s-totala}) is the polarisation tensor of scalar QED, which for equilibrium plasma was discussed in {\it e.g.} \cite{Kraemmer:1994az} using the imaginary-time formalism. Since there are two selectron fields in ${\cal N} =1$ SUSY QED, the expression (\ref{Pi-k-s-totala}) should be multiplied by a factor of 2 to get the complete selectron contribution to the polarisation tensor.

\begin{center}
{\it  Final result}
\end{center}
\label{fr-photon-sqed}

Combining the electron (\ref{Pi-k-e-finala}) and selectron (\ref{Pi-k-s-totala}) contributions, we get the final expression of the polarisation tensor
\ba
\label{Pi-k-finala}
\Pi^{\mu \nu}(k) 
= 4e^2  \int \frac{d^3p}{(2\pi )^3} \, \frac{f_e({\bf p})+f_s({\bf p})}{E_p} \,
\frac{k^2 p^\mu p^\nu - \big(p^\mu k^\nu + k^\mu p^\nu - g^{\mu \nu} (k \cdot p)\big)(k \cdot p)}
{(k \cdot p + i0^+)^2}.
\ea
As seen, $\Pi^{\mu \nu}(k)$ vanishes in the vacuum limit when $f_e , f_s \rightarrow 0$. This is a nice feature of supersymmetric plasma. In the nonsupersymmetric counterpart, the polarisation tensor is given by Eq.~(\ref{Pi-k-e-finala}) where, as already discussed, the vacuum contribution diverges and it requires a special treatment. Up to the vacuum contribution, the polarisation tensor of supersymmetric plasma and of its non-supersymmetric counterpart has the same structure. 

\begin{center}
{\bf Electron self-energy}
\end{center}
\label{electron-se-sqed}

\begin{figure}[!h]
\centering
\includegraphics*[width=0.45\textwidth]{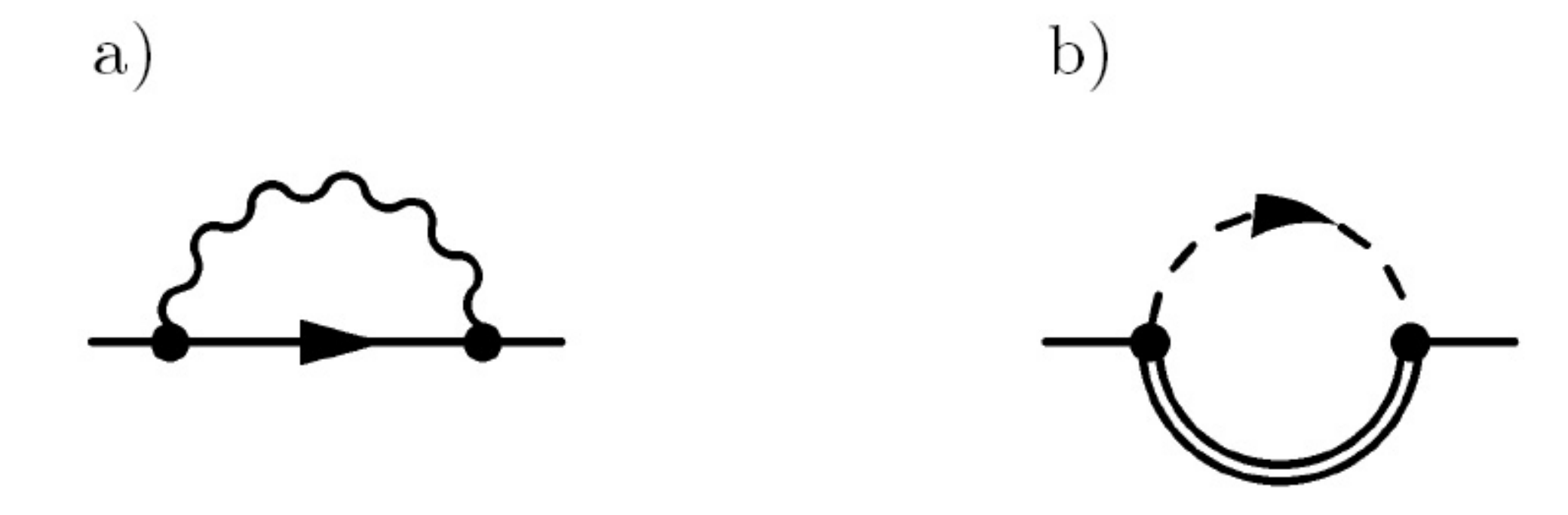}
\caption{Contributions to the electron self-energy. }
\label{fig-electron}
\end{figure}

The electron self-energy $\Sigma$ can be defined by means of the Dyson-Schwinger equation 
\ba
\label{DSE-electron}
i{\cal S} (k) = i S (k) + i S(k) \, \big(-i\Sigma (k) \big) \, i{\cal S}(k) , 
\ea 
where ${\cal S}$ and $S$ is the interacting and free electron propagator, respectively. The lowest order contributions to $\Sigma$ are given by two diagrams shown in Fig.~\ref{fig-electron}. The solid, wavy, dashed and double-solid lines denote, respectively, the electron, photon, selectron and photino fields.

\begin{center}
{\it  Electron-photon loop}
\end{center}
\label{e-p-loop-sqed}

The contribution to the electron self-energy corresponding to the graph depicted in Fig.~\ref{fig-electron}a is given by an appropriately modified Eq.~(\ref{Pi-k-e-1a}) that is 
\ba
\label{Si-k-a-1aa}
-i {_{(a)} \Sigma} (k) = - \frac{e^2}{2} \int \frac{d^4p}{(2\pi )^4}
\Big[\gamma^\mu iS^+(p+k) \gamma^\nu iD^{\rm sym}_{\mu \nu}(p)
+ \gamma^\mu  iS^{\rm sym}(p)  \gamma^\nu iD^-_{\mu \nu}(p-k) \Big],
\ea
where $S^+$ and $S^{\rm sym}$ are electron free propagators and $D_{\mu\nu}^-$ and $D_{\mu\nu}^{\rm sym}$ are photon free propagators. Plugging the functions $D^{-}_{\mu \nu}$, $D^{\rm sym}_{\mu \nu}$ and $S^{+}$, $S^{\rm sym}$ given by Eqs.~(\ref{D-adv}), (\ref{D-sym}), (\ref{S-ret}), and (\ref{S-sym}), respectively, in Eq.~(\ref{Si-k-a-1aa}), one finds
\ba
\label{Si-k-a-3a} 
_{(a)} \Sigma (k) 
\!\!\!&=& \!\!\!
\frac{e^2}{2} \int \frac{d^3p}{(2\pi )^3 E_p} 
\\ [2mm] \nn
&& \!\!\!\!\!\!
\times \bigg\{ \bigg[ \frac{p\sla+k\sla }{(p+k)^2 + i\, {\rm sgn}\big((p+k)_0\big)0^+}
- \frac{p\sla - k\sla}{(p-k)^2 - i\, {\rm sgn}\big((p-k)_0\big)0^+}\bigg]  \big(2 f_\gamma ({\bf p}) +1\big)
\\ [2mm] \nn
&& \!\!\!\!\!\!
-\bigg[ \frac{p\sla }{(p-k)^2 - i\, {\rm sgn}\big((p-k)_0\big)0^+} 
- \frac{p\sla }{(p+k)^2 + i\, {\rm sgn}\big((p+k)_0\big)0^+} \bigg] \big(2 f_e({\bf p}) - 1\big)  \bigg\}.
\ea
Applying the hard-loop approximation, one obtains
\ba
\label{Si-k-a-4}
_{(a)} \Sigma (k) &=& e^2  \int \frac{d^3p}{(2\pi )^3} \,
 \frac{ f_\gamma ({\bf p}) +  f_e ({\bf p})}{E_p}  \, \frac{p\sla}{k\cdot p + i 0^+} ,
\ea
which is the well-known form of the self-energy of electrons and of quarks in ultrarelativistic plasmas, see {\it e.g.} the review \cite{Blaizot:2001nr}.

\begin{center}
{\it  Selectron-photino loop}
\end{center}
\label{s-po-loop-sqed}

Since there are two selectron fields in  ${\cal N} =1$ SUSY QED there are two contributions to the electron self-energy corresponding to the graph depicted in Fig.~\ref{fig-electron}b. The first one corresponding to the left selectron field equals 
\ba
\label{Si-k-b-1a}
i {_{(bL)}\Sigma} (k) = e^2 \int \frac{d^4p}{(2\pi )^4} \Big[ i \Delta^+(p+k) P_L i S^{\rm sym} (p) P_R
+ i\Delta^{\rm sym} (p)  P_L i S^-(p-k) P_R \Big],
\ea
where $P_L$ and $P_R$ are the projection operators defined by (\ref{projection-op}), $\Delta^+$ and $\Delta^{\rm sym}$ are free Green functions of the selectron field, which is a scalar one, and $S^-$ and $S^{\rm sym}$ are free Green functions of the photino field, which is a fermion one. Subsequently, we substitute the functions $\Delta^+$, $\Delta^{\rm sym}$ and $S^-$, $S^{\rm sym}$ given by Eqs.~(\ref{Del-ret}), (\ref{Del-sym}), (\ref{S-adv}), and (\ref{S-sym}), respectively, into Eq.~(\ref{Si-k-b-1a}). One needs to remember that the distribution functions of selectrons $f_s$ and photinos $f_{\tilde \gamma}$ should be implemented in these formulas. Applying the hard-loop approximation, one finds
\ba
\label{Si-k-b-3}
_{(bL)} \Sigma (k) = e^2  \int \frac{d^3p}{(2\pi )^3} \,
\frac{ f_{\tilde \gamma} ({\bf p}) +  f_s ({\bf p})}{E_p} \, \frac{P_L  p\sla P_R}{k\cdot p + i 0^+}.
\ea

Computing the contribution  corresponding to the graph depicted in Fig.~\ref{fig-electron}b with the right selectron field, we get
\ba
\label{Si-k-b-3-R}
_{(bR)} \Sigma (k) = e^2  \int \frac{d^3p}{(2\pi )^3} \,
 \frac{ f_{\tilde \gamma} ({\bf p}) +  f_s ({\bf p})}{E_p} \, \frac{P_R p\sla P_L}{k\cdot p + i 0^+} .
\ea
Because $P_L  p\sla P_R + P_R p\sla P_L = p\sla$, the total contribution given by the graph from Fig.~\ref{fig-electron}b equals
\ba
\label{Si-k-b-total}
_{(b)} \Sigma (k) = e^2 \int \frac{d^3p}{(2\pi )^3} \,
\frac{ f_{\tilde \gamma} ({\bf p}) +  f_s ({\bf p})}{E_p} \, \frac{ p\sla }{k\cdot p + i 0^+} .
\ea

\begin{center}
{\it  Final result}
\end{center}
\label{fr-electron-sqed}

The sum of expressions (\ref{Si-k-a-4}) and (\ref{Si-k-b-total}) gives the complete electron self-energy
\ba
\label{Si-k-finala}
\Sigma (k) = e^2  \int \frac{d^3p}{(2\pi )^3} \;
 \frac{ f_\gamma ({\bf p}) +  f_e ({\bf p}) + f_{\tilde \gamma} ({\bf p}) + f_s ({\bf p})}{E_p}  \,
\frac{ p\sla }{k\cdot p + i 0^+} .
\ea
As seen, the electron self-energy has the same structure for the supersymmetric plasma and for its nonsupersymmetric counterpart. 

\begin{center}
{\bf Photino self-energy}
\end{center}
\label{photino-se-sqed}

\begin{figure}[!h]
\centering
\includegraphics*[width=0.17\textwidth]{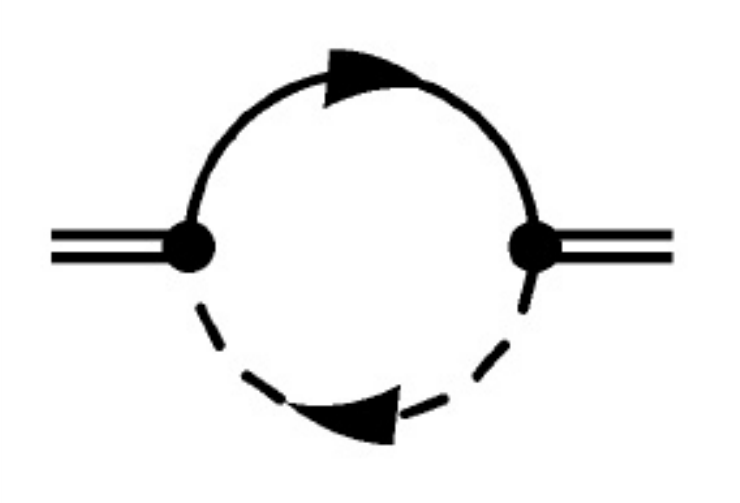}
\caption{Contribution to the photino self-energy.}
\label{fig-photino}
\end{figure}

The photino self-energy $\tilde \Pi$ can be also defined by means of the Dyson-Schwinger equation 
\ba
i{\cal S} (k) = i S(k) + i S(k) \, \big(- i\tilde \Pi(k) \big) \, i {\cal S}(k) ,
\ea 
where ${\cal S}$ and $S$ is the interacting and free photino propagator, respectively. The lowest order contribution to $\tilde \Pi$ is given by the diagram shown in Fig.~\ref{fig-photino}. The solid, dashed and double-solid lines denote, respectively, the electron, selectron and photino fields.  

Since there are two selectron fields in ${\cal N} = 1 $ SUSY QED there are two contributions represented by the diagram corresponding to the left and right selectrons. An appropriate modification of Eq.~(\ref{Pi-k-e-1a}) leads us to
\ba
\label{Pi-k-f-1} 
i {_{(L)}} \tilde \Pi(k) = e^2 \int \frac{d^4p}{(2\pi )^4} 
\Big[P_R iS^+(p+k)P_L i \Delta^{\rm sym}(p) + P_R i S^{\rm sym}(p) P_L i \Delta^-(p-k) \Big], 
\ea
where the contribution from left selectrons is taken into account and $P_L$ and $P_R$ are the projection operators defined by (\ref{projection-op}). Now one substitutes the functions $S^+$, $S^{\rm sym}$ and $\Delta^-$, $\Delta^{\rm sym}$ given by Eqs.~(\ref{S-ret}), (\ref{S-sym}), (\ref{Del-adv}), and (\ref{Del-sym}), respectively, into Eq.~(\ref{Pi-k-f-1}). The propagators should be taken with the electron and selectron distribution functions, respectively. Performing the integration over $p_0$ and changing ${\bf p}$ into $-{\bf p}$ in the terms representing antiparticles, we obtain
\ba 
\label{Pi-k-f-4} 
 {_{(L)}} \tilde \Pi(k)
\!\!\! &=& \!\!\!
\frac{e^2}{2} \int \frac{d^3p}{(2\pi )^3E_p} 
\\ \nn  
&& \!\!\!\!
\times \bigg\{\bigg[\frac{P_R (p\sla + k\sla) P_L } {(p+k)^2 + i\, {\rm sgn}\big((p+k)_0\big)0^+}
-\frac{P_R (p\sla -k\sla) P_L} {(p-k)^2 - i\,{\rm sgn}\big((p-k)_0\big)0^+}\bigg] \big( 2f_s({\bf p})+1\big)
\\ \nn 
&& \!\!\!\!
+\bigg[\frac{P_R p\sla P_L } {(p+k)^2 +i \, {\rm sgn}\big((p+k)_0\big)0^+} 
- \frac{P_R p\sla P_L} {(p-k)^2 - i\, {\rm sgn}\big((p-k)_0\big)0^+}\bigg] \big( 2f_e({\bf p}) - 1\big).
\bigg\},
\ea
Adopting the hard-loop approximation one gets
\ba 
\label{Pi-k-f-6}  
{_{(L)}} \tilde \Pi(k) = e^2
\int \frac{d^3p}{(2\pi )^3} \; \frac{f_s({\bf p})+f_e({\bf p})}{E_p} \;
 \frac{P_R p\sla P_L }{k\cdot p + i 0^+} . 
\ea

Since the contribution to the photino self-energy coming from right selectrons, which is obtained in the same way, reads 
\ba
\label{Pi-k-f-7}  
{_{(R)}} \tilde \Pi(k) = e^2 \int \frac{d^3p}{(2\pi )^3} \; \frac{f_s({\bf p})+f_e({\bf p})}{E_p} \;
\frac{P_L p\sla P_R }{k\cdot p + i 0^+} ,
\ea
one finds, using the well-known identity $P_R p\sla P_L + P_L p\sla P_R = p\sla$, the complete photino self-energy as
\ba
\label{Pi-k-f-final}
\tilde \Pi(k) = e^2 \int \frac{d^3p}{(2\pi )^3} \; 
\frac{f_s({\bf p})+f_e({\bf p})}{E_p} \; \frac{ p\sla }{k\cdot p + i 0^+} .
\ea
As seen, the structure of the photino self-energy (\ref{Pi-k-f-final}) comes as no surprise as it is the same as the electron self-energy (\ref{Si-k-finala}).

\begin{center}
{\bf Selectron self-energy}
\end{center}
\label{selectron-se-sqed}

\begin{figure}[!h]
\centering
\includegraphics*[width=0.9\textwidth]{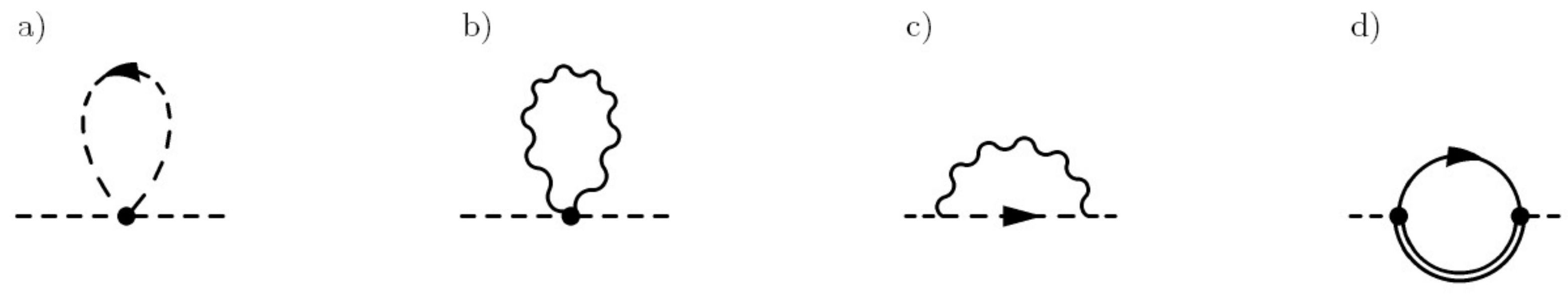}
\caption{Contributions to the selectron self-energy.}
\label{fig-selectron}
\end{figure}

The selectron self-energy $\tilde \Sigma$ is also defined by means of the Dyson-Schwinger equation 
\ba
\label{DSE-scalar}
i\tilde \Delta (k) = i \Delta (k) + i \Delta (k) \, i\tilde \Sigma (k)  \, i\tilde \Delta(k) , 
\ea 
where $\tilde \Delta$ and $\Delta$ is the interacting and free propagator, respectively. The lowest order contributions to $\tilde \Sigma$ are given by four diagrams shown in Fig.~\ref{fig-selectron}. The solid, wavy, dashed and double-solid lines denote, respectively, electron, photon, selectron and photino fields. Below we compute the self-energy of the left selectron. The result for the right selectron is the same.

\begin{center}
{\it  Selectron tadpole}
\end{center}
\label{selectron-tadpole-sqed}

There are two contributions represented by the graph depicted in Fig.~\ref{fig-selectron}a, as the tadpole  line corresponds to either left or right selectron. In the first case we have
\ba
\label{Si-k-s-t-1} 
i {_{(aL)}}\tilde \Sigma_L(k) = -2i e^2  \int \frac{d^4p}{(2\pi )^4}  i\Delta^<(p) .
\ea
Substituting the function $\Delta^<$ given by Eq.~(\ref{Del-<}) into Eq.~(\ref{Si-k-s-t-1}), one finds
\ba 
\label{Si-k-s-t-2}
{_{(aL)}}\tilde \Sigma_L(k) = - e^2  \int \frac{d^3p}{(2\pi )^3}  \,
\frac{2f_s({\bf p})+1}{E_p}  , 
\ea
where the equality $\bar f_s({\bf p}) = f_s({\bf p})$ is assumed. The second contribution corresponding to the right-selectron field equals
\ba
\label{Si-k-s-t-3} 
i {_{(aR)}}\tilde \Sigma_L (k) = i e^2
\int \frac{d^4p}{(2\pi )^4}  i\Delta^<(p) ,
\ea
and it gives
\ba
\label{Si-k-s-t-4} 
_{(aR)}\tilde \Sigma_L (k) = \frac{e^2}{2} \int \frac{d^3p}{(2\pi )^3}  \,
\frac{2f_s({\bf p})+1}{E_p}  . 
\ea
Summing up the contributions (\ref{Si-k-s-t-2}) and (\ref{Si-k-s-t-4}), one finds the following complete result of the  selectron tadpole
\ba
\label{Si-k-s-t-5} 
_{(a)}\tilde \Sigma_L (k) = - \frac{e^2}{2} \int \frac{d^3p}{(2\pi )^3}  \,
\frac{2f_s({\bf p}) +1}{E_p}  . 
\ea

\begin{center}
{\it  Photon tadpole}
\end{center}
\label{photon-tadpole-sqed}

The contribution to the selectron self-energy coming from the photon tadpole shown in Fig.~\ref{fig-selectron}b equals
\ba
\label{Si-k-s-d-1} 
i {_{(b)}}\tilde \Sigma_L (k) = i e^2 g^{\mu \nu} \int \frac{d^4p}{(2\pi )^4}  i D^<_{\mu \nu}(p) ,
\ea
where the symmetry factor $1/2$ is included. Eq.~(\ref{Si-k-s-d-1}) gives 
\ba 
\label{Si-k-s-d-2} 
_{(b)}\tilde \Sigma_L (k) =  - 2 e^2 \int \frac{d^3p}{(2\pi )^3}  \,
\frac{2f_\gamma({\bf p})+1}{E_p} ,
\ea
when the function $D^<_{\mu \nu}$ (\ref{D-<}) is substituted into Eq.~(\ref{Si-k-s-d-1}).

\begin{center}
{\it  Selectron-photon loop}
\end{center}
\label{s-p-loop-sqed}

The contribution represented by the graph depicted in Fig.~\ref{fig-selectron}c equals
\ba 
\label{Si-k-s-p-1} 
i {_{(c)}}\tilde \Sigma_L (k) 
&=& 
-\frac{e^2}{2} \int \frac{d^4p}{(2\pi )^4} 
\Big[(p+2k)^\mu i D^+_{\mu \nu}(p+k) \, (p+2k)^\nu i \Delta^{\rm sym}(p) 
\\ [2mm] \nn
&& \qquad\qquad\qquad\qquad\qquad\qquad\qquad\qquad
+ (p+k)^\mu i D^{\rm sym}_{\mu \nu}(p) \, (p+k)^\nu i \Delta^-(p-k) \Big] ,
\ea
which after the substitution of the functions $D^+_{\mu \nu}, \, D^{\rm sym}_{\mu \nu}$ and $\Delta^-, \, \Delta^{\rm sym}$ in the forms (\ref{D-ret}), (\ref{D-sym}), (\ref{Del-adv}), and (\ref{Del-sym}), respectively, leads to
\ba  
\label{Si-k-s-p-3} 
{_{(c)}}\tilde \Sigma_L(k) 
\!\!\! & = & \!\!\!
\frac{e^2}{4} \int \frac{d^3p}{(2\pi )^3 E_p} 
\\ [2mm] \nn
&& \!\!\!\! 
\times \bigg[ \bigg(\frac{(p+2k)^2 } {(p+k)^2 + i\, {\rm sgn}\big((p+k)_0\big)0^+} 
+ \frac{(p-2k)^2 } {(p-k)^2 - i\, {\rm sgn}\big((p-k)_0\big)0^+} \bigg) \big(2f_s({\bf p}) +1\big) 
\\ [2mm] \nn 
&& \!\!\!\! 
+ \bigg(\frac{(p+k)^2} {(p-k)^2 - i\, {\rm sgn} \big((p-k)_0\big)0^+}
+ \frac{(p-k)^2} {(p+k)^2 + i\, {\rm sgn}\big((p+k)_0\big)0^+}\bigg) 
\big(2 f_\gamma({\bf p}) + 1\big)
\bigg] , 
\ea
where we have assumed that $\bar f_s({\bf p}) = f_s({\bf p})$.  Within the hard-loop approximation, one obtains
\ba 
\label{Si-k-s-p-4} 
{_{(c)}}\tilde \Sigma_L (k) = \frac{e^2}{2} \int \frac{d^3p}{(2\pi )^3}
\frac{4f_\gamma({\bf p}) -2f_s({\bf p})+1}{E_p} .
\ea
We note that the sum of the contributions (\ref{Si-k-s-d-2}) and (\ref{Si-k-s-p-4}), which equals
\ba 
\label{Si-scalar-QED} 
{_{(b+c)}}\tilde \Sigma_L (k) = - \frac{e^2}{2} \int \frac{d^3p}{(2\pi )^3}
\frac{4f_\gamma({\bf p}) +2f_s({\bf p})+3}{E_p} ,
\ea
represents the scalar self-energy of scalar QED which for equilibrium plasma was discussed in {\it e.g.} \cite{Kraemmer:1994az} within the imaginary-time formalism.

\begin{center}
{\it  Electron-photino loop}
\end{center}
\label{e-p-loop-sqed}

The graph depicted in Fig.~\ref{fig-selectron}d provides
\ba
\label{Si-k-r-1} 
i _{(d)} \tilde \Sigma_L (k) &=& 
e^2 \! \int \frac{d^4p}{(2\pi )^4}
{\rm Tr}\Big[ P_R \, i S_e^+(p+k) \, P_L \, i S_{\tilde \gamma}^{\rm sym} (p) 
\\ [2mm] \nn 
&& \qquad \qquad \qquad  \qquad \qquad \qquad
+ P_R \, iS_e^{\rm sym} (p) \,  P_L \, i S_{\tilde \gamma}^-(p-k)  \Big],
\ea
where two types of the fermion propagators enter. We have used the indices to differentiate them, namely, $S_e^+$ and $S_e^{\rm sym}$ denote the electron Green functions and $S_{\tilde \gamma}^-$ and $S_{\tilde \gamma}^{\rm sym}$ denote the photino ones. Substituting these functions given by Eqs.~(\ref{S-ret}), (\ref{S-adv}), and (\ref{S-sym}) with the electron and photino distribution functions into Eq.~(\ref{Si-k-r-1}) and repeating the same steps which have been made in the previous subsections, we find in the hard-loop approximation the following expression
\ba
\label{Si-k-r-3} 
_{(d)} \tilde \Sigma_L (k) = -2 e^2  \int \frac{d^3p}{(2\pi )^3} \,
\frac{ f_{\tilde \gamma} ({\bf p}) + f_e ({\bf p})-1}{E_p}, 
\ea
where we have assumed that $f_e({\bf p}) = \bar f_e({\bf p})$.

\begin{center}
{\it Final result}
\end{center}
\label{fr-selectron-sqed}

The sum of contributions (\ref{Si-k-s-t-5}), (\ref{Si-k-s-d-2}), (\ref{Si-k-s-p-4}), and (\ref{Si-k-r-3}) gives the complete self-energy of the left selectron 
\ba
\label{Si-k-r-final}  
\tilde \Sigma (k) = -2 e^2 \int \frac{d^3p}{(2\pi )^3} \,
\frac{f_e ({\bf p}) + f_\gamma ({\bf p}) +  f_s ({\bf p})+  f_{\tilde \gamma} ({\bf p}) }{E_p}, 
\ea
which equals the complete self-energy of right selectron. For this reason the index $L$ is dropped. As seen, the self-energy (\ref{Si-k-r-final}) is independent of $k$ and because of supersymmetry it vanishes in the vacuum limit when all the distribution functions are zero. This is also effect of the supersymmetry that the distribution functions of electrons and of selectrons enter  the formula (\ref{Si-k-r-final}) with the coefficients equal to each other. The same is true for the distribution functions of photons and of photinos.

Before we conclude the results of this section let us discuss the self-energies of the supersymmetric Yang-Mills plasma.

\subsection{${\cal N} =4$ super Yang-Mills plasma}
\label{ssec-super-ym}

In this subsection the self-energies of all fields occuring in the system described by $\mathcal{N}=4$ super Yang-Mills are computed\footnote{This subsection is based on our work published in \cite{Czajka:2012gq}.}. The plasma, as in case of supersymmetric QED one, is assumed to be homogeneous but the momentum distribution is, in general, different from equilibrium one. Therefore, we use the same techniques based on the Keldysh-Schwinger formalism and performing the calculations within the hard-loop approximation. Since $\mathcal{N}=4$ super Yang-Mills is a conformal theory, all fields are massless. The charges of SYMP are assumed to vanish. The constituents of SYMP carry colour charges but we assume that the plasma is globally and locally colourless. 

We start our consideration with writing down the dispersion equations into which the respective self-energies enter. Then, we derive the self-energies corresponding to all fields of SYMP.

\subsubsection{Dispersion equations}
\label{sssec-dis-eqs}

Since the super Yang-Mills plasma is constituted by gluon, fermion, and scalar fields here we write down, in analogy to supersymmetric QED plasma, the dispersion equations of quasi-gluons, quasi-fermions, and quasi-scalars, which are as follows
\ba
\label{dis-photon-1}
{\rm det}\big[ k^2 g^{\mu \nu} -k^{\mu} k^{\nu} - \Pi^{\mu \nu}(k) \big] = 0 ,
\\ [2mm]
\label{dis-electron-1}
 {\rm det}\big[ k\sla  - \Sigma (k) \big]  = 0 ,
\\ [2mm]
\label{dis-selectron}
k^2 + P(k) = 0 ,
\ea
where $\Pi^{\mu \nu}(k)$, $\Sigma(k)$, and $P(k)$ are the self-energies of gluon, fermion, and scalar fields, respectively.

\subsubsection{Self-energies}
\label{sssec-sym-se}

We compute here the self-energies which enter the dispersion equations (\ref{dis-photon-1}), (\ref{dis-electron-1}), and (\ref{dis-selectron}). The vertices of ${\cal N} = 4$ super Yang-Mills, which are used in our perturbative calculations, are listed in Appendix \ref{appendix-FR}. Since the plasma is assumed to be homogeneous, all the Green functions of the Keldysh-Schwinger formalism which are going to be used here and have been derived in Sec. \ref{sec-KS-form}, are independent of a position $X$. The self-energies derived here can be defined by the Dyson-Schwinger equation as in case of supersymmetric QED. Namely Eqs. (\ref{DSE-photon}), (\ref{DSE-electron}), and (\ref{DSE-scalar}) define the self-energies of gluon, fermion and scalar fields, respectively.

\begin{center}
{\bf Polarisation tensor}
\end{center}
\label{polar-tensor-sym}

\begin{figure}[!h]
\centering
\includegraphics*[width=0.7\textwidth]{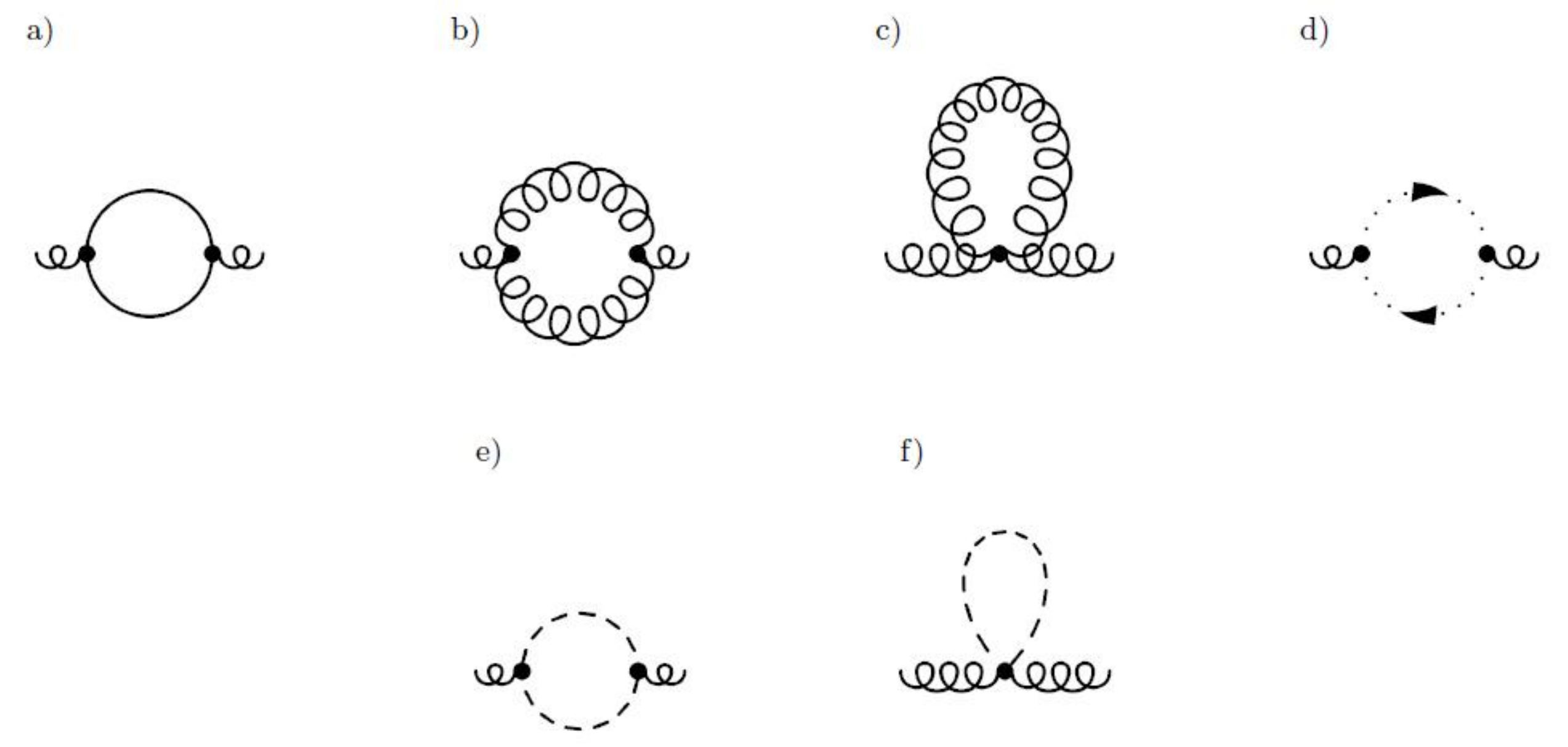}
\caption{Contributions to the gluon self-energy. }
\label{fig-gluon}
\end{figure}

The lowest order contributions to the gluon polarisation tensor are given by six diagrams shown in Fig.~\ref{fig-gluon}. The curly, plain, dotted and dashed lines denote, respectively, gluon, fermion, ghost, and scalar fields.

\begin{center}
{\it Fermion loop}
\end{center}
\label{fermion-loop-sym}

The contribution to the gluon polarisation tensor coming from the fermion loop is depicted in the graph in Fig.~\ref{fig-gluon}a. The respective formula is obtained in the same way as that of the electron loop of photon polarisation tensor (\ref{Pi-k-e-finala}). For this reason we omit some steps of the derivation. Using the vertices given in Appendix~\ref{appendix-FR}, the contribution to the contour polarisation tensor is immediately written down in the coordinate space as
\ba
\label{contour-Pi}
_{(a)}\Pi^{\mu \nu}_{ab}(x,y)
= -ig^2 N_c \delta_{ab} {\rm Tr} \big[\gamma^\mu S_{ij}(x,y) \gamma^\nu S_{ji}(y,x)\big] .
\ea
where the trace is taken over spinor indices; the indices $i,j$ correspond to types of fermions as defined in Table \ref{table-field-content}. The factor $(-1)$ due to the fermion loop is included and the relation $f_{acd }f_{bcd}=\delta_{ab} N_c$ is used here. 

The retarded polarisation tensor is obtained in terms of the formula (\ref{pi-1a-1}) and after some elementary manipulations it may be written down in the form
\ba
\label{Pi-x-1}
\big({_{(a)} \Pi^+(x)} \big)^{\mu \nu}_{ab}
= -\frac{ig^2}{2} N_c \delta_{ab} {\rm Tr}\Big[\gamma^\mu S^+_{ij}(x) \gamma^\nu S^{\rm sym}_{ji}(-x)
+ \gamma^\mu S^{\rm sym}_{ji}(x) \gamma^\nu S^-_{ij}(-x) \Big],
\ea
which in the momentum space reads
\ba
\label{Pi-k-e-1}
\big({_{(a)} \Pi^+(k)} \big)^{\mu \nu}_{ab} 
&=& 
-\frac{ig^2}{2} N_c \delta_{ab} \int \frac{d^4p}{(2\pi)^4}
\\ [2mm] \nn
&& \qquad
\times {\rm Tr} \Big[\gamma^\mu S^+_{ij}(p+k) \gamma ^\nu S^{\rm sym}_{ji}(p)
+ \gamma^\mu S^{\rm sym}_{ji}(p) \gamma^\nu S^-_{ij}(p-k) \Big].
\ea

Further on the index $+$ is dropped and $\Pi^+$ is denoted as  $\Pi$, as only the retarded polarisation tensor is discussed. Substituting the functions $S^{+}, S^- S^{\rm sym}$ given by Eqs.~(\ref{S-ret}), (\ref{S-adv}), and (\ref{S-sym}), respectively, into the formula (\ref{Pi-k-e-1}), one finds
\ba
\label{Pi-k-e-4}
_{(a)}\Pi^{\mu \nu}_{ab}(k)
&=& 
- 4 g^2 N_c  \delta_{ab} \int \frac{d^3p}{(2\pi)^3} \, \frac{2f_f({\bf p})-1}{E_p} 
\\ [2mm] \nn
&& 
\times \bigg(\frac{2p^\mu p^\nu + k^\mu p^\nu + p^\mu k^\nu - g^{\mu \nu} (k \cdot p)}
{(p+k)^2 + i\, {\rm sgn}\big((p+k)_0\big)0^+}
+ \frac{2p^\mu p^\nu - k^\mu p^\nu - p^\mu k^\nu + g^{\mu \nu} (k \cdot p)}
{(p-k)^2 - i\, {\rm sgn}\big((p-k)_0\big)0^+} \bigg) ,
\ea
where the traces of gamma matrices are computed and it is taken into account that $p^2 =0$.

In the hard-loop approximation, when $p \gg k$, the relations (\ref{HLA-plus}) and (\ref{HLA-minus}) are used and the polarisation tensor gets the form
\ba
\label{Pi-k-e-final}
_{(a)} \Pi^{\mu \nu}_{ab}(k)
&=& 
4 g^2 N_c \delta_{ab} \int \frac{d^3p}{(2\pi)^3} \, \frac{2f_f({\bf p})-1}{E_p} 
\\ [2mm] \nn
&& \qquad\qquad
\times \frac{k^2 p^\mu p^\nu - \big(k^\mu p^\nu + p^\mu k^\nu
- g^{\mu \nu} (k \cdot p) \big) (k \cdot p)} {(k\cdot p + i 0^+)^2} ,
\ea
which has the well-known structure of the polarisation tensor of gauge bosons in ultrarelativistic QED and QCD plasmas. As seen, $\Pi(k)$ is symmetric with respect to Lorentz indices $ {_{(a)}\Pi}^{\mu \nu}_{ab}(k) =  {_{(a)}\Pi}^{\nu \mu}_{ab}(k)$ and transverse $k_\mu  {_{(a)} \Pi}^{\mu \nu}_{ab}(k) = 0$, as required by the gauge invariance. In the vacuum limit, when the fermion distribution function $f_f({\bf p})$ vanishes, the polarisation tensor (\ref{Pi-k-e-final}) is still nonzero (actually infinite). As we will see, the vacuum contribution to the complete polarisation tensor exactly vanishes due to the supersymmetry.

\begin{center}
{\it Gluon loop}
\end{center}
\label{gluon-loop-sym}

In analogy to the fermion-loop expression (\ref{Pi-k-e-1}), one finds the gluon-loop contribution to the retarded polarisation tensor shown in Fig.~\ref{fig-gluon}b as
\ba
\label{Pi-gluon-loop-1}
_{(b)}\Pi^{\mu \nu}_{ab}(k) 
\!\!\! &=& \!\!\!
\frac{ig^2}{4} \int \frac{d^4p}{(2\pi )^4}  \int \frac{d^4q}{(2\pi )^4} 
\\ [2mm] \nn
&& \!\!\!\!
\times \Big[ (2\pi)^4 \delta^{(4)}(k+p-q) \Gamma^{\mu \sigma \rho}_{acd} (k,-q,p)
\big( D^+(q)  \big)^{cc'}_{\sigma \sigma'}  \big( D^{\rm sym}(p) \big)^{dd'}_{\rho \rho'}  
\Gamma^{\sigma' \nu \rho'}_{c'bd'} (q,-k,-p)
\\ [2mm] \nn
&& \!\!\!\!
+ (2\pi)^4 \delta^{(4)}(k-p+q) \Gamma^{\mu \sigma \rho}_{acd} (k,q,-p)
\big( D^-(q)  \big)^{cc'}_{\sigma \sigma'}  \big( D^{\rm sym}(p) \big)^{dd'}_{\rho \rho'}  
\Gamma^{\sigma' \nu \rho'}_{c'bd'} (-q,-k,p) \Big],
\ea
where the combinatorial factor $1/2$ is included and $\Gamma^{\mu \nu \rho}_{abc} (k,p,q)$ is the three-gluon vertex function  
\ba
\label{3-g-vertex-1}
\Gamma^{\mu \nu \rho}_{abc} (k,p,q) \equiv f_{abc} \Gamma^{\mu \nu \rho} (k,p,q) 
\ea
with
\ba
\label{3-g-vertex-2}
\Gamma^{\mu \nu \rho} (k,p,q) \equiv g^{\mu \nu }(k-p)^\rho+g^{\nu \rho}(p-q)^\mu +g^{\rho \mu}(q-k)^\nu .
\ea
The vertex function usually includes the factor $g$ which is dropped here but it is included in the formula of polarisation tensor.

Substituting the gluon Green functions (\ref{D-ret}), (\ref{D-adv}), and (\ref{D-sym}) into Eq.~(\ref{Pi-gluon-loop-1}), we get
\ba
\label{Pi-gluon-loop-2}
_{(b)}\Pi^{\mu \nu}_{ab}(k) 
\!\!\! &=& \!\!\!
- \frac{ig^2}{4} N_c \delta^{ab} \int \frac{d^4p}{(2\pi )^4}  \int d^4q \Delta^{\rm sym}(p)
\\ [2mm] \nn
&& \!\!\!\!
\times \Big[ \delta^{(4)}(k+p-q) M^{\mu \nu} (k,q,p) \Delta^+(q)
+ \delta^{(4)}(k-p+q)  M^{\mu \nu} (k,-q,-p) \Delta^-(q) \Big],
\ea
where the identity $f^{acd}f^{cbd} = - N_c \delta^{ab}$ has been used and the tensor $M^{\mu\nu}$ equals
\be
\label{tensor-M-def}
M^{\mu \nu} (k,q,p) \equiv \Gamma^{\mu \sigma \rho} (k,-q,p)
g_{\sigma \sigma'} g_{\rho \rho'} \Gamma^{\sigma' \nu \rho'} (q,-k,-p) .
\ee
The tensor (\ref{tensor-M-def}) is computed as
\ba
\label{tensor-M-F}
M^{\mu \nu} (k,q,p) 
&=& 
g^{\mu \nu}\big[(k+q)^2 + (k-p)^2 \big] - 2k^\mu k^\nu + 2q^\mu q^\nu + 2p^\mu p^\nu  
\\ [2mm] \nn
&& \qquad\qquad\qquad
- (k^\mu q^\nu + q^\mu k^\nu) + (k^\mu p^\nu + p^\mu k^\nu) + 3  (q^\mu p^\nu + q^\mu k^\nu) 
\ea
and for $q= p+k$ we have
\ba
\label{tensor-M-F-2}
M^{\mu \nu} (k,p+k,p) = g^{\mu \nu}\big[5k^2 + 2(k\cdot p) + 2p^2 \big] 
- 2k^\mu k^\nu + 10 p^\mu p^\nu + 5(k^\mu p^\nu + p^\mu k^\nu).
\ea
Within the hard-loop approximation, when $p \gg k$, we obtain
\ba
\label{tensor-M-F-HL}
M^{\mu \nu} (k,p+k,p) \approx 2 g^{\mu \nu} (k\cdot p)
+ 10 p^\mu p^\nu + 5(k^\mu p^\nu + p^\mu k^\nu),
\ea
where we have taken into account that $p^2=0$. Using Eq.~(\ref{tensor-M-F}) one immediately finds
\ba
\label{tensor-M-F-3}
M^{\mu \nu} (k,-q,-p) 
&=& 
g^{\mu \nu}\big[(k-q)^2 + (k+p)^2 \big] - 2k^\mu k^\nu + 2q^\mu q^\nu + 2p^\mu p^\nu 
\\ [2mm] \nn
&& \qquad\qquad\qquad
+ (k^\mu q^\nu + q^\mu k^\nu) - (k^\mu p^\nu + p^\mu k^\nu) + 3  (q^\mu p^\nu + q^\mu k^\nu) ,
\ea
and for $q = p - k$ we have
\ba
\label{tensor-M-F-4}
M^{\mu \nu} (k,-p+k,-p) = g^{\mu \nu}\big[5k^2 - 2(k\cdot p) + 2p^2 \big] 
-2 k^\mu k^\nu + 10p^\mu p^\nu (k^\mu p^\nu + p^\mu k^\nu) ,
\ea
and finally the hard-loop approximate expression reads
\ba
\label{tensor-M-F-HL-2}
M^{\mu \nu} (k,p+k,p) \approx - 2 g^{\mu \nu} (k\cdot p)
+ 10 p^\mu p^\nu - 5(k^\mu p^\nu + p^\mu k^\nu).
\ea

Substituting the expressions (\ref{tensor-M-F-HL}), and (\ref{tensor-M-F-HL-2}) into Eq.~(\ref{Pi-gluon-loop-2}), we get
\ba
\label{Pi-gluon-loop-3}
_{(b)}\Pi^{\mu \nu}_{ab}(k) 
&=& 
- \frac{ig^2}{4} N_c \delta^{ab} \int \frac{d^4p}{(2\pi )^4} \; \Delta^{\rm sym}(p)
\\ [2mm] \nn
&& 
\times \Big[ \big(2 g^{\mu \nu} (k\cdot p)
+ 5(k^\mu p^\nu + p^\mu k^\nu)+ 10 p^\mu p^\nu  \big) \Delta^+(p+k) 
\\ [2mm] \nn
&& 
- \big(2 g^{\mu \nu} (k\cdot p)
+ 5(k^\mu p^\nu + p^\mu k^\nu) - 10 p^\mu p^\nu  \big) \Delta^-(p-k) \Big].
\ea
With the explicit form of the functions $\Delta^+$, $\Delta^-$ and $\Delta^{\rm sym}$ given by the formulas (\ref{Del-ret}), (\ref{Del-adv}), and (\ref{Del-sym}), respectively, Eq.~(\ref{Pi-gluon-loop-2}) equals
\ba
\label{Pi-gluon-l-4a}
_{(b)}\Pi^{\mu \nu}_{ab}(k) 
\!\!\! &=& \!\!\!
- \frac{g^2}{2} N_c \delta^{ab} \int \frac{d^3p}{(2\pi )^3} \frac{2 f_g({\bf p}) +1}{2E_p}
\\ [2mm] \nn
&& \!\!\!\!\!\!\!\!\!\!\!\!\!\!\!\!\!\!\!\!\!\!\!\!
\times \bigg[ \frac{2 g^{\mu \nu} (k\cdot p)
+ 5(k^\mu p^\nu + p^\mu k^\nu)+ 10 p^\mu p^\nu }{(p+k)^2 + i{\rm sgn}\big((p+k)_0 \big) 0^+}
+ \frac{-2 g^{\mu \nu} (k\cdot p)
- 5(k^\mu p^\nu + p^\mu k^\nu)+ 10 p^\mu p^\nu }{(p-k)^2 - i{\rm sgn}\big((p-k)_0 \big) 0^+}
\bigg].
\ea
Adopting the hard-loop approximation (\ref{HLA-plus}), and (\ref{HLA-minus}) to Eq.~(\ref{Pi-gluon-l-4a}) we finally get
\ba
\label{Pi-gluon-loop-5}
_{(b)}\Pi^{\mu \nu}_{ab}(k) 
&=& 
\frac{g^2}{2} N_c \delta^{ab} \int \frac{d^3p}{(2\pi )^3} 
\\ [2mm] \nn
&& \qquad
\times \frac{2 f_g({\bf p}) +1}{2E_p} \frac{- 2 g^{\mu \nu} (k\cdot p)^2
- 5(k^\mu p^\nu + p^\mu k^\nu) (k\cdot p) + 5k^2 p^\mu p^\nu}{(k\cdot p + i 0^+)^2} .\;
\ea

\begin{center}
{\it Gluon tadpole}
\end{center}
\label{gluon-tadpole-sym}

The gluon-tadpole contribution to the retarded polarisation tensor, which shown in Fig.~\ref{fig-gluon}c,
equals
\ba
\label{Pi-gluon-tadpole-1}
_{(c)}\Pi^{\mu \nu}_{ab}(k) = - \frac{g^2}{2} \int \frac{d^4p}{(2\pi )^4}
\Gamma^{\mu \nu \rho}_{abcc \rho} \Delta^<(p)  ,
\ea
where the combinatorial factor $1/2$ is included. In Eq. (\ref{Pi-gluon-tadpole-1}) we have inserted the gluon Green function $D_{\mu\nu}^>$ in the form (\ref{D->-Del}) in order to factor out the Lorentz and colour structure. These structures are included in the function $\Gamma^{\mu \nu \rho \sigma }_{abcd}$ which equals
\ba
\label{4-g-vertex}
\Gamma^{\mu \nu \rho \sigma }_{abcd} 
&\equiv&
f_{abe}f_{ecd}(g^{\mu \sigma} g^{\nu \rho} - g^{\mu \rho} g^{\nu \sigma})
+ f_{ace}f_{edb}(g^{\mu \rho} g^{\nu \sigma} - g^{\mu \nu} g^{\rho \sigma})
\\ [2mm] \nn
&& \qquad\qquad
+ f_{ade}f_{ebc}(g^{\mu \nu} g^{\rho \sigma} - g^{\mu \sigma} g^{\nu \rho}).
\ea
With the explicit form of the function $\Delta^<(p)$ given by Eq.~(\ref{Del-<}), the formula (\ref{Pi-gluon-tadpole-1}) provides 
\ba
_{(c)}\Pi^{\mu \nu}_{ab}(k) = \frac{3g^2}{2} N_c \, \delta_{ab} g^{\mu \nu}
\int \frac{d^3p}{(2\pi )^3} \frac{2 f_g({\bf p}) +1}{E_p} .
\ea

\begin{center}
{\it Ghost loop}
\end{center}
\label{ghost-loop-sym}

The ghost-loop contribution to the retarded polarisation tensor, which is shown in Fig.~\ref{fig-gluon}d, equals
\ba
\label{Pi-ghost-loop-2}
_{(d)}\Pi^{\mu \nu}_{ab}(k) 
&=& 
- \frac{ig^2}{2} f^{acd} f^{ebf}  \int \frac{d^4p}{(2\pi )^4} \; \Delta_{ce}^{\rm sym}(p)
\\ [2mm] \nn
&& \qquad \qquad
\times \Big[ (p+k)^\mu p^{\nu} \Delta_{df}^+(p+k) + p^\mu (p-k)^\nu \Delta_{df}^-(p-k) \Big], \nn
\ea
where the factor $(-1)$ is included as we deal with the fermion loop. Using the explicit forms of the ghost functions $\Delta_{ab}^+$, $\Delta^-_{ab}$ and $\Delta_{ab}^{\rm sym}$ derived in Sec. \ref{sec-ghosts-KS} and given by Eqs.~(\ref{Del-g-ret}), (\ref{Del-g-adv}), and (\ref{Del-g-sym}), respectively, the formula (\ref{Pi-ghost-loop-2}) is manipulated to
\ba
\label{Pi-ghost-loop-4}
 _{(d)}\Pi^{\mu \nu}_{ab}(k) 
= -\frac{g^2}{4} N_c \delta_{ab} \int \frac{d^3p}{(2\pi )^3} \; \frac{2f_g({\bf p})+1}{E_p}
\frac{k^2 p^\mu p^\nu - (k^\mu p^\nu + p^\mu k^\nu) (k\cdot p)}{(k\cdot p + i 0^+)^2},
\ea
where the identity $f^{aed}f^{ebd}=-\delta^{ab}N_c$ is used. The expression (\ref{Pi-ghost-loop-4}) holds in the hard-loop approximation.

As already mentioned, the fermion-loop contribution to the polarisation tensor is symmetric and transverse with respect to Lorentz indices. The same holds for the sum of gluon-loop, gluon-tadpole and ghost-loop contributions which gives the gluon polarisation tensor in pure gluodynamics (QCD with no quarks).  The sum of the three contributions equals
\ba
\label{Pi-b-c-d}
 _{(b)+(c)+(d)}\Pi^{\mu \nu}_{ab}(k)
&=& 
g^2 N_c \delta_{ab} \int \frac{d^3p}{(2\pi )^3} \frac{2f_g({\bf p})+ 1}{E_p}
\\ [2mm] \nn
&& \qquad \qquad
\times \frac{k^2 p^\mu p^\nu + g^{\mu \nu} (k\cdot p)^2
- (k^\mu p^\nu + p^\mu k^\nu) (k\cdot p)}{(k\cdot p + i 0^+)^2}.
\ea
To our best knowledge this is the first computation of the gluon polarisation tensor in the hard-loop approximation performed in the Keldysh-Schwinger (real time) formalism which explicitly demonstrates the transversality of the tensor. In Refs.~\cite{Weldon:1982aq,Mrowczynski:2000ed}, where the equilibrium and non-equilibrium anisotropic plasmas were considered, respectively, the transversality of $\Pi^{\mu \nu}(k)$ was actually assumed.  In the case of imaginary time formalism, the computation of the gluon polarisation tensor in the hard-loop approximation is the textbook material \cite{LeBellac,Kapusta-Gale}. 

\begin{center}
{\it Scalar loop}
\end{center}
\label{scalar-loop-sym}

The contribution to the polarisation tensor coming from the scalar loop depicted in Fig.~\ref{fig-gluon}e is given by
\ba
\label{Pi-s-l-1}
 _{(e)}\Pi^{\mu \nu}_{ab}(k) 
&=& - \frac{ig^2}{2} N_c \delta_{ab} \delta^{AA} \int \frac{d^4p}{(2\pi)^4}
\Big[(2p+k)^\mu (2p+k)^\nu  \Delta^+ (p+k)  \Delta^{\rm sym}(p)
\\ [2mm] \nn
&& \qquad\qquad\qquad\qquad\qquad\qquad
+ (2p-k)^\mu (2p-k)^\nu  \Delta^{\rm sym} (p) \Delta^-(p-k) \Big], 
\ea
which changes into 
\ba
\label{Pi-s-l-3}
_{(e)}\Pi^{\mu \nu}_{ab}(k) =  3g^2 N_c \delta_{ab}
\int \frac{d^3p}{(2\pi )^3} \,\frac{2f_s({\bf p})+1}{E_p} \,
\frac{k^2 p^\mu p^\nu - (p^\mu k^\nu + k^\mu p^\nu)(k \cdot p)}{(k \cdot p + i0^+)^2}
\ea
when the functions $\Delta^{+}$, $\Delta^-$, and $\Delta^{\rm sym}$ given by Eqs.~(\ref{Del-ret}), (\ref{Del-adv}), and (\ref{Del-sym}), respectively are used and the hard-loop approximation is adopted.

\newpage
\begin{center}
{\it Scalar tadpole}
\end{center}
\label{scalar-tadpole-sym}

The contribution to the polarisation tensor coming from the scalar tadpole depicted in Fig.~\ref{fig-gluon}f is
\ba
\label{Pi-s-t-1}
_{(f)}\Pi^{\mu \nu}_{ab}(k) = - ig^2 N_c \delta^{ab} \delta^{AA} g^{\mu \nu}
\int \frac{d^4p}{(2\pi)^4} \Delta^<(p) ,
\ea
where the combinatorial factor $1/2$ is included. With the function $\Delta^<$ given by Eq.~(\ref{Del-<}) we have
\ba
\label{Pi-s-t-2}
_{(f)}\Pi^{\mu \nu}_{ab}(k) = 3 g^2 N_c \delta_{ab} g^{\mu \nu}
\int \frac{d^3p}{(2\pi )^3}  \, \frac{2f_s({\bf p})+1}{E_p}  .
\ea

We get the complete contribution from a scalar field to the polarisation tensor by summing up the scalar loop and scalar tadpole. Thus, one finds
\ba
\label{Pi-s-total}
_{(e+f)}\Pi^{\mu \nu}_{ab}(k) 
&=&  
3 g^2 N_c \delta_{ab} \int \frac{d^3p}{(2\pi )^3} \frac{2f_s({\bf p})+1}{E_p}
\\ [2mm] \nn
&& \qquad \qquad
\times \frac{k^2 p^\mu p^\nu - \big(p^\mu k^\nu + k^\mu p^\nu - g^{\mu \nu} (k \cdot p)\big)(k \cdot p)}
{(k \cdot p + i0^+)^2},
\ea
which has the structure corresponding to the scalar QED.  Then, it is not a surprise that the polarisation tensor (\ref{Pi-s-total}) is symmetric and transverse. 

\begin{center}
{\it Final result}
\end{center}
\label{fr-pol-tensor-sym}

After summing up all contributions, we get the final expression of the gluon polarisation tensor
\be
\label{Pi-k-final-s}
\Pi^{\mu \nu}_{ab}(k)= g^2 N_c \delta_{ab} \int \frac{d^3p}{(2\pi)^3}\frac{f({\bf p})}{E_p} 
\frac{k^2 p^\mu p^\nu - (k^\mu p^\nu + p^\mu k^\nu - g^{\mu \nu} (k\cdot p))
(k\cdot p)}{(k\cdot p + i 0^+)^2},
\ee
where 
\be
\label{f-def}
f({\bf p}) \equiv 2f_g({\bf p}) + 8f_f({\bf p}) + 6f_s({\bf p})
\ee
is the effective distribution function of the plasma constituents.  We observe that the coefficients in front of the distributions functions $f_g({\bf p})$,   $f_f({\bf p})$, $f_s({\bf p})$ equal the numbers of degrees of freedom (except colours) of, respectively, gauge bosons, fermions and scalars, {\it cf.} Table \ref{table-field-content}. This is obviously a manifestation of supersymmetry. Another effect of the supersymmetry is vanishing of the tensor (\ref{Pi-k-final-s}) in the vacuum limit when $f({\bf p})  = 0$. Needless to say, the polarisation tensor (\ref{Pi-k-final-s}) is symmetric and transverse in Lorentz indices and thus it is gauge independent.

In the case of QCD plasma, one gets the polarisation tensor of the form (\ref{Pi-k-final-s}) after the vacuum contribution is subtracted. For the QGP with the number $N_f$ of massless flavors, the effective distribution function equals
\be
\label{f-def-QGP}
f_{\rm QGP}({\bf p}) \equiv 2f_g({\bf p}) +  \frac{N_f}{N_c} \big(f_q({\bf p}) + f_{\bar q}({\bf p}) \big) ,
\ee
where $f_q({\bf p})$, $f_{\bar q}({\bf p})$ are the distribution functions of quarks and antiquarks which contribute differently to the polarisation tensor than fermions of the ${\cal N} = 4$ super Yang-Mills. This happens because (anti-)quarks of QCD belong to the fundamental representation of ${\rm SU}(N_c)$ while the fermions of SYMP belong to the adjoint representation. 

\begin{center}
{\bf Fermion self-energy}
\end{center}
\label{fermion-se-sym}

\begin{figure}[!h]
\centering
\includegraphics*[width=0.7\textwidth]{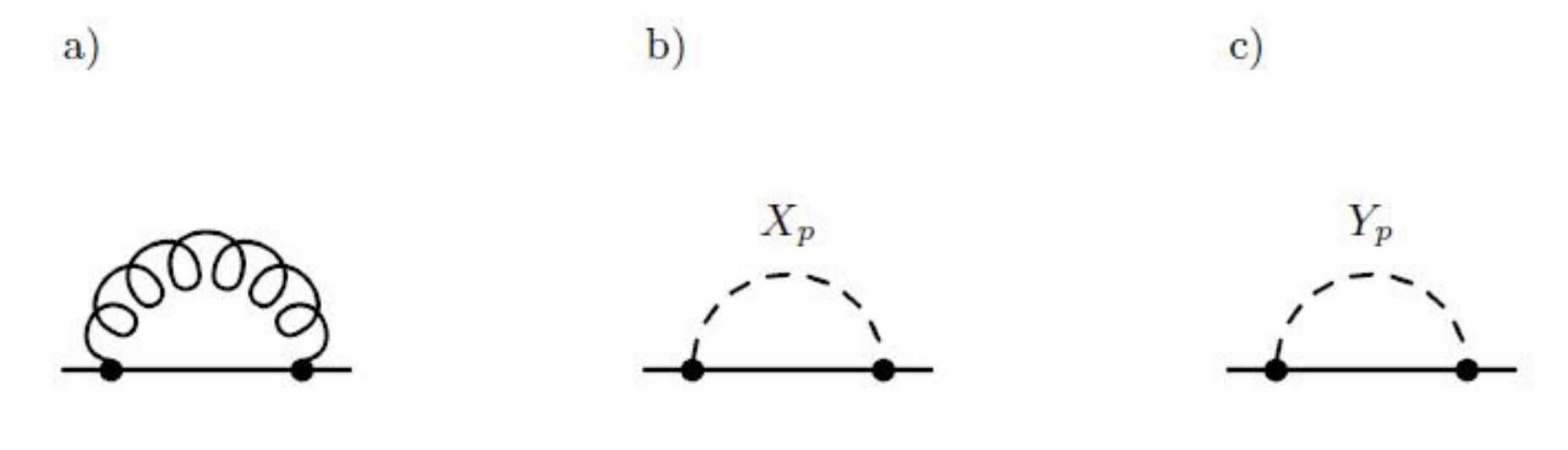}
\caption{Contributions to the fermion self-energy. }
\label{fig-fermion}
\end{figure}

The lowest order contributions to the fermion self-energy are given by diagrams shown in Fig.~\ref{fig-fermion}. The curly, plain, and dashed lines denote, respectively, gluon, fermion, and scalar fields.

\begin{center}
{\it Fermion-gluon loop}
\end{center}
\label{q-g-loop-sym}

The contribution to the fermion self-energy corresponding to the graph depicted in Fig.~\ref{fig-fermion}a is given by
\ba
\label{Si-k-a-1}
_{(a)}\Sigma^{ij}_{ab}(k) = \frac{ig^2}{2} N_c \delta_{ab} \delta^{ij}
\! \int \! \frac{d^4p}{(2\pi )^4}
\Big[\gamma^\mu S^+(p+k) \gamma^\nu D_{\mu\nu}^{\rm sym}(p)
+ \gamma^\mu S^{\rm sym}(p) \gamma^\nu D_{\mu\nu}^-(p-k) \Big].\quad
\ea
With the functions $D^-_{\mu\nu}$, $D_{\mu\nu}^{\rm sym}$ and $S^+$, $S^{\rm sym}$ given by Eqs.~(\ref{D-adv}), (\ref{D-sym}), (\ref{S-ret}), and (\ref{S-sym}), respectively, one obtains
\ba
\label{Si-k-a-3}
_{(a)}\Sigma^{ij}_{ab}(k) = g^2 N_c \delta_{ab} \delta^{ij}
\int \frac{d^3p}{(2\pi )^3} \, \frac{ f_g ({\bf p}) +  f_f ({\bf p})}{E_p}  \,
\frac{p\sla}{k\cdot p + i 0^+} ,
\ea
where the traces over gamma matrices are computed and the hard-loop approximation is applied.
Eq.~(\ref{Si-k-a-3}) has the well-known form of the electron self-energy in QED.

\begin{center}
{\it Fermion-scalar loops}
\end{center}
\label{q-s-loop-sym}

Since there are scalar and pseudoscalar fields $X_p$ and $Y_p$, there are two contributions to the fermion self-energy corresponding to the graphs depicted in Figs.~\ref{fig-fermion}b, \ref{fig-fermion}c. The first one corresponding to the $X_p$ field equals
\ba
\label{Si-k-b-1}
{_{(b)}\Sigma}^{ij}_{ab}(k) =  \frac{ig^2}{2} N_c \delta^{ab} \alpha^p_{ik}\alpha^p_{kj}
\int \frac{d^4p}{(2\pi )^4} \Big[ S^+(p+k)  \Delta^{\rm sym}(p) + S^{\rm sym}(p)  \Delta^-(p-k) \Big].
\ea
Because of the relations (\ref{alpha-beta-relations}), one finds that $\alpha^p_{ik}\alpha^p_{kj}=-3\delta_{ij}$.
Using the result and substituting the fermion functions $S^+$, $S^{\rm sym}$ and the scalar  ones $\Delta^-$, $\Delta^{\rm sym}$ given by Eqs.~(\ref{S-ret}), (\ref{S-sym}), (\ref{Del-adv}), and (\ref{Del-sym}) into Eq.~(\ref{Si-k-b-1}), one obtains the following result
\ba
\label{Si-k-b-2}
{_{(b)}\Sigma}^{ij}_{ab} (k) = \frac{3g^2}{2} N_c \delta_{ab} \delta^{ij}
\int \frac{d^3p}{(2\pi)^3} \frac{ f_f ({\bf p}) +  f_s ({\bf p})}{E_p}  \,
\frac{p\sla}{k\cdot p + i 0^+} ,
\ea
which holds in the hard-loop approximation.

The contribution due to the pseudoscalar field $Y_p$ is
\ba
\label{Si-k-c-1}
{_{(c)}\Sigma}^{ij}_{ab}(k) 
&=& 
\frac{ig^2}{2} N_c \delta^{ab} \beta^p_{ik}\beta^p_{kj}
\int \frac{d^4p}{(2\pi )^4} 
\\ [2mm] \nn
&& \qquad \qquad
\times \Big[ \gamma_5 S^+(p+k) \gamma_5 \Delta^{\rm sym}(p) 
+ \gamma_5 S^{\rm sym}(p) \gamma_5 \Delta^-(p-k) \Big].
\ea
Because $\beta^p_{ik}\beta^p_{kj}=-3\delta_{ij}$, $\gamma_\mu \gamma_5=-\gamma_5 \gamma_\mu$, and $\gamma_5^2 = 1$, we again obtain the result (\ref{Si-k-b-2}).

\begin{center}
{\it Final result}
\end{center}
\label{fr-fermion-sym}

Summing up all the contributions, we get the final expression for the fermion self-energy
\ba
\label{Si-k-final}
\Sigma^{ij}_{ab}(k) = \frac{g^2}{2} \, N_c \delta_{ab}\delta^{ij}
\int \frac{d^3p}{(2\pi )^3}
\frac{f({\bf p})}{E_p}  \, \frac{p\sla}{k\cdot p + i 0^+}.
\ea
which, as the polarisation tensor (\ref{Pi-k-final-s}), depends on the effective distribution function (\ref{f-def}).

\begin{center}
{\bf Scalar self-energy}
\end{center}
\label{scalar-se-sym}

\begin{figure}[!h]
\centering
\includegraphics*[width=0.7\textwidth]{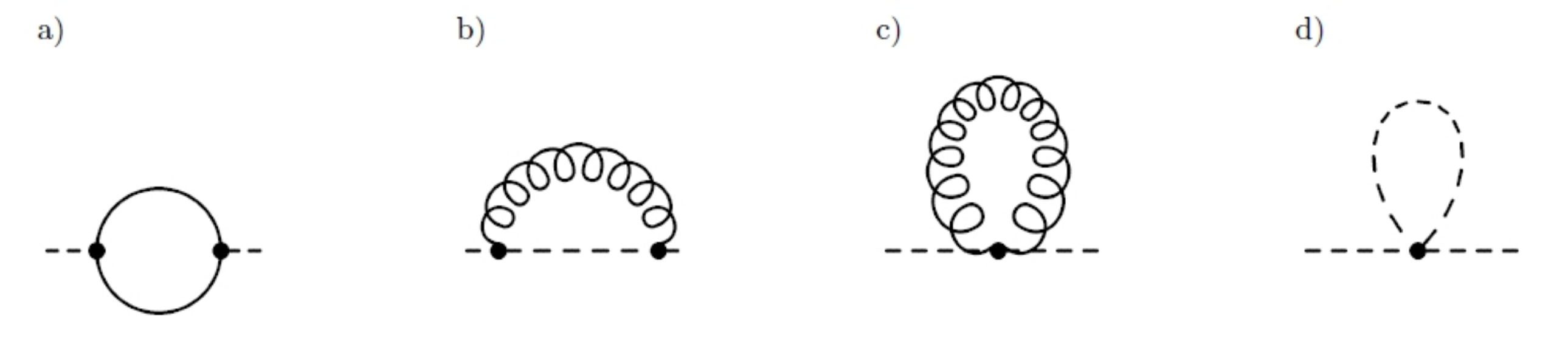}
\caption{Contributions to the scalar self-energy. }
\label{fig-scalar}
\end{figure}

The lowest order contributions to the scalar self-energy are given by the diagrams shown in Fig.~\ref{fig-scalar}. The curly, plain, and dashed lines denote, respectively, gluon, fermion, and scalar fields.

Since there are scalar ($X_p$) and pseudoscalar ($Y_p$) fields, we have to consider separately the self-energies of  $X_p$ and $Y_p$. However, one observes that only the coupling of scalars to fermions differs for $X_p$ and $Y_p$. The self-interaction and the coupling to the gauge field are the same. Therefore, only the fermion-loop contribution to the scalar self-energy, which is shown in the diagram Fig.~\ref{fig-scalar}a, needs to be computed separately for the $X_p$ and $Y_p$ fields.

\begin{center}
{\it Fermion loop}
\end{center}
\label{quark-loop-sym}

In the case of the scalar $X_p$ field, the diagram  Fig.~\ref{fig-scalar}a provides
\ba
\label{P-k-a-1}
_{(a)} P^{pq}_{ab}(k) = \frac{ig^2}{4} \, N_c \delta^{ab} \alpha^p_{ij} \alpha^q_{ji} 
\int \frac{d^4p}{(2\pi)^4}
{\rm Tr}\Big[   S^+ (p+k)  S^{\rm sym} (p) + S^{\rm sym} (p)   S^- (p-k) \Big].
\ea
where the symmetry factor $1/2$ and the extra minus sign due to the fermionic character of the loop are included. With the explicit form of the functions $S^{+}$, $S^-$, and $S^{\rm sym}$  given by Eqs.~(\ref{S-ret}), (\ref{S-adv}), and (\ref{S-sym}) and the identity
$\alpha^p_{ij}\alpha^q_{ji}=-4\delta^{pq}$ which follows from the relations (\ref{alpha-beta-relations}), one finds
\ba
\label{P-k-a-2}
_{(a)}P^{pq}_{ab}(k) = -4 g^2 N_c \delta_{ab} \delta^{pq}
\int \frac{d^3p}{(2\pi )^3} \, \frac{ 2f_f({\bf p})-1}{E_p}.
\ea
The result holds in the hard-loop approximation. For the pseudoscalar $Y_p$ we obtain the same expression because
$\beta^p_{ij}\beta^q_{ji}=-4\delta^{pq}$, $\gamma_5 \gamma_\mu = -\gamma_\mu \gamma_5$ and $\gamma_5^2 = 1$.
Therefore, we replace the indices $p,q$ by $A,B$ and we write down the result (\ref{P-k-a-2}) as 
\ba
\label{P-k-a-5}
_{(a)}P^{AB}_{ab}(k) = -4 g^2 N_c \delta_{ab} \delta^{AB}
\int \frac{d^3p}{(2\pi )^3} \, \frac{ 2f_f({\bf p})-1}{E_p}.
\ea

\begin{center}
{\it Gluon-scalar loop}
\end{center}
\label{g-s-loop-sym}

The contribution represented by the graph depicted in Fig.~\ref{fig-scalar}b equals
\ba
\label{P-k-b-1}
{_{(b)}}P^{AB}_{ab}(k) 
&=& 
- \frac{ig^2}{2} N_c \delta_{ab} \delta^{AB} \int \frac{d^4p}{(2\pi)^4}
\Big[(p+2k)^\mu (p+2k)^\nu \Delta^+(p+k) \,  D_{\mu\nu}^{\rm sym}(p)\qquad
\\ [2mm] \nn
&& \qquad\qquad\qquad\qquad\qquad\qquad
+ (p+k)^\mu (p+k)^\nu \Delta^{\rm sym}(p) \, D_{\mu\nu}^-(p-k) \Big] ,
\ea
which after the substitution of the functions $D^-_{\mu\nu}, \, D_{\mu\nu}^{\rm sym}$
and $\Delta^{+}, \Delta^{\rm sym}$ in the form (\ref{D-adv}), (\ref{D-sym}), (\ref{Del-ret}), and 
(\ref{Del-sym}) leads to
\ba
\label{P-k-b-3}
_{(b)}P^{AB}_{ab}(k) = \frac{g^2}{2} N_c \delta_{ab} \delta^{AB}
\int \frac{d^3p}{(2\pi )^3} \frac{4f_g({\bf p}) - 2f_s({\bf p}) + 1}{E_p} ,
\ea
which is valid within the hard-loop approximation.

\begin{center}
{\it Gluon and scalar tadpoles}
\end{center}
\label{gluon-scalar-tadpole-sym}

The contributions coming from the gluon tadpole shown in Fig.~\ref{fig-scalar}c and the scalar tadpole from Fig.~\ref{fig-scalar}d equal, respectively, 
\ba
\label{P-k-c-2}
_{(c)}P^{AB}_{ab}(k) 
&=& 
- 2 g^2 N_c \delta_{ab} \delta^{AB}
\int \frac{d^3p}{(2\pi)^3} \frac{2f_g({\bf p})+1}{E_p},
\\ [2mm]
\label{P-k-d-2}
_{(d)}P^{AB}_{ab}(k) 
&=& 
- 5g^2 N_c \delta_{ab} \delta^{AB}
\int \frac{d^3p}{(2\pi)^3} \frac{2f_s({\bf p})+1}{2E_p}.
\ea
In both cases the symmetry factor $1/2$ is included. 

\begin{center}
{\it Final result}
\end{center}
\label{fr-scalar-sym}

Summing up all contributions, we obtain the final formula of scalar self-energy
\be
\label{P-k-final}
P^{AB}_{ab}(k) = - g^2 N_c \delta_{ab} \delta^{AB}
\int \frac{d^3p}{(2\pi)^3} \frac{f({\bf p})}{E_p},
\ee
which depends, as $\Pi$ and $\Sigma$, on the effective distribution function (\ref{f-def}).

Having obtained the self-energies of both ${\mathcal N}=1$ SUSY QED and ${\mathcal N}=4$ super Yang-Mills, we are obviously in the offing to extract spectra of collective excitations. Before finding them let us first sum up and discuss the consequences of our findings.

\subsection{Universality of the self-energies}
\label{ssec-universality-se}

In this part we are going to summarise our results derived in the subsections \ref{ssec-susy-qed} and \ref{ssec-super-ym} so as that one can easily observe arising universality of the self-energies. 

In the Tables \ref{tab-gluons}, \ref{tab-fermions}, and \ref{tab-scalars} we present all, discussed before, diagrams of the lowest order (one loop) contributions to the self-energies of gauge bosons, fermions, and scalars, respectively, for all studied theories. The curly, plain, dotted, and dashed lines denote, respectively, the gauge, fermion, ghost, and scalar fields.

\begin{table}[!h]
\caption{\label{tab-gluons} The diagrams of the lowest order contributions to the polarisation tensors.}
\centering
\begin{tabular}{m{5cm} m{8cm}}
\hline \hline
\vspace{1mm}
Plasma system & Lowest order diagrams
\\
\hline \hline
\\
QED & \includegraphics[scale=0.2]{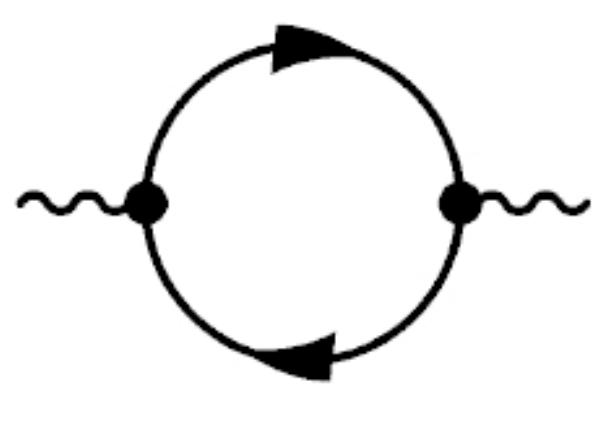}
\\
scalar QED & \includegraphics[scale=0.2]{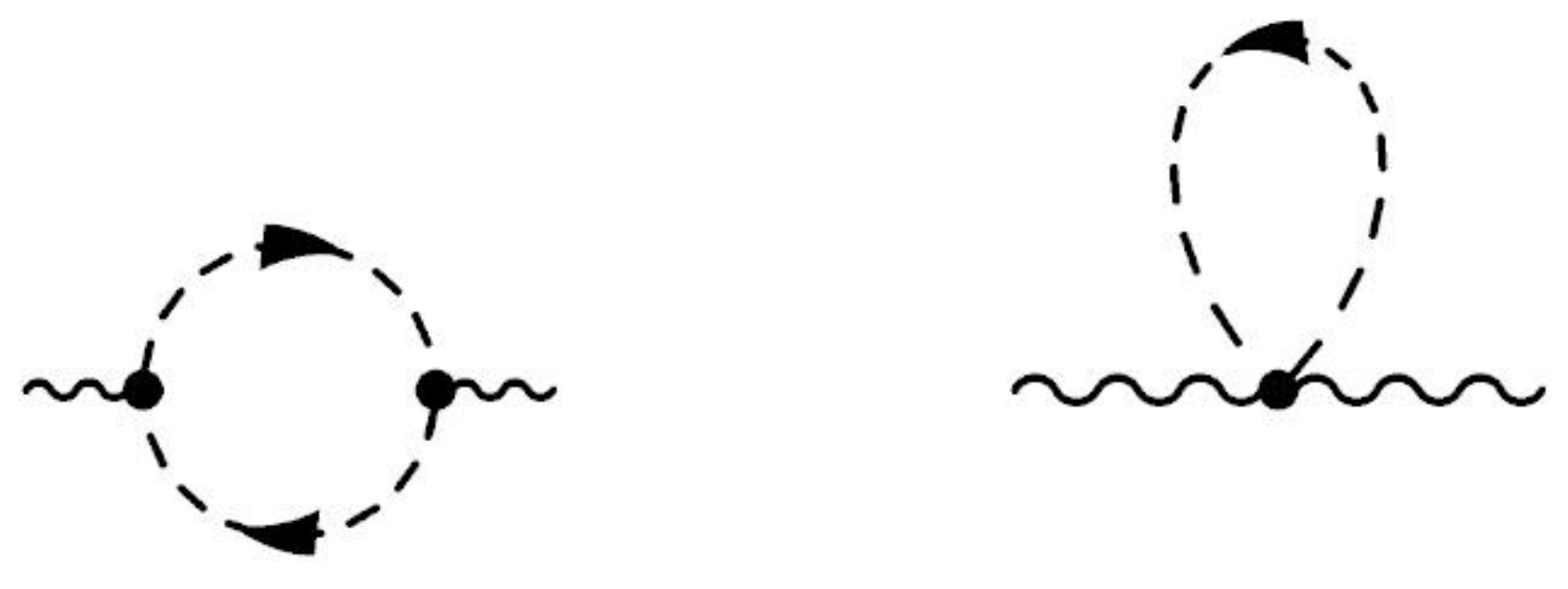}
\\
${\cal N}=1$ SUSY QED & \includegraphics[scale=0.2]{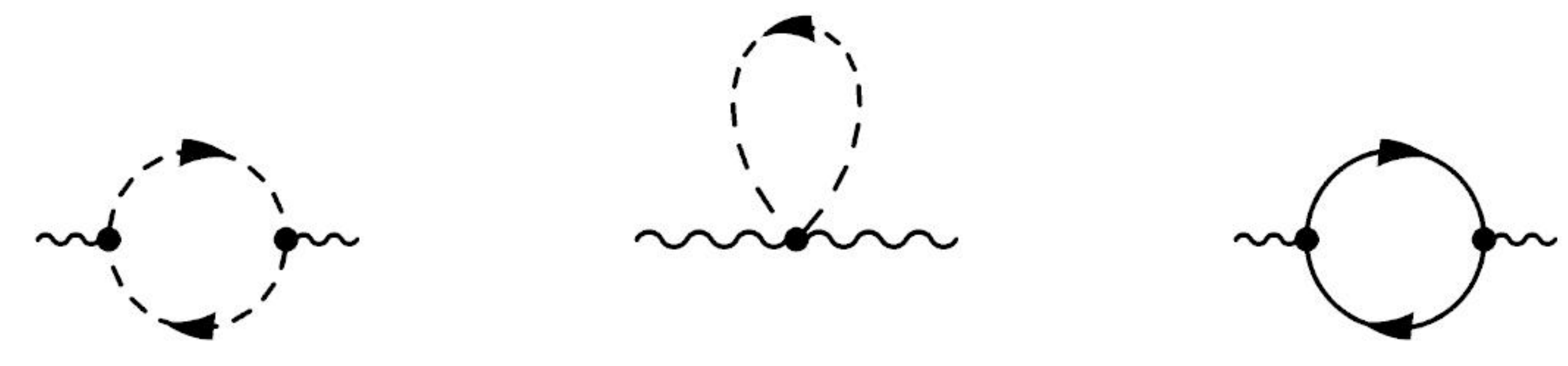}
\\
Yang-Mills & \includegraphics[scale=0.2]{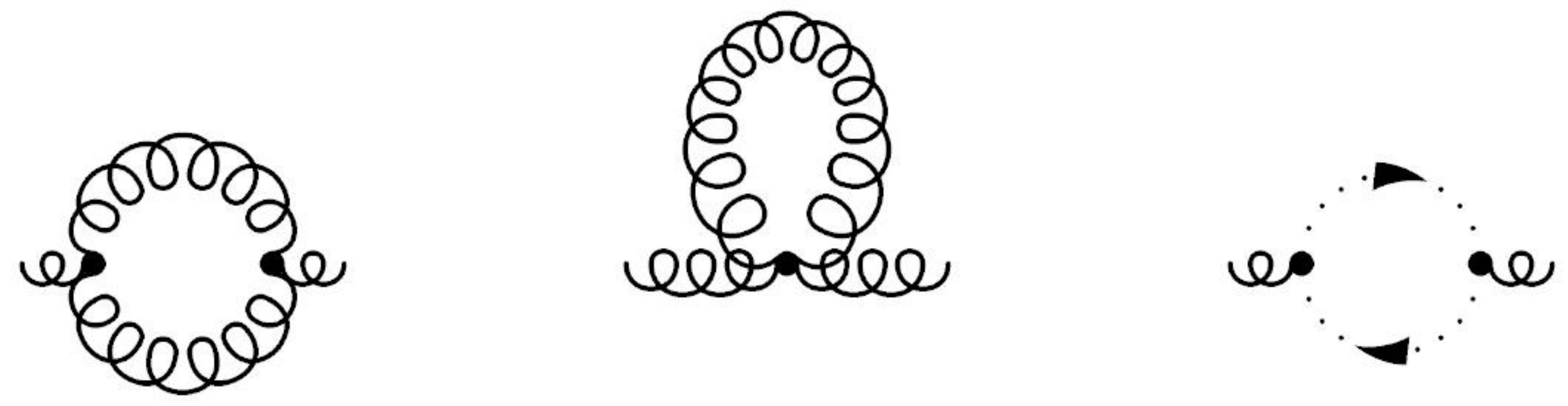}
\\
QCD & \includegraphics[scale=0.2]{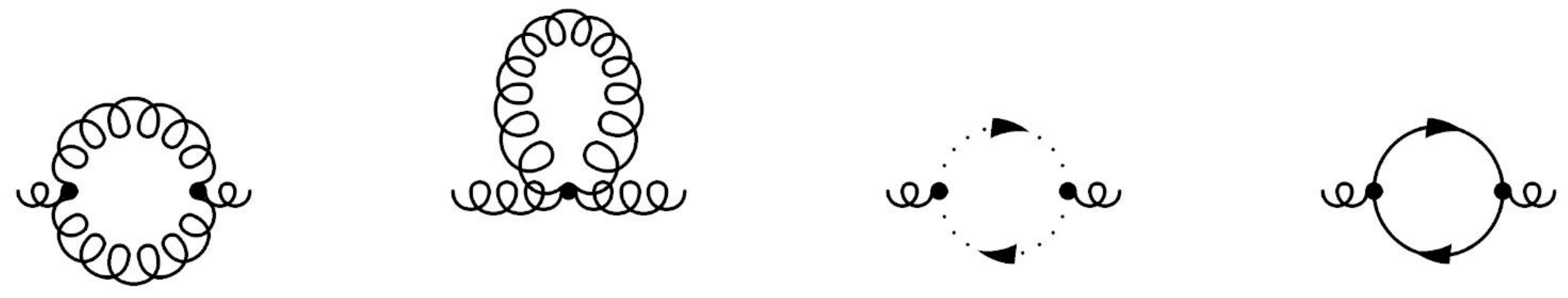}
\\
${\cal N}=4$ super Yang-Mills & \includegraphics[scale=0.2]{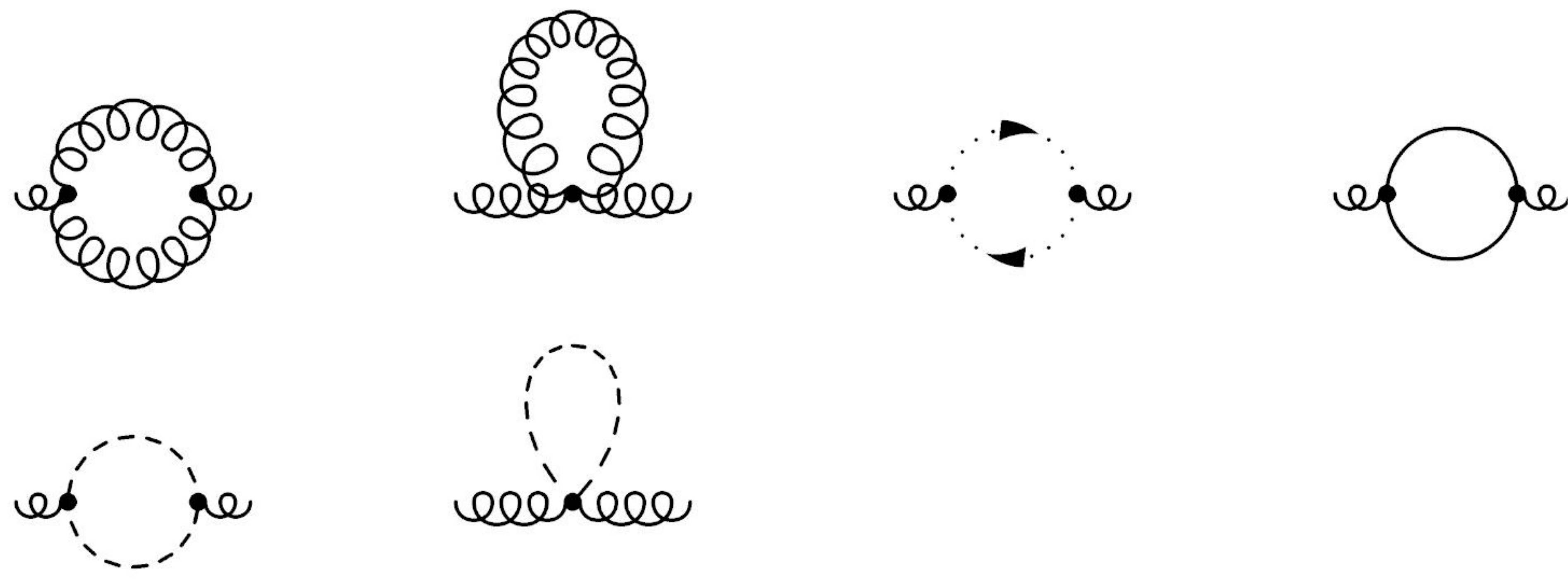}
\\
\hline \hline
\end{tabular}
\end{table}

As seen in Table \ref{tab-gluons}, both the number of diagrams contributing to the polarisation tensor and their forms are different for each theory. We have the fermion, scalar and gluon loops and the scalar and gluon tadpoles which differently depend on the external momentum. Accordingly, there is no surprise that the polarisation tensors $\Pi^{\mu \nu}(k)$ are quite different for each theory. However, when the external momentum $k$ is much smaller than the internal momentum $p$, which flows along the loop and is carried by a plasma constituent, that is when the long-wavelength limit is taken, we get a very striking result: the (retarded) polarisation tensors of all theories are of the same form 
\ba
\label{Pi-k-final}
\Pi^{\mu \nu}(k)= C_{\Pi}\int \frac{d^3p}{(2\pi)^3} \frac{f_{\Pi}({\bf p})}{E_p} 
\frac{k^2 p^\mu p^\nu - (k^\mu p^\nu + p^\mu k^\nu - g^{\mu \nu} (k\cdot p))
(k\cdot p)}{(k\cdot p + i 0^+)^2},
\ea
where $C_{\Pi}$ is the factor and $f_{\Pi}({\bf p})$ the effective distribution function of plasma constituents which are both given in Table \ref{tab-factors-gluons} for each plasma system.

\begin{table}[!h]
\caption{\label{tab-factors-gluons} The factors entering the polarisation tensors.} \centering
\begin{tabular}{lcc}
\hline \hline
Plasma system & $C_{\Pi}$ &  $f_{\Pi}({\bf p})$
\\
\hline \hline
\\
\vspace{1mm}
QED & $e^2$ &  $2 f_e({\bf p}) + 2 \bar f_e({\bf p})$
\\
\vspace{1mm}
scalar QED & $e^2$ &  $f_s({\bf p}) + \bar f_s({\bf p})$
\\
\vspace{1mm}
${\cal N}=1$ SUSY QED & $e^2$ &  $2 f_e({\bf p}) + 2 \bar f_e({\bf p}) + 2f_s({\bf p}) + 2\bar f_s({\bf p}) $
\\
\vspace{1mm}
Yang-Mills & $~~~g^2 N_c  \delta^{ab}~~~$ & $2 f_g ({\bf p})$
\\
\vspace{1mm}
QCD &  $g^2 N_c  \delta^{ab}$ & $2 f_g ({\bf p}) + \frac{N_f}{N_c} \big(f_q ({\bf p}) + \bar f_q ({\bf p}) \big) $
\\
\vspace{1mm}
${\cal N}=4$ super Yang-Mills &  $g^2 N_c  \delta^{ab}$ & $2f_g ({\bf p}) + 8f_f ({\bf p}) + 6f_s ({\bf p}) $
\\
\hline \hline
\end{tabular}
\end{table}

The functions $f_e({\bf p})$ and $\bar f_e({\bf p})$ denote, as previously, the electron and, respectively, positron distribution functions. The meaning of other functions can be easily guessed. All functions are normalized in such a way that
\ba
\rho_f = \int \frac{d^3p}{(2\pi)^3} f_f ({\bf p})
\ea
is density of particles of a type $f$ of a given spin and colour, if any. Particles of the same type but different spin and/or colour are assumed to have the same momentum distribution. The left and right selectrons in ${\cal N}=1$ SUSY QED have the same  momentum distribution as well. It is also assumed that quarks of all flavors, similarly as all fermions and all scalars in the $\mathcal{N}=4$ super Yang-Mills plasma, have the same momentum distribution. In case of non-supersymmetric plasmas, there is subtracted from the formula (\ref{Pi-k-final}) the (infinite) vacuum contribution which otherwise survives when $f_{\Pi}({\bf p})$ is sent to zero. The subtraction is not needed in the supersymmetric theories where the vacuum effect cancels out.  The polarisation tensor (\ref{Pi-k-final}), which is chosen to obey the retarded initial condition, is symmetric in Lorentz indices, $\Pi^{\mu\nu}(k) = \Pi^{\nu\mu}(k)$, and transverse, $k_\mu \Pi^{\mu\nu}(k)=0$, which guarantees its gauge independence. We repeat that the transversality of $\Pi^{\mu\nu}(k)$ is not an assumption but it automatically results from the calculations. 

One wonders how the universality of the polarisation tensor emerges. This is not the case that every one-loop contribution behaves in the same way in the long-wavelength limit. Just the opposite, the fermion loops contribute differently than boson ones, and the tadpoles are different than the loops. However, every subset of diagrams which is, as a sum of the diagrams, gauge independent, has the same  long-wavelength limit. For example, in the $\mathcal{N}=4$ super Yang-Mills theory we have three such subsets. The first one is simply the fermion loop, the second one is the sum of the scalar loop and scalar tadpole, and the third gauge independent subset is the sum of the gluon loop, the gluon tadpole and the ghost loop. 

Let us now discuss the fermion self-energies. In Table \ref{tab-fermions} there are listed the lowest order contributions to the fermion self-energies of every theory.
\begin{table}[!h]
\caption{\label{tab-fermions} The diagrams of the lowest order contributions to the fermion self-energies.}\centering
\begin{tabular}{m{6cm} m{7cm}}
\hline \hline
Plasma system & Lowest order diagrams
\\
\hline \hline
\\
QED & \includegraphics[scale=0.2]{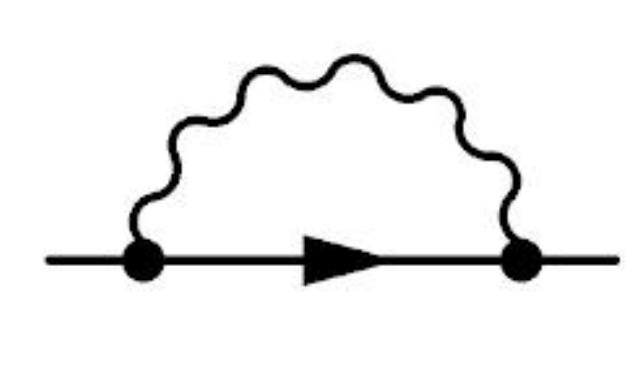}
\\
electron in ${\cal N}=1$ SUSY QED & \includegraphics[scale=0.2]{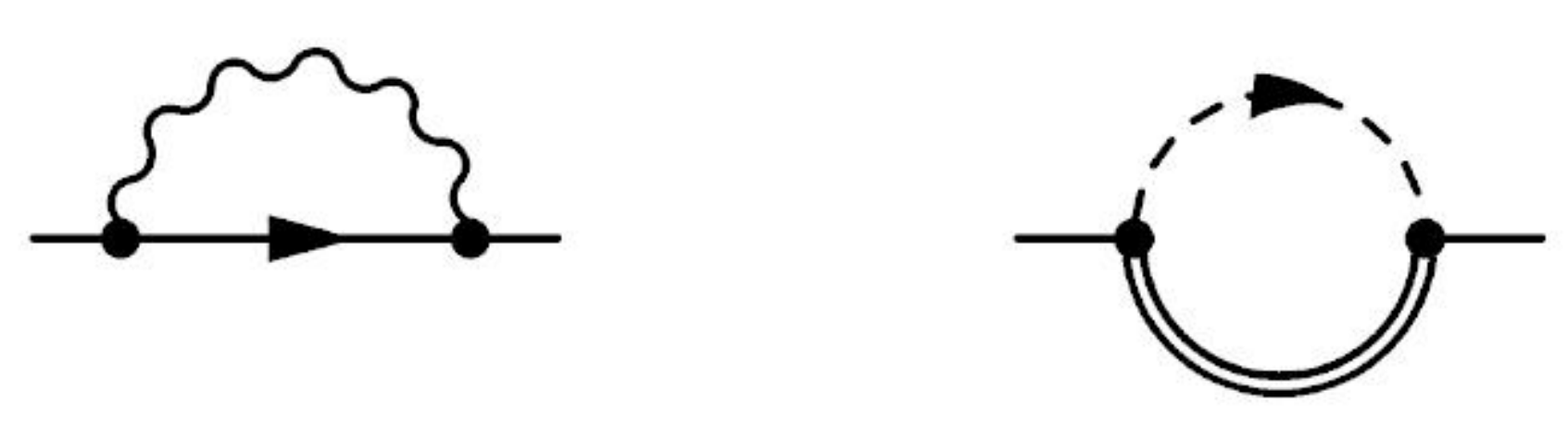}
\\
photino in ${\cal N}=1$ super QED & \includegraphics[scale=0.2]{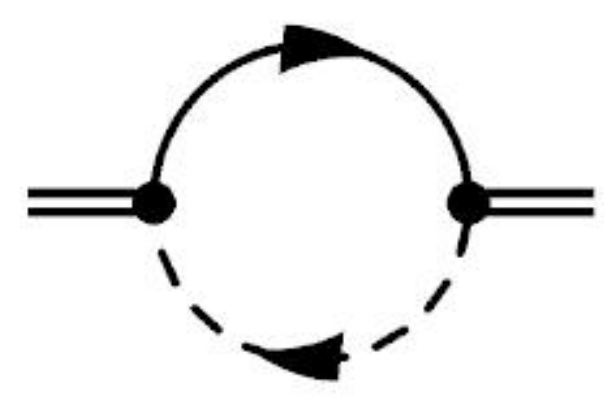}
\\
QCD & \includegraphics[scale=0.2]{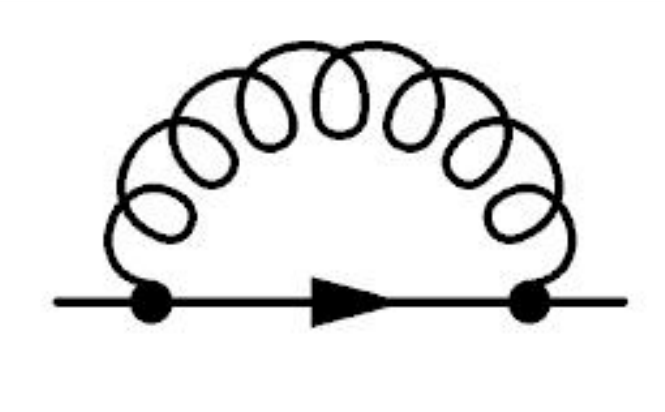}
\\
${\cal N}=4$ super Yang-Mills & \includegraphics[scale=0.2]{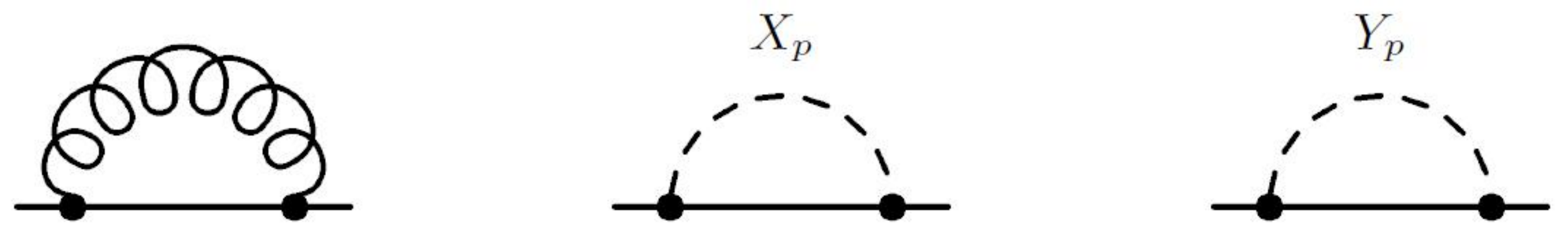}
\\
\hline \hline
\end{tabular}
\end{table}
In case of the ${\cal N}=1$ SUSY QED, there are the Dirac fermions and Majorana fermions which have to be treated differently. As in case of the polarisation tensor, the fermion self-energies $\Sigma(k)$ are quite different for each theory. However, when the hard-loop approximation is applied, the (retarded) self-energies of all theories are of the same form 
\ba
\label{Si-k-final}
\Sigma(k) = C_{\Sigma} \int \frac{d^3p}{(2\pi )^3}
\frac{f_{\Sigma}({\bf p})}{E_p}  \, \frac{p\sla}{k\cdot p + i 0^+},
\ea
where $C_{\Sigma}$ and $f_{\Sigma}({\bf p})$ are both given in Table \ref{tab-factors-fermions} for each plasma system. The indices $m, n = 1, 2, \dots N_c$ label quark colours in the fundamental representation of ${\rm SU}(N_c)$ group.
\begin{table}[!h]
\caption{\label{tab-factors-fermions} The factors entering the fermion self-energies.}\centering
\begin{tabular}{lcc}
\hline \hline
Plasma system & $C_{\Sigma}$ &  $f_{\Sigma}({\bf p})$
\\
\hline \hline
\\
\vspace{1mm}
QED &  $\frac{e^2}{2}$ &  $ 2 f_\gamma ({\bf p}) + f_e({\bf p}) + \bar f_e({\bf p}) $
\\
\vspace{1mm}
electron in ${\cal N}=1$ SUSY QED &  $\frac{e^2}{2}$ &  $ 2 f_\gamma ({\bf p}) + f_e({\bf p}) + \bar f_e({\bf p}) + 2 f_{\tilde\gamma} ({\bf p})$ 
\\
\vspace{1mm}
 &  &  $ + f_s({\bf p}) + \bar f_s({\bf p}) $
\\
\vspace{1mm}
photino in ${\cal N}=1$ SUSY QED &  $\frac{e^2}{2}$ &  $f_e({\bf p}) + \bar f_e({\bf p}) + f_s({\bf p}) + \bar f_s({\bf p}) $
\\
\vspace{1mm}
QCD &  $~~~ \frac{g^2}{2} \frac{N_c^2-1}{2 N_c} \delta^{m n} \delta^{ij}~~~$ & $2 f_g ({\bf p}) + N_f \big(f_q ({\bf p}) + \bar f_q ({\bf p}) \big) $
\\
\vspace{1mm}
${\cal N}=4$ super Yang-Mills & $\frac{g^2}{2} N_c  \delta^{ab} \delta^{ij}$ & $2f_g ({\bf p}) + 8f_f ({\bf p}) + 6f_s ({\bf p}) $
\\
\hline \hline
\end{tabular}
\end{table}

Finally we discuss the scalar self-energies. Table \ref{tab-scalars} shows the diagrams of the lowest order contributions to the scalar self-energy of three theories where scalars occur.  As in case of the polarisation tensors and fermion self-energies, the self-energy of scalars $P(k)$ are quite different for each theory. However, within the hard-loop approximation we obtain the amazingly repetitive result - the scalar self-energies of all theories have the same form 
\ba
\label{P-k-final}
P(k) = - C_P \int \frac{d^3p}{(2\pi)^3} \frac{f_P({\bf p})}{E_p},
\ea
where $C_P$ and $f_P({\bf p})$ are both given in Table \ref{tab-factors-scalars} for each plasma system. As seen, the self-energy (\ref{P-k-final}) is real, negative and it is independent of the wavevector $k$.

\begin{table}[!h]
\caption{\label{tab-scalars} The diagrams of the lowest order contributions to the scalar self-energies.}\centering
\begin{tabular}{m{5cm} m{8cm}}
\hline \hline
Plasma system & Lowest order diagrams
\\
\hline \hline
\\
scalar QED &  \includegraphics[scale=0.2]{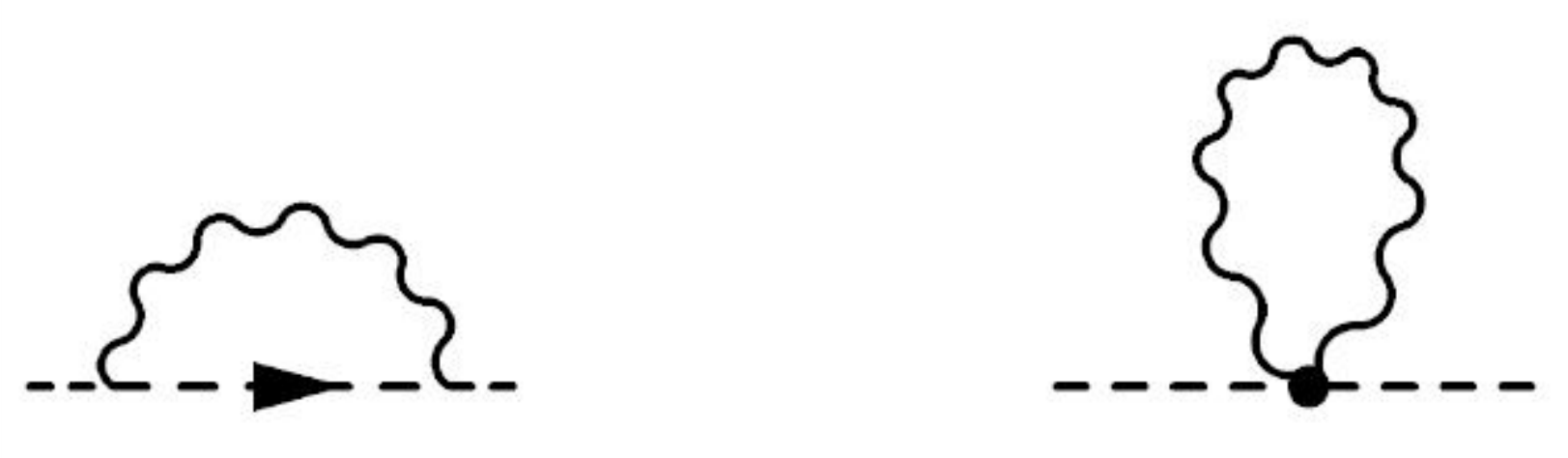}
\\
${\cal N}=1$ SUSY QED & \includegraphics[scale=0.2]{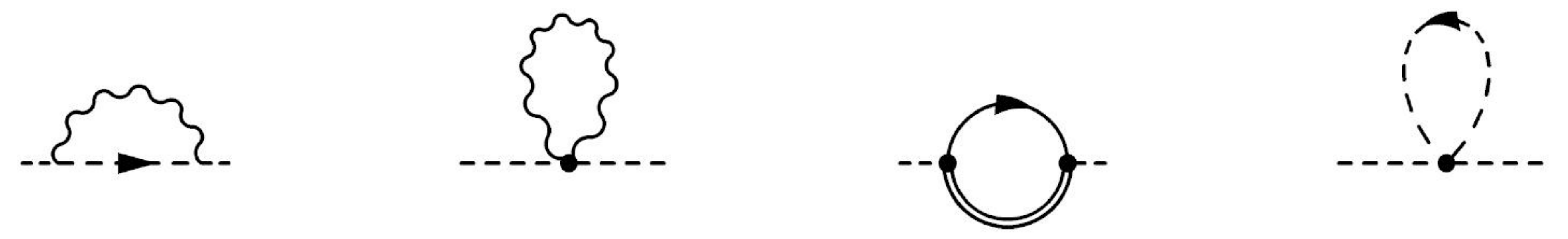}
\\
${\cal N}=4$ super Yang-Mills & \includegraphics[scale=0.2]{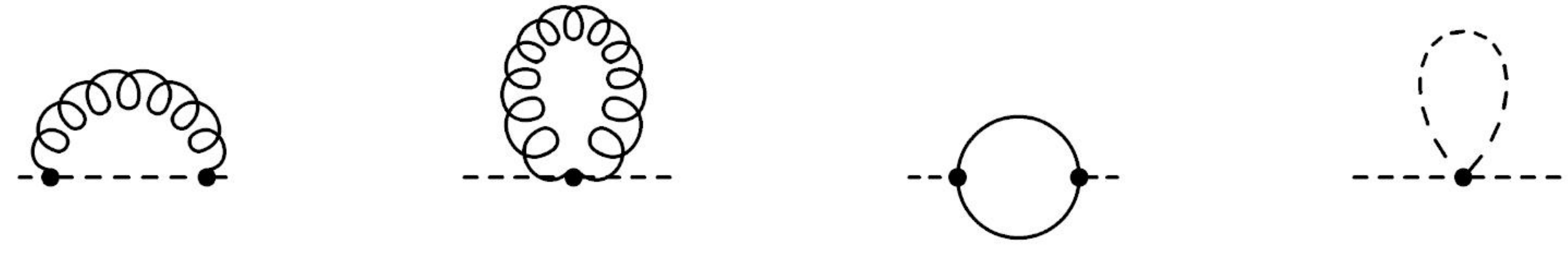}
\\
\hline \hline
\end{tabular}
\end{table}

\begin{table}[!h]
\caption{\label{tab-factors-scalars} The factors entering the scalar self-energies.}\centering
\begin{tabular}{lcc}
\hline \hline
Plasma system &  $C_P$ &  $f_P({\bf p})$
\\
\hline \hline
\\
\vspace{1mm}
scalar QED & $e^2$ & $ 2 f_\gamma ({\bf p}) + f_s({\bf p}) + \bar f_s({\bf p}) $
\\
\vspace{1mm}
${\cal N}=1$ SUSY QED & $e^2$ & $ 2 f_\gamma ({\bf p}) + f_e({\bf p}) + \bar f_e({\bf p}) + 2 f_{\tilde\gamma} ({\bf p}) $
\\
 &  &  $+ f_s({\bf p}) + \bar f_s({\bf p}) $
\\
\vspace{1mm}
${\cal N}=4$ super Yang-Mills & $~~~ g^2 N_c  \delta^{ab} \delta^{AB}~~~$ & $2f_g ({\bf p}) + 8f_f ({\bf p}) + 6f_s ({\bf p}) $
\\
\hline \hline
\end{tabular}
\end{table}
   
The universal expressions of the self-energies (\ref{Pi-k-final}), (\ref{Si-k-final}), and (\ref{P-k-final}) have been obtained in the limit when the external momentum $k$ is much smaller than the internal momentum $p$ which is carried by a plasma constituent. However, it appears that the self-energies (\ref{Pi-k-final}), (\ref{Si-k-final}), and (\ref{P-k-final}) are valid when the external momentum $k$ is not too small. It is most easily seen in case of the fermion self-energy (\ref{Si-k-final}) which diverges as $k \rightarrow 0$. When we deal with an equilibrium (isotropic) plasma of the temperature $T$, the characteristic momentum of (massless) plasma constituents is of the order $T$. One observes that if the external momentum $k$ is of the order $g^2 T$, which is the ultrasoft scale, the self-energy (\ref{Si-k-final}) is not perturbatively small as it is of the order ${\cal O}(g^0)$. Therefore, the expression (\ref{Si-k-final}) is meaningless for $k \le g^2 T$.  Since $k$ must be much smaller than $p \sim T$, one arrives at the well-known conclusion that the self-energy (\ref{Si-k-final}) is valid at the soft scale - when $k$ is of the order $gT$.  Analysing higher order corrections to the self-energies (\ref{Pi-k-final}), (\ref{Si-k-final}), (\ref{P-k-final}), one shows that they are indeed valid for $k \sim gT$ and they break down at the magnetic scale because of the infrared problem of gauge theories, see {\it e.g.} \cite{Lebedev:1989ev} or the review \cite{Kraemmer:2003gd}. When the momentum distribution of plasma particles is anisotropic, instead of the temperature $T$, we have a characteristic four-momentum ${\cal P}^\mu$ of plasma constituents and the hard-loop approximation requires that ${\cal P}^\mu \gg k^\mu$ which should be understood as a set of four conditions for each component of the four-momentum $k^\mu$. Validity of the self-energies (\ref{Pi-k-final}), (\ref{Si-k-final}), and (\ref{P-k-final}) is then limited to $k^\mu \sim g {\cal P}^\mu$.

\subsection{Effective action}
\label{ssec-eff-action}

Since the calculated self-energies appeared to be universal, one can ask whether they could be of any other form. For this reason let us consider an effective action, the most fundamental quality encoding the dynamics of a given system\footnote{This subsection is based on our work published as \cite{Czajka:2014gaa}.}.

The hard-loop effective action was first derived for equilibrium and nonequilibrium systems in \cite{Taylor:1990ia,Frenkel:1991ts,Braaten:1991gm} and \cite{Mrowczynski:2004kv}, respectively. Let us add that the introduction of the effective action makes the formulation of the hard-loop approach gain some elegance and concision.

Since a self-energy of a given field is the second functional derivative of the action $S$ with respect to the field, it strongly constrains a possible form of the respective action. Consequently, the self-energies of gauge boson, fermion, and scalar fields equal
\ba
\label{se-Pi}
\Pi^{\mu \nu}(x,y) &=& \frac{\delta^2 S}{\delta A_\mu(x) \,\delta A_\nu(y)}, 
\\ [2mm]
\label{se-Sigma}
\Sigma (x,y) &=& \frac{\delta^2 S}{\delta \bar\Psi (x) \,\delta \Psi(y)}, 
\\ [2mm]
\label{se-P}
P(x,y) &=& \frac{\delta^2 S}{\delta \phi^*(x) \,\delta \phi (y)},
\ea
where the field indices, which are different for different theories under consideration, are suppressed. The action will be obtained by integrating the formulas (\ref{se-Pi})-(\ref{se-P}) over the respective fields. 

\subsubsection{Universality of the hard-loop action}
\label{sssec-universality-se}

Having the self-energies $\Pi^{\mu\nu}(k),\; \Sigma(k)$, and $P(k)$ given by Eqs.~(\ref{Pi-k-final}), (\ref{Si-k-final}), and (\ref{P-k-final}), respectively, we can reconstruct the effective action. Integrating the formulas (\ref{se-Pi})-(\ref{se-P}) over the respective fields, we obtain the Lagrangian densities
\ba
\label{action-A-1}
{\cal L}^{A}_2(x) &=&
\frac{1}{2} \int d^4y \; A_\mu(x) \Pi^{\mu \nu}(x-y) A_\nu(y) ,
\\ [2mm]
\label{action-Psi-1}
{\cal L}^{\Psi}_2(x) &=&
\int d^4y \; \bar\Psi(x) \Sigma (x-y) \Psi(y) ,
\\ [2mm]
\label{action-Phi-1}
{\cal L}^{\Phi}_2(x) &=&
\int d^4y \; \phi^*(x) P(x-y) \phi(y) .
\ea
In case of ${\cal N}=4$ super Yang-Mills, where the scalar fields are real, there is an extra factor 1/2 in the r.h.s of Eq.~(\ref{action-Phi-1}). The subscript `2' indicates that the above effective actions generate only two-point functions. We omit the field indices in Eqs.~(\ref{action-A-1})-(\ref{action-Phi-1}) to keep the expressions applicable to all considered theories. The action is obviously related to the Lagrangian density as $S = \int d^4x \, {\cal L}$. Using the explicit expressions of  the self-energies (\ref{Pi-k-final}), (\ref{Si-k-final}), and (\ref{P-k-final}),  the Lagrangians (\ref{action-A-1})-(\ref{action-Phi-1}) can be manipulated, as first shown in \cite{Braaten:1991gm}, to the forms
\ba
\label{action-A-2}
{\cal L}^{A}_2(x) &=& C_\Pi \int \frac{d^3p}{(2\pi )^3} \,
\frac{f_\Pi ({\bf p})}{E_p} \,
F_{\mu \nu} (x) {p^\nu p^\rho \over (p \cdot \partial)^2} F_\rho^{\mu} (x) ,
\\ [2mm]
\label{action-Psi-2}
{\cal L}^{\Psi}_2(x) &=& C_\Sigma
\int \frac{d^3p}{(2\pi )^3} \, \frac{ f_\Sigma({\bf p})}{E_p} \,
\bar{\Psi}(x) {p \cdot \gamma \over p\cdot \partial} \Psi(x) ,
\\ [2mm]
\label{action-Phi-2}
{\cal L}^{\Phi}_2(x) &=& - C_P
\int \frac{d^3p}{(2\pi )^3} \, \frac{f_P({\bf p})}{E_p} \;
\phi^*(x) \phi(x) ,
\ea
where the operator inverse to ${p \cdot \partial}$ acts as
\ba
\frac{1}{p \cdot \partial} \Psi(x) \equiv i\int \frac{d^4k}{(2\pi)^4} \frac{e^{i k \cdot x}}{p \cdot k} \Psi(k).
\ea
The operator $({p \cdot \partial})^{-2}$ is defined analogously. 

The $n-$point functions with $n > 2$, which are generated by the actions (\ref{action-A-2})-(\ref{action-Phi-2}), identically vanish, as the actions are quadratic in fields. We also observe that the action of scalars (\ref{action-Phi-2}) is gauge invariant for every theory which includes the scalar field. Moreover, the gauge boson action (\ref{action-A-2}) is invariant as well but only in the Abelian theories.  The fermion action is gauge dependent in all theories under consideration. Therefore, the fermion action and, in general, the gauge boson action need to be modified to comply with the principle of gauge invariance. This is achieved by simply replacing the usual derivative $\partial^\mu$ by the covariant one $D^\mu$ in Eqs.~(\ref{action-A-2}) and (\ref{action-Psi-2}). Thus, we obtain
\ba
\label{action-A-HL}
{\cal L}^{A}_{\rm HL}(x) &=& C_\Pi \int \frac{d^3p}{(2\pi )^3} \,
\frac{f_\Pi({\bf p})}{E_p} \,
F_{\mu \nu} (x) {p^\nu p^\rho \over (p \cdot D)^2} F_\rho^{\mu} (x) ,
\\ [2mm]
\label{action-Psi-HL}
{\cal L}^{\Psi}_{\rm HL}(x) &=& C_\Sigma
\int \frac{d^3p}{(2\pi )^3} \, \frac{ f_\Sigma({\bf p})}{E_p} \,
\bar{\Psi}(x) {p \cdot \gamma \over p\cdot D} \Psi(x) ,
\\ [2mm]
\label{action-Phi-HL}
{\cal L}^{\Phi}_{\rm HL}(x) &=& - C_P
\int \frac{d^3p}{(2\pi )^3} \, \frac{f_P({\bf p})}{E_p} \;
\phi^*(x) \phi(x) .
\ea
The forms of covariant derivatives present in Eqs.~(\ref{action-A-HL}) and (\ref{action-Psi-HL}) depend on the theory under consideration. In the electromagnetic theories, the derivative in the gauge boson action (\ref{action-A-HL}) is, as already mentioned,  the usual derivative while that in the fermion action  (\ref{action-Psi-HL}) is $D^\mu = \partial^\mu - ie A^\mu$. The operator $(p \cdot D)^{-1}$ acts as 
\ba
\label{inv-cov}
\frac{1}{p \cdot D} \Psi(x) \equiv \frac{1}{p \cdot \partial} \sum_{n=0}^\infty
\Big(- ie p \cdot A(x) \frac{1}{p \cdot \partial}\Big)^n  \Psi(x).
\ea
In the ${\cal N}=4$ super Yang-Mills the covariant derivatives in Eqs.~(\ref{action-A-HL}) and (\ref{action-Psi-HL})  are both in the adjoint representation of ${\rm SU}(N_c)$ gauge group. The formula (\ref{inv-cov}) should be then appropriately modified. In QCD, the covariant derivative in Eq.~(\ref{action-A-HL}) is in the adjoint representation but that in Eq.~(\ref{action-Psi-HL}) is in the fundamental one. As already mentioned,  there is an extra factor 1/2 in the r.h.s of Eq.~(\ref{action-Phi-HL}) in case of ${\cal N}=4$ super Yang-Mills.

The hard-loop actions (\ref{action-A-HL}), (\ref{action-Psi-HL}), and (\ref{action-Phi-HL}) are all of the universal form for a whole class of gauge theories. However, the case of Abelian fields differs from that of nonAbelian ones. In the electromagnetic theories the gauge boson and scalar actions are quadratic in fields.  Therefore, the $n-$point functions generated by these actions vanish for $n>2$. Only the fermion action generates the non-trivial three-point and higher functions. The action (\ref{action-Psi-HL}) is, in particular, responsible for a modification of the electromagnetic vertex. In the theories, both the gauge boson and fermion actions generate  the non-trivial three-point and higher functions. Therefore, the gluon-fermion, three-gluon, and four-gluon couplings are all modified. 

\subsubsection{Discussion}
\label{sssec-discussion}

We have shown that the hard-loop self-energies of gauge, fermion, and scalar fields are of the universal structures and so are the effective actions of QED, scalar QED, $\mathcal{N}=1$ SUSY QED, Yang-Mills, QCD, and $\mathcal{N}=4$ super Yang-Mills. One asks why the universality occurs physically. Taking into account a diversity of the theories - various field content and microscopic interactions - the uniqueness of the hard-loop effective action is rather surprising. 

To better understand the problem in physical terms, let us consider the QED plasma of spin 1/2 electrons and positrons and the scalar QED plasma of spin 0 particles and antiparticles. The universality of hard-loop action means that neither effects of quantum statistics of plasma constituents are observable nor the differences in elementary interactions which govern the dynamics of the two systems. Both facts can be understood as follows. The hard-loop approximation requires that the momentum at which a plasma is probed, that is the wavevector $k$, is much smaller than the typical momentum of a plasma constituent $p$. Therefore, the length scale, at which the plasma is probed, $1/k$, is much greater than the characteristic de Broglie wavelength of plasma particle, $1/p$. The hard-loop approximation thus corresponds to the classical limit where fermions and bosons of the same masses and charges are not distinguishable. The fact that the differences in elementary interactions are not seen results from the very nature of gauge theories - the gauge symmetry fully controls the interaction. And the hard-loop effective actions obey the gauge symmetry. 

The universality of the hard-loop actions has far-reaching physical consequences: the characteristics of all plasma systems under consideration, which occur at the soft scale, are qualitatively the same. In particular, spectra of collective excitations of gauge, fermion, and scalar fields are the same. Therefore, if the electromagnetic plasma with a given momentum distribution is, say, unstable, the quark-gluon plasma with this momentum distribution is unstable as well. We conclude that in spite of all differences, the plasma systems under consideration are very similar to each other at the soft scale. However, the hard-loop approach breaks down for the momenta at and below the magnetic sale. Then, systems governed by different theories can behave very differently. In particular, the QED plasma is very different from the QCD one, as in the latter case effects of confinement apparently appear at the magnetic scale. Recently, there have been undertaken several efforts to extend methods of the hard-loop approach to the ultrasoft scale \cite{Maas:2011se,Gao:2014rqa,Hidaka:2011rz,Blaizot:2014hka,Su:2014rma}. These efforts explicitly show limitations of the universality we have elaborated on here.

\subsection{Spectrum of collective modes}
\label{ssec-modes}

When the self-energies computed in Secs.~\ref{ssec-susy-qed} and \ref{ssec-super-ym} are substituted into the respective dispersion equations, collective modes can be found as solutions of the equations. Below we briefly discuss the gluon, fermion, and scalar excitations.

\begin{center}
{\it Spectrum of gauge boson collective modes}
\end{center}
\label{spec-gb}

As already discussed, the structure of polarisation tensor (\ref{Pi-k-final}) is universal. Therefore, the spectrum of collective excitations of gauge bosons is the same in all considered systems.

The polarisation tensor in equilibrium plasma, which is isotropic, can be decomposed into two components transverse and parallel to the wavevector ${\bf k}$, $\Pi_T$ and $\Pi_L$, which have the forms
\ba
\label{pi-tr}
\Pi_T(\omega,{\bf k} )&=&\frac{m_D^2}{2}\Bigg[\frac{\omega^2}{{\bf k}^2}+\frac{\omega}{2} \frac{{\bf k}^2-\omega^2}{|{\bf k}|^3} \ln\bigg(\frac{\omega+|{\bf k}|}{\omega-|{\bf k}|} \bigg)  \Bigg],\\
\label{pi-par}
\Pi_L(\omega,{\bf k})&=&m_D^2\frac{\omega^2}{{\bf k}^2}\Bigg[-1+\frac{\omega}{2|{\bf k}|} \ln\bigg(\frac{\omega+|{\bf k}|}{\omega-|{\bf k}|} \bigg) \Bigg],
\ea
where $m_D$ is the Debye mass, see, {\it e.g.}, the textbook \cite{LeBellac}. The Debye mass depends on the plasma under consideration and, for example, for QED plasma of massless electrons and positrons it equals $m_D=gT/\sqrt{6}$. For other considered systems the Debye mass is similar, the only difference lies in the numerical factors as discussed, {\it e.g.}, in Sec. \ref{basics} of this thesis. The decomposition of the polarisation tensor leads to the splitting of the dispersion equation (\ref{dis-eq-A}) into the transverse dispersion equation
\be
\label{dis-tr}
\omega^2 - {\bf k}^2-\Pi_T(\omega,{\bf k})=0
\ee
and the longitudinal one
\be
\label{dis-lon}
\omega^2 -\Pi_L(\omega,{\bf k})=0.
\ee

Eqs. (\ref{dis-tr}) and (\ref{dis-lon}) have analytical solutions only in special cases. In the long-wavelength limit, that is when $\omega \gg |{\bf k}|$, the dispersion relations are given in the approximate forms
\be
\label{sol-tr}
\omega^2_T({\bf k}) \approx \omega_p^2 + \frac{6}{5}{\bf k}^2,
\ee
\be
\label{sol-lon}
\omega^2_L({\bf k}) \approx \omega_p^2 + \frac{3}{5}{\bf k}^2.
\ee
where $\omega_p$ is the plasma (Langmuir) frequency giving the minimal oscillation frequency. This frequency is also called the thermal mass of a gauge boson and is related to the Debye mass as $\omega_p=m_D/\sqrt{3}$. As seen, in any considered plasma in equilibrium, there are two modes: a transverse mode and longitudinal one, a plasmon. A mode is called transverse when the electric field is transverse to the wavevector and a longitudinal one when the electric field is parallel to the wavevector. As the Maxwell equations show, the transverse modes, sometimes called magnetic, appear as the consequence of electric current oscillations and the longitudinal ones, called also electric, are associated with electric charge fluctuations. In the limit of vanishing wavevector ${\bf k}$, the transversal and longitudinal modes are not differentiable from each other and they propagate with the lowest frequency of plasma oscillations $\omega_p$.  

\begin{figure}[!h]
\centering
\includegraphics*[width=0.45\textwidth]{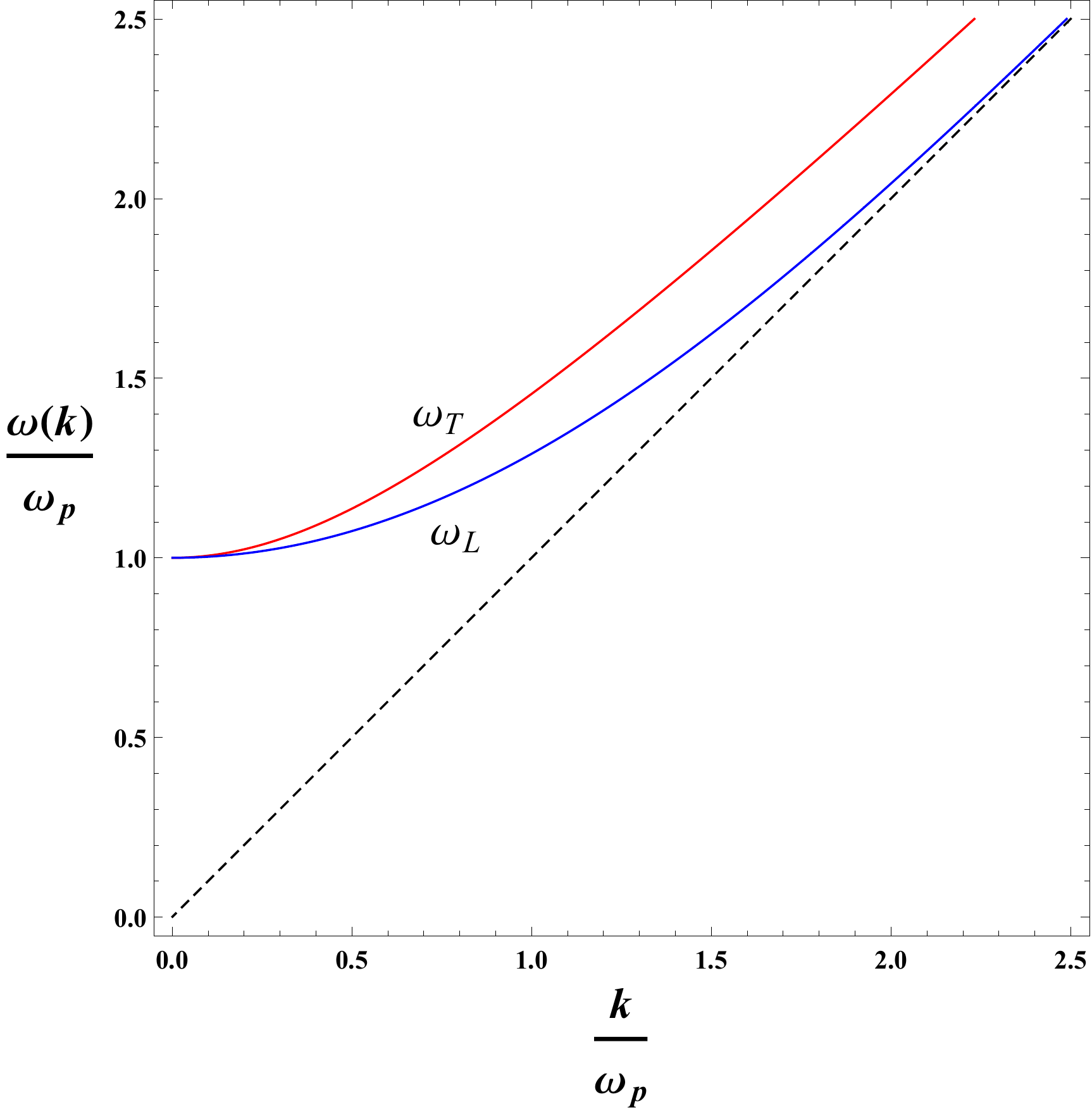}
\caption{Dispersion relations for transverse and longitudinal gauge bosons. }
\label{fig-coll-photons}
\end{figure}

Having given the transverse and longitudinal components of the polarisation tensor (\ref{pi-tr}) and (\ref{pi-par}) one also solves Eqs. (\ref{dis-tr}) and (\ref{dis-lon}) numerically. These numerical results are plotted in Fig. \ref{fig-coll-photons}, where the dispersion laws are expressed in units of $\omega_p$.

As seen, both modes lie above the light cone given by the relation $\omega=|{\bf k}|$. This means that there in no Landau damping in the limit of massless constituents of the plasma. At large momentum the transverse excitation behaves as a massive real gauge boson while the plasmon, as a pure effect of the medium, is exponentially suppressed. 

In case of the plasmas out of equilibrium we deal with a wide variety of possible collective excitations. In particular, there are unstable modes, see  {\it e.g.} the review  \cite{Mrowczynski:2007hb}, which exponentially grow in time and strongly influence the system's dynamics. An illuminating analysis of such modes is also shown in \cite{Carrington:2014bla}.

\begin{center}
{\it Spectrum of fermion collective modes}
\end{center}
\label{spec-fer}

The form of Majorana fermion self-energy (\ref{Si-k-final}), which holds for photinos of supersymmetric electromagnetic plasma and fermions of super Yang-Mills, happens to be the same as the Dirac fermion self-energy corresponding to electrons of QED and SUSY QED and quarks in QCD plasma. Therefore, we have an identical spectrum of excitations of fermions in all these systems. 

As discussed in, {\it e.g.}, \cite{LeBellac}, in equilibrium plasma there are two fermionic modes of opposite helicity over chirality ratio. Their forms are delivered by the following dispersion equations
\ba
\label{sig-solutions}
\omega \mp |{\bf k}| -\frac{m_f^2}{2|{\bf k}|} \Bigg[ \bigg( 1 \mp \frac{\omega}{|{\bf k}|} \bigg) \ln\frac{\omega+|{\bf k}|}{\omega-|{\bf k}|} \pm 2  \Bigg] =0,
\ea
where $m_f$ is an effective thermal mass gained by a fermion interacting with its environment. For small values of momentum ${\bf k}$, Eqs. (\ref{sig-solutions}) have the following approximate solutions
\be
\label{spect-fermions}
\omega_\pm({\bf k}) \approx m_f \pm \frac{1}{3}|{\bf k}|,
\ee
where the signs $\pm$ correspond to positive or negative helicity over chirality ratio $\pm \chi$. Since in the vacuum positive energy fermions posses  $+ \chi$, $\omega_+$ is related to the mode corresponding to elementary fermions. Another mode, a plasmino, is a specific medium effect with $- \chi$. When $|{\bf k}|=0$ the helicity is not defined, one obtains $\omega_+(0)=\omega_-(0)$.

\begin{figure}[!h]
\centering
\includegraphics*[width=0.45\textwidth]{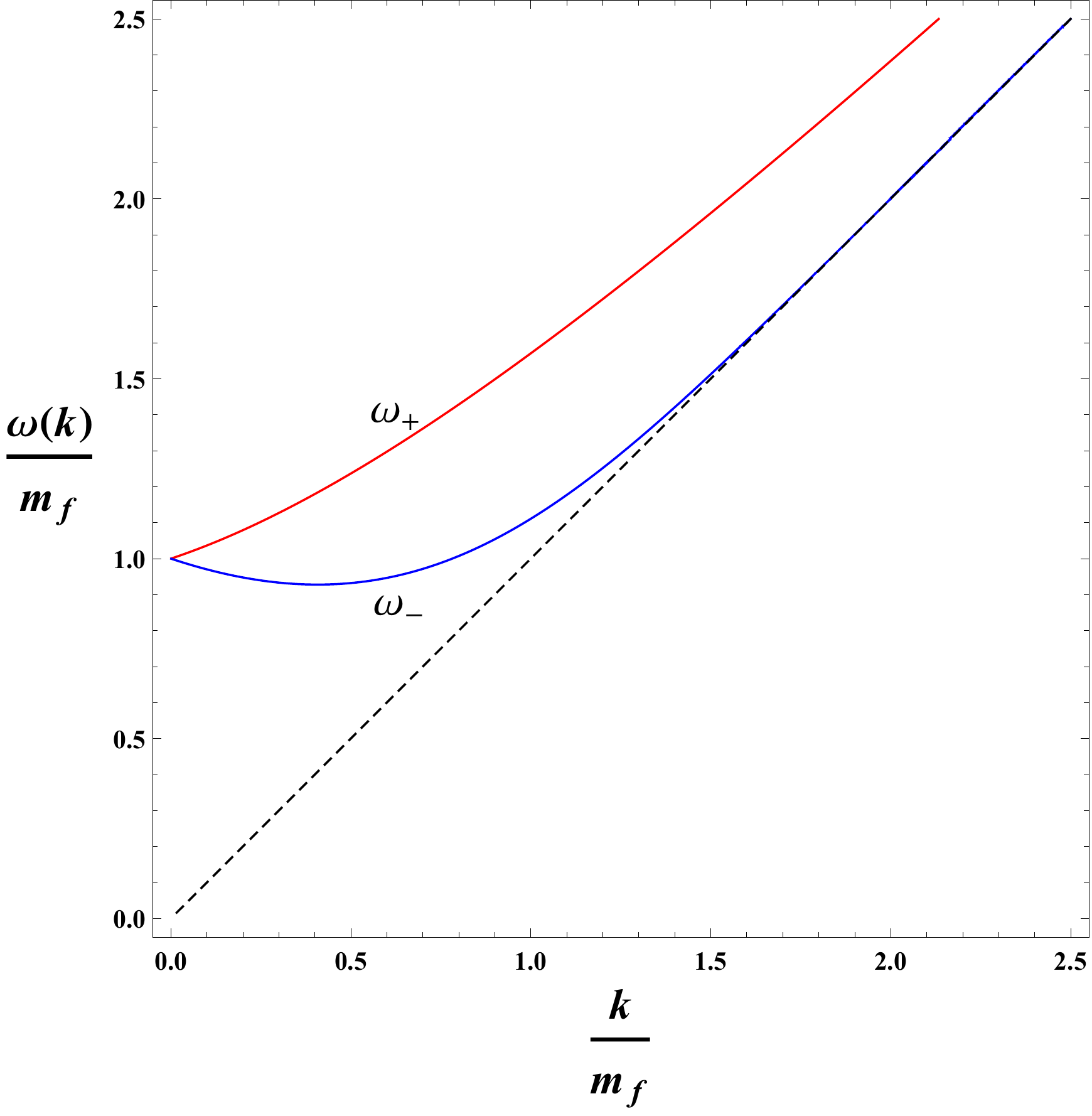}
\caption{Dispersion relations for the fermionic excitations. }
\label{fig-coll-electrons}
\end{figure}

The dispersion equations (\ref{sig-solutions}) for arbitrary momentum ${\bf k}$ can be solved numerically and they are plotted in Fig. \ref{fig-coll-electrons}. As seen, at large momentum $\omega_+({\bf k})$ describes elementary fermions whereas the plasmino, which has a shallow minimum at $|{\bf k}| \approx m_f$, is of no physical meaning.
  
In non-equilibrium plasma the spectrum of fermion collective excitations changes but no unstable modes have been found even for an extremely anisotropic momentum distribution \cite{Mrowczynski:2001az,Schenke:2006fz}. Then the spectra of fermion excitations are only slightly different against the equilibrium scenario. One claims that supersymmetry does not change anything.

\begin{center}
{\it Spectrum of scalar collective modes}
\end{center}
\label{spec-sca}

The scalar self-energy (\ref{P-k-final}) is  independent of momentum, it is negative and real. Therefore,  $P(k)$ can be written as $P(k) = - m^2_{\rm eff}$ where $m_{\rm eff}$ is the effective scalar mass. Then, the solutions of dispersion equation (\ref{dis-eq-selectron}) are $E_p = \pm \sqrt{m^2_{\rm eff} + {\bf p}^2}$.

We conclude this section by saying that the gauge boson and fermion excitations of supersymmetric plasma systems are the same as in their nonsupersymmetric counterparts. The scalar excitations are of the form of a free massive relativistic particle.

\newpage 
\thispagestyle{plain}

\section{Transport phenomena}
\label{sec-tr-charact}

Transport properties of any plasma system are related to the second order of perturbation theory and thus they might look as only corrections to more quintessential collective excitations discussed in the previous chapter. However, since the collective excitations result from a mean field dynamics, there is no entropy production, no dissipation. It is just transport phenomena that lead to the increase in an entropy and therefore they play a central role in a process of equilibration.

Transport coefficients of quark-gluon plasma include baryon and strangeness diffusion, electric charge and heat conductivity, shear and bulk viscosity and colour conductivity, and they have been studied in detail in weak coupling regime, see \cite{Arnold:2000dr,Arnold:2003zc,Arnold:2006fz,Arnold:1998cy} and references therein. Microscopically, the coefficients come from collisions of plasma constituents. As transport properties of a medium are controlled by elementary processes occurring in the system, one has to specify them and calculate the respective cross sections. 

Since the temperature is the only dimensional parameter, which characterises the equilibrium plasma of massless constituents, the parametric form of majority of transports coefficients can be determined. For example, the shear viscosity $\eta$ must be proportional to $T^3$ and the colour conductivity $\sigma_c$ to $T$. At the leading order $\eta \sim T^3/g^4\ln g^{-1}$ and $\sigma_c \sim T/\ln g^{-1}$ as shown in \cite{Arnold:2000dr} and \cite{Arnold:1998cy}. The dominant contributions to both transport coefficients of the quark-gluon plasma come from the binary collisions driven by a one-gluon exchange which corresponds to the matrix elements squared diverging as $t^{-2}$ for $t \rightarrow 0$. The factor $1/\ln g^{-1}$ appears due to the infrared singularity of the Coulomb-like interaction which is regulated by the gluon self-energy. Actually the physics behind the formulas of $\eta$ and $\sigma_c$ is rather different. The viscosity is governed by collisions with the momentum transfer of the order $gT$ while for the colour conductivity the softer collisions with the momentum transfer of the order $g^2T$ play a crucial role. 

One expects the same parametric form of $\eta$, $\sigma_c$ and other transport coefficients in the QCD and $\mathcal{N}=4$ super Yang-Mills plasma. The reason lies in the masslessness of the plasma constituents and Coulomb-like nature of interaction which dominates the dynamics of the system. As a result it does not much matter that the sets of elementary processes in the two plasma systems are different. The analysis of the shear viscosity of the $\mathcal{N}=4$ super Yang-Mills plasma is given in \cite{Huot:2006ys} and indeed it proves that the shear viscosity coefficients of the quark-gluon and super Yang-Mills plasmas differ only by numerical factors which mostly reflect different numbers of degrees of freedom in the two plasmas. It is worth noting that since the super Yang-Mills system is exactly conformal the bulk viscosity is identically zero.

In this section we focus on such characteristics which are not constrained by the dimensional analysis and can strongly depend on a specific scattering process under consideration. Namely, we consider collisional and radiative energy losses of a high-energy test particle traveling through hot plasmas. In this case not only the temperature but also the energy of the test particle may alter the characteristics. We start, however, with the analysis of all binary processes of $\mathcal{N}=1$ SUSY QED plasma and next of $\mathcal{N}=4$ super Yang-Mills one providing the respective cross sections.

\subsection{Cross sections of binary interactions}
\label{sssec-cross-sqed}

Here we derive and discuss the cross sections of all binary interactions of both the $\mathcal{N}=1$ SUSY QED and $\mathcal{N}=4$ super Yang-Mills plasmas which occur at the order of $\alpha^2$ ($\alpha \equiv g^2/4\pi$).

The Feynman diagrams corresponding to a given process $1+2 \rightarrow 3+4 $ are drawn in such a way that the particle $1$ comes from the upper left corner of the diagram, the particle $2$ comes from the lower left corner, the particle $3$ goes to the upper right, and the particle $4$ goes to the lower right corner. The four-momenta of incoming particles $1, 2$, respectively, are denoted as $p_1=(E_1, {\bf p}_1)$, $p_2=(E_2, {\bf p}_2)$ and these of outgoing particles $3, 4$ as $p_3=(E_3, {\bf p}_3)$, $p_4=(E_4, {\bf p}_4)$, respectively, regardless of whether it is a scalar, fermion or gauge boson particle. We follow here the convention used in the textbook by Bechler \cite{Bechler}.

The cross section of two interacting bosons is defined in the following way
\ba
\label{cs-def}
d\sigma'= \frac{1}{|{\bf v}_1-{\bf v}_2|} \frac{1}{2E_1} \frac{1}{2E_2} |\mathcal{M}|^2
\frac{d^3p_3}{(2\pi)^3 2E_3} \frac{d^3p_4}{(2\pi)^3 2E_4} (2\pi)^4 
\delta^{(4)}(p_1+p_2-p_3-p_4),
\ea
where $|{\bf v}_1-{\bf v}_2|$ is a relative velocity of incoming particles, the delta expresses the conservation of four-momenta and $\mathcal{M}$ is the Lorentz-invariant scattering amplitude of a given process. In case of interacting fermions we should make the replacement $\frac{1}{2E}\rightarrow\frac{m}{E}$ for every fermion in the formula (\ref{cs-def}). Hereinafter we premise that the incoming particles move parallel to each other, and therefore 
\ba
\label{factor-v}
E_1 E_2 |{\bf v}_1-{\bf v}_2| = \sqrt{(p_1 \cdot p_2)^2 - m_1^2 m_2^2}.
\ea
All the cross sections computed here are averaged over internal degrees of freedom of initial state particles $N^{\rm dof}_1$ and $N^{\rm dof}_2$ and summed over internal degrees of freedom of final state particles $N^{\rm dof}$. So, one may write 
\ba
\label{cs-def-pol}
|\overline{\mathcal{M}}|^2 = \frac{1}{N^{\rm dof}_1}\frac{1}{N^{\rm dof}_2} \sum_{N^{\rm dof}} |\mathcal{M}|^2.
\ea
Then the cross section is given as
\ba
\label{cross-sec}
d\sigma= \frac{1}{4} \frac{|\overline{\mathcal{M}}|^2}{\sqrt{(p_1 \cdot p_2)^2 - m_1^2 m_2^2}} 
\frac{d^3p_3}{(2\pi)^3 2E_3} \frac{d^3p_4}{(2\pi)^3 2E_4} (2\pi)^4 
\delta^{(4)}(p_1+p_2-p_3-p_4).
\ea

The computed cross sections are expressed through the Mandelstam invariants $s,t$ and $u$ defined in the standard way. For a process $1+2 \longrightarrow 3+4$, we have
\ba
\label{s}
s &\equiv& (p_1+p_2)^2 = (p_3+p_4)^2 ,
\\ [2mm]
t &\equiv& (p_1 - p_3)^2 = (p_2 - p_4)^2 ,
\\ [2mm]
\label{u}
u &\equiv& (p_1 - p_4)^2 = (p_2 - p_3)^2,
\ea
which satisfy the condition
\ba
\label{mand-var}
s+t+u=m_1^2+m_2^2+m_3^2+m_4^2.
\ea 

The cross section (\ref{cross-sec}) expressed by the Mandelstam variables gets the form
\ba
\label{cs-mand}
-\frac{d\sigma}{dt} = \frac{1}{16\pi}\frac{|\overline{\mathcal{M}}|^2}{s(s - m_1^2 m_2^2)} .
\ea
The matrix elements $\mathcal{M}$ is computed for every single process of a given system by means of the Feynman rules collected in Appendix \ref{appendix-FR}.

Further on all cross sections of processes occurring in both supersymmetric systems are given.
 
\subsubsection{Cross sections of the $\mathcal{N}=1$ SUSY QED plasma}
\label{ssec-cross-sec-qed}

We start with the processes occurring in the $\mathcal{N}=1$ SUSY QED plasma of electrons, selectrons, photons and photinos\footnote{The results presented here come from our paper \cite{Czajka:2011zn}.}. They are denoted as $e$, $\tilde{e}$, $\gamma$, $\tilde{\gamma}$, respectively. Since the procedure of derivation of any cross section of interest is well known and easily accessible, see the handbooks \cite{Bjorken,Bechler}, only the cross section of the Compton scattering on selectron is derived in detail. The remaining cross sections are simply listed. The Compton scattering is chosen as it is important in the context of the further discussion.

\begin{center}
{\it Compton scattering on selectron}
\end{center}
\label{compton-scat}

\begin{figure}[!h]
\centering
\includegraphics[scale=.45]{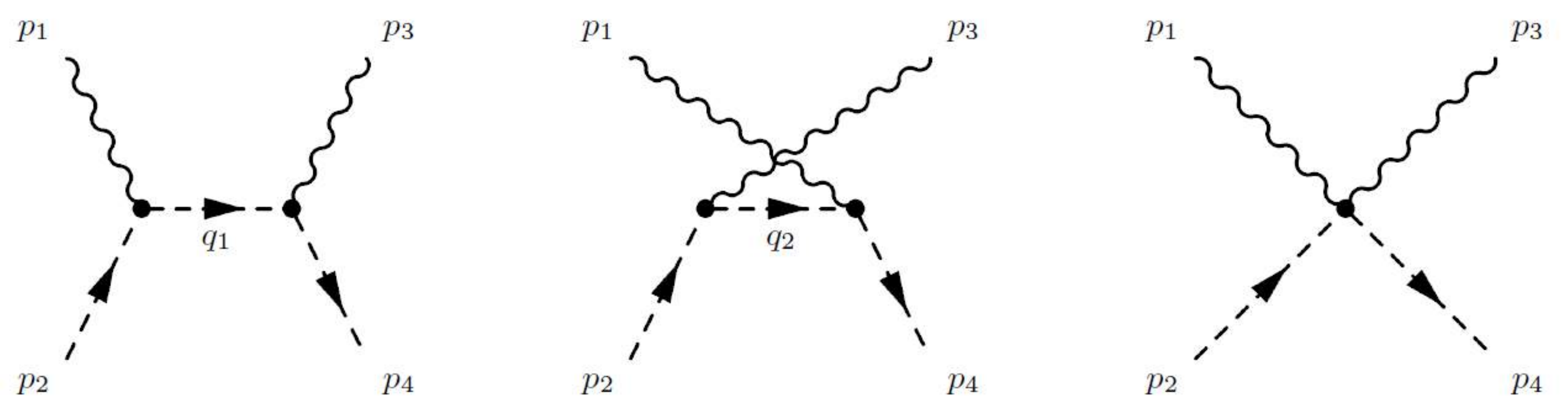}
\caption{Compton scattering on selectron. }
\label{compton-selectron}
\end{figure}

We consider here the process of a photon interacting with a left-handed selectron of negative charge $\gamma \tilde{e}^-_L \longrightarrow \gamma \tilde{e}^-_L$. The Feynamn diagrams corresponding to this process are shown in Fig~\ref{compton-selectron}.

The scattering amplitude is written as a sum of the three contributions as
\ba
\label{m-full}
\mathcal{M} = \mathcal{M}_1+\mathcal{M}_2+\mathcal{M}_3.
\ea
Then, following the Feynman rules collected in Appendix \ref{appendix-FR} we write down the respective contributions as
\ba
\label{m-1}
\mathcal{M}_1 &=& \varepsilon^\nu ({\bf p}_3,\lambda') \, (-ie(q_1+p_4)_\nu) \, i\Delta(q_1) \,
(-ie(q_1+p_2)_\mu) \, \varepsilon^\mu ({\bf p}_1,\lambda), 
\\ [2mm]
\label{m-2}
\mathcal{M}_2 &=& \varepsilon^\nu ({\bf p}_3,\lambda') \, (-ie(q_2+p_2)_\nu) \, i\Delta(q_2) \,
(-ie(q_2+p_4)_\mu) \, \varepsilon^\mu ({\bf p}_1,\lambda), 
\\ [2mm]
\label{m-3}
\mathcal{M}_3 &=& \varepsilon^\nu ({\bf p}_3,\lambda') \, 2ie^2 g_{\mu\nu} \, \varepsilon^\mu ({\bf p}_1,\lambda) ,
\ea
where $\varepsilon^\mu ({\bf p},\lambda)$ is a polarisation vector of a photon, $\Delta(q)$ is a scalar propagator with $q$ being $q_1=p_1+p_2$ or $q_2=p_2-p_3$. The expressions (\ref{m-1})-(\ref{m-3}) can be simplified to
\ba
\label{m-11}
\mathcal{M}_1 &=&\frac{-ie^2}{q_1^2} \, \varepsilon^\nu_{3,\lambda'} \, (q_1+p_4)_\nu \,
(q_1+p_2)_\mu \, \varepsilon^\mu_{1,\lambda}, 
\\ [2mm]
\label{m-22}
\mathcal{M}_2 &=& \frac{-ie^2}{q_2^2} \, \varepsilon^\nu_{3,\lambda'} \, (q_2+p_2)_\nu \,
(q_2+p_4)_\mu \, \varepsilon^\mu_{1,\lambda}, 
\\ [2mm]
\label{m-33}
\mathcal{M}_3 &=& 2ie^2 \, \varepsilon_{\mu,3,\lambda'} \, \varepsilon^\mu_{1,\lambda} ,
\ea
where we have introduced the notation $\varepsilon^\mu ({\bf p}_1,\lambda) \equiv \varepsilon^\mu_{1,\lambda}$ and  $\varepsilon^\nu ({\bf p}_3,\lambda') \equiv \varepsilon^\nu_{3,\lambda'}$ and substituted the explicit expression of the scalar propagator (\ref{scalar-prop}), where the masses are neglected as we study an ultrarelativistic plasma. Subsequently, we obtain the scattering amplitude squared as
\ba
\label{m-all}
|\mathcal{M}|^2= \mathcal{M}^* \mathcal{M}
&=&
\frac{e^4}{q_1^4}  \, (q_1+p_2)_\mu \, (q_1+p_4)_\nu \, \varepsilon^\nu_{3,\lambda'} \, 
\varepsilon^\sigma_{3,\lambda'} \, (q_1+p_4)_\sigma \, (q_1+p_2)_\rho \, 
\varepsilon^\rho_{1,\lambda} \, \varepsilon^\mu_{1,\lambda} \qquad
\\ [2mm] \nn
& + & \frac{e^4}{q_2^4}  \, (q_2+p_4)_\mu \,(q_2+p_2)_\nu \, \varepsilon^\nu_{3,\lambda'} \, 
\varepsilon^\sigma_{3,\lambda'} \, (q_2+p_2)_\sigma \, (q_2+p_4)_\rho \, \varepsilon^\rho_{1,\lambda} \, \varepsilon^\mu_{1,\lambda}  
\\ [2mm] \nn
& + & 
\frac{2e^4}{q_1^2 q_2^2}  \, (q_2+p_4)_\mu \,(q_2+p_2)_\nu \, \varepsilon^\nu_{3,\lambda'} \, \varepsilon^\sigma_{3,\lambda'} \, (q_1+p_4)_\sigma \, (q_1+p_2)_\rho \, \varepsilon^\rho_{1,\lambda} \, \varepsilon^\mu_{1,\lambda}  
\\ [2mm] \nn
& - & 
\frac{4e^4}{q_1^2} \, \varepsilon_{\mu,3,\lambda'} \, \varepsilon^\nu_{3,\lambda'} \, 
(q_1+p_4)_\nu \, (q_1+p_2)_\sigma \, \varepsilon^\sigma_{1,\lambda} \, \varepsilon^\mu_{1,\lambda} 
\\ [2mm] \nn
& - & 
\frac{4e^4}{q_2^2}  \, \varepsilon_{\mu,3,\lambda'} \, \varepsilon^\nu_{3,\lambda'} \, 
(q_2+p_2)_\nu \, (q_2+p_4)_\sigma \, \varepsilon^\sigma_{1,\lambda} \, \varepsilon^\mu_{1,\lambda} 
\\ [2mm] \nn
& + & 
4e^4  \, \varepsilon_{\mu,3,\lambda'} \, \varepsilon_{\nu,3,\lambda'} \, 
\varepsilon^\nu_{1,\lambda} \, \varepsilon^\mu_{1,\lambda}.
\ea
Next, we average over the polarisations of the initial photon and sum over the polarisations of the final one, making use of the formulas (\ref{cs-def-pol}) and next of (\ref{summation-bosons}). Therefore, we get
\ba
\label{m-all-av}
|\overline{\mathcal{M}}|^2 
&=&
\frac{e^4}{2}\bigg[\frac{1}{q_1^4}  \, (q_1+p_2)^2 (q_1+p_4)^2
 + \frac{1}{q_2^4}  \, (q_2+p_4)^2 \,(q_2+p_2)^2  
\\ [2mm] \nn
&& \qquad 
+ \frac{2}{q_1^2 q_2^2}  \, (q_2+p_4) \cdot (q_1+p_2) \,(q_2+p_2) \cdot (q_1+p_4)  
\\ [2mm] \nn
&& \qquad 
- \frac{4}{q_1^2} \, (q_1+p_4) \cdot (q_1+p_2) - \frac{4}{q_2^2} \, (q_2+p_2) \cdot (q_2+p_4)
+ 16 \bigg]. 
\ea
Taking into account that $q_1=p_1+p_2$ and $q_2=p_2-p_3$ and the mass-shell condition, $p_i^2=0$, we obtain
\ba
\label{m-all-av-1}
|\overline{\mathcal{M}}|^2 
\!\!&=&\!\!
\frac{e^4}{2}\bigg[
\frac{8 p_1 \cdot p_2 (p_1 \cdot p_2 + p_1 \cdot p_4 + p_2 \cdot p_4)}{(2 p_1 \cdot p_2)^2}
+ \frac{8 p_2 \cdot p_3 (p_2 \cdot p_3 + p_3 \cdot p_4 + p_2 \cdot p_4)}{(-2 p_2 \cdot p_3)^2} \qquad
\\ [2mm] \nn
&& 
+ \frac{(2p_1 \cdot p_2 + 2p_4 \cdot p_2 - p_3 \cdot p_4 - p_1 \cdot p_3 - p_2 \cdot p_3)}
{(2p_1 \cdot p_2) (-2p_2 \cdot p_3)} 
\\ [2mm] \nn
&& \qquad \qquad
\times (3p_2 \cdot p_4 +2p_1 \cdot p_4 - 2p_2 \cdot p_3 - p_1 \cdot p_3)
\\ [2mm] \nn
&& 
- \frac{4(3 p_1 \cdot p_2 + 2p_2 \cdot p_4 + p_1 \cdot p_4)}{2p_1 \cdot p_2} 
- \frac{4(3p_2 \cdot p_3 - 2p_2 \cdot p_4 + p_3 \cdot p_4)}{2p_2 \cdot p_3} 
+ 16 \bigg].
\ea
For massless particles the Mandelstam variables, defined by (\ref{s})-(\ref{u}), equal
\ba
s &=& 2p_1 \cdot p_2 = 2p_3 \cdot p_4, 
\\ [2mm]
t &=& - 2p_1 \cdot p_3 =-2p_2 \cdot p_4,
\\ [2mm]
u &=& - 2p_1 \cdot p_4 = - 2p_2 \cdot p_3.
\ea
The scattering amplitude expressed through $s$, $t$, and $u$ yields
\ba
\label{m-all-av-stu}
|\overline{\mathcal{M}}|^2 =
\frac{e^4}{2}\bigg[ \frac{4s^2}{s^2} + \frac{4u^2}{u^2} - \frac{t^2}{us} 
- \frac{6s - 4t - 2u}{s} - \frac{6u - 4t -2s}{u} + 16 \bigg],
\ea
which simplifies to
\ba
\label{m-all-av-stu-all}
|\overline{\mathcal{M}}|^2 = 4e^4.
\ea
With the scattering matrix (\ref{m-all-av-stu-all}) the cross section of the Compton scattering on the left selectron (\ref{cs-mand}) is given by
\ba
\label{cs-compton}
-\frac{d\sigma}{dt}= \frac{4\pi \alpha^2}{s^2},
\ea
where $\alpha=e^2/4\pi$. The cross section (\ref{cs-compton}) holds for selectrons of both positive and negative charge and of both left and right type. As seen, this cross section is independent of momentum transfer, that is, it does not depend on $t$ nor $u$. It implies that the scattering is isotropic in the centre-of-mass frame which is extremely different than scattering caused by one-photon-exchange interaction characteristic of the electromagnetic plasma. Further consequences of this property are discussed in the subsection \ref{ssec-tr-coef}.

All other processes together with corresponding Feynman diagrams and cross sections of the $\mathcal{N}=1$ SUSY QED plasma are listed in Table~\ref{tab-corr-binray}.
\begin{table}
\caption{\label{tab-corr-binray} Elementary processes in the $\mathcal{N}=1$ SUSY QED plasma.}
\begin{tabular}{m{.5cm} m{3.5cm} m{5cm} m{5cm}}
\hline \hline
$n^0$ & \hspace{0.4cm} Process & \hspace{0.9cm} Diagrams 
& \hspace{0.7cm} Cross section $-\frac{d\sigma}{dt}$
\\ \hline \hline \\
         1& $e^{\mp}e^{\mp} \longrightarrow e^{\mp}e^{\mp}$ &
        \includegraphics[scale=.25]{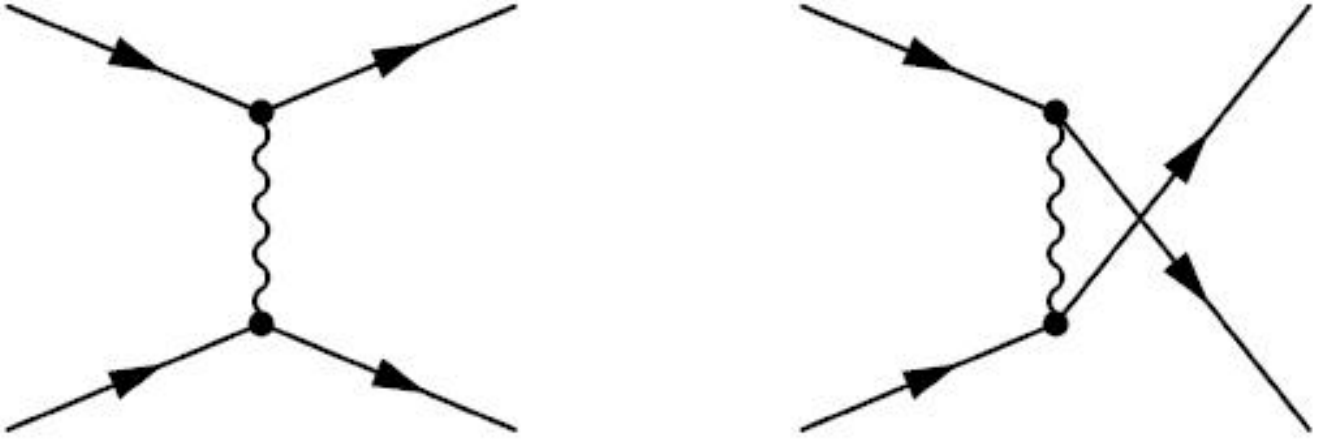} &
        $\frac{2\pi \alpha^2}{s^2} \big(\frac{s^2+u^2}{t^2}+\frac{s^2+t^2}{u^2}+\frac{2s^2}{tu}
        \big)$
        \\ \\
        2& $e^\pm e^\mp \longrightarrow  e^\pm e^\mp$ &
        \includegraphics[scale=.25]{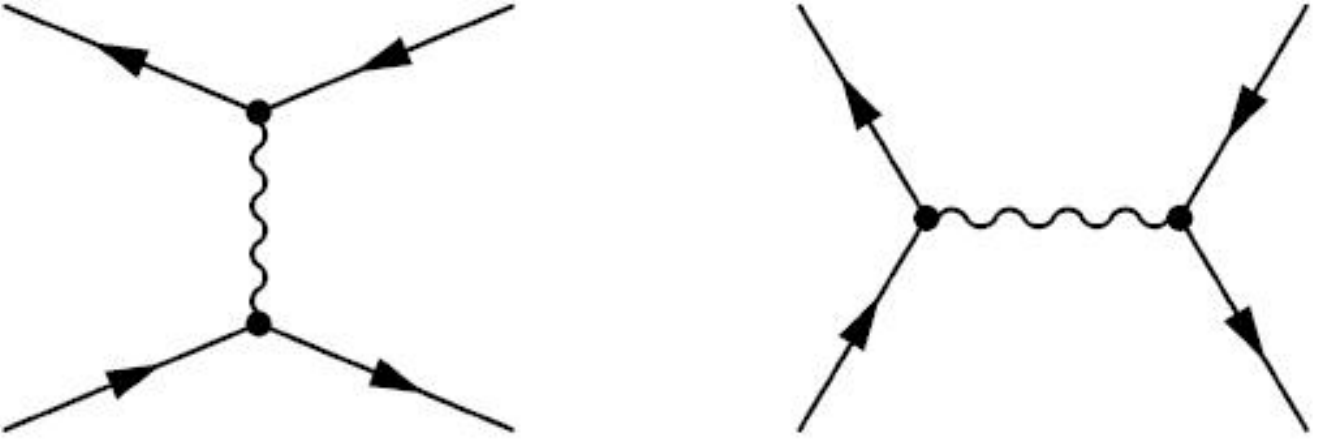} &
        $\frac{2\pi \alpha^2}{s^2} \big(\frac{s^2+u^2}{t^2}+\frac{u^2+t^2}{s^2}+\frac{2u^2}{ts}
        \big)$
        \\ \\
        3& $\gamma e^\mp \longrightarrow  \gamma e^\mp$ &
        \includegraphics[scale=.25]{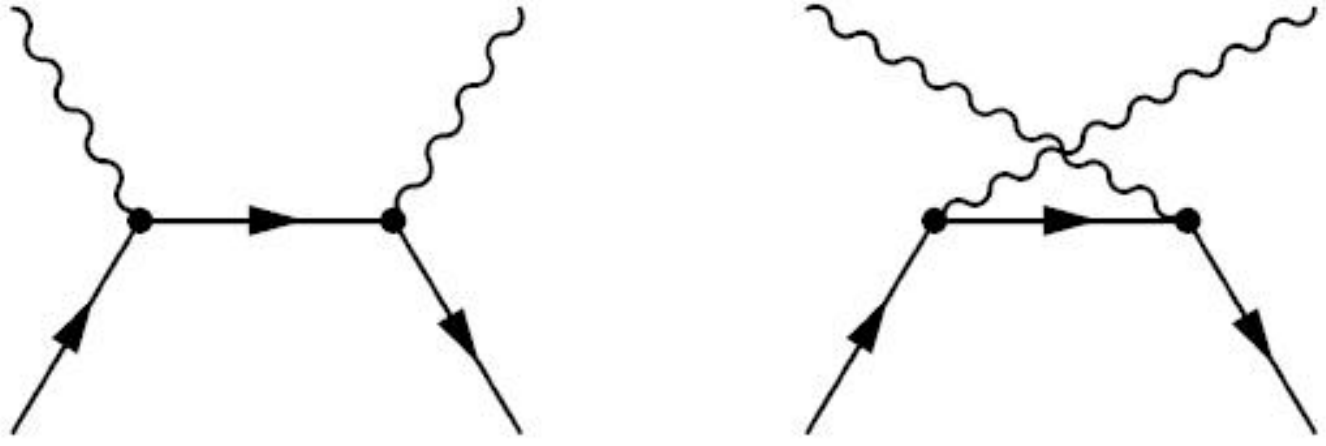}&
        $-\frac{2\pi \alpha^2}{s^2} \big(\frac{s}{u}+\frac{u}{s}
        \big)$
        \\ \\
        4& $e^\pm e^\mp \longrightarrow  \gamma \gamma $ &
        \includegraphics[scale=.25]{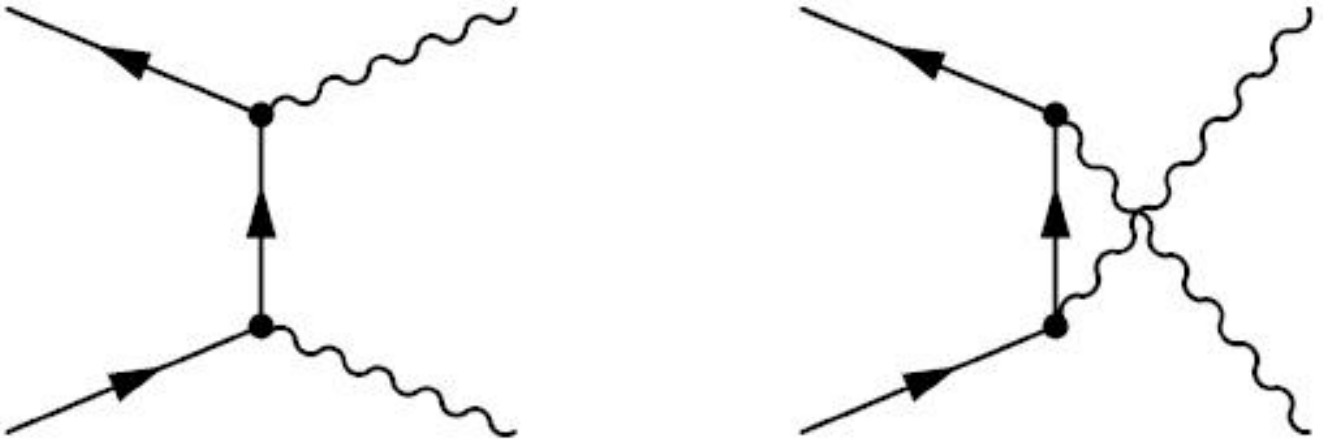}&
        $\frac{2\pi \alpha^2}{s^2} \big(\frac{t}{u}+\frac{u}{t}
        \big)$
        \\ \\
        5& $\gamma \gamma  \longrightarrow e^\mp e^\pm $ &
        \includegraphics[scale=.25]{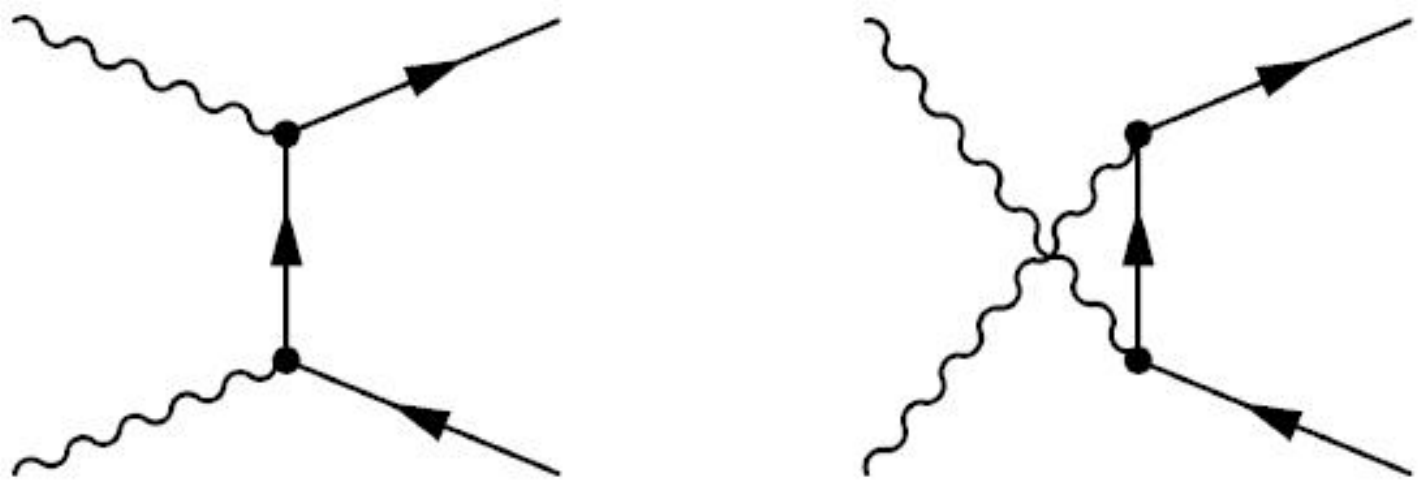}&
        $\frac{2\pi \alpha^2}{s^2} \big(\frac{t}{u}+\frac{u}{t}
        \big)$
        \\ \\
        6& $\tilde{\gamma} e^{\mp}  \longrightarrow \tilde{\gamma} e^{\mp}$ &
        \includegraphics[scale=.25]{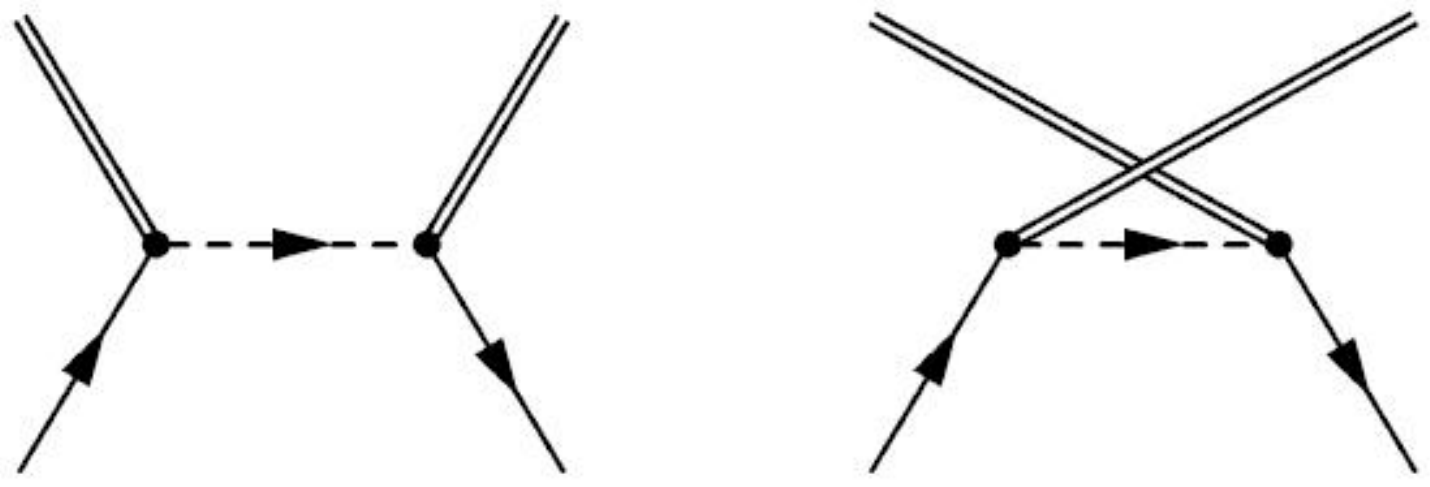}&
        $\frac{4\pi \alpha^2}{s^2}$
        \\ \\
        7& $e^{\pm} e^{\mp} \longrightarrow \tilde{\gamma}\tilde{\gamma}$ &
        \includegraphics[scale=.25]{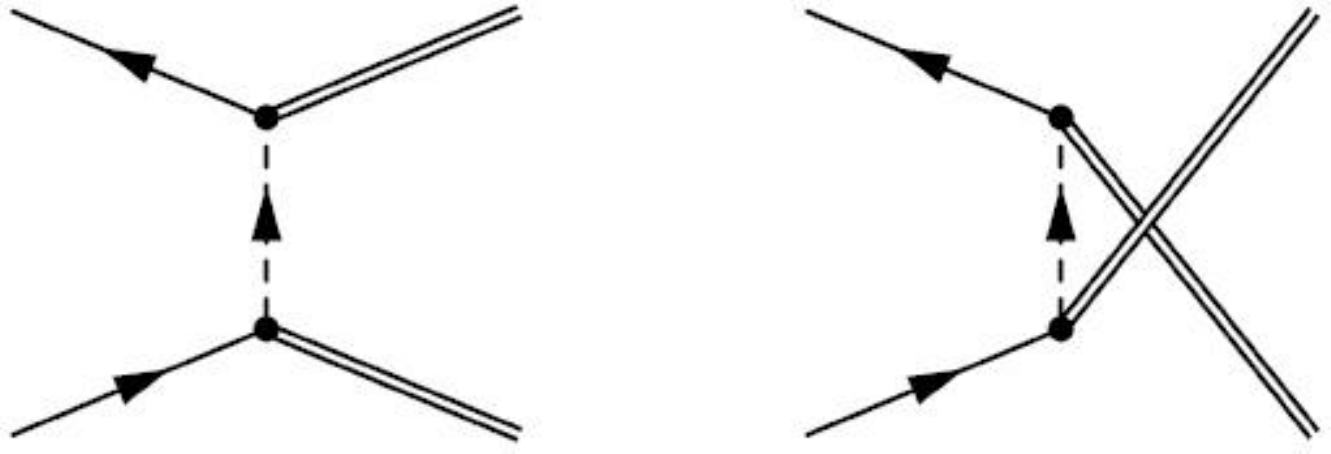}&
        $\frac{4\pi \alpha^2}{s^2}$
        \\ \\
        8& $\tilde{\gamma} \tilde{\gamma} \longrightarrow e^{\mp} e^{\pm}$ &
        \includegraphics[scale=.25]{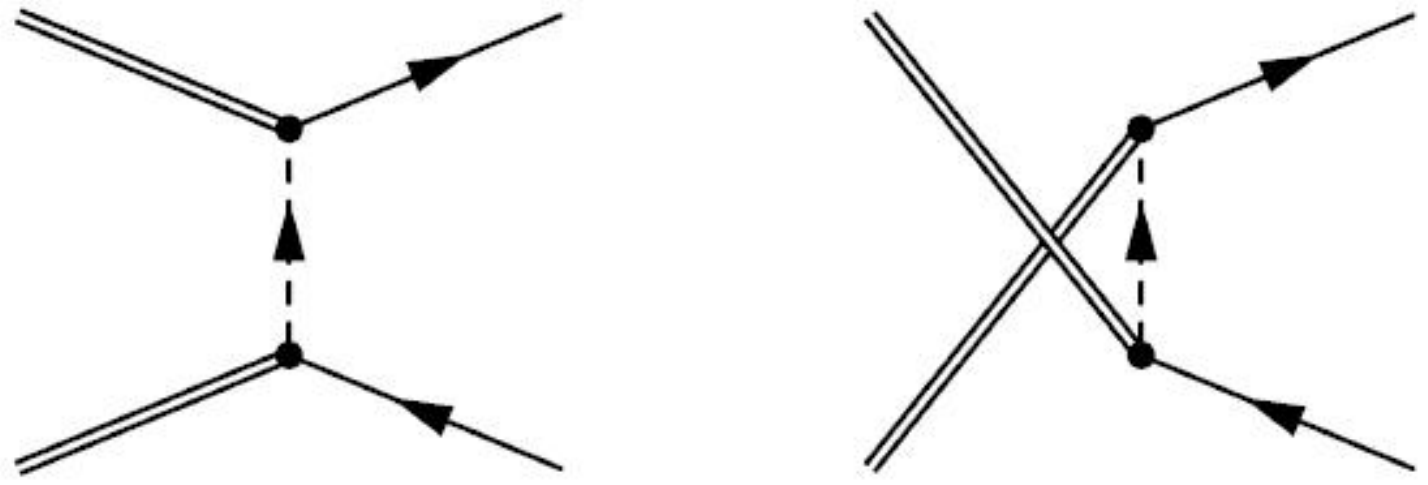}&
        $\frac{4\pi \alpha^2}{s^2}$
        \\ \\
        9& $\tilde{\gamma} e^{\mp} \longrightarrow \gamma \tilde{e}^{\mp}_{L,R} $ &
        \includegraphics[scale=.25]{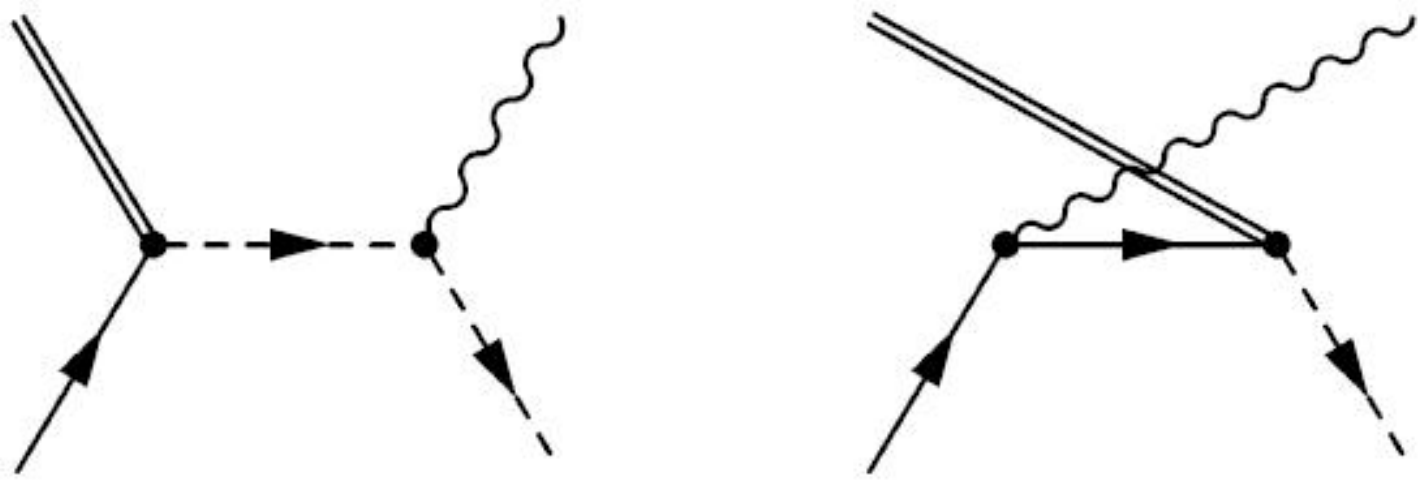}&
        $\frac{\pi \alpha^2}{s^2} \frac{t}{u}$
        \\ \\
        10& $ \gamma \tilde{e}^{\mp}_{L,R}  \longrightarrow  \tilde{\gamma} e^{\mp}$ &
        \includegraphics[scale=.25]{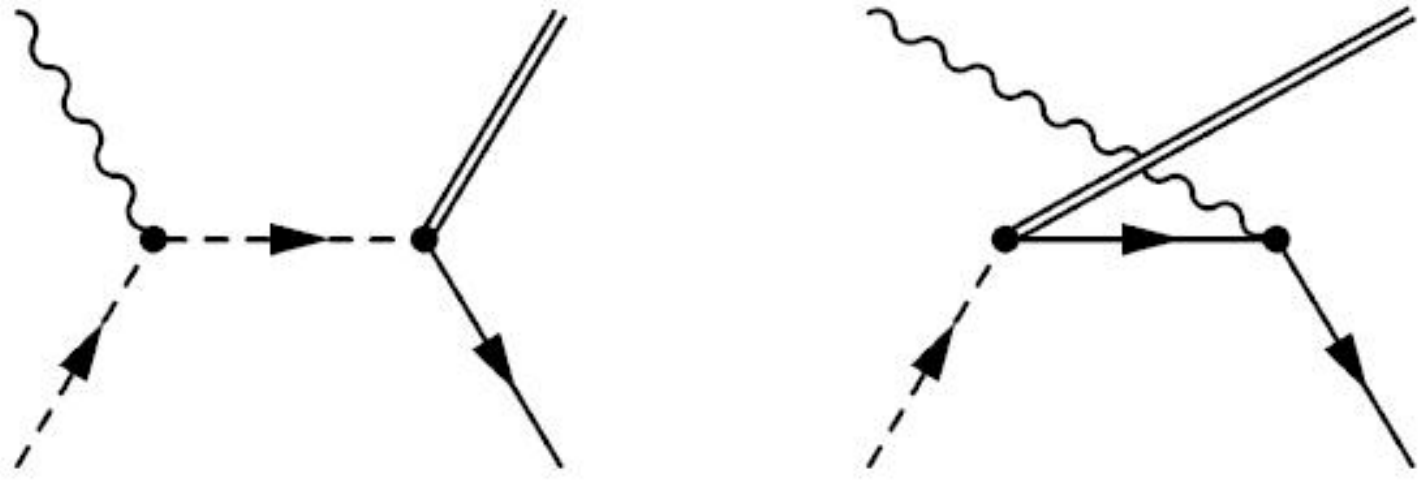}&
        $\frac{2\pi \alpha^2}{s^2} \frac{t}{u}$
         \\ \\
        11& $\gamma  e^{\mp} \longrightarrow \tilde{\gamma} \tilde{e}^{\mp}_{L,R} $ &
        \includegraphics[scale=.25]{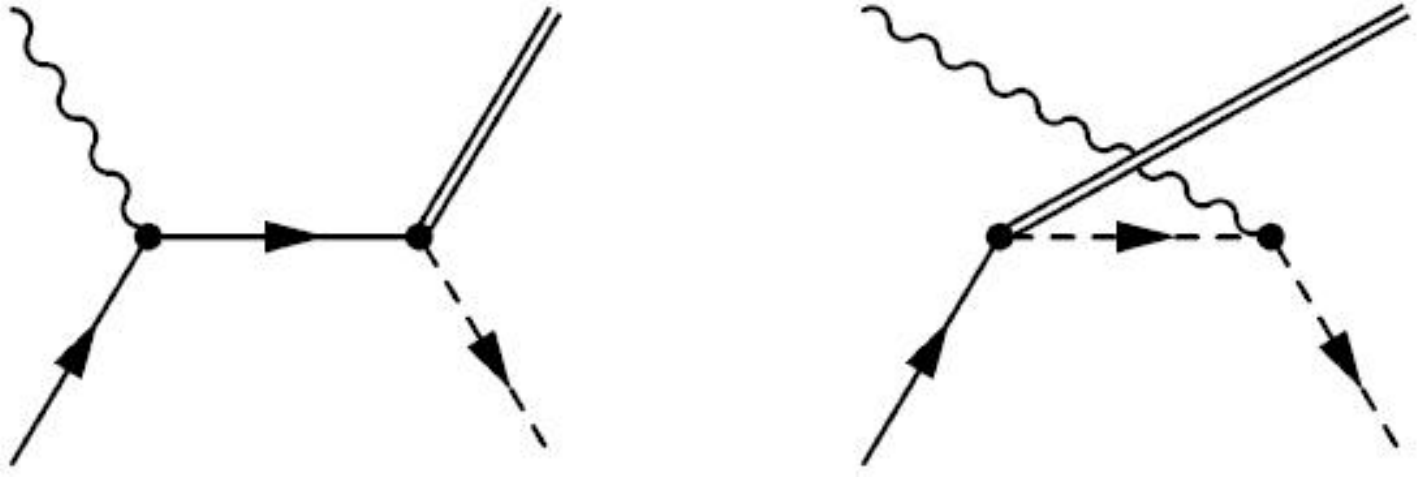}&
        $-\frac{2\pi \alpha^2}{s^2} \frac{t}{s}$
        \\ \\
        12& $\tilde{\gamma} \tilde{e}^{\mp}_{L,R}  \longrightarrow  \gamma e^{\mp} $ &
        \includegraphics[scale=.25]{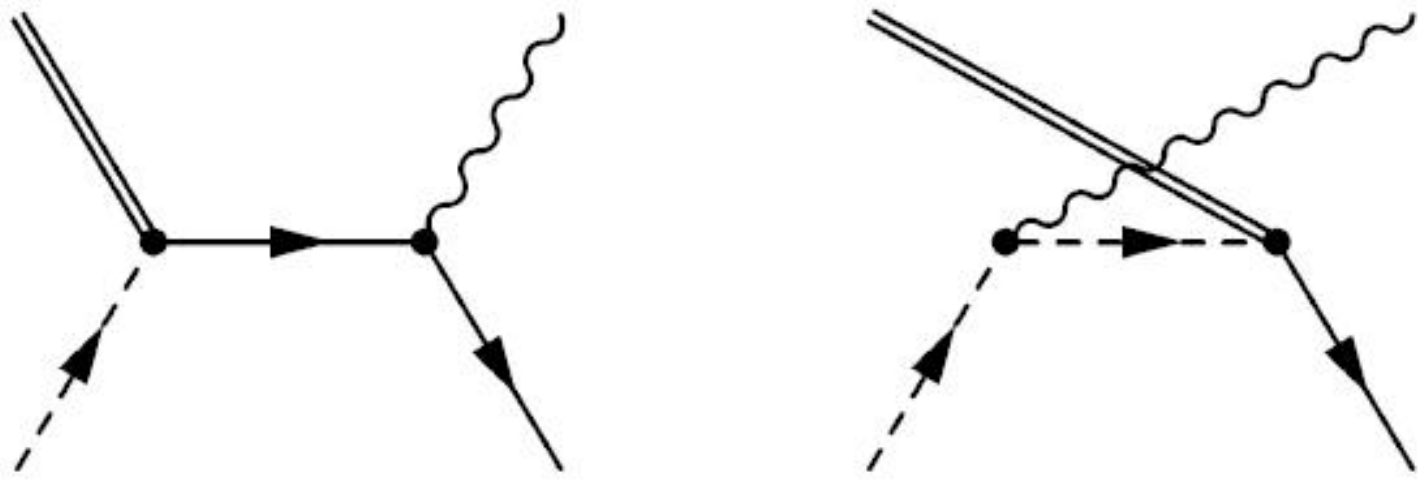}&
        $-\frac{2\pi \alpha^2}{s^2} \frac{t}{s}$
        \\ 
\hline\hline
\end{tabular}
\end{table}
\begin{table}
\begin{tabular}{m{.5cm} m{4cm} m{5.5cm} m{4cm}}
\hline \hline
$n^0$ & \hspace{0.4cm} Process & \hspace{0.9cm} Diagrams 
& \hspace{0.4cm} Cross section $-\frac{d\sigma}{dt}$
\\ \hline \hline \\
        13& $\tilde{e}^{\pm}_{L,R} e^{\mp} \longrightarrow \tilde{\gamma} \gamma $ &
        \includegraphics[scale=.25]{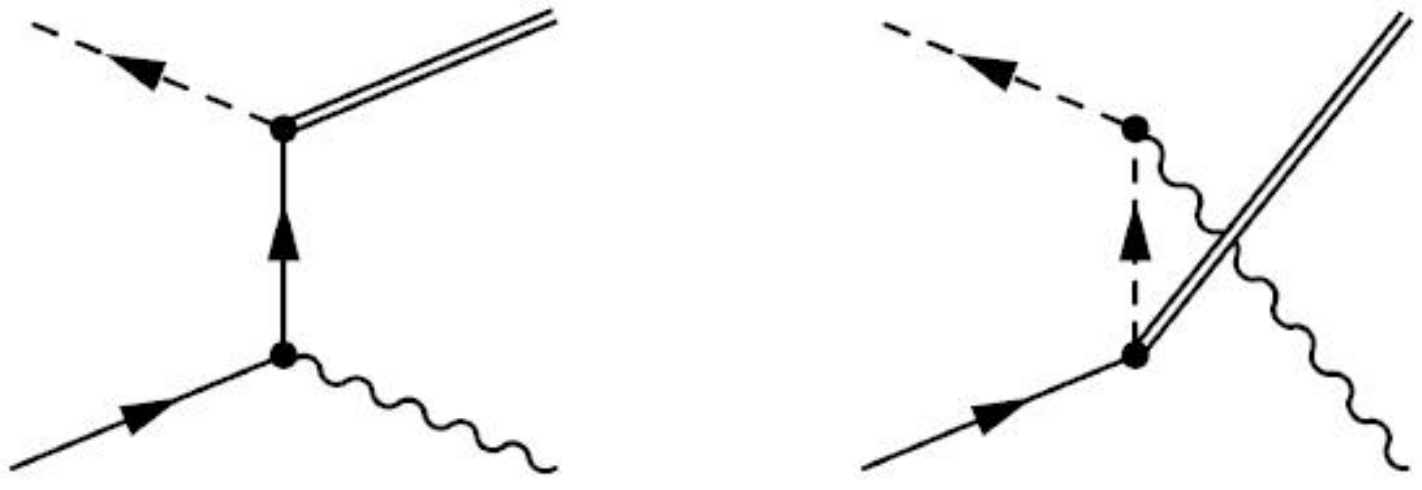}&
        $-\frac{2\pi \alpha^2}{s^2} \frac{s}{t}$
        \\ \\
        14& $\tilde{\gamma} \gamma \longrightarrow \tilde{e}^{\mp}_{L,R} e^{\pm} $ &
        \includegraphics[scale=.25]{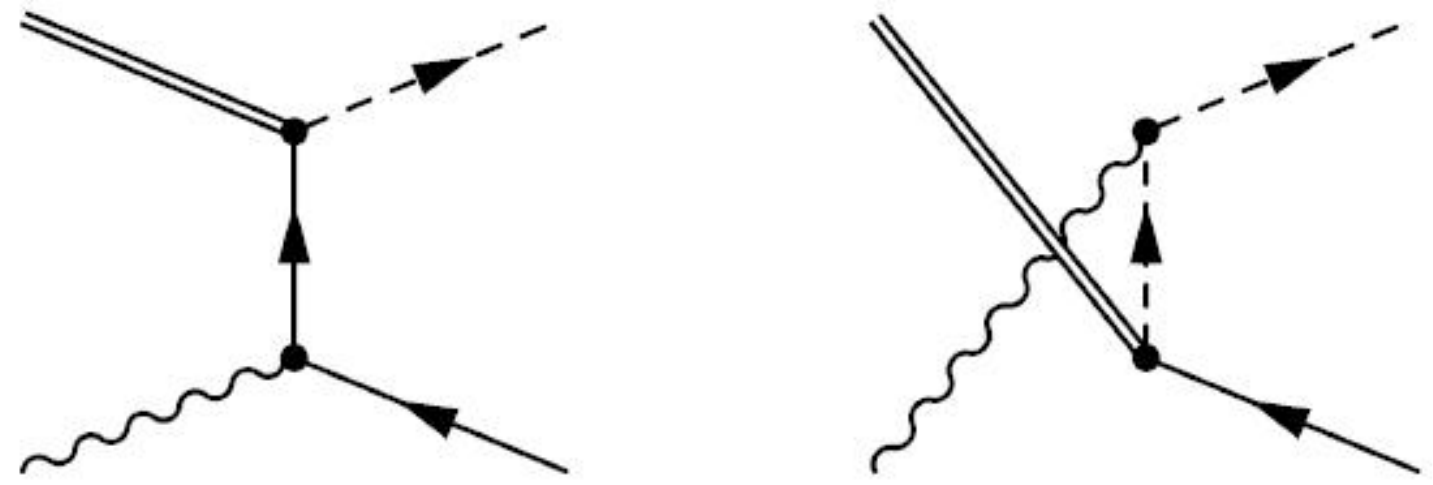}&
        $-\frac{\pi \alpha^2}{s^2} \frac{s}{t}$
         \\ \\
        15& $\tilde{e}^{\mp}_{L,R}e^{\mp} \longrightarrow \tilde{e}^{\mp}_{L,R} e^{\mp}$ &
        \includegraphics[scale=.25]{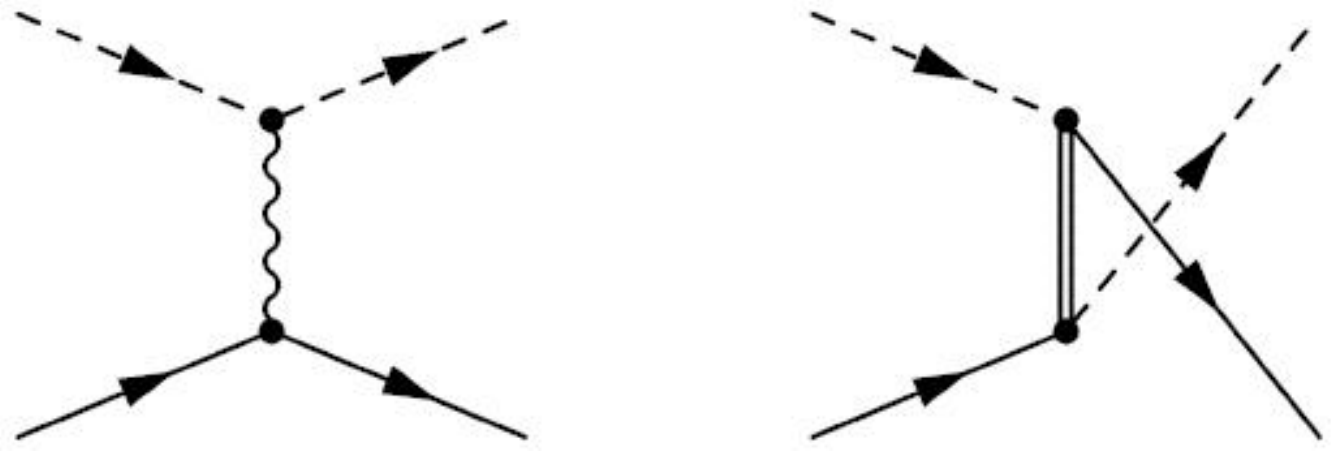}&
        $-\frac{2\pi \alpha^2}{s^2} \frac{s(s^2+u^2)}{ut^2}$
        \\ \\
        16& $\tilde{e}^{\pm}_{L,R} e^{\mp} \longrightarrow \tilde{e}^{\pm}_{L,R} e^{\mp}$ &
        \includegraphics[scale=.25]{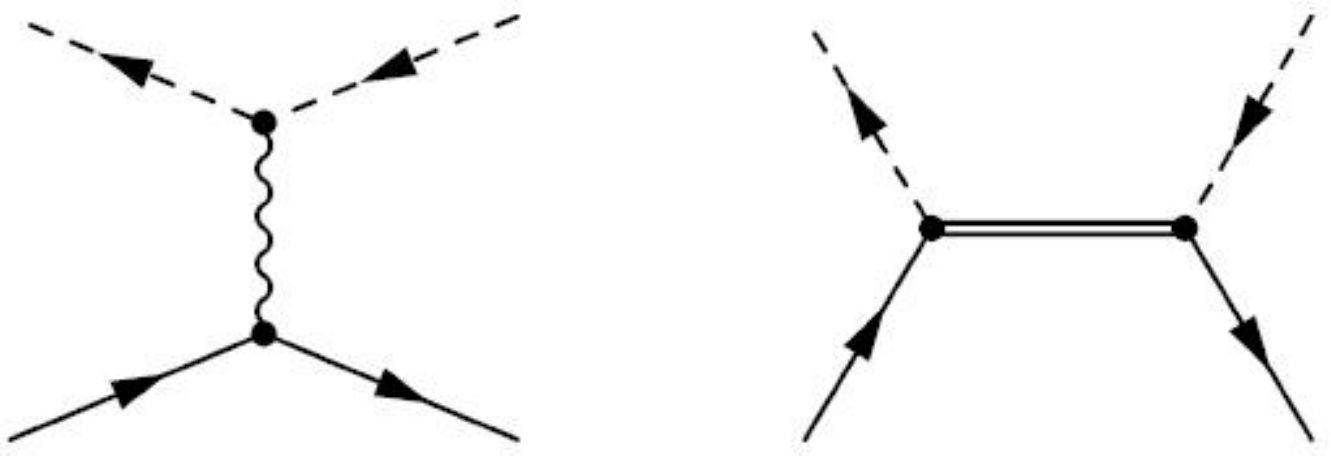}&
        $-\frac{2\pi \alpha^2}{s^2} \frac{u(s^2+u^2)}{st^2}$
        \\ \\
        17& $e^{\pm} e^{\mp} \longrightarrow \tilde{e}^{\pm}_{L,R} \tilde{e}^{\mp}_{L,R}$ &
        \includegraphics[scale=.25]{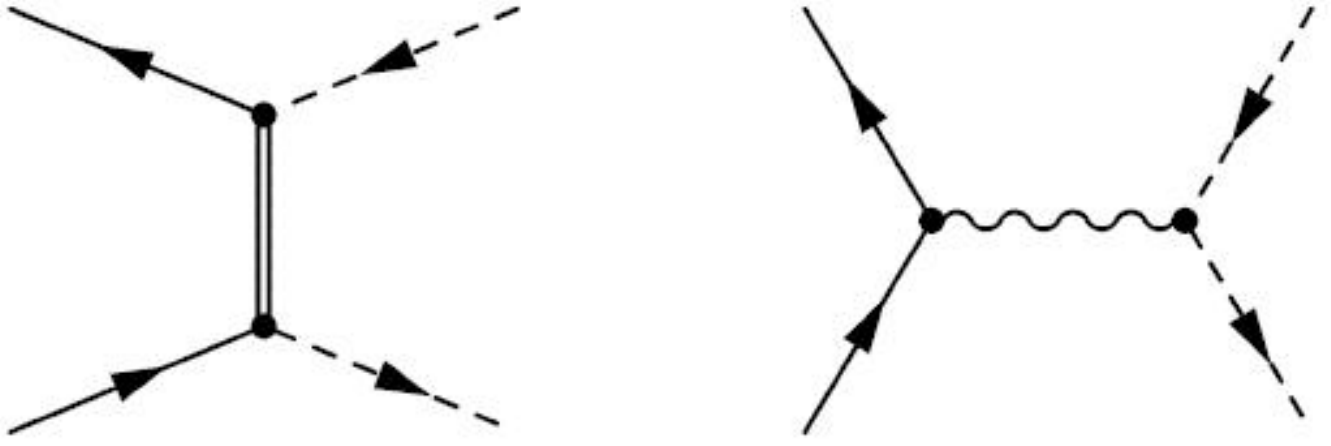}&
        $\frac{\pi \alpha^2}{s^2} \frac{u(t^2+u^2)}{ts^2}$
        \\ \\
        18& $\tilde{e}^{\pm}_{L,R} \tilde{e}^{\mp}_{L,R} \longrightarrow e^{\pm} e^{\mp}$ &
        \includegraphics[scale=.25]{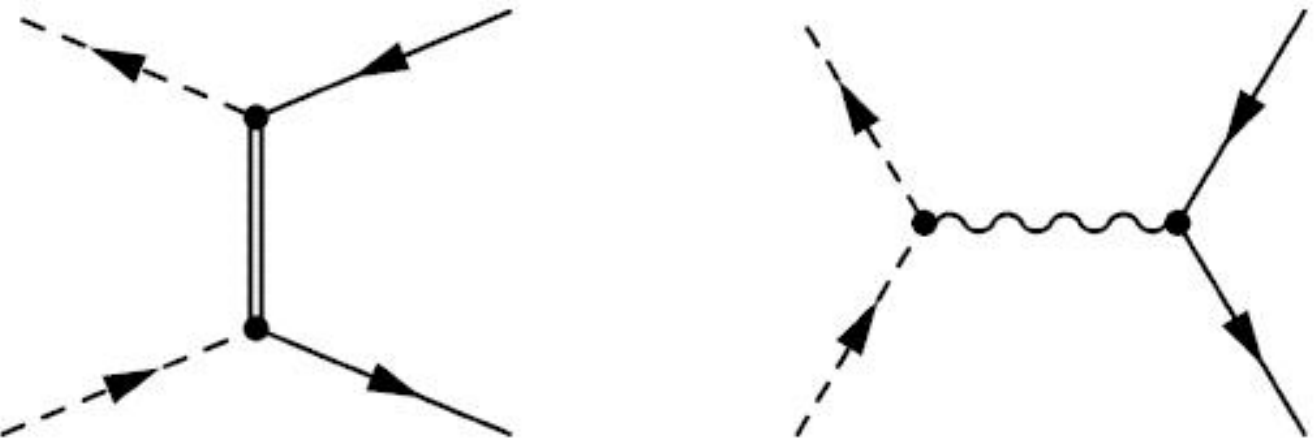}&
        $\frac{4\pi \alpha^2}{s^2} \frac{u(t^2+u^2)}{ts^2}$
        \\ \\
        19& $\tilde{e}^{\mp}_{L,R} \tilde{e}^{\mp}_{L,R} \longrightarrow \tilde{e}^{\mp}_{L,R} \tilde{e}^{\mp}_{L,R}$ &
        \includegraphics[scale=.25]{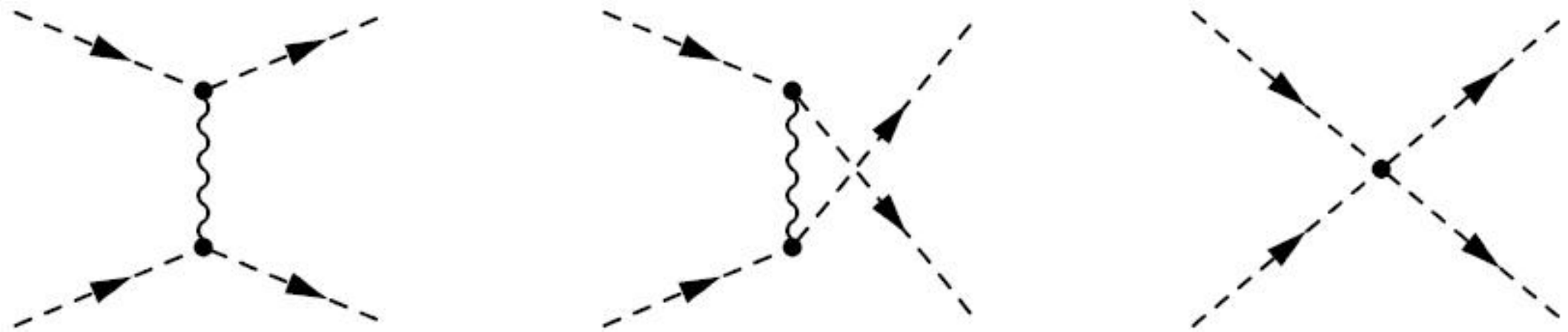}&
        $\frac{4\pi \alpha^2}{s^2}\big(\frac{u}{t} + \frac{t}{u}\big)^2$
        \\ \\
        20& $\tilde{e}^{\pm}_{L,R} \tilde{e}^{\mp}_{L,R} \longrightarrow \tilde{e}^{\pm}_{L,R} \tilde{e}^{\mp}_{L,R}$ &
        \includegraphics[scale=.25]{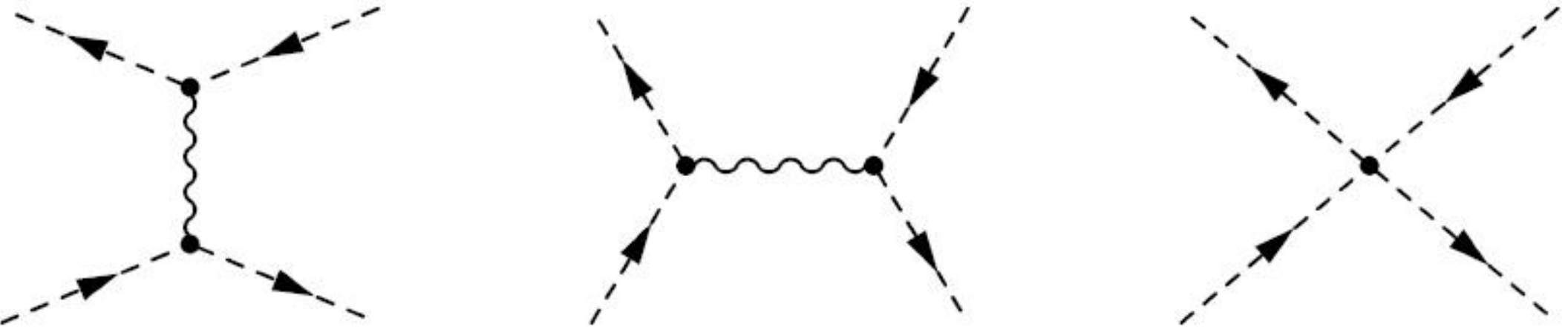}&
         $\frac{4\pi \alpha^2}{s^2}\big(\frac{s}{t}+\frac{t}{s} \big)^2$
        \\ \\
        21& $\tilde{e}^{\mp}_{L,R} \tilde{e}^{\mp}_{R,L} \longrightarrow \tilde{e}^{\mp}_{L,R} \tilde{e}^{\mp}_{R,L}$ &
        \includegraphics[scale=.25]{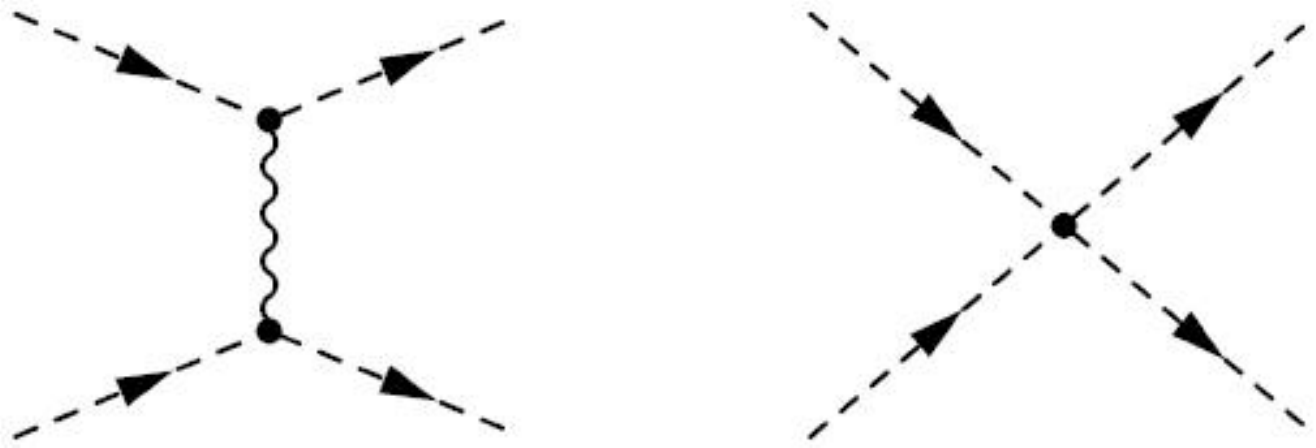}&
        $\frac{4\pi \alpha^2}{t^2}$
        \\ \\
        22& $\tilde{e}^{\pm}_{L,R} \tilde{e}^{\mp}_{R,L} \longrightarrow \tilde{e}^{\pm}_{L,R} \tilde{e}^{\mp}_{R,L}$ &
        \includegraphics[scale=.25]{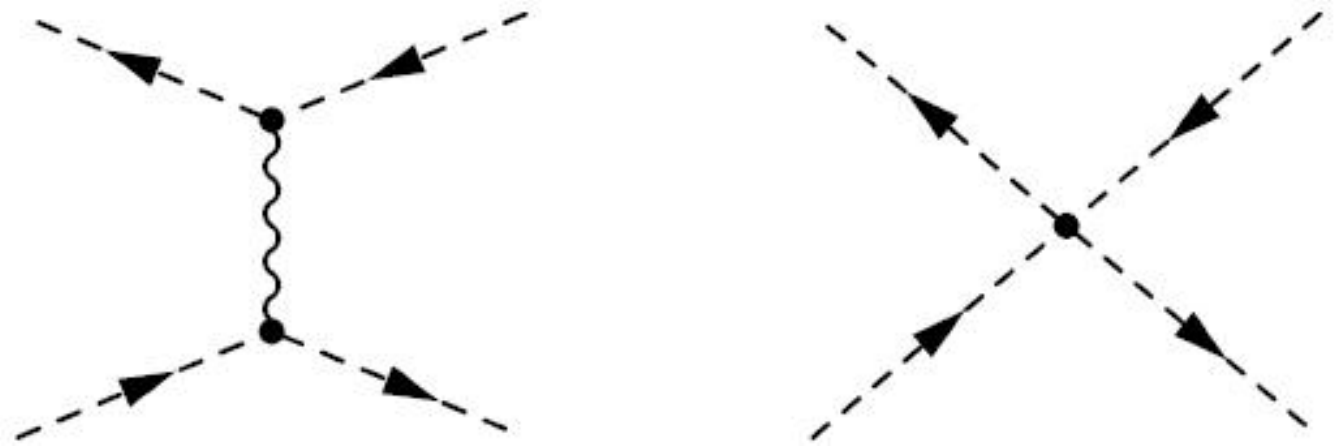}&
        $\frac{4\pi \alpha^2}{s^2} \frac{u^2}{t^2}$
        \\ \\
        23& $\tilde{e}^{\pm}_{L,R} \tilde{e}^{\mp}_{R,L} \longrightarrow \tilde{e}^{\pm}_{R,L} \tilde{e}^{\mp}_{L,R}$ &
        \includegraphics[scale=.25]{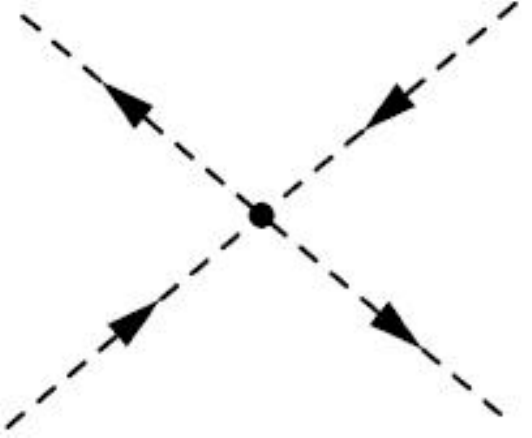} &
        $\frac{\pi \alpha^2}{s^2}$
        \\ \\
        24& $\tilde{e}^{\mp}_{L,R} \tilde{e}^{\mp}_{L,R}
        \longrightarrow \tilde{e}^{\mp}_{R,L} \tilde{e}^{\mp}_{R,L}$ &
        \includegraphics[scale=.25]{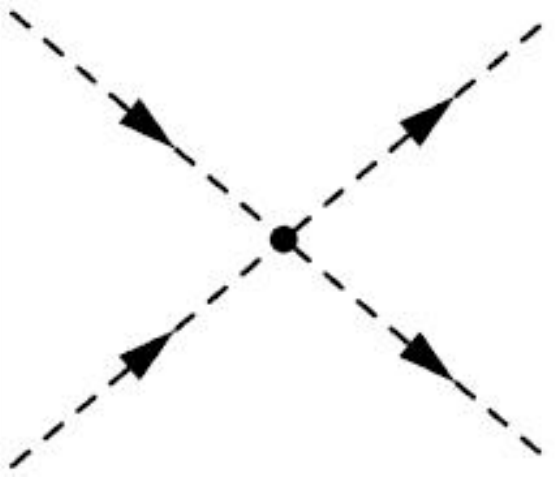}&
        $\frac{\pi \alpha^2}{s^2}$
        \\ 
        \hline\hline
\end{tabular}
\end{table}

\begin{table}[!h]
\begin{tabular}{m{.5cm} m{3.5cm} m{6cm} m{5cm}}
\hline \hline
$n^0$ & \hspace{0.4cm} Process & \hspace{0.9cm} Diagrams 
& \hspace{0.4cm} Cross section $-\frac{d\sigma}{dt}$
\\ \hline \hline \\
        25& $\gamma \tilde{e}^{\mp}_{L,R} \longrightarrow \gamma \tilde{e}^{\mp}_{L,R}$ &
        \includegraphics[scale=.25]{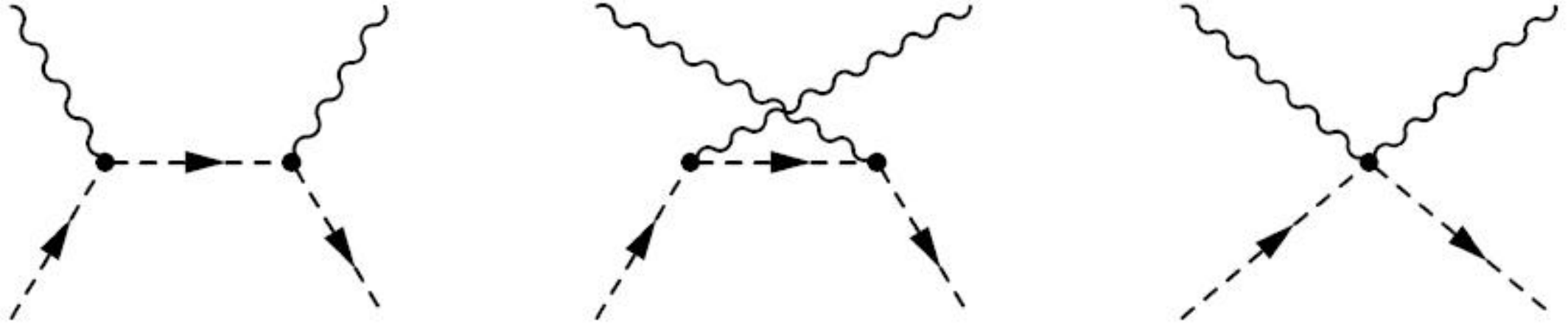}&
        $\frac{4\pi \alpha^2}{s^2}$
        \\ \\
        26& $\tilde{e}^{\pm}_{L,R} \tilde{e}^{\mp}_{L,R} \longrightarrow \gamma \gamma$ &
        \includegraphics[scale=.25]{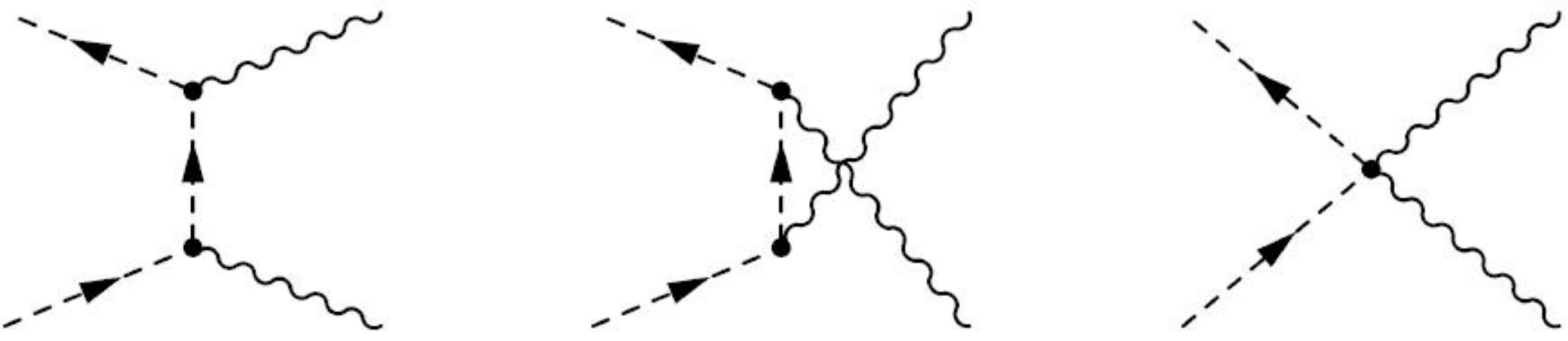}&
        $\frac{8\pi \alpha^2}{s^2}$
        \\ \\
        27& $\gamma \gamma \longrightarrow \tilde{e}^{\mp}_{L,R} \tilde{e}^{\pm}_{L,R}$ &
        \includegraphics[scale=.25]{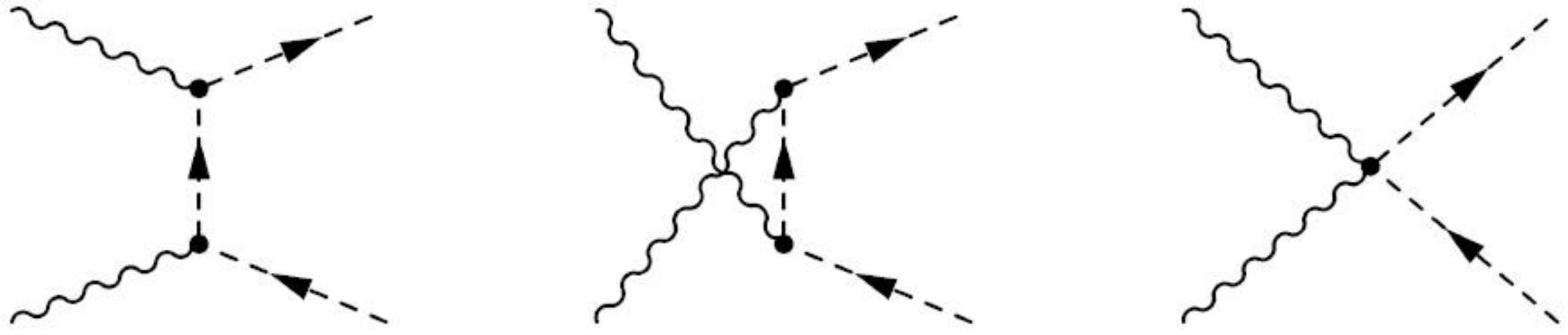}&
        $\frac{2\pi \alpha^2}{s^2}$
        \\ \\
        28& $\tilde{\gamma} \tilde{e}^{\mp}_{L,R} \longrightarrow \tilde{\gamma} \tilde{e}^{\mp}_{L,R} $ &
        \includegraphics[scale=.25]{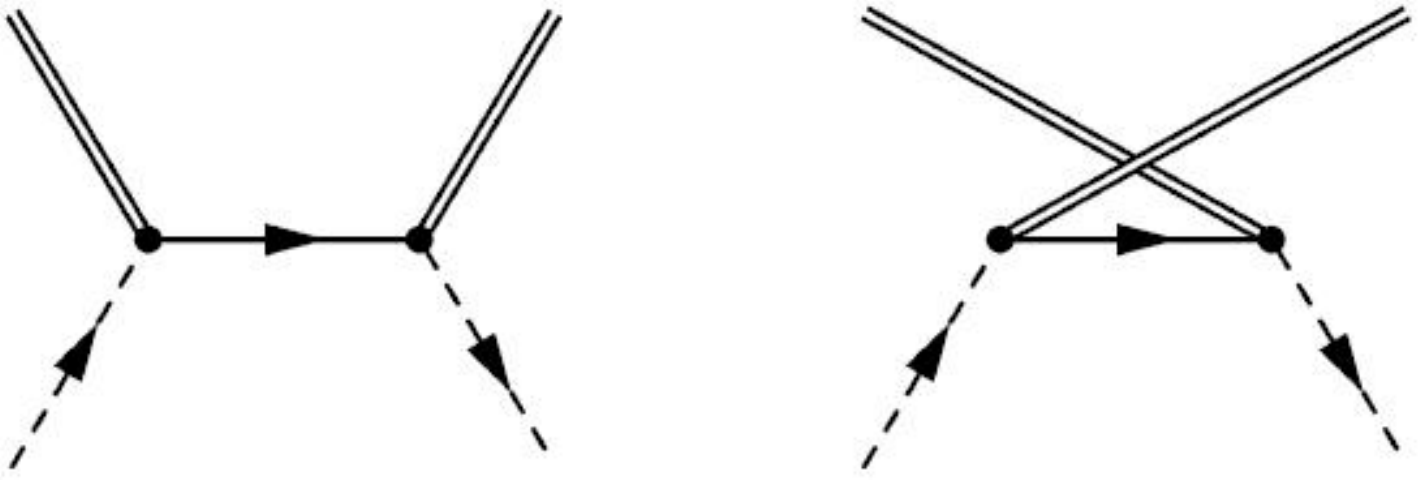}&
        $-\frac{2\pi \alpha^2}{s^2} \big(\frac{u}{s}+\frac{s}{u}
        \big)$
        \\ \\
        29& $\tilde{e}^{\pm}_{L,R} \tilde{e}^{\mp}_{L,R} \longrightarrow \tilde{\gamma} \tilde{\gamma}$ &
        \includegraphics[scale=.25]{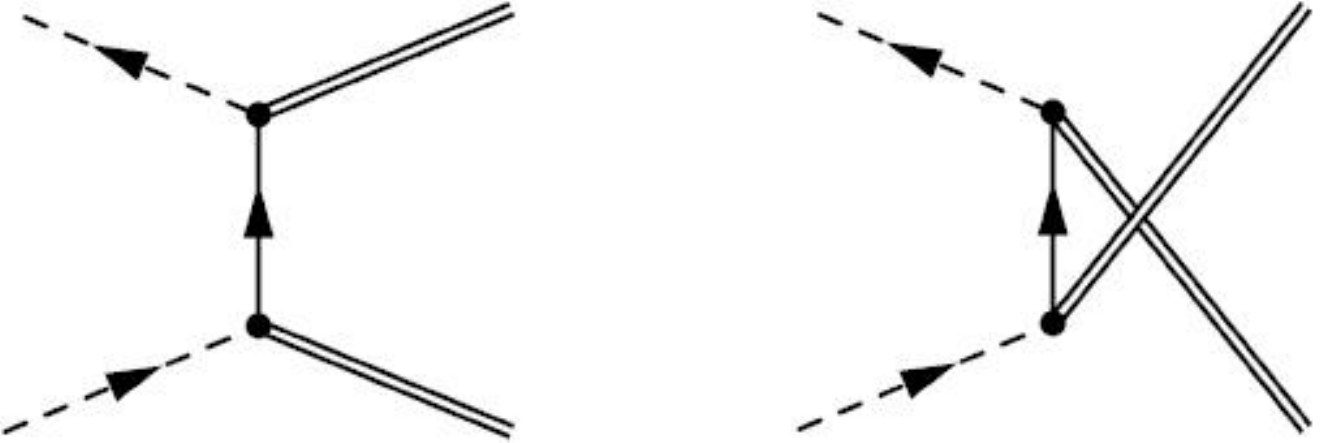}&
        $\frac{4\pi \alpha^2}{s^2} \big(\frac{u}{t}+\frac{t}{u}
        \big)$
        \\ \\
        30& $\tilde{\gamma} \tilde{\gamma} \longrightarrow \tilde{e}^{\pm}_{L,R} \tilde{e}^{\mp}_{L,R}$ &
        \includegraphics[scale=.25]{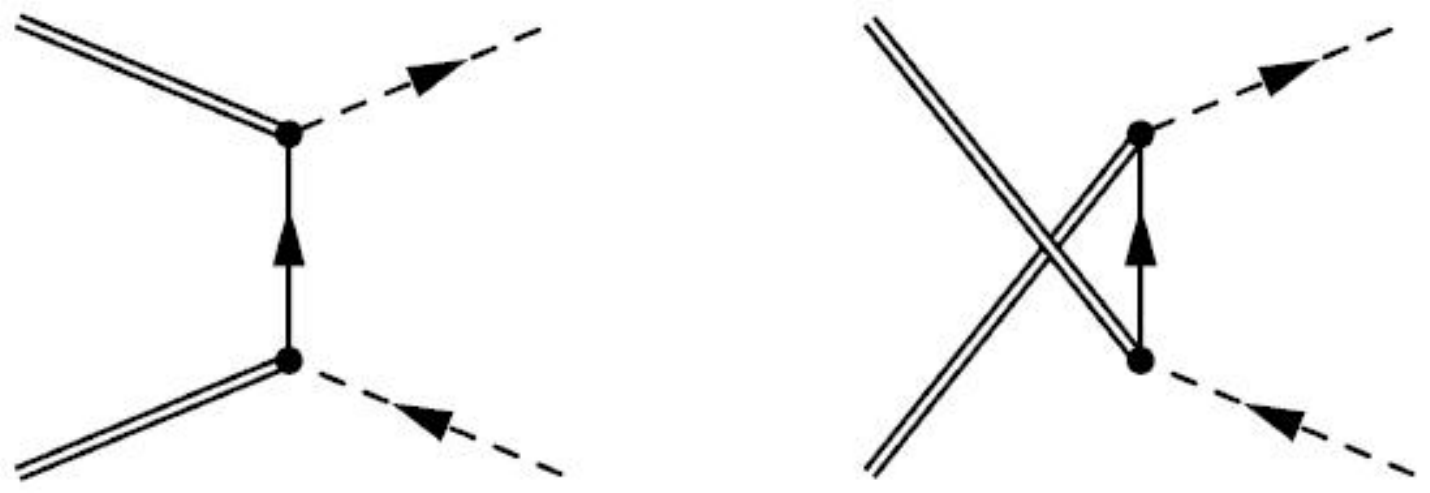}&
        $\frac{\pi \alpha^2}{s^2} \big(\frac{u}{t}+\frac{t}{u}
        \big)$
        \\ \\
        31& $\tilde{e}^{\pm}_{L,R} e^{\mp} \longrightarrow e^{\pm} \tilde{e}^{\mp}_{R,L}$ &
        \includegraphics[scale=.25]{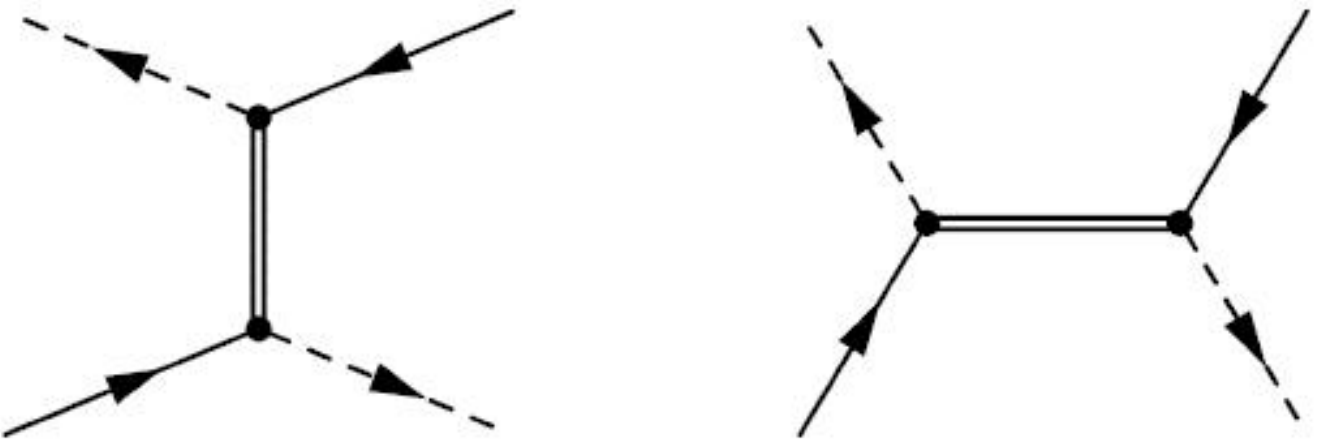}&
        $-\frac{2\pi \alpha^2}{s^2}\big( \frac{s}{t} + \frac{t}{s} \big)$
        \\ \\
        32& $e^{\mp} e^{\mp} \longrightarrow \tilde{e}^{\mp}_{L,R} \tilde{e}^{\mp}_{R,L} $ &
        \includegraphics[scale=.25]{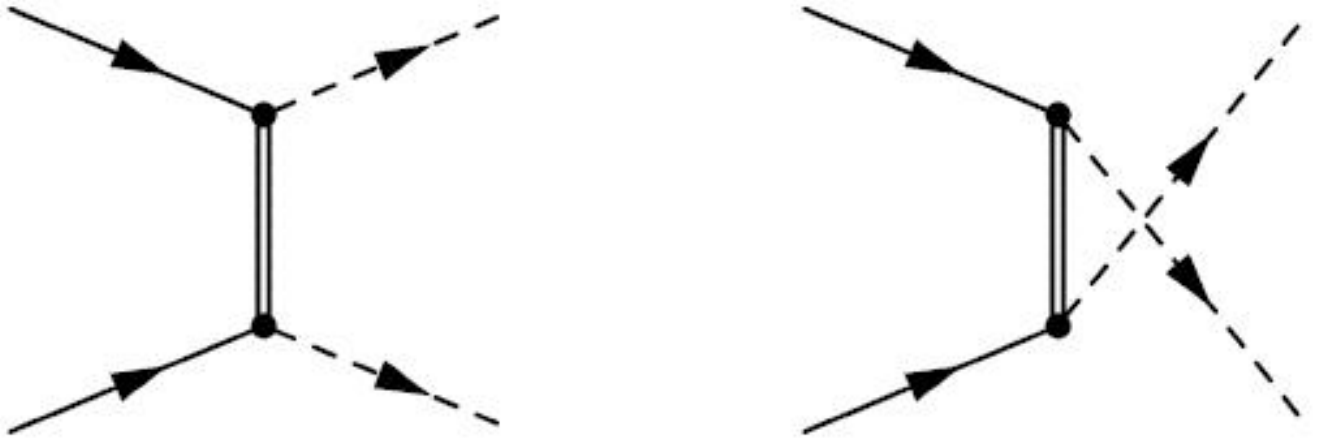} &
        $\frac{\pi \alpha^2}{s^2}\big( \frac{u}{t} + \frac{t}{u} \big)$
        \\ \\
        33& $\tilde{e}^{\mp}_{L,R} \tilde{e}^{\mp}_{R,L} \longrightarrow e^{\mp} e^{\mp}$ &
        \includegraphics[scale=.25]{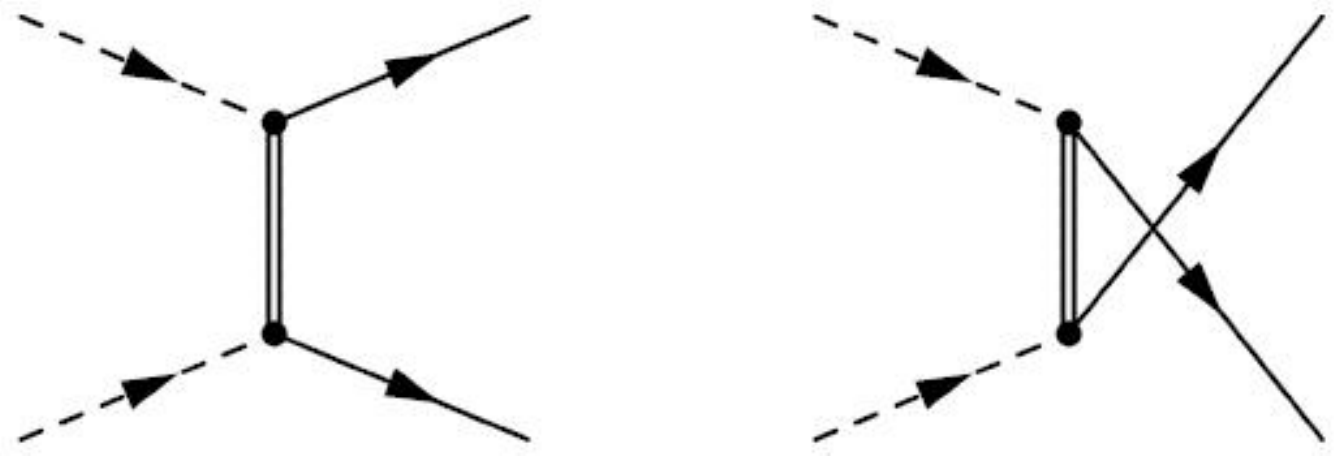}&
        $\frac{4\pi \alpha^2}{s^2}\big( \frac{u}{t} + \frac{t}{u} \big)$
        \\ 
        \hline\hline
\end{tabular}
\end{table}

In every process the initial and/or final state particles can carry positive or negative charge; selectrons can be additionally `$R$' (right) or `$L$' (left). For each listed reaction the Feynman diagrams from the third column correspond to the upper combination of charges of interacting particles. For example, the Feynman diagrams of the process $e^{\mp}e^{\mp} \longrightarrow e^{\mp}e^{\mp}$ actually represent the scattering $e^- e^- \longrightarrow e^- e^-$. The solid, dashed, wavy and double-solid lines correspond to electrons, selectrons, photons and photinos, respectively.  

As seen from Table~\ref{tab-corr-binray}, the first five processes occur in the supersymmetric QED plasma and in an electromagnetic plasma of electrons, positrons and photons. The remaining processes are characteristic of the  ${\cal N} =1$ SUSY QED  plasma. As one observes, the processes no. 6 -- 8 and 23 -- 26 are independent of momentum transfer. As a result, the scattering is isotropic in the centre-of-mass frame of colliding particles. Such processes are qualitatively different from those in an electromagnetic plasma (processes no. 1 -- 5) which are dominated by an interaction with small momentum transfer and, therefore, the scattering mostly occurs at small scattering angles. We also observe that for each plasma particle $e$, $\gamma$, $\tilde{e}$, $\tilde{\gamma}$ there is a process in which the cross section is independent of momentum transfer. Since the collisional processes determine transport characteristics, the discussion on to what extent the isotropic scattering gives rise to qualitative differences of supersymmetric QED plasma against usual QED one is provided later on.

\subsubsection{Cross sections of the ${\mathcal N} =4$ super Yang-Mills plasma}
\label{sssec-cross-sec-sym}

Here, we list the elementary processes which occur at the lowest nontrivial order in SYMP. The cross sections of the processes are presented in Table~\ref{table-collisions}. $G, F, S$ denote a gluon, fermion and scalar, respectively. The matrix elements, which were first computed in \cite{Huot:2006ys}, are expressed through the Mandelstam invariants $s,t$ and $u$ defined in the standard way, Eqs. (\ref{s})-(\ref{u}). For a given process, the differential cross section is computed through the formula (\ref{cs-mand}) where $N^{\rm dof}_1$ and $N^{\rm dof}_2$ are given in Table~\ref{table-field-content}. 
\begin{table}[!h]
\caption{\label{table-collisions} Elementary processes in the ${\cal N} =4$ super Yang-Mills plasma.}
\begin{center}
\begin{tabular}{m{1cm} m{4cm} m{7cm}}
\hline\hline
$n^0$ &  Process & $\frac{1}{g^4} \frac{1}{N_c^2(N_c^2 -1)}\sum |M|^2$
\\ \hline\hline \\
1& $GG \leftrightarrow GG$ & $8 \big(\frac{s^2 + u^2}{t^2} + \frac{u^2 + t^2}{s^2}+\frac{t^2 + s^2}{u^2} +3 \big)$
\\[2mm]
2 & $GF \leftrightarrow GF$ &  $32 \big(\frac{s^2 + u^2}{t^2} - \frac{u}{s} - \frac{s}{u} \big)$
\\[2mm]
3 & $GG \leftrightarrow FF$ &  $32 \big(\frac{t^2 + u^2}{s^2} - \frac{u}{t} - \frac{t}{u} \big)$
\\[2mm]
4 & $GS \leftrightarrow GS$ & $24 \big(\frac{s^2 + u^2}{t^2} + 1 \big)$
\\[2mm]
5 & $GG \leftrightarrow SS$ &  $24 \big(\frac{t^2 + u^2}{s^2} + 1 \big)$
\\[2mm]
6 & $GF \leftrightarrow SF$ &  $- 96 \big( \frac{u}{s} + \frac{s}{u} +1 \big)$
\\[2mm]
7 & $GS \leftrightarrow FF$ & $- 96 \big( \frac{u}{t} + \frac{t}{u} +1 \big)$
\\[2mm]
8 & $FS \leftrightarrow FS$ &  $- 96 \big[ \frac{2us}{t^2} + 3 \big( \frac{u}{s} + \frac{s}{u} \big) + 1 \big]$
\\[2mm]
9 & $SS \leftrightarrow FF$ & $- 96 \big[ \frac{2ut}{s^2} + 3 \big( \frac{u}{t} + \frac{t}{u} \big) + 1 \big]$
\\[2mm]
10 & $SS \leftrightarrow SS$ & $72 \big(\frac{s^2 + u^2}{t^2} + \frac{u^2 + t^2}{s^2}+\frac{t^2 + s^2}{u^2} +3 \big)$
\\[2mm]
11 & $FF \leftrightarrow FF$ & $128 \big(\frac{s^2 + u^2}{t^2} + \frac{u^2 + t^2}{s^2}+\frac{t^2 + s^2}{u^2} +3 \big)$
\\[2mm]
\hline\hline
\end{tabular}
\end{center}
\end{table}

\subsection{Transport coefficients}
\label{ssec-tr-coef}

As already discussed, the temperature is the only dimensional parameter of an ultrarelativistic equilibrium system. Accordingly, the parametric structure of most transport coefficients is uniquely determined by dimensional arguments. Possible differences between the coefficients of a supersymmentric and corresponding nonsupersymmetric plasmas lie in numerical factors. However, there are some other characteristics that are not so strongly constrained by dimensional arguments. These are, in particular, a collisional energy loss and momentum broadening. The latter determines a magnitude of radiative energy loss of a highly energetic particle in a plasma \cite{Baier:1996sk}. Both characteristics may depend not only on the plasma temperature but also on the energy of a test particle traversing the medium. Moreover, these characteristics seem to strongly depend on a specific process under consideration.

We have found that there is a group of processes in ${\cal N} =1$ SUSY QED plasma, as the Compton scattering on a selectron, whose cross sections are qualitatively different from those of the usual QED plasma. Namely, the processes are independent of momentum transfer. Thus, one may expect that energy losses and momentum broadening caused by theses processes are different from the ones caused by the Coulomb-like interactions dominated by small momentum transfers. 

Below we calculate both the energy loss and momentum broadening which occur due to the Compton scattering on selectron, whose cross section is given by (\ref{cs-compton}).

\subsubsection{Energy loss}
\label{sssec-e-loss}

Let us consider a high-energy selectron traversing an equilibrium ${\cal N} =1$ SUSY QED plasma. The selectron interacts with plasma particles of all types but we take into account only scattering on photons. This is only for demonstration but in general all processes, which contribute additively to the energy loss, must be included. More physically, one can think of a selectron flying across a photon gas.

To abide by the convention used in the whole chapter, the initial plasma photon four-momentum is denoted as $p^\mu_1 = (E_1, {\bf p}_1)$ with $E_1 \equiv |{\bf p}_1|$ and that of a selectron as $p^\mu_2 = (E_2, {\bf p}_2)$. The final four-momenta of the photon and selectron are, respectively, $p^\mu_3 = (E_3, {\bf p}_3)$ and  ${p}^\mu_4 = (E_4, {\bf p}_4)$. The energy loss of the selectron per unit length is then
\ba
\label{e-loss-def}
\frac{dE}{dx}= -\int d\Gamma (E_2 -  E_4),
\ea
where $d\Gamma$ is the interaction rate given as
\ba
\label{rate-def}
d\Gamma = |\mathcal{M}|^2 \frac{f_\gamma(\mathbf{p}_1) }{16 E_1 E_2 E_3 E_4} 
\, \frac{d^3p_1}{(2\pi)^3} \, \frac{d^3p_4}{(2\pi)^3} \,
 \frac{d^3p_3}{(2\pi)^3}\, (2\pi)^4 \,   \delta^{(4)}(p_1 + p_2 -p_3 - p_4) ;
\ea
$f_\gamma(\mathbf{p}_1)$ is the distribution function of plasma photons and $\mathcal{M}$ denotes, as previously, the scattering amplitude. We have neglected here the quantum factor $f_\gamma(\mathbf{p}_3) +1$ which is important when the momentum of final state photon is of order of plasma temperature. Since the scattering process under consideration leads to a sizeable momentum transfer and we are mostly interested in energy loss of a highly energetic particle, the factor can be safely  ignored. 

When $\mathcal{M}$ describes a scattering driven by one photon exchange, the formula (\ref{e-loss-def}) with the rate (\ref{rate-def}) leads to an infinite result due to the long range nature of electromagnetic interaction. The problem is cured by including the effect of screening in a plasma medium. In the case of photon-selectron scattering the matrix element equals $|\mathcal{M}|^2 = 4 e^4$  and it does not need any modification to provide a finite energy loss. 

Substituting the interaction rate (\ref{rate-def}) with $|\mathcal{M}|^2 = 4 e^4$ into Eq.~(\ref{e-loss-def}) and performing the trivial integration over ${\bf p}_3$, we obtain 
\ba
\label{e-loss1}
\frac{dE}{dx}= -\frac{e^4}{4}\int \frac{d^3p_1}{(2\pi)^3}
\frac{d^3q}{(2\pi)^3} \frac{f_\gamma(\mathbf{p}_1)(E_2-E_4)}{E_2^2 E_1 E_3 q} 
\: 2\pi \delta(\cos\theta - \overline{\cos \theta} ) ,
\ea
where ${\bf q} \equiv {\bf p}_4 - {\bf p}_2={\bf p}_1 - {\bf p}_3$ is the momentum transfer and $q\equiv |{\bf q}|$; $\theta$ is the angle between the vectors ${\bf p}_2$ and ${\bf q}$ and $\overline{\cos \theta}$ is the solution of the energy conservation equation
\ba
\label{cos-bar}
\overline{\cos \theta} = \frac{(E_2+E_1 - E_3)^2 - E_2^2 - q^2}{2 q E_2},
\ea
provided $-1 \le \overline{\cos \theta} \le 1$.

Now we make use of the assumption that the plasma is in thermal equilibrium and therefore it is isotropic. As a result, the momentum distribution of plasma photons depends on ${\bf p}_1$ only through $E_1$ and we write it as $f_\gamma(E_1)$. Consequently, the energy loss is independent of the orientation of the momentum $\mathbf{p}_2$. Therefore, following \cite{Braaten:1991jj} we average the formula (\ref{e-loss1}) over the orientation of $\mathbf{p}_2$ with respect to ${\bf q}$ and we get 
\ba
\label{e-loss2}
\frac{dE}{dx}= \int \frac{d\Omega}{4\pi} \frac{dE}{dx}
= -\frac{e^4}{2^8 \pi^5}\int d^3p_1 d^3q 
\, \frac{f_\gamma(E_1) (E_2-E_4)}{E_2^2 E_1 E_3 q}.
\ea
We write down the integral over ${\bf q}$ in spherical coordinates where the axis $z$ is along the momentum ${\bf p}_1$. Then, the integral over orientation of ${\bf p}_1$ is trivial and one obtains
\ba
\label{e-loss-3}
\frac{dE}{dx}= -\frac{e^4}{2^5 \pi^3}
\int_0^\infty dE_1 E_1^2
\int_{q_{\rm min}}^{q_{\rm max}} dq q^2 
\int_{(\cos \theta_1)_{\rm min}}^{(\cos \theta_1)_{\rm max}}
d(\cos\theta_1)
\frac{f_\gamma(E_1)\, (E_2 - E_4)}{E_2^2 E_1 E_3 q},
\ea
where $\theta_1$ is the angle between the vectors ${\bf p}_1$ and ${\bf q}$. The integration limits must be chosen in such a way that the energy conservation is satisfied. Instead of $\cos \theta_1$ it appears more convenient to use the variable $\omega \equiv E_2 -E_4 = \sqrt{E_1^2 -2 E_1 q \cos \theta_1 +q^2} - E_1$. Then, the expression (\ref{e-loss-3}) can be written in the form
\ba
\label{e-loss-4}
\frac{dE}{dx} = -\frac{e^4}{2^5 \pi^3 E^2}\int_0^\infty d E_1 f_\gamma(E_1) 
\int_{q_{\rm min}}^{q_{\rm max}}dq 
\int_{\omega_{\rm min}}^{\omega_{\rm max}}
d\omega \, \omega .
\ea

To find the integration limits we express $\overline{\cos \theta}$, which is given by Eq.~(\ref{cos-bar}), through the variable $\omega$ and we demand that $-1 \le \overline{\cos \theta} \le 1$ keeping in mind that $-E_1 \le \omega \le E_2$. Then, a somewhat lengthy but elementary analysis leads to the expression
\ba
\label{e-loss-5}
\frac{dE}{dx} 
\!\!\!&=& \!\!\!
-\frac{e^4}{2^5 \pi^3 \, E^2}  \Bigg\{ \int_0^{E_2} {\rm d}E_1 f_\gamma(E_1) 
\\ [2mm] \nn 
&& \qquad
\times\Bigg[\int_0^{E_1} d q \int_{-q}^{q} d \omega \, \omega
+  \int_{E_1}^{E_2} d q \int_{q -2E_1}^{q}  d\omega \, \omega 
+  \int_{E_2}^{E_2+E_1} dq \int_{q -2E_1}^{2E_2-q} d \omega \, \omega \Bigg]
\\ [2mm] \nn 
&+& 
 \int_{E_2}^{\infty} d E_1 f_\gamma(E_1)
 \\ [2mm] \nn 
&& \qquad
\times\Bigg[ \int_0^{E_2} d q \int_{-q}^{q} d \omega \, \omega 
+  \int_{E_2}^{E_1} dq \int_{-q}^{2E_2 - q} d\omega \, \omega
+ \int_{E_1}^{E_2+E_1} d q \int_{q -2E_1}^{2E_2 - q} d \omega \, \omega 
\Bigg] \Bigg\},\nn
\ea
which after performing simple integrations over $q$ and $\omega$ gives
\ba
\label{e-loss-n(E)}
\frac{d E}{dx} = -\frac{e^4}{2^5 \pi^3 E}
\int_0^{\infty}  d E_1 f_\gamma(E_1) \big( E_2 E_1 - E_1^2\big).
\ea

To check correctness of rather complicated integration domain in Eq.~(\ref{e-loss-n(E)}), one observes that the integral (\ref{e-loss-5}) becomes simple when $\omega \equiv E_2 -E_4$ is replaced by unity. Then, the integral 
\ba
\int \frac{d^3p_4}{(2\pi)^3 2E_4} \,
 \frac{d^3p_3}{(2\pi)^3 2E_3}\, (2\pi)^4 \,   \delta^{(4)}(p_1 + p_2 -p_3 - p_4)  = \frac{1}{8\pi}
\ea
is Lorentz invariant and can be easily computed in the centre-of-mass frame. We have reproduced this result in the frame, where the energy loss is computed, performing the integration over the domains in $q-\omega$ space from Eq.~(\ref{e-loss-5}).

The energy distribution of photons in equilibrium plasma is of Bose-Einstein form
\ba
\label{B-E}
f_\gamma(E) = \frac{2}{e^{\frac{E}{T}}-1},
\ea
where the factor of 2 takes into account two photon polarisations and $T$ is the plasma temperature. Substituting the distribution (\ref{B-E}) into Eq.~(\ref{e-loss-n(E)}), one finds 
\ba
\label{e-loss-final}
\frac{dE}{d x} = -\frac{e^4}{2^5 3 \pi}\, T^2 
\bigg[1 - \frac{12 \zeta(3)}{\pi^2}  \frac{T}{E} \bigg] ,
\ea
where $\zeta(z)$ is the zeta Riemann function that $\zeta(3) \approx 1.202$, and we omitted the index `2' as only the energy of the incoming particle enters the formula.

As far as one is interested in the jet suppression observed in relativistic heavy-ion collisions, the limit $E \gg T$ is worth consideration. In the limit, we have the result
\ba
\label{e-loss-BigE}
\frac{d E}{d x}=   -\frac{e^4}{2^5 3 \pi}\, T^2 , 
\ea
which should be confronted with the energy loss of an energetic muon in ultrarelativistic electromagnetic plasma of electrons, positrons and photons \cite{Braaten:1991jj}
\ba
\label{e-loss-QED}
\frac{dE}{dx}= -\frac{e^4}{48\pi^3}\, T^2 \Big( \ln\frac{E}{eT}+2.031\Big).
\ea
As seen, the Coulomb energy-loss formula (\ref{e-loss-QED}) differs from (\ref{e-loss-BigE}) by the logarithm term which comes from the integration over the momentum transfer from the minimal ($q_\textrm{min}$) to maximal ($q_\textrm{max}$) value. The latter one is of the order of the energy of the test particle ($q_\textrm{max} \sim E$). In vacuum $q_\textrm{min}=0$ and consequently the integral, which equals $\ln(q_\textrm{max}/q_\textrm{min})$, diverges. In a plasma medium the long range Coulomb forces are screened and $q_\textrm{min}$ is of the order of Debye mass which in ultrarelativistic plasma is roughly $eT$. Thus, the logarithm term gets the form as in Eq. (\ref{e-loss-QED}). That said, the similarity of formulas (\ref{e-loss-QED}) and (\ref{e-loss-BigE}) is rather surprising if one realizes how different are the differential cross sections responsible for the energy losses. 

The energy loss can be estimated using more qualitative arguments. Then the energy loss may be evaluated as $\frac{dE}{d x} \sim \langle \Delta E \rangle / \lambda$, where $\langle \Delta E \rangle$ is the typical change of particle's energy in a single collision and $\lambda$ is the particle's mean free path given as $\lambda^{-1} = \rho \, \sigma$ with $\rho \sim T^3$ being the density of scatterers and $\sigma$ denoting the cross section. For the differential cross section $\frac{ d\sigma}{dt} \sim e^4/s^2$, the total cross section is $\sigma \sim e^4/s$. When a highly energetic particle with energy $E$ scatters on massless plasma particle, $s \sim ET$ and consequently  $\sigma \sim e^4/(ET)$. The inverse mean free path is thus estimated as $\lambda^{-1} \sim e^4 T^2/E$.  When the scattering process is independent of momentum transfer, $\langle \Delta E \rangle$ is of order $E$ and we finally find $-\frac{dE}{d x} \sim e^4 T^2$. When compared to the case of Coulomb scattering, the energy transfer in a single collision is much bigger but the cross section is smaller in the same proportion. Consequently, the two interactions corresponding to very different differential cross sections lead to very similar energy losses. The authors of \cite{Dusling:2009df} arrived to the analogous conclusion discussing viscous corrections to the distribution function caused by the collisions driven by a one-gluon exchange or by a $\phi^4$ interaction.

Since two extremely different cross sections of the processes of the ${\cal N} =1$ SUSY QED plasma lead to qualitatively the same energy losses one can expect that the same amount of energy is lost per length unit in the QCD and ${\cal N} =4$ super Yang-Mills plasmas.

\subsubsection{Broadening of transverse momentum}
\label{sssec-qhat-sqed}

We consider here the second transport characteristic of equilibrium ${\cal N} =1$ SUSY QED plasma which is the momentum broadening of an energetic selectron due to its interaction with plasma photons. The quantity, which is usually denoted as $\hat q$, determines the magnitude of radiative energy loss of a highly energetic particle in a plasma medium \cite{Baier:1996sk}. It is defined  as
\ba
\label{qhat-def}
\hat q = \int d\Gamma \, q_T^2,
\ea
where $d\Gamma$ is, as previously, the interaction rate and $q_T$ is the momentum transfer to the selectron which is perpendicular to the selectron initial momentum. 

Since $\hat q$ is computed in exactly the same way as the energy loss, it can be obtained by replacing $E_4 - E_2$ by $q_T^2$ in the formulas from the previous section. Then, instead of equation (\ref{e-loss-n(E)}) one finds
\ba
\label{qhat-n(E)}
\hat{q} = \frac{e^4}{2^4 3 \pi^3 E}
\int_0^{\infty}  d E_1 f_\gamma(E_1) \bigg[E_2 E_1^2 + \frac{2}{3}E_1^3 \bigg].
\ea

With the momentum distribution of plasma photons of the Bose-Einstein form (\ref{B-E}), Eq.~(\ref{qhat-n(E)}) gives
\ba
\label{qhat-final}
\hat{q} = \frac{e^4}{12 \pi^3}\, T^3 
\bigg[ \zeta(3)  + \frac{\pi^4}{45}  \frac{T}{E}\bigg] .
\ea

When the momentum broadening is caused by scattering driven by one-photon exchange, $\hat{q}$ is of the order $e^4 \ln(1/e) \, T^3$ \cite{Baier:2008js}. Therefore, we conclude that the momentum broadening, and consequently the radiative energy loss, of a highly energetic particle in the SUSY QED plasma is rather similar to that in the electromagnetic plasma of electrons, positrons and photons.

We expect an analogous situation in super Yang-Mills system. There are various elementary process but the  momentum broadening of highly energetic particles does not much differ from that in QGP.

We conclude this section by saying that despite remarkable differences between the sets of microscopical processes
occurring in supersymmetric plasmas and their nonsupersymmetric counterparts the physical characteristics are very similar in both systems. Specifically, the energy loss caused by different interactions is qualitatively the same. And the same holds for the momentum broadening.

\newpage
\thispagestyle{plain}

\section{Conclusions}
\label{conclusions}

\vspace{-1cm}

\begin{tabular}{ m{5cm} l }
& \emph{The more he looked inside}
\\
& \emph{the more Piglet wasn't there.} 
\end{tabular}

\vspace{.3cm}

\begin{tabular}{ m{5cm} r }
& The House at Pooh Corner, Alan Alexander Milne
\end{tabular}

\vspace{2cm}

In this work we systematically compared supersymmetric plasma systems with their nonsupersymmetric counterparts and the more we tried to find differences between them the more they were not visible. We compared the ${\cal N} =1$ SUSY QED with QED plasma and the ${\cal N} =4$ super Yang-Mills with QCD system. The main motivation for this work was the AdS/CFT correspondence which provides us with the method to investigate strongly coupled conformal systems via weakly coupled gravity. Thus, the properties of strongly coupled system described by the ${\cal N} =4$ super Yang-Mills theory can be extracted. However, the reliability of the information about quark-gluon plasma obtained from the ${\cal N} =4$ super Yang-Mills plasma is still in question. Here we carried out the comparative analysis of supersymmetric and nonsupersymmetric plasmas not in a strong, but in weak coupling regime so that perturbative methods have been applied.

The study of all the systems was performed within the Keldysh-Schwinger formalism which is applicable to both equilibrium and nonequilibrium many-body systems. However, to study nonAbelian gauge theories such as QCD or ${\cal N} =4$ super Yang-Mills in covariant gauges, which make the calculations much simpler than in physical gauges, one needs to introduce ghost fields. In the framework of functional methods we found the way how to implement ghosts into the Keldysh-Schwinger formulation of nonAbelian gauge theories. First, the generating functional was constructed and then the general Slavnov-Taylor identity in the Feynman gauge was derived. From the general form of the identity one is able to find its specific forms which establish relations between different Green functions. Therefore, the specific identity fulfilled by the gluon Green functions was found. The identity reveals the relationship between the gluon propagator and the free propagator of ghosts. It appears that the Green functions of the unphysical ghost field are determined by a distribution function of physical gluons. These ghost Green functions are an essential element of the perturbative series and they are also used in this thesis in order to compute the polarisation tensor of gluons. What is even more important, the introduction of ghosts into Keldysh-Schwinger formalism opens up an opportunity of making perturbative computations in nonAbelian gauge theories in the Feynman gauge.

After these rather formal considerations, an effort was put to derive some physical characteristics of supersymmetric plasma which were next compared to those of their nonsupersymmetric counterparts. We started with a comparison of the ${\cal N} =1$ SUSY QED plasma with the QED plasma of electrons, positrons, and photons which are described by Abelian theories. Then, we went to confront the ${\cal N} =4$ super Yang-Mills plasma to the QCD plasma of quarks and gluons, which are governed by more complex nonAbelian theories. Since all the systems were treated as ultrarelativistic ones the masses of all particles were neglected in the computations. At the beginning the collective excitations were considered. The dispersion relations of all constituents of a given plasma system, that are gauge bosons, fermions, and scalars, were found. In order to find the respective collective modes, the self-energies of all these fields were computed in the hard-loop approximation, that is when the wavevector of a mode is much smaller than typical momentum of a plasma constituent. 

The polarisation tensor of photons of both ${\cal N} =1$ SUSY QED and usual QED was found to be of the same structure. On calculating the polarisation tensor of both ${\cal N} =4$ super Yang-Mills and QCD we worked in the Feynman gauge and among others the ghost loop contributed to it. Thus, we used the Green functions of ghosts derived in this thesis. The final structure of the gluon polarisation tensor of ${\cal N} =4$ super Yang-Mills is the same as that of QCD and also as these of ${\cal N} =1$ SUSY QED and usual QED. The only differences lie in numerical factors reflecting different numbers of degrees of freedom. The polarisation tensors are transverse and symmetric with respect to the Lorentz indices. Their transversality, which appears automatically, guarantees its invariance due to the gauge symmetry. When the distribution functions entering the polarisation tensors are sent to zero, that is, when the vacuum limit is taken, the polarisation tensors of supersymmetric theories vanish, which is a manifestation of supersymmetry. In case of the QED and QCD polarisation tensors the vacuum contribution is nonzero.

The self-energies of fermions are again found to be of the same structure in all considered theories. It appears that irrespective of the fact that a fermion is a Dirac or Majorana one, the structure of fermionic self-energy does not change. And the same holds for scalars, which occur in the supersymmetric plasmas. The scalar self-energy of both the ${\cal N} =1$ SUSY QED and ${\cal N} =4$ super Yang-Mills plasmas is independent of the wavevector, it is negative and real.

Irrespective of a diversity of field theories considered, the self-energies of a given field turned out to have unique forms in the hard-loop approximation. Having obtained them it was possible to construct the effective hard-loop action. Since any self-energy is the second functional derivative of an action with respect to the field, the possible form of the action is strongly constrained. By integrating the self-energies and taking into account some arguments of gauge symmetries, we obtained the hard-loop actions which are all of universal forms for a whole class of gauge theories: QED, scalar QED, ${\cal N} =1$ SUSY QED, Yang-Mills, QCD, and ${\cal N} =4$ super Yang-Mills. Making allowance for various field contents and microscopic interactions, the universality of hard-loop actions is rather surprising. Nevertheless it is understandable if one bears in mind that the hard-loop approximation corresponds to the classical limit where effects of quantum statistics are not observable and therefore fermions and bosons are not distinguishable. Besides that, differences in elementary interactions are not seen as the gauge symmetry fully controls the interaction.

The universality of self-energies and accordingly of the hard-loop action has far-reaching consequences as the properties of all plasma systems studied here, which occur at the soft scale, are qualitatively the same. Especially, spectra of collective excitations of gauge, fermion, and scalar fields are the same in all the plasmas. In case of gauge bosons in equilibrium systems there are always two modes: longitudinal and transverse. When a plasma is out of equilibrium there is a whole variety of possible collective excitations and unstable modes occur as well. As far as fermionic modes are considered, there are two such modes of opposite helicity over chirality ratio in equilibrium plasma. In nonequilibrium plasma the spectrum changes but no unstable fermionic modes are seen. Supersymmetry does not change anything here. The scalar modes in turn behave as relativistic massive particles. 

Subsequently, the transport properties of supersymmetric plasmas in equilibrium were analysed. Since the temperature is the only dimensional parameter, which characterises the equilibrium plasma of massless constituents, the parametric structure of most of the transport coefficients is determined. For this reason we concentrated on such quantities which are not limited by dimensional arguments and can strongly depend on a specific process under consideration. Thus, we derived exact formulas of collisional energy loss and momentum broadening of highly energetic particle traversing a hot medium. Prior to that we found that among 33 binary processes occurring in the supersymmetric electromagnetic plasma there are a few, in particular the Compton scattering on selectron, which are isotropic in the centre-of-mass frame. The cross sections of them are independent of momentum transfer and therefore they are qualitatively different than these of the usual QED plasma whose scattering relies mostly on one-photon exchange and thus it is dominated by small angle deflections. We found that for the processes independent of momentum transfer the collisional energy loss and momentum broadening are very similar to the respective characteristics of the usual QED plasma when the limit of high energy of the particle traversing the medium is taken. In other words, two extremely different cross sections lead to similar amount of energy losses. When the process is independent of the momentum transfer an energy transfer is large but the collisions are rare and in the other case an energy transfer is small but the collisions are frequent. The situation in the super Yang-Mills plasma is expected to be similar.

To conclude, the performed analysis of a whole set of gauge theories allows to state that both collective phenomena and transport properties in supersymmetric and nonsupersymmetric plasma systems are astonishingly similar to each other. In particular the plasma described by the ${\cal N} =4$ super Yang-Mills theory is qualitatively very similar to the quark-gluon plasma when the coupling constant is small. The differences lie in numerical coefficients which reflect different numbers of degrees of freedom. Although our findings do not fully justify using the AdS/CFT duality to infer information about the strongly coupled QGP, they make the approach more plausible.

\newpage




\newpage
\thispagestyle{plain}
 
\appendix
\textcolor[rgb]{1.00,1.00,1.00}{.}
\vspace{8cm}

\addappheadtotoc
\begin{center}
\Huge \textbf{Appendices}
\end{center}


\newcommand{\mat}[4]{\left(\begin{array}{cc} #1 & #2 \\ #3 & #4
                           \end{array}\right)}

\newpage
\section{Supersymmetry}
\label{susy}
\thispagestyle{plain}
\vspace{-1cm}

Since we consider in this thesis the plasma systems governed by supersymmetric theories it is in order to explain a very idea of supersymmetry. Supersymmetry is a transformation that relates fermion degrees of freedom to boson ones and \emph{vice versa}, so that each particle of a given type is associated with its superpartner of an another type. An illuminating introduction to supersymmetric theories can be found in \cite{Sohnius:1985qm} or \cite{Signer:2009dx}. 

Studying supersymmetries it is natural to work on the superspace spanned by the coordinates $X=(x^\mu, \theta^\alpha, \bar \theta^{\dot \alpha})$. The variables $\theta^\alpha$ and $\bar \theta^{\dot \alpha}=(\theta^\alpha)^\dagger$, with $\alpha, \dot\alpha =1,2$ denoting indices of Weyl spinors, are the Grassmann variables whose properties are discussed in detail in Appendix~\ref{appendix-Grassmann-algebra}. Let us stress that it is a matter of notation that the dotted indices hereinafter label the adjoint variables. The supersymmetry is generated by sets of fermionic generators $Q_\alpha$, called also supercharges, that change the spin of a state by $1/2$ as follows
\ba
\label{susy-gen}
Q_\alpha |bos \rangle = |ferm \rangle_\alpha,  \qquad\qquad\qquad   Q_\alpha |ferm \rangle^\alpha = |bos \rangle.
\ea
If there is only one pair of the generators then we have ${\cal N}\!=\!1$ supersymmetry. Then the (anti-)commutation relations among the fermionic operators $Q_\alpha$, and bosonic ones $P^\rho$ and $M^{\rho\sigma}$, which generate translations and Lorentz transformations (rotations and boosts), respectively, constitute the Poincar\'{e} superalgebra, which determines symmetries of the theory
\ba
\label{algPP}
\left[P^\rho,P^\sigma\right] &=& 0 \, , 
\\ [2mm]
\left[P^\rho,M^{\nu\sigma}\right] &=& i (g^{\rho\nu} P^\sigma - g^{\rho\sigma} P^\nu) \, , 
\\ [2mm]
\label{alg-pp2}
\left[M^{\mu\nu}, M^{\rho\sigma}\right]&=& -i ( g^{\mu\rho}
M^{\nu\sigma}+g^{\nu\sigma} M^{\mu\rho}- g^{\mu\sigma} M^{\nu\rho}-g^{\nu\rho} M^{\mu\sigma}) \, ,
\\ [2mm]
\label{alg-pp-3}
\left\{Q_\alpha,Q_\beta\right\} &=& \left\{{\bar{Q}}_{\dot{\alpha}},{\bar{Q}}_{\dot{\beta}}\right\} = 0 \, ,
\\ [2mm]
\left\{Q_\alpha,\bar{Q}_{\dot{\beta}}\right\} &=& 2 (\sigma^\rho)_{\alpha\dot{\beta}} P_\rho \, ,
\\ [2mm]
\left[Q_\alpha,P^\rho\right] &=& 0 \, ,
\\ [2mm]
\label{MQ}
\left[M^{\rho\sigma},Q_\alpha\right] &=& -i (\sigma^{\rho\sigma})_\alpha^{\ \beta} Q_\beta \, ,
\ea
where $\sigma^\rho$ are the Pauli matrices. In these formulas the bosonic operators are defined as $P^\mu \equiv i\partial^\mu$ and $M^{\mu\nu}=i(x^\mu \partial^\nu-x^\nu \partial^\mu)$. The operator $M^{\mu\nu}$ has the additional element $\frac{i}{4}[\gamma^\mu,\gamma^\nu]$ in case of fermions with $\gamma^\mu$ being the Dirac matrices. The fermionic operators are as follows
\ba
\label{fermionic-operators}
Q_\alpha=i\partial_\alpha - \sigma^\mu_{\alpha \dot \alpha} \bar \theta^{\bar \alpha} \partial_\mu, \qquad\qquad
\bar Q_{\dot \alpha}=i\bar\partial_{\dot \alpha} - \theta^{\alpha} \sigma^\mu_{\alpha \dot \alpha} \partial_\mu,
\ea
where $\partial_\alpha$ and $\bar\partial_{\dot \alpha}$ are the derivatives over the Grassmann variables. The relations (\ref{algPP})-(\ref{alg-pp2}) define the Poincar\'{e} group which is the group of space-time symmetries of any relativistic field theory. The relations (\ref{alg-pp-3})-(\ref{MQ}) are characteristic of the supersymmetry.

The supersymmetric algebra of the $\mathcal{N} = 4$ super Yang-Mills theory, which involves four pairs of supercharges, is much richer but its structure is rather irrelevant for the considerations presented in this thesis.

\newpage
\section{Derivation of real-time Green functions in equilibrium}
\label{appendix-FG}
\thispagestyle{plain}
\vspace{-1cm}

\subsection{Derivation of the functions of the scalar field}
\label{appendix-scalars}

Here we derive the Green functions of real-time arguments for the scalar field in equilibrium. We start with $\Delta^>(x,y)$ and $\Delta^<(x,y)$ which are
defined by
\ba
\label{bigger-GF-SF}
i \Delta^{>}(x,y) 
& = & 
\frac{{\rm Tr} \big[\hat \rho \; \phi(x) \phi(y) \big]}{{\rm Tr}[\hat \rho]}, 
\\ [2mm]
\label{smaller-GF-SF}
i \Delta^{<}(x,y) 
& = & 
\frac{{\rm Tr} \big[\hat \rho \; \phi(y) \phi(x) \big]}{{\rm Tr}[\hat \rho]},
\ea
where the field operator is given by
\ba
\label{field-op}
\phi(x)=\int \frac{d^3 k}{(2\pi)^3\sqrt{2\omega_k }}
\Big( \hat a({\bf k}) e^{-i kx} +\hat a^\dagger({\bf k}) e^{i kx}\Big)
\ea
with $k^\mu = (\omega_{\bf k}, {\bf k})$ and $\omega_{\bf k} \equiv \sqrt{ {\bf k}^2 + m^2 }$. The density operator in
equilibrium equals
\ba
\label{density-matrix-SF}
\hat \rho = \exp{(-\beta \hat H)},
\ea
where $\beta \equiv 1/T$ and $\hat H$ is a Hamiltonian of the system of the form
\ba
\label{hamiltonian}
\hat H= \int \frac{d^3k}{(2\pi)^3}\frac{\omega({\bf k})}{2} 
\Big(\hat a^\dagger({\bf k}) \hat a({\bf k})  + \hat a({\bf k}) \hat a^\dagger({\bf k}) \Big),
\ea
where $\hat a({\bf k})$ is an annihilation operator and $\hat a^\dagger({\bf k})$ - a creation operator.

To manage with calculations, we discretize a continuous momentum space so that we have the set of discrete momenta $\{{\bf k}_1,{\bf k}_2,{\bf k}_3, \ldots \}$, and thus the field is given as the sum
\ba
\label{field-op-dis}
\phi(x)=\sqrt{\Delta_{\bf k}}\sum_i \frac{1}{\sqrt{2\omega_i }}
\Big( \hat a_i e^{-i k_ix} +\hat a^\dagger_i e^{i k_ix}\Big),
\ea
where $\Delta_{\bf k}$ is the volume of a momentum-space cell. Then, the Hamiltonian can be written as a sum of energies of independent oscillators
\ba
\label{hamiltonian-discrete}
\hat H= \sum_i \frac{\omega_i}{2} 
\Big(\hat a^\dagger_i \hat a_i  + \hat a_i \hat a^\dagger_i \big),
\ea
where $\hat a_i\equiv \sqrt{\Delta_{\bf k}}\hat a({\bf k}_i)$ and $\hat a^\dagger_i\equiv \sqrt{\Delta_{\bf k}}\hat a^\dagger({\bf k}_i)$. The operators $\hat a_i$ and $\hat a^\dagger_i$ are dimensionless and satisfy the following commutation relations
\ba
\label{op-an-kr-comm}
\big[\hat a_i,\hat a_j^\dagger\big] &=& \delta^{ij}, 
\\ [2mm]
\big[\hat a_i,\hat a_j\big] &=& 0, 
\\ [2mm]
\label{op-an-kr-comm1}
\big[\hat a_i^\dagger,\hat a_j^\dagger\big] &=& 0.
\ea
Using the relations (\ref{op-an-kr-comm})-(\ref{op-an-kr-comm1}), the Hamiltonian (\ref{hamiltonian-discrete}) gets the form
\ba
\label{hamiltonian-discrete-1}
\hat H= \sum_i \omega_i
\bigg(\hat a^\dagger_i \hat a_i  + \frac{1}{2} \bigg)
\ea
and after applying the normal ordering of the creation and annihilation operators,  we write
\ba
\label{hamiltonian-normal-ord}
\hat H= \sum_i \omega_i \hat a^\dagger_i \hat a_i = \sum_i \omega_i \hat n_i,
\ea
where $\hat n_i =\hat a^\dagger_i \hat a_i$ is the operator of particle number with the momentum ${\bf k}_i$.
The Fock space is built of the mutually orthogonal states $|n_1,n_2,n_3, \ldots \rangle$, and the action of the
annihilation and creation operators is defined in the following way
\ba
\label{a-many-s}
\hat a_i |n_1,n_2, \ldots,n_i,\ldots \rangle &=&\sqrt{n_i}\,|n_1,n_2,\ldots,n_i-1,\ldots \rangle,
\\ [2mm]
\label{a+many-s}
\hat a_i^\dagger|n_1,n_2, \ldots,n_i,\ldots \rangle &=&\sqrt{n_i+1}\,|n_1,n_2,\ldots,n_i+1,\ldots\rangle,
\ea
The states $|n_1,n_2,n_3, \ldots \rangle$ are the eigenstates of the energy operator $\hat H$ and also of the operator of
number of particles $\hat n$, so that
\ba
\label{eigenvalue-H}
\hat H |n_1,n_2, \ldots,n_i,\ldots \rangle &=& \sum_i \omega_i n_i |n_1,n_2, \ldots,n_i, \ldots \rangle, 
\\ [2mm]
\hat n |n_1,n_2, \ldots,n_i, \ldots \rangle &=& \sum_i  n_i |n_1,n_2, \ldots,n_i, \ldots \rangle.
\ea

Keeping in mind these elementary facts we write the Green function $\Delta^>(x,y)$ as
\ba
\label{bigger-GF-SF-1}
\Delta^{>}(x,y) 
& = & 
Z^{-1} \int \frac{d^3 k d^3 p}{2 (2\pi)^3 \sqrt{\omega_k \omega_p}} 
\sum_{n_1}  \sum_{n_2} \sum_{n_3} \ldots  
\\ [2mm] \nn
&& 
\times  \big\langle n_1,n_2, \ldots \big| \exp{(-\beta \hat H)} 
\Big( \hat a({\bf k}) e^{-i kx} +\hat a^\dagger({\bf k}) e^{i kx}\Big) 
\\ [2mm] \nn
&& \qquad\qquad \qquad\qquad\qquad
\times \Big( \hat a({\bf p}) e^{-i py} +\hat a^\dagger({\bf p}) e^{i py}\Big) \big| n_1,n_2, \ldots  \big\rangle ,
\ea
where we denoted the denominator of (\ref{bigger-GF-SF}), which is a partition function, as
\ba
\label{Z}
Z={\rm Tr}\big[\hat \rho \big] =  {\rm Tr}\big[e^{-\beta \hat H}\big].
\ea
Let us first find an exact form of the partition function which can be written as
\ba
\label{Z-1}
Z= \sum_{n_1}\sum_{n_2}\sum_{n_3} \dots \langle n_1,n_2,n_3, \ldots | e^{-\beta \hat H} |n_1,n_2,n_3, \ldots \rangle.
\ea
Using Eq.~(\ref{eigenvalue-H}) we find the following equality
\ba
\label{identity-1}
&&\exp{(-\beta \hat H)} |n_1,n_2,n_3, \ldots  \rangle 
\\ [2mm] \nn
&& \qquad\qquad
=\sum_{k=0}^\infty \frac{(-\beta)^k}{k!} \hat H^k  |n_1,n_2,n_3, \ldots \rangle 
= \sum_{k=0}^\infty \frac{(-\beta)^k}{k!}  \bigg(\sum_i \omega_i n_i\bigg)^k |n_1,n_2,n_3, \ldots \rangle 
\\ [2mm]\nn
&& \qquad\qquad
=\sum_{k=0}^\infty \sum_i \frac{(-\beta )^k}{k!}  \big(\omega_i n_i\big)^k |n_1,n_2,n_3, \ldots \rangle 
=\exp{\Big(-\beta \sum_i\omega_i n_i\Big)} |n_1,n_2,n_3, \ldots \rangle .
\ea
Inserting the result (\ref{identity-1}) into (\ref{Z-1}), we find the following formula of the partition function
\ba
\label{Z-2}
Z &=& \sum_{n_1}\sum_{n_2}\sum_{n_3} \dots \langle n_1,n_2,n_3, \ldots | 
e^{-\beta \hat H} |n_1,n_2,n_3, \ldots \rangle 
\\ [2mm] \nn
&=& 
\sum_{n_1}\sum_{n_2}\sum_{n_3} \dots  e^{-\beta \sum_i\omega_i n_i}
= \sum_{n_1}\sum_{n_2}\sum_{n_3} \dots  e^{-\beta \omega_1 n_1} e^{-\beta \omega_2 n_2}e^{-\beta \omega_3 n_3} \dots 
\\ [2mm] \nn
&=& 
\sum_{n_1} e^{-\beta \omega_1 n_1} \sum_{n_2} e^{-\beta \omega_2 n_2} \sum_{n_3} e^{-\beta \omega_3 n_3} \dots 
\ea
For every sum in (\ref{Z-2}) we use the identity
\ba
\label{geometric-series}
\sum_{n=0}^\infty q^n= \frac{1}{1-q}
\ea
and then we obtain
\ba
\label{Z-3}
Z = \frac{1}{1-e^{-\beta \omega_1}} \frac{1}{1-e^{-\beta \omega_2}} \frac{1}{1-e^{-\beta \omega_3}} \dots
\ea
The result (\ref{Z-3}) can be also rewritten as
\ba
\label{Z-4}
Z = \exp \bigg[ \sum_i {\rm ln}\Big( \frac{1}{1-e^{-\beta \omega_i}} \Big)\bigg]
= \exp \bigg[ - \sum_i {\rm ln}\Big(1-e^{-\beta \omega_i} \Big)\bigg],
\ea
The formula (\ref{Z-4}) is especially useful if we would like to come back to the continuous momentum space. The
partition function is then given as
\ba
\label{Z-cont}
Z = \exp \bigg[ - \frac{1}{\Delta_{\bf k}} \int \frac{d^3 k}{(2\pi)^3} {\rm ln}\Big(1-e^{-\beta \omega_{\bf k}} \Big)\bigg]
= \exp \bigg[ - V \int \frac{d^3 k}{(2\pi)^3} {\rm ln}\Big(1-e^{-\beta \omega_{\bf k}} \Big)\bigg],
\ea
where the inverse volume of the momentum cell $\Delta_{\bf k}$ is replaced by the system's volume $V$.

Continuing with the consideration of $\Delta^>(x,y)$, given by the formula~(\ref{bigger-GF-SF-1}), we perform
discretization of momenta and manipulate the expression to the form
\ba
\label{bigger-GF-SF-2}
\Delta^{>}(x,y) 
& = & 
Z^{-1} \Delta_{\bf k}
\sum_{n_1} \sum_{n_2} \sum_{n_3} \ldots 
e^{-\beta \omega_1 n_1} e^{-\beta \omega_2 n_2} e^{-\beta \omega_3 n_3} \dots
\sum_i \sum_j \frac{1}{2 \sqrt{\omega_i \omega_j}} \qquad \qquad
\\ [2mm] \nn
& & 
\times\Big[ e^{-i(k_i x+p_j y)}\big\langle n_1, \dots, n_i, \dots \big| 
\hat a_i \hat a_j \big| n_1, \dots, n_j, \dots \big\rangle 
\\ [2mm] \nn
&& 
+e^{-i(k_i x-p_j y)} \big\langle n_1, \dots, n_i, \dots \big| 
\hat a_i \hat a_j^\dagger \big|n_1, \dots, n_j, \dots \big\rangle 
\\ [2mm] \nn
&& 
+e^{i(k_i x-p_j y)} \big\langle n_1, \dots, n_i, \dots \big| 
\hat a_i^\dagger \hat a_j \big|n_1, \dots, n_j, \dots \big\rangle 
\\ [2mm] \nn
&& 
+e^{i(k_i x+p_j y)} \big\langle n_1, \dots, n_i, \dots \big| 
\hat a_i^\dagger \hat a_j^\dagger \big|n_1, \dots, n_j, \dots \big\rangle \Big],
\ea
where $k_i^\mu = (\omega_i, {\bf k}_i)$. The action of the creation and annihilation operators, as dictated by the definitions
(\ref{a-many-s}) and (\ref{a+many-s}), gives us the following contributions to $\Delta^>(x,y)$
\ba
\label{1-contr}
&&
\big\langle n_1, \dots, n_i, \dots \big| 
\hat a_i \hat a_j \big| n_1, \dots, n_j, \dots \big\rangle
\\ [2mm] \nn
&& \qquad\qquad\qquad\qquad
= \sqrt{(n_i+1)n_j} \big\langle n_1, \dots, n_i +1, \dots \big| 
n_1, \dots, n_j - 1, \dots \big\rangle =0,
\\ [2mm] 
&& 
\label{2-contr}
\big\langle n_1, \dots, n_i, \dots \big| 
\hat a^\dagger_i \hat a^\dagger_j \big| n_1, \dots, n_j, \dots \big\rangle
\\ [2mm] \nn
&& \qquad\qquad\qquad\qquad
= \sqrt{n_i (n_j+1)} \big\langle n_1, \dots, n_i -1, \dots \big| 
n_1, \dots, n_j + 1, \dots \big\rangle =0,
\ea
\ba
\label{3-contr}
\big\langle n_1, \dots, n_i, \dots \big| 
\hat a_i \hat a_j^\dagger \big|n_1, \dots, n_j, \dots \big\rangle
&=& \sqrt{(n_i+1)(n_j+1)} \; \delta^{ij}, 
\\ [2mm]
\label{4-contr}
\big\langle n_1, \dots, n_i, \dots \big| 
\hat a_i^\dagger \hat a_j \big|n_1, \dots, n_j, \dots \big\rangle
&=&\sqrt{n_i n_j}\; \delta^{ij}.
\ea
As seen, the action of the operators in formulas (\ref{1-contr}) and (\ref{2-contr}) produces in result orthogonal states, which is why their product vanish. Including the results (\ref{3-contr}) and (\ref{4-contr}) in the formula (\ref{bigger-GF-SF-2}), we find
\ba
\label{bigger-GF-SF-3}
\Delta^{>}(x,y) 
& = & 
Z^{-1} \Delta_{\bf k} \sum_{n_1} \sum_{n_2} \sum_{n_3} \dots 
e^{-\beta \omega_1 n_1} e^{-\beta \omega_2 n_2} e^{-\beta \omega_3 n_3} \dots 
\\ [2mm] \nn
& & \qquad\qquad\qquad\qquad
\times \sum_i \frac{1}{2 \omega_i }
\Big[(n_i +1) e^{-i k_i (x-y)} + n_i e^{i k_i (x-y)} \Big],
\ea
where we summed over all momenta indexed by $j$. The formula (\ref{bigger-GF-SF-3}) is equivalent to
\ba
\label{bigger-GF-SF-4}
\Delta^{>}(x,y) 
& = & 
Z^{-1} \Delta_{\bf k} \sum_{n_1} \sum_{n_2} \sum_{n_3} \dots 
e^{-\beta \omega_1 n_1} e^{-\beta \omega_2 n_2} e^{-\beta \omega_3 n_3} \dots 
\\ [2mm] \nn
& & \qquad\qquad\qquad
\times \sum_i \frac{1}{2 \omega_i }
\Big[n_i \Big(e^{-i k_i (x-y)} + e^{i k_i (x-y)}\Big) + e^{-i k_i (x-y)} \Big].
\ea
To proceed with the calculations we consider now only the sum over $n_1$. It helps us find a general rule how to perform all summations in (\ref{bigger-GF-SF-4}). So let us calculate the following expression
\ba
\label{sum1}
\frac{1}{2 \omega_1 } \sum_{n_1} 
\bigg[n_1  e^{-\beta \omega_1 n_1} \Big(e^{-i k_1 (x-y)} + e^{i k_1 (x-y)}\Big) 
+  e^{-\beta \omega_1 n_1} e^{-i k_1 (x-y)} \bigg].
\ea
The following identities are needed to solve the problem:
\ba
\label{math-trick-SF-1}
\sum_{n=0}^\infty q^n= \frac{1}{1-q},
\ea
which we use for the second term in the brackets in (\ref{sum1}) and for every other sum over $n_i$ with $i>1$, and
\ba
\label{math-trick-SF-2}
\sum_{n=0}^\infty n q^n= q \frac{d}{dq} \sum_{n=0}^\infty q^n =q \frac{d}{dq} \frac{1}{1-q}=\frac{q}{(1-q)^2}
\ea
which we use for the first term in the bracket. Accordingly, in place of (\ref{sum1}) we get
\ba
\label{sum1-a}
\frac{1}{2 \omega_1 } 
\bigg[\frac{e^{-\beta \omega_1}}{(1-e^{-\beta \omega_1})^2}  \Big(e^{-i k_1 (x-y)} + e^{i k_1 (x-y)}\Big) 
+\frac{1}{1-e^{-\beta \omega_1}} e^{-i k_1 (x-y)} \bigg].
\ea
Since
\ba
\frac{e^{-\beta \omega_1}}{(1-e^{-\beta \omega_1})^2}  \Big(e^{-i k_1 (x-y)} + e^{i k_1 (x-y)}\Big) 
+\frac{1}{1-e^{-\beta \omega_1}} e^{-i k_1 (x-y)} \qquad \qquad \qquad \qquad 
\\ [2mm] \nn
=\frac{1}{1-e^{-\beta \omega_1}}
\bigg[\frac{e^{-\beta \omega_1}}{1-e^{-\beta \omega_1}}  \Big(e^{-i k_1 (x-y)} + e^{i k_1 (x-y)}\Big) 
+ e^{-i k_1 (x-y)} \bigg] ,
\ea
in the expression (\ref{bigger-GF-SF-4}) the partition function $Z$ factors out, and we obtain the following form of $\Delta^>(x,y)$
\ba
\label{bigger-GF-SF-5}
i\Delta^{>}(x,y) = \Delta_{\bf k} \sum_i \frac{1}{2\omega_i}
\bigg[\frac{e^{-\beta \omega_i}}{1-e^{-\beta \omega_i}} 
\Big(e^{-ik_i(x-y)}+e^{ik_i(x-y)}\Big) +e^{-ik_i(x-y)} \bigg].
\ea
Changing a discrete momentum space into a continuous one, we can write
\ba
\label{bigger-GF-SF-6}
i\Delta^{>}(x,y)  = 
\int \frac{d^3 k}{(2\pi)^3} \frac{1}{2\omega_{\bf k}} 
\bigg[\frac{e^{-\beta \omega_{\bf k}}}{1-e^{-\beta \omega_{\bf k}}} 
\Big(e^{-ik(x-y)}+e^{ik(x-y)}\Big) + e^{-ik(x-y)} \bigg]
\ea
and pointing out that
\ba
\label{distr-fun-SF}
f(\omega_{\bf k})=\frac{1}{e^{\beta \omega_{\bf k}}-1} 
= \frac{e^{-\beta \omega_{\bf k}}}{1-e^{-\beta \omega_{\bf k}}},
\ea
we simply get
\ba
\label{bigger-GF-SF-7}
i\Delta^{>}(x,y) = \int \frac{d^3 k}{(2\pi)^3} \frac{1}{2\omega_{\bf k}} 
\bigg[f(\omega_{\bf k}) \Big(e^{-ik(x-y)}+e^{ik(x-y)}\Big)+e^{-ik(x-y)} \bigg],
\ea
or equivalently
\ba
\label{bigger-GF-SF-8}
i\Delta^{>}(x,y)  = \int \frac{d^3 k}{(2\pi)^3} \frac{1}{2\omega_{\bf k}} 
\bigg[(f(\omega_{\bf k})+1) e^{-ik(x-y)}+ f(\omega_{\bf k}) e^{ik(x-y)}\big)\bigg].
\ea
Repeating the same steps, we find $\Delta^<(x,y)$ as
\ba
\label{smaller-GF-SF}
i\Delta^{<}(x,y)  = \int \frac{d^3 k}{(2\pi)^3} \frac{1}{2\omega_{\bf k}} 
\Big[f(\omega_{\bf k}) e^{-ik(x-y)}+ (f(\omega_{\bf k})+1) e^{ik(x-y)}\big)\Big].
\ea
Let us now find the Green function $\Delta^>(x,y)$ in the momentum space. Performing the Fourier transform of
(\ref{bigger-GF-SF-8}), where we put $y=0$, we get
\ba
\label{bigger-GF-SF-FT}
\Delta^{>}(p) =
- \frac{i}{2} \int dt \, d^3 x \; \int \frac{d^3 k}{(2\pi)^3 \omega_{\bf k}} 
\Big[(f(\omega_{\bf k})+1) e^{-i(k-p)x}+ f(\omega_{\bf k}) e^{i(k+p)}\big)\Big].
\ea
Using the following identity 
\ba
\label{delta}
\int dt \, d^3 x\; e^{i(k-p)x}= (2\pi)^4 \delta(\omega_{\bf k} - p_0) \delta^{(3)}({\bf k} - {\bf p})
\ea
and performing the integration over three-momenta, we find
\ba
\label{bigger-GF-SF-FT-1}
\Delta^{>}(p)  =  -\frac{i\pi}{\omega_{\bf p}} 
\Big[ \delta(\omega_{\bf p} - p_0) (f(\omega_{\bf p})+1)
+  \delta(\omega_{\bf p} + p_0) f(\omega_{\bf p})\Big].
\ea
Analogously we get
\ba
\label{smaller-GF-SF-FT}
\Delta^{<}(p) & = & -\frac{i\pi}{\omega_{\bf p}} 
\Big[ \delta(\omega_{\bf p} - p_0) f(\omega_{\bf p})
+  \delta(\omega_{\bf p} + p_0)( f(\omega_{\bf p})+1)\Big].
\ea
We see that both $\Delta^>(p)$ and $\Delta^<(p)$ are non zero only for $p^2=m^2$. As one can notice that the Kubo-Martin-Schwinger boundary condition (KMS condition), which is discussed in detail in Appendix \ref{appendix-KMS-condition} and has the form
\ba
\label{KMS-cond-SC}
\Delta^<(p)=e^{-\beta p_0} \Delta^>(p)
\ea
is fulfilled.

Let us now compute the Green functions $\Delta^c(x,y)$ and $\Delta^a(x,y)$. Due to the relation (\ref{rel-2}), we have 
\ba
\label{causal-GF-SF-1}
\Delta^{c} (x,y)
&=&
-\frac{i}{2} \bigg\{\Theta (x_0-y_0) \int \frac{d^3 k}{(2\pi)^3 \omega_{\bf k}} 
\bigg[(f(\omega_{\bf k})+1) e^{-ik(x-y)}+ f(\omega_{\bf k}) e^{ik(x-y)}\big)\bigg] \quad \quad 
\\ [2mm] \nn
&& \quad \;\; 
+ \Theta (y_0-x_0) \int \frac{d^3 k}{(2\pi)^3 \omega_{\bf k}} 
\bigg[f(\omega_{\bf k}) e^{-ik(x-y)}+ (f(\omega_{\bf k})+1) e^{ik(x-y)}\big)\bigg] \bigg\},
\ea
\ba
\label{anticausal-GF-SF-1}
\Delta^{a} (x,y)
&=&
-\frac{i}{2} \bigg\{\Theta (x_0-y_0) \int \frac{d^3 k}{(2\pi)^3 \omega_{\bf k}} 
\Big[f(\omega_{\bf k}) e^{-ik(x-y)}+ (f(\omega_{\bf k})+1) e^{ik(x-y)}\big)\Big] \quad\quad 
\\ [2mm] \nn
&& \quad \;\; 
+ \Theta (y_0-x_0)  \int \frac{d^3 k}{(2\pi)^3 \omega_{\bf k}} 
\Big[(f(\omega_{\bf k})+1) e^{-ik(x-y)}+ f(\omega_{\bf k}) e^{ik(x-y)}\big)\Big] \bigg\}.
\ea
Calculating the function $\Delta^c$ in the momentum space we obtain
\ba
\label{causal-GF-SF-FT-2}
\Delta^{c} (p)
&=&
-\frac{i}{2 \omega_{\bf p}} 
\bigg\{\int_{-\infty}^\infty dt \;
f(\omega_{\bf p}) \Big[e^{i(p_0-\omega_{\bf p})t} + e^{i(p_0+\omega_{\bf p})t} \Big] 
\\ [2mm] \nn
&& \qquad \qquad \qquad \qquad  
+ \; \int_0^{\infty} dt \;  e^{i(p_0-\omega_{\bf p})t}
+ \int_{-\infty}^0 dt \; e^{i(p_0+\omega_{\bf p})t} \bigg\}.
\ea
The first term of (\ref{causal-GF-SF-FT-2}) can be immediately written as
\ba
\label{first-t-chronological}
\int_{-\infty}^\infty dt \; f(\omega_{\bf p}) \Big[e^{i(p_0-\omega_{\bf p})t} + e^{i(p_0+\omega_{\bf p})t}\Big]
= 2\pi f(\omega_{\bf p}) \big[ \delta(p_0-\omega_{\bf p}) + \delta(p_0+\omega_{\bf p}) \big].
\ea
The second and third term is ill defined, so that we change $\omega_{\bf p} \rightarrow \omega_{\bf p}-i0^+$ to make the limit $t \rightarrow \pm \infty$ meaningful. Then, we have
\ba
\label{second-t-chronological}
\int_0^\infty dt\; e^{i(p_0-\omega_{\bf p}+i0^+) t} 
=\frac{-ie^{i(p_0-\omega_{\bf p}+i0^+)t}}{p_0-\omega_{\bf p}+i0^+}\bigg|_0^\infty 
=\frac{i}{p_0-\omega_{\bf p}+i0^+}
\ea
and
\ba
\label{third-t-chronological}
\int_{-\infty}^0 dt\; e^{i(p_0+\omega_{\bf p}-i0^+) t} 
=\frac{-ie^{i(p_0+\omega_{\bf p}-i0^+)t}}{p_0+\omega_{\bf p}-i0^+}\bigg|_{-\infty}^0 
=\frac{-i}{p_0+\omega_{\bf p}-i0^+}.
\ea
After adding all terms we find the final result of $\Delta^c(p)$ as
\ba
\label{causal-GF-SF-FT-3}
\Delta^{c} (p)&=&\frac{1}{p^2-m^2+i0^+} -
\frac{i\pi}{\omega_{\bf p}} f(\omega_{\bf p}) \Big[\delta(p_0 - \omega_{\bf p})+\delta(p_0 + \omega_{\bf p}) \Big]
\ea
and that of $\Delta^a(p)$ as
\ba
\label{anticausal-GF-SF-FT-3}
\Delta^{a} (p)&=&-\frac{1}{p^2 -m^2- i0^+} - 
\frac{i\pi}{\omega_{\bf p}} f(\omega_{\bf p}) \Big[\delta(p_0 - \omega_{\bf p})+\delta(p_0 + \omega_{\bf p}) \Big].
\ea
In the first term of Eq.~(\ref{causal-GF-SF-FT-3}) one recognizes the usual Feynman propagator. The second term
represents the effect of a medium. The first term is non-zero for any $p$ while the second one only for $p^2=m^2$.

Let us now calculate the retarded and advanced Green functions, $\Delta^+(p)$ and $\Delta^-(p)$. For this purpose we use the relations (\ref{rel-5}) which involve the functions derived $\Delta^>, \Delta^<$, and $\Delta^c$. The
retarded Green function is given as
\ba
\label{retarded-GF}
\Delta^+(p)= \Delta^c(p)-\Delta^<(p)
\ea
and after inserting the formulas (\ref{causal-GF-SF-FT-3}) and (\ref{smaller-GF-SF-FT}) we get
\ba
\label{retarded-GF-1}
\Delta^+(p)=\frac{i\pi}{\omega_{\bf p}} \delta(p_0+\omega_{\bf p}) 
+ \frac{1}{p^2 - m^2 + i0^+}.
\ea
Applying the identity 
\ba
\label{identity-P}
\frac{1}{x \pm i0^+}= \mathcal{P}\frac{1}{x}\mp i\pi \delta(x),
\ea
we obtain
\ba
\label{retarded-GF-2}
\Delta^+(p)=\mathcal{P} \frac{1}{p^2 - m^2} 
- \frac{i\pi}{2\omega_{\bf p}} \Big[\delta(p_0-\omega_{\bf p}) - \delta(p_0+\omega_{\bf p})\Big].
\ea
The retarded Green function (\ref{retarded-GF-2}) can be written using the sign function, which, depending on the
sign of $p_0$, selects the right delta function. Consequently $\Delta^+ (p)$ can be expressed as
\ba
\label{retarded-GF-3}
\Delta^+(p)=\mathcal{P} \frac{1}{p^2 - m^2} - \frac{i\pi}{2\omega_{\bf p}} \textrm{sgn}(p_0)
\Big[\delta(p_0+\omega_{\bf p}) +\delta(p_0-\omega_{\bf p})\Big]
\ea
and thus recalling Eq. (\ref{identity-P}) we can write the final form of the retarded Green function as
\ba
\label{retarded-GF-SF-FT}
\Delta^{+} (p)= \frac{1}{p^2-m^2+i {\rm sgn}(p_0)0^+},
\ea
Analogously, the advanced Green function is defined by
\ba
\label{advanced-GF}
\Delta^-(p)= \Delta^c(p)-\Delta^>(p)
\ea
and performing analogous steps as in the case of $\Delta^+$ we find its exact form as
\ba
\label{advanced-GF-SF-FT}
\Delta^{-} (p)= \frac{1}{p^2-m^2 -i {\rm sgn}(p_0)0^+}.
\ea
It is worth noting that the propagators $\Delta^\pm(p)$ given by (\ref{retarded-GF-SF-FT}) and (\ref{advanced-GF-SF-FT})
coincide with the ground state propagators, as there is no contribution from the distribution function $f(\omega_{\bf p})$.
  
\subsection{Derivation of the functions of electromagnetic field}
\label{appendix-photons}

Below we derive the Green functions of real-time arguments for the electromagnetic field in equilibrium. Since the
procedure of their derivation is analogous to that one presented in the previous section about the scalar field, we
only expose the differences occurring in case of the electromagnetic field. We start with $D_{\mu\nu}^>(x,y)$ and
$D_{\mu\nu}^<(x,y)$ which are defined by
\ba
\label{bigger-GF-EF}
i D_{\mu\nu}^{>}(x,y) & = & 
\frac{{\rm Tr} \big[\hat \rho \; A_\mu(x) A_\nu(y) \big]}{{\rm Tr}[\hat \rho]}, 
\\ [2mm]
\label{smaller-GF-EF}
i D_{\mu\nu}^{<}(x,y) & = & 
\frac{{\rm Tr} \big[\hat \rho \; A_\mu(y) A_\nu(x) \big]}{{\rm Tr}[\hat \rho]},
\ea
where the field operator is given by
\ba
\label{field-op-em}
{\bf A} (x)=\sum_{\lambda=1}^2 \int \frac{d^3 k}{(2\pi)^3\sqrt{2\omega_k }} {\bf \epsilon}({\bf k},\lambda)
\Big( \hat a({\bf k}, \lambda) e^{-i kx} +\hat a^\dagger({\bf k}, \lambda) e^{i kx}\Big)
\ea
with $k^\mu = (\omega_{\bf k}, {\bf k})$ where $\omega_{\bf k} \equiv |{\bf k}|$, and ${\bf \epsilon}({\bf k},\lambda)$ is
the real (${\bf \epsilon}({\bf k},\lambda)={\bf \epsilon}^*({\bf k},\lambda)$) polarization vector of unit length (${\bf \epsilon}^2({\bf k},\lambda)=1$). As we work in the Coulomb gauge where
\ba
\label{coulomb-gauge-cond}
\nabla \cdot {\bf A}=0,
\ea
the polarization vector satisfies the condition
\ba
\label{cond-gauge-pol}
{\bf k} \cdot {\bf \epsilon}({\bf k},\lambda) = 0.
\ea
Since the electromagnetic field has two polarization states there are two polarization vectors which we denote as $\lambda=\pm1$. The vectors ${\bf \epsilon}({\bf
k},+1)$, ${\bf \epsilon}({\bf k},-1)$ and $|{\bf k}|/{\bf k}$ form the orthonormal base in 3-dimensional space, what
means that
\ba
\label{pol-vec}
{\bf \epsilon}({\bf k},\lambda)\cdot{\bf \epsilon}({\bf k},\lambda')=\delta^{\lambda\lambda'}.
\ea
The density operator in equilibrium equals
\ba
\label{density-matrix-EF}
\hat \rho = \exp{(-\beta \hat H)},
\ea
where $\beta \equiv 1/T$ and $\hat H$ is the normally ordered Hamiltonian of the system
\ba
\label{hamiltonian-em}
\hat H= \sum_{\pm\lambda} \int \frac{d^3k}{(2\pi)^3}\omega({\bf k}) \; 
\hat a^\dagger({\bf k}, \lambda) \hat a({\bf k}, \lambda).
\ea
$\hat a({\bf k},\lambda)$ and $\hat a^\dagger({\bf k},\lambda)$ are the annihilation and creation operators, respectively. To proceed with the calculations, we discretize a continuous momentum space so that we have a set of discrete momenta $\{{\bf k}_1,{\bf k}_2,{\bf k}_3, \ldots \}$, and thus the Hamiltonian can be written as
\be
\label{hamiltonian-discrete-em}
\hat H= \sum_{\pm\lambda} \sum_i \omega_i \hat a^\dagger_{i,\lambda} \hat a_{i,\lambda} 
= \sum_{\pm\lambda} \sum_i \omega_i \hat n_{i,\lambda},
\ee
where $\hat a_{i,\lambda}\equiv \sqrt{\Delta_{\bf k}}\hat a({\bf k}_i, \lambda)$, $\hat a^\dagger_{i,\lambda}\equiv
\sqrt{\Delta_{\bf k}}\hat a^\dagger({\bf k}_i, \lambda)$ and $\Delta_{\bf k}$ is the volume of a momentum space cell.
$\hat n_{i,\lambda} =\hat a^\dagger_{i,\lambda} \hat a_{i,\lambda}$ is the operator of a particle number with a momentum ${\bf k}_i$. The operators $\hat a_{i,\lambda}$ and $\hat a^\dagger_{i,\lambda}$ are dimensionless and satisfy the following commutation relations
\ba
\label{op-an-kr-comm-em}
[\hat a_{i,\lambda},\hat a_{j,\lambda'}^\dagger] &=& \delta^{\lambda\lambda'}\delta^{ij}, 
\\ [2mm]
[\hat a_{i,\lambda},\hat a_{j,\lambda'}] &=& 0, 
\\ [2mm]
[\hat a_{i,\lambda}^\dagger,\hat a_{j,\lambda'}^\dagger] &=& 0.
\ea
The Fock space is built of the mutually orthogonal states $|n_{1,+}, n_{1,-},n_{2,+},n_{2,-},
\ldots\rangle$, where `$+$' and `$-$' denote two different spin states of electromagnetic field and the action of the
annihilation operator and the creation one is defined in the following way
\ba
\label{a-many}
&&\hat a_{i,\lambda} |n_{1,+},n_{1,-},n_{2,+},n_{2,-}, \ldots,n_{i,\lambda},\ldots \rangle 
\\ [2mm] \nn
&& \qquad\qquad\qquad\qquad
=\sqrt{n_{i,\lambda}} \,|n_{1,+},n_{1,-},n_{2,+},n_{2,-}, \ldots,n_{i,\lambda}-1,\ldots \rangle,
\\  [2mm]
&&
\label{a+many}
\hat a_{i,\lambda}^\dagger|n_{1,+},n_{1,-},n_{2,+},n_{2,-}, \ldots,n_{i,\lambda},\ldots \rangle 
\\ [2mm] \nn
&& \qquad\qquad\qquad\qquad
=\sqrt{n_{i,\lambda}+1}\,|n_{1,+},n_{1,-},n_{2,+},n_{2,-},\ldots,n_{i,\lambda}+1,\ldots\rangle,\;\,
\ea
The states $|n_{1,+}, n_{1,-},n_{2,+},n_{2,-}, \ldots \rangle$ are the eigenstates of the energy operator
$\hat H$ and also of the operator of number of particles $\hat n_{i,\lambda}$, so that
\ba
\label{eigenvalue-H-em}
&& \hat H |n_{1,+},n_{1,-},n_{2,+},n_{2,-}, \ldots,n_{i,\lambda},\ldots \rangle  
\\ [2mm] \nn
&&  \qquad\qquad\qquad\qquad
= \sum_i \sum_\lambda \omega_i n_{i,\lambda} 
|n_{1,+},n_{1,-},n_{2,+},n_{2,-}, \ldots,n_{i,\lambda},\ldots \rangle, 
\\ [2mm]
\label{kkhh}
&& 
\hat n_{i,\lambda} |n_{1,+},n_{1,-},n_{2,+},n_{2,-}, \ldots,n_{i,\lambda},\ldots \rangle  
\\ [2mm] \nn
&& \qquad\qquad\qquad\qquad
= \sum_i \sum_\lambda n_{i,\lambda} |n_{1,+},n_{1,-},n_{2,+},n_{2,-}, \ldots,n_{i,\lambda},\ldots \rangle,  
\ea
where $\sum_{\lambda}\ldots \equiv \sum_{\lambda=1}^2 \ldots$. The Green function $D_{ij}^{>}(x,y)$ can be then written as
\ba
\label{bigger-GF-EF-1}
D_{ij}^{>}(x,y) 
& = & 
Z^{-1} \sum_{\lambda',\lambda''} \int \frac{d^3 k d^3 p}{2 (2\pi)^3 \sqrt{\omega_k \omega_p}} 
\sum_{n_{1,+}}  \sum_{n_{1,-}} \sum_{n_{2,+}} \sum_{n_{2,-}} \ldots
{\bf \epsilon}^i({\bf k},\lambda') \cdot {\bf \epsilon}^j({\bf p},\lambda'') \qquad
\\ [2mm] \nn
&& 
\times  \big\langle n_{1,+},n_{1,-},n_{2,+},n_{2,-}, \ldots,n_{i,\lambda},\ldots| \exp{(-\beta \hat H)} 
\\ [2mm] \nn
&&
\times \Big( \hat a({\bf k},\lambda') e^{-i kx} +\hat a^\dagger({\bf k},\lambda') e^{i kx}\Big)
\Big( \hat a({\bf p},\lambda'') e^{-i py} +\hat a^\dagger({\bf p},\lambda'') e^{i py}\Big)
\\ [2mm] \nn
&& \qquad\qquad\qquad\qquad\qquad\qquad\qquad
\times \big|n_{1,+},n_{1,-},n_{2,+},n_{2,-}, \ldots,n_{i,\lambda},\ldots \rangle, 
\ea
where we denoted the denominator of (\ref{bigger-GF-EF-1}), which is a partition function, as
\ba
\label{Z-em}
Z={\rm Tr}\big[\hat \rho \big] =  {\rm Tr}\big[e^{-\beta \hat H}\big].
\ea
The partition function is found analogously to that of the scalar field and it equals
\ba
\label{Z-3-em}
Z =\bigg(\frac{1}{1-e^{-\beta \omega_1}}\bigg)^2 \bigg(\frac{1}{1-e^{-\beta \omega_2}}\bigg)^2 
\bigg(\frac{1}{1-e^{-\beta \omega_3}}\bigg)^2 \dots
\ea
where the powers of 2 have appeared as for every momentum there exist two polarization states. The result
(\ref{Z-3-em}) can be also rewritten in the other way
\ba
\label{Z-4-em}
Z = \exp \bigg[ 2 \sum_i {\rm ln}\Big( \frac{1}{1-e^{-\beta \omega_i}} \Big)\Big]
= \exp \Big[ - 2 \sum_i {\rm ln}\Big(1-e^{-\beta \omega_i} \Big)\bigg],
\ea
In the continuous momentum space the partition function is given by
\ba
\label{Z-cont-em}
Z = \exp \bigg[ - \frac{2}{\Delta_{\bf k}} \int \frac{d^3 k}{(2\pi)^3} {\rm ln}\Big(1-e^{-\beta \omega_{\bf k}} \Big)\Big]
= \exp \Big[ -2 V \int \frac{d^3 k}{(2\pi)^3} {\rm ln}\Big(1-e^{-\beta \omega_{\bf k}} \Big)\bigg].
\ea
Continuing with the consideration of $D_{ij}^>(x,y)$, given by the formula~(\ref{bigger-GF-EF-1}), we manipulate the expression to get
\ba
\label{bigger-GF-EF-2}
D_{ij}^{>}(x,y) 
& = & 
Z^{-1} \Delta_{\bf k}
\sum_{n_{1,+}} \sum_{n_{1,-}} \sum_{n_{2,+}} \sum_{n_{2,-}} \ldots 
e^{-\beta \omega_1 n_{1,+}}  e^{-\beta \omega_1 n_{1,-}} 
e^{-\beta \omega_2 n_{2,+}} e^{-\beta \omega_2 n_{2,-}} \dots
\\ [2mm] \nn
&& 
\times \sum_{\lambda'} \sum_{\lambda''} \sum_k \sum_l \frac{1}{2 \sqrt{\omega_k \omega_l}} 
\epsilon_{k,\lambda'}^i \cdot \epsilon_{l,\lambda''}^j 
\\ [2mm] \nn
&&
\times \bigg[ e^{-i(k_k x+p_l y)}\big\langle n_{1,+},n_{1,-}, \dots, n_{k,\lambda}, \dots \big| 
\hat a_{k,\lambda'} \hat a_{l,\lambda''} \big| n_{1,+},n_{1,-}, \dots, n_{l,\lambda}, \dots \big\rangle 
\\ [2mm] \nn
&&
+e^{-i(k_k x-p_l y)} \big\langle n_{1,+},n_{1,-}, \dots, n_{k,\lambda}, \dots \big| 
\hat a_{k,\lambda'} \hat a_{l,\lambda''}^\dagger \big|n_{1,+},n_{1,-}, \dots, n_{l,\lambda}, \dots \big\rangle
\\ [2mm] \nn
&& 
+e^{i(k_k x-p_l y)} \big\langle  n_{1,+},n_{1,-}, \dots, n_{k,\lambda}, \dots \big| 
\hat a_{k,\lambda'}^\dagger \hat a_{l,\lambda''} \big|n_{1,+},n_{1,-}, \dots, n_{l,\lambda}, \dots \big\rangle 
\\ [2mm] \nn
&&
+e^{i(k_k x+p_l y)} \big\langle  n_{1,+},n_{1,-}, \dots, n_{k,\lambda}, \dots \big| 
\hat a_{k,\lambda'}^\dagger \hat a_{l,\lambda''}^\dagger \big|n_{1,+},n_{1,-}, \dots, n_{l,\lambda}, \dots \big\rangle
\bigg],
\ea
where $k_i^\mu = (\omega_i, {\bf k}_i)$. Next, using the results analogous to (\ref{1-contr})-(\ref{4-contr}), we find
\ba
\label{bigger-GF-EF-3}
D_{ij}^{>}(x,y) 
& = & 
Z^{-1} \Delta_{\bf k} \sum_{n_{1,+}} \sum_{n_{1,-}} \sum_{n_{2,+}} \sum_{n_{2,-}} \ldots 
e^{-\beta \omega_1 n_{1,+}}  e^{-\beta \omega_1 n_{1,-}} 
e^{-\beta \omega_2 n_{2,+}} e^{-\beta \omega_2 n_{2,-}} \dots \qquad 
\\ [2mm] \nn
& & 
\times \sum_{\lambda'} \sum_l \frac{1}{2 \omega_l } \epsilon_{l,\lambda'}^i \cdot \epsilon_{l,\lambda'}^j
\Big[(n_{l,\lambda'} +1) e^{-i k_l (x-y)} + n_{l,\lambda'} e^{i k_l (x-y)} \Big]. 
\ea
The formula (\ref{bigger-GF-EF-3}) is equivalent to
\ba
\label{bigger-GF-EF-4}
D_{ij}^{>}(x,y) 
& = & 
Z^{-1} \Delta_{\bf k} \sum_{n_{1,+}} \sum_{n_{1,-}} \sum_{n_{2,+}} \sum_{n_{2,-}} \ldots 
e^{-\beta \omega_1 n_{1,+}}  e^{-\beta \omega_1 n_{1,-}} 
e^{-\beta \omega_2 n_{2,+}} e^{-\beta \omega_2 n_{2,-}} \dots \qquad 
\\ [2mm] \nn
& & 
\times \sum_{\lambda'}\sum_l \frac{1}{2 \omega_l }\epsilon_{l,\lambda'}^i \cdot \epsilon_{l,\lambda'}^j
\Big[n_{l,\lambda'} (e^{-i k_l (x-y)} + e^{i k_l (x-y)}) + e^{-i k_l (x-y)} \Big],
\ea
which is converted to
\ba
\label{bigger-GF-EF-5}
iD_{ij}^{>}(x,y) = \Delta_{\bf k} \sum_{\lambda'} \sum_l \frac{1}{\omega_l}
\epsilon_{l,\lambda'}^i \cdot \epsilon_{l,\lambda'}^j
\bigg[\frac{e^{-\beta \omega_l}}{1-e^{-\beta \omega_l}} 
\big(e^{-ik_l(x-y)}+e^{ik_l(x-y)}\big) +e^{-ik_l(x-y)} \bigg].
\ea

Changing a discrete momentum space into a continuous one, we can write
\ba
\label{bigger-GF-EF-6}
iD_{ij}^{>}(x,y) 
& = & 
\sum_{\lambda'}
\int \frac{d^3 k}{(2\pi)^3} \frac{1}{\omega_{\bf k}} 
\epsilon^i ({\bf k},\lambda') \cdot \epsilon^j ({\bf k},\lambda') 
\\ [2mm] \nn
&& \qquad\qquad\qquad
\times\bigg[\frac{e^{-\beta \omega_{\bf k}}}{1-e^{-\beta \omega_{\bf k}}} 
\big(e^{-ik(x-y)}+e^{ik(x-y)}\big) + e^{-ik(x-y)} \bigg] \nn
\ea
and noting that
\ba
\label{distr-fun-EF}
f(\omega_{\bf k})=\frac{1}{e^{\beta \omega_{\bf k}}-1} 
= \frac{e^{-\beta \omega_{\bf k}}}{1-e^{-\beta \omega_{\bf k}}} 
\ea
we simply get
\ba
\label{bigger-GF-EF-7}
iD_{ij}^{>}(x,y) = \int \frac{d^3 k}{(2\pi)^3} \frac{1}{\omega_{\bf k}} 
\bigg( \delta^{ij} -\frac{k^ik^j}{{\bf k}^2} \bigg)
\bigg[f(\omega_{\bf k}) \big(e^{-ik(x-y)}+e^{ik(x-y)}\big)+e^{-ik(x-y)} \bigg],
\ea
where the following identity was applied
\ba
\label{pol-vec-sum}
\sum_{\lambda'}\epsilon^i({\bf k},\lambda') \cdot \epsilon^j({\bf k},\lambda')
= \delta^{ij} -\frac{k^ik^j}{{\bf k}^2}.
\ea
The formula (\ref{bigger-GF-EF-7}) can be rewritten as
\ba
\label{bigger-GF-EF-8}
iD_{ij}^{>}(x,y) = \int \frac{d^3 k}{(2\pi)^3} \frac{1}{\omega_{\bf k}} 
\bigg( \delta^{ij} -\frac{k^ik^j}{{\bf k}^2} \bigg)
\bigg[(f(\omega_{\bf k})+1) e^{-ik(x-y)}+ f(\omega_{\bf k}) e^{ik(x-y)}\bigg],
\ea
which gives
\ba
\label{bigger-GF-EF-FT-1}
D_{ij}^{>}(p)  =  -\frac{2i\pi}{\omega_{\bf p}} 
\bigg( \delta^{ij} -\frac{p^ip^j}{{\bf p}^2} \bigg)
\bigg[ \delta(\omega_{\bf p} - p_0) (f(\omega_{\bf p})+1)
+  \delta(\omega_{\bf p} + p_0) f(\omega_{\bf p})\bigg].
\ea
Analogously, we get
\ba
\label{smaller-GF-EF-FT}
D_{ij}^{<}(p)  =  -\frac{2i\pi}{\omega_{\bf p}} 
\bigg( \delta^{ij} -\frac{p^ip^j}{{\bf p}^2} \bigg)
\bigg[ \delta(\omega_{\bf p} - p_0) f(\omega_{\bf p})
+  \delta(\omega_{\bf p} + p_0)( f(\omega_{\bf p})+1) \bigg].
\ea
We see that both $D_{ij}^>(p)$ and $D_{ij}^<(p)$ are non zero only for $p^2=0$.

Let us now compute the Green functions $D_{ij}^c(x,y)$ and $D_{ij}^a(x,y)$ which are defined by
\ba
\label{causal-GF-EF}
D_{ij}^c(x,y)&=&\Theta (x_0-y_0)D_{ij}^> (x,y)+\Theta (y_0-x_0)D_{ij}^< (x,y),
\\ [2mm]
\label{anticausal-GF-EF}
D_{ij}^a(x,y)&=&\Theta (x_0-y_0)D_{ij}^< (x,y)+\Theta (y_0-x_0)D_{ij}^> (x,y).
\ea
These functions are found in the phase space as
\ba
\label{causal-GF-EF-FT-3-e}
D_{ij}^{c} (p)= \bigg( \delta^{ij} -\frac{p^ip^j}{{\bf p}^2} \bigg)
\bigg[\frac{1}{p^2+i0^+} -
\frac{i\pi}{\omega_{\bf p}} f(\omega_{\bf p}) \big(\delta(p_0 - \omega_{\bf p})+\delta(p_0 + \omega_{\bf p}) \big) \bigg] \qquad
\ea
and that of $D_{ij}^a(p)$ as
\ba
\label{anticausal-GF-EF-FT-3-e}
D_{ij}^{a} (p)=-\bigg( \delta^{ij} -\frac{p^ip^j}{{\bf p}^2} \bigg)
\bigg[\frac{1}{p^2 - i0^+} - 
\frac{i\pi}{\omega_{\bf p}} f(\omega_{\bf p}) \big(\delta(p_0 - \omega_{\bf p})+\delta(p_0 + \omega_{\bf p}) \big)\bigg].\qquad
\ea
One recognizes in the first term of Eq.~(\ref{causal-GF-EF-FT-3-e}), the usual Feynman propagator in the Coulomb gauge. The second term represents the effect of a medium. The first term is non-zero for any $p$ while the second one only for $p^2=0$.

The analysis of retarded and advanced Green functions follows that of the scalar field. So, here we only present the final formulas of $D_{ij}^+(p)$ and $D_{ij}^-(p)$, that are
\ba
\label{retarded-GF-EF-FT}
D_{ij}^{+} (p)&=&\frac{\delta^{ij} -\frac{p^ip^j}{{\bf p}^2} }{p^2+i {\rm sgn}(p_0)0^+},
\\ [2mm] 
\label{advanced-GF-EF-FT}
D_{ij}^{-} (p)&=& \frac{\delta^{ij} -\frac{p^ip^j}{{\bf p}^2} }{p^2 -i {\rm sgn}(p_0)0^+}.
\ea
As seen, the retarded and advanced propagators coincide with the vacuum ones.

\newpage
\section{Kubo-Martin-Schwinger condition}
\label{appendix-KMS-condition}
\thispagestyle{plain}
\vspace{-1cm}

Here we derive the Kubo-Martin-Schwinger (KMS) boundary condition for the scalar field Green functions
$\Delta^<(x,y)$ and $\Delta^>(x,y)$ of equilibrium system.

The Green functions $\Delta^>(x,y)$ and $\Delta^<(x,y)$ are defined as
\ba
\label{d1}
i\Delta^>(x,y) &\equiv& Z^{-1} \textrm{Tr}\Big[ \hat \rho \phi(x)  \phi(y)  \Big], 
\\ [2mm]
\label{d2}
i\Delta^<(x,y) &\equiv& Z^{-1} \textrm{Tr}\Big[ \hat \rho \phi(y)  \phi(x)  \Big],
\ea
where $ Z \equiv \textrm{Tr}[ \hat \rho(t_0)]= \textrm{Tr}[e^{-\beta \hat H}]$. Let us write the functions as follows
\ba
\label{d11}
i\Delta^>(x,y) &\equiv& \frac{1}{Z} \textrm{Tr}\Big[ \hat \rho \phi(t_1,{\bf x})  \phi(t_2,{\bf y})  \Big], 
\\ [2mm]
\label{d22}
i\Delta^<(x,y) &\equiv& \frac{1}{Z} \textrm{Tr}\Big[ \hat \rho \phi(t_2,{\bf y})  \phi(t_1,{\bf x})  \Big],
\ea

The field operators are given in the Heisenberg picture as
\ba
\label{heis-pic}
\phi(x)=\phi(t_1,{\bf x}) = e^{i \hat H t} \phi({\bf x}) e^{-i \hat H t}.
\ea
Then the trace in Eq. (\ref{d22}) can be written as
\ba
\textrm{Tr}\Big[ \hat \rho \phi(t_2,{\bf y})  \phi(t_1,{\bf x})  \Big] = 
\textrm{Tr}\Big[  \phi(t_1,{\bf x}) \hat \rho \phi(t_2,{\bf y})  \Big]= 
\textrm{Tr}\Big[  e^{-\beta \hat H} e^{\beta \hat H} \phi(t_1,{\bf x})  e^{-\beta \hat H} \phi(t_2,{\bf y})  \Big].
\ea
Using the relation (\ref{heis-pic}), one finds that
\ba
 e^{\beta \hat H} \phi(t_1,{\bf x}) e^{-\beta \hat H} = \phi(t_1 -i\beta,{\bf x}),
\ea
and thus we have the following relation
\ba
\label{KMS-condition}
\Delta^<(x,y)=\Delta^>(t_1-i\beta,{\bf x},t_2,{\bf y}),
\ea
which is the known Kubo-Martin-Schwinger condition.

The equilibrium Green functions $\Delta^>(x,y)$ and $\Delta^<(x,y)$ depend on their arguments only through their
difference, so that
\ba
\label{KMS-1}
\Delta^<(x,y) & = & \Delta^<(x-y) = \Delta^<(t_1-t_2,{\bf x}-{\bf y}) = \Delta^<(t,{\bf x}),
\ea
where we have put $t_2=0,{\bf y}=0$ and then $t_1 \rightarrow t$. Subsequently, we have
\ba
\Delta^>(t_1-i\beta,{\bf x},t_2,{\bf y}) & = & \Delta^>(t_1-t_2-i\beta,{\bf x}-{\bf y}) = \Delta^>(t-i\beta,{\bf x})
\ea
and thus the formula (\ref{KMS-condition}) can be rewritten as
\ba
\label{KMS-condition-1}
\Delta^<(t, {\bf x})=\Delta^>(t-i\beta,{\bf x}).
\ea

\begin{figure}[th!]
\centering
\includegraphics[scale=0.4]{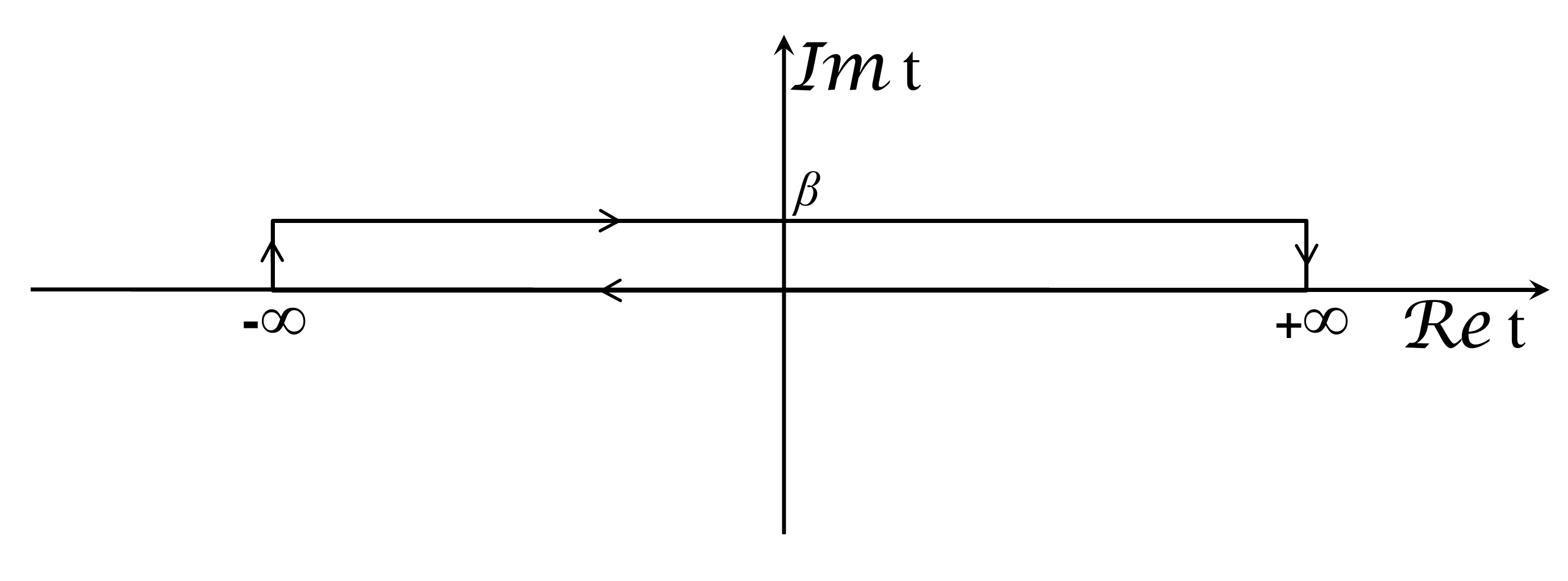}
\caption{The contour.}
\label{contour}
\end{figure}

Performing the Fourier transformation of Eq. (\ref{KMS-condition-1}), we get
\ba
\label{KMS-condition-FT}
\Delta^<(p)=\int_{-\infty}^\infty d^4x \; e^{i px} \Delta^<(x) = \int_{-\infty}^\infty d^4 x \; e^{ipx}
\Delta^>(t-i\beta,{\bf x}).
\ea
Changing the variable $t-i\beta \rightarrow t'$, we have
\ba
\label{KMS-condition-FT-1}
\Delta^<(p) 
&=&
\int_{-\infty+i\beta}^{\infty+i\beta} dt' \int d^3x \;
e^{i\omega (t' +i\beta)} e^{-i{\bf p}\cdot{\bf x}} \Delta^>(t',{\bf x}) 
\\ [2mm] \nn
&=& 
e^{-\beta \omega} \int_{-\infty+i\beta}^{\infty+i\beta} dt' \int d^3x \;
e^{i\omega t'} e^{-i{\bf p}\cdot{\bf x}} \Delta^>(t',{\bf x}).
\ea
If the function $e^{i\omega t'} e^{-i{\bf p}\cdot{\bf x}} \Delta^>(t',{\bf x})$ is analytical for $\beta \geq \Im \; t \geq 0$, that is, inside the contour shown in Fig.\ref{contour}, then due to the Cauchy theorem we conclude that
\ba
\label{KMS-condition-FT-help}
\int_{-\infty+i\beta}^{\infty+i\beta} dt \int d^3x \; e^{i\omega t'} e^{-i{\bf p}\cdot{\bf x}} 
\Delta^>(t',{\bf x}) = 
\int_{-\infty}^{\infty} dt \int d^3x \; e^{i\omega t'} e^{-i{\bf p}\cdot{\bf x}} \Delta^>(t',{\bf x}),
\ea
as the integrals over the vertical parts of the contour vanish. Thus the KMS condition is of the form
\ba
\label{KMS-condition-FT-2}
\Delta^<(p)= e^{-\beta \omega} \Delta^>(p).
\ea

\newpage
\section{Calculus of Grassmann variables}
\label{appendix-Grassmann-algebra}
\thispagestyle{plain}
\vspace{-1cm}

\subsection{Grassmann algebra}
An algebra whose generators $\theta_1, \theta_2, \ldots, \theta_n$ anticommute with each other that is 
\be
\{\theta_i, \theta_j  \} \equiv \theta_i \theta_j +\theta_j \theta_i = 0,
\ee
is called the Grassmann algebra with $n$ generators and it is denoted as ${\cal G}_n$. The squares and higher powers of the generators vanish because $\theta_i^2= \theta_i \theta_i = - \theta_i \theta_i=0$. The Grassmann algebra ${\cal G}_n$ is the $2^n-$dimensional linear space (over a field of numbers) which  is generated by the following monomials
\ba
&1,& 
\\ [2mm] \nn
&\quad \qquad \qquad \theta_1, \theta_2, \ldots, \theta_n,&
\\ [2mm] \nn
&\quad \qquad \qquad \qquad \qquad \theta_1 \theta_2, \theta_2 \theta_3, \ldots, \theta_{n-1}\theta_n, &
\\ [2mm] \nn
&\vdots &
\\ [2mm] \nn
&\qquad \qquad\theta_1\theta_2 \ldots \theta_n .&
\ea
Elements of  ${\cal G}_n$ obviously commute with numbers belonging to the field.

Every element of the Grassmann algebra ${\cal G}_n$ can be expressed as a linear combination of the monomials
\ba
\label{decomp}
f(\theta)=f_0+\sum_k f_1(k)\theta_k + \sum_{k_1,k_2} f_2(k_1,k_2) \theta_{k_1}\theta_{k_2} + \ldots
+\sum_{k_i}f_n(k_1, \ldots, k_n) \theta_{k_1} \ldots \theta_{k_n} ,
\ea
where the coefficients $f_i$ are antisymmetric with respect to interchange of any two arguments. If a coefficient $f_i$ is not fully antisymmetric, the symmetric part does not contribute to $f(\theta)$, as squares of all generators vanish.

\subsection{Derivatives}

A left-side and right-side derivatives of a monomial are defined as 
\ba
\frac{\partial}{\partial \theta_p}\theta_1 \ldots \theta_s 
&\equiv& 
\delta_{1p}\theta_2 \ldots \theta_s - \delta_{2p} \theta_1\ldots\theta_s + \ldots
+ (-1)^{s-1} \delta_{sp}\theta_1 \ldots \theta_{s-1},
\\ [2mm]
\theta_1 \ldots \theta_s \frac{\partial}{\partial \theta_p} 
&\equiv& 
\delta_{sp}\theta_1 \ldots \theta_{s-1} - \delta_{(s-1)p} \theta_1 \ldots \theta_{s-2}\theta_s  + \ldots
+ (-1)^{s-1} \delta_{1p}\theta_2 \ldots \theta_{s}.
\ea

\subsection{Integrals}

To define the integral over Grassmann variables, one first introduces the symbol of differentials $d\theta_i$ which satisfy the anticommutation relations
\ba
\{d\theta_i, d\theta_j \}=\{ \theta_i, d\theta_j \}=0.
\ea
The integrals, also called the Berezin integrals, are defined by the equalities
\ba
\label{Berezin-int}
\int d\theta_j = 0, \;\;\;\;\;\;\;\;\; \int \theta_i d\theta_j = \delta_{ij}.
\ea
Variables are changed in the integrals on the Grassmann algebra in a specific way. In case of linear change $\{\chi_1, \chi_2, \dots \chi_n\} \rightarrow \{\theta_1, \theta_2, \dots \theta_n\}$ we write
\ba
\theta_m=\sum_l a_{ml}\chi_l,
\ea
where $a_{ml}$ is a non-singular matrix, and the differentials  transform as
\ba
d\theta_n=\sum_k b_{nk} d\chi_k.
\ea
Let us compute the matrix $b_{nk}$. The definition of integrals (\ref{Berezin-int}) implies 
\ba
\delta_{mn} = \int \theta_m d\theta_n = \int \sum_{kl} a_{ml} b_{nk} \chi_l d\chi_k 
= \sum_{kl} a_{ml} b_{nk} \delta_{lk} 
= \int \sum_{k} a_{mk} b_{nk} .
\ea
Therefore, $b_{nk} = (a^{-1})_{kn}$ and consequently $ d\theta_n=\sum_k (a^{-1})_{kn} d\chi_k$.
One further finds that
\ba
\int f(\theta) d\theta_n \ldots d\theta_1 = \textrm{det}(a^{-1})\int f \big(\theta(\chi)\big) d\chi_n \ldots d\chi_1.
\ea

\subsection{Dirac delta function}

The function 
\ba
\label{Dirac-delta-def}
\delta^{(n)}(\theta_1, \theta_2, \dots \theta_n) \equiv \theta_1 \theta_2 \dots \theta_n
\ea
plays a role of the Dirac delta-like function in the Grassmann algebra as
\ba
\int f(\theta) \,  \delta^{(n)}(\theta_1, \theta_2, \dots \theta_n) \, d\theta_n \dots d\theta_1 = f_0 ,
\ea
where $f_0$ is defined through the decomposition (\ref{decomp}).

It appears that the function $\delta^{(n)}(\phi_1, \phi_2, \dots \phi_n)$ can be expressed as the integral
\ba
\delta^{(n)}(\phi_1, \phi_2, \dots \phi_n) = \int 
\exp\Big[- \sum_{k=1}^n \pi_k\phi_k\Big] \; d\pi_n \,d\pi_{n-1} \dots d\pi_1.
\ea
Indeed, expanding the exponential and taking into account the $n-$th term, which is the only one contributing to the integral, one finds 
\ba
\int \exp\Big[- \sum_{k=1}^n \pi_k\phi_k\Big] \; d\pi_n \,d\pi_{n-1} \dots d\pi_1
&=&
\frac{(-1)^n}{n!}  \int \Big[\sum_{k=1}^n \pi_k\phi_k\Big]^n 
d\pi_n \,d\pi_{n-1} \dots d\pi_1
\\[2mm] \nn
&=&
(-1)^n \int  \pi_1\phi_1 \, \pi_2\phi_2 \dots \pi_n\phi_n \;
d\pi_n \,d\pi_{n-1} \dots d\pi_1
\\[2mm] \nn
&=& 
\phi_1 \phi_2 \dots \phi_n = \delta^{(n)}(\phi_1, \phi_2, \dots \phi_n) ,
\ea
where the factor $(-1)^n$ was compensated by the same factor resulting from the interchanges of $\pi_i$ and $\phi_j$.

\newpage
\section{Feynman rules}
\label{appendix-FR}
\thispagestyle{plain}
\vspace{-1cm}

We collect here the Feynman rules which are used in our calculations. We follow the convention used in the textbook \cite{Bechler}.

In order to compute a cross section of a given scattering process, one has to calculate the scattering amplitude $\mathcal{M}$. The construction of the scattering amplitude starts with drawing all possible Feynman diagrams corresponding to a given process. The diagrams consist of external and internal lines. The former represent particles of initial and final states and the latter describe virtual particles wich are responsible for interaction. Herein, the particles on the left hand side of a diagram are initial particles and these on the righ hand side are the ones of the final state. Particles carrying a charge are denoted by lines with arrows. All lines correspond to a given factor and these correspondences are shown in Tables \ref{table-external} and \ref{table-internal}. The lines meet in vertices whose forms depend on a sort of interaction and they are given in Tables \ref{table-sqed-vertex} and \ref{table-sym-vertex}.

\begin{table}[!h]
\caption{\label{table-external}External particles and the respective factors.}
\centering
\begin{tabular}{l m{5cm} m{3cm}}
\hline \hline
External Particle & \hspace{1cm}Graphical line & Counterpart
\\
\hline \hline
\\ \vspace{2mm}
ingoing fermion & \centering \includegraphics[scale=0.4]{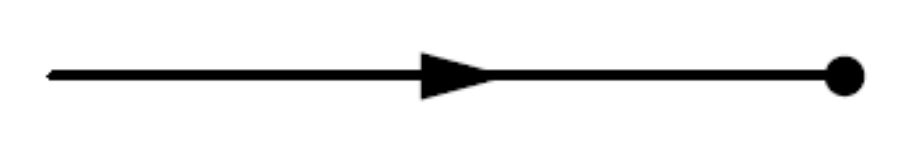} & $u({\bf p},s)$\\ \vspace{2mm}
outgoing fermion & \centering \includegraphics[scale=0.4]{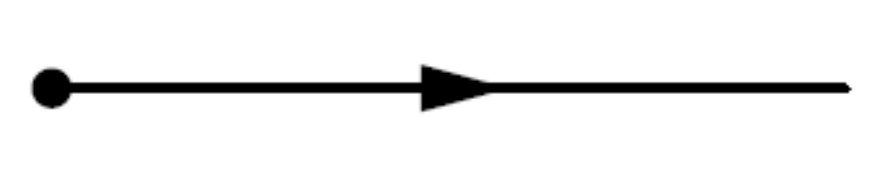} & $\bar{u}({\bf p},s)$\\ \vspace{2mm}
ingoing antifermion & \centering \includegraphics[scale=0.4]{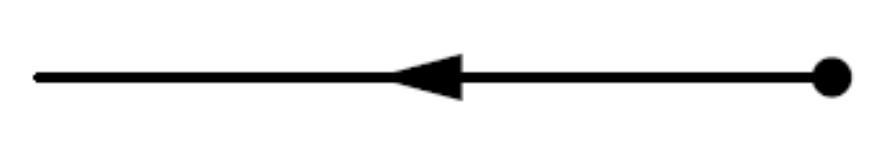} & $\bar{\upsilon}({\bf p}, s)$\\ \vspace{2mm}
outgoing antifermion & \centering \includegraphics[scale=0.4]{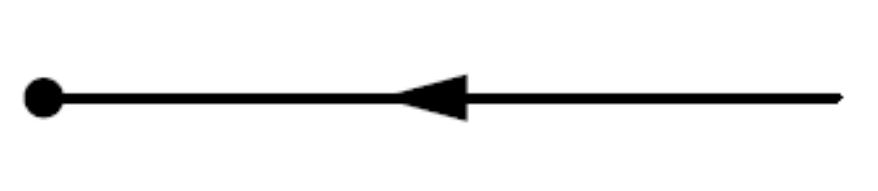} & $\upsilon({\bf p}, s)$\\ \vspace{2mm}
ingoing photino & \centering \includegraphics[scale=0.4]{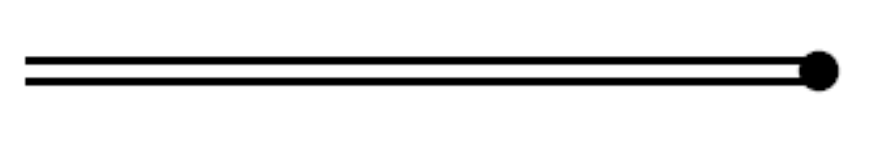} & $u({\bf p},s)$ \\ \vspace{2mm}
outgoing photino & \centering \includegraphics[scale=0.4]{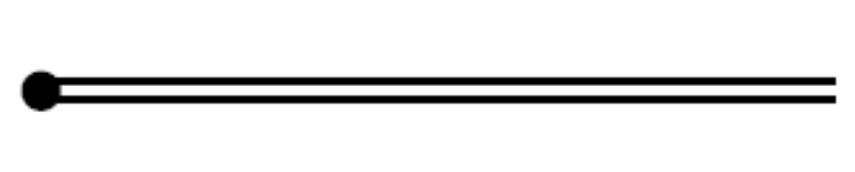} & $\bar{u}({\bf p},s)$\\ \vspace{2mm}
ingoing scalar & \centering \includegraphics[scale=0.4]{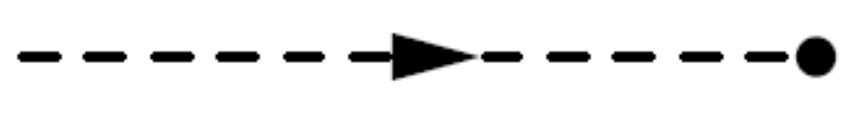} & 1\\ \vspace{2mm}
outgoing scalar & \centering \includegraphics[scale=0.4]{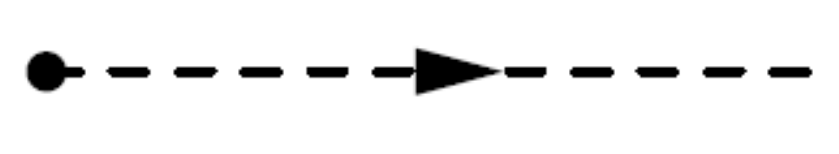} & 1\\ \vspace{2mm}
ingoing photon & \centering \includegraphics[scale=0.4]{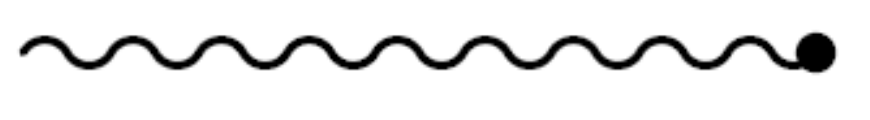} & $\varepsilon^\mu({\bf k}, \lambda)$\\ \vspace{2mm}
outgoing photon & \centering \includegraphics[scale=0.4]{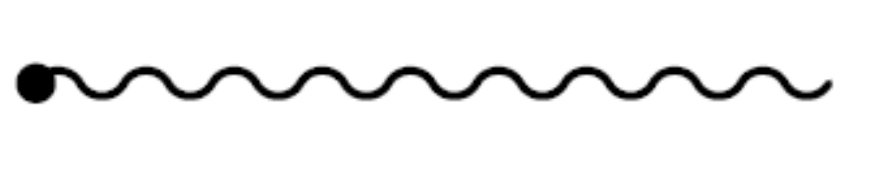} & $\varepsilon^{\mu}({\bf k}, \lambda)$\\ \vspace{2mm}
ingoing gluon & \centering \includegraphics[scale=0.4]{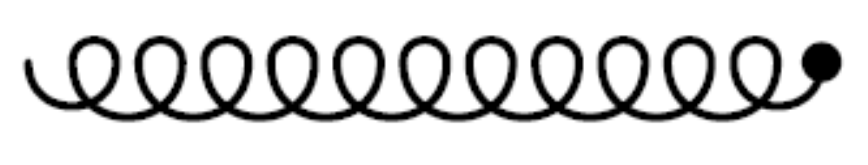} & $\varepsilon^\mu({\bf k}, \lambda)$\\ \vspace{2mm}
outgoing gluon & \centering \includegraphics[scale=0.4]{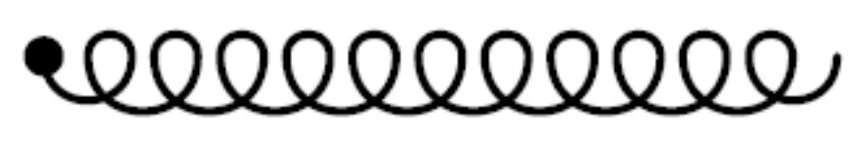} & $\varepsilon^{\mu}({\bf k}, \lambda)$\\
\hline \hline
\end{tabular}
\end{table}

\begin{table}[!h]
\caption{\label{table-internal} Internal particles and the respective factors.}
\centering
\begin{tabular}{m{3.5cm} c m{3.5cm} c m{6cm}}
\hline \hline
External Particle && \hspace{.5cm}Graphical line && \hspace{1cm}Counterpart
\\
\hline  \hline
\\ 
        \centering fermion &&
        \centering \includegraphics[scale=0.4]{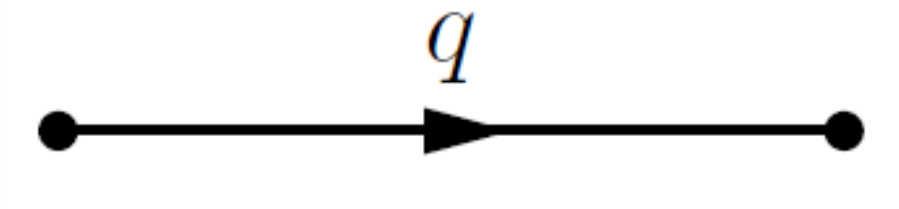}&&
        \be iS(q)=\frac{i(q\sla + m)}{q^2-m^2+i0^+} \ee
        \\ 
        \centering photino &&
        \centering \includegraphics[scale=.4]{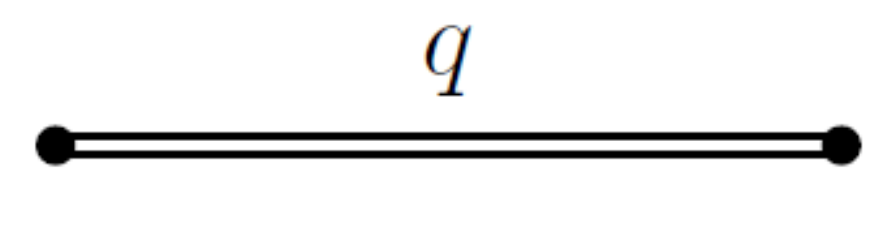}&&
        \be iS(q)=\frac{i(q\sla + m)}{q^2-m^2+i0^+} \ee
        \\ 
        \centering scalar &&
        \centering \includegraphics[scale=.4]{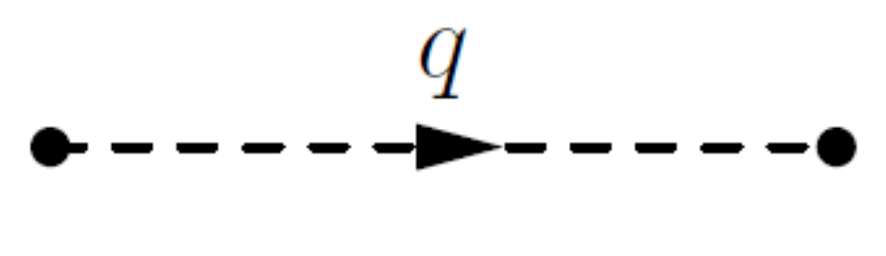}&&
        \be \label{scalar-prop} i\Delta(q)=\frac{i}{q^2-m^2+i0^+} \ee
        \\ 
        \centering photon &&
        \centering \includegraphics[scale=.4]{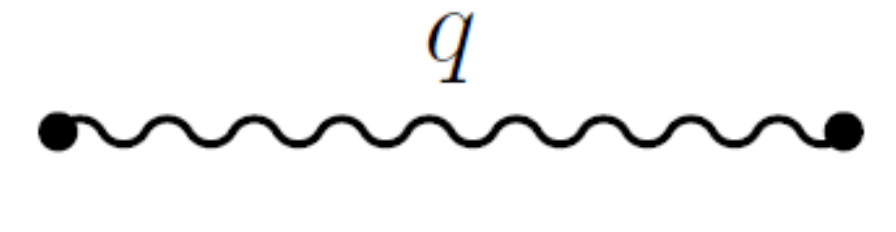}&&
        \be iD^{\mu\nu}(q)=\frac{-ig^{\mu\nu}}{q^2+i0^+} \ee
        \\ 
        \centering gluon &&
        \centering \includegraphics[scale=.4]{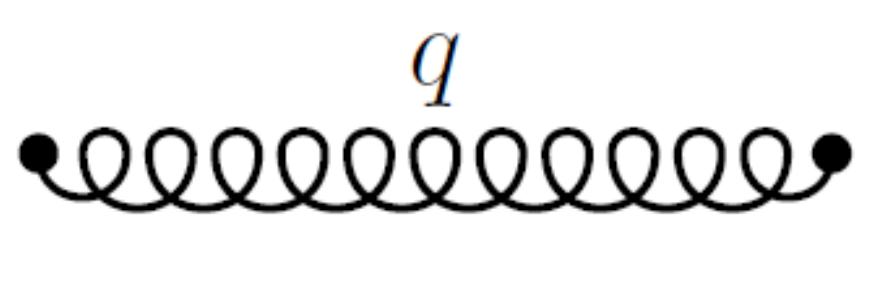}&&
        \be iD_{ab}^{\mu\nu}(q)=\frac{-i\delta^{ab} g^{\mu\nu} }{q^2+i0^+} \ee
        \\
\hline \hline
\end{tabular}
\end{table}

\vspace{1cm}

The factors $u({\bf p}, \, s)$ and $\upsilon({\bf p}, \, s)$, which correspond to the external fermion lines, {\it cf.} Tab. \ref{table-external}, are dependent on the momentum ${\bf p}$ and spin $s$ of a particle, are spinors which satisfy the free Dirac equation:
\be
\label{Dirac-spinors}
[p\sla -m]u({\bf p}, \,s)=0,    \qquad    [p\sla +m]\upsilon({\bf p}, \,s)=0,
\ee
where $p\sla \equiv p_\mu \gamma^\mu = E\gamma^0-\mathbf{p}\cdot \boldsymbol{\gamma}$.
$\bar{u}({\bf p}, \, s)$ and $\bar{\upsilon}({\bf p}, \, s)$ are the conjugate spinors to $u({\bf p}, \, s)$ and $\upsilon({\bf p}, \, s)$
\be
\label{spinors-conj}
\bar{u}\equiv u^\dag \gamma^0\,, \qquad   \bar{\upsilon}\equiv \upsilon^\dag\gamma^0\,.
\ee
The spinors fulfil the following normalisation conditions
\ba
\label{norm-cond}
\bar{u}({\bf p}, \, s) u({\bf p}, \, s') &=& - \bar{\upsilon}({\bf p}, \, s)\upsilon({\bf p}, \, s')= \delta^{ss'}, \\
\bar{u}({\bf p}, \, s) \upsilon({\bf p}, \, s') &=& \bar{\upsilon}({\bf p}, \, s) u({\bf p}, \, s')= 0
\ea
and the completeness relations
\be
\label{complet-rel-1}
\sum_{\pm s} u_\alpha({\bf p},s) \bar{u}_\beta ({\bf p},s) =
\bigg( \frac{p\sla +m}{2m} \bigg)_{\alpha\beta}
\ee
\be
\label{complet-rel-2}
\sum_{\pm s} \upsilon_\alpha({\bf p},s) \bar{\upsilon}_\beta({\bf p},s)=
\bigg( \frac{p\sla - m}{2m} \bigg)_{\alpha\beta}.
\ee
Subtracting the relation (\ref{complet-rel-2}) from (\ref{complet-rel-1}) we get the standard completeness relation
\be
\label{complet-rel-1}
\sum_{\pm s} \bigg( u_\alpha({\bf p},s) \bar{u}_\beta ({\bf p},s) 
- \upsilon_\alpha({\bf p},s) \bar{\upsilon}_\beta({\bf p},s) \bigg) =\delta_{\alpha\beta}
\ee

Photons and gluons are characterised by the real polarisation vector $\varepsilon^\mu({\bf k},\lambda)$ with the momentum ${\bf k}$ and the polarisation valued by $\lambda=1,2$. Due to the Lorentz condition $(\partial^\mu A_\mu=0)$, the polarisation vector is transverse
\be
\label{pol-vec}
\varepsilon_\mu({\bf k},\lambda) k^\mu=0 ,
\ee
where $k^\mu=(\omega_{\bf k}, {\bf k})$. In the arbitrary Lorentz frame the polarisation vector $\varepsilon^\mu({\bf k},\lambda)$ is space-like and normalised
\be
\label{cond-norm}
\varepsilon_\mu ({\bf k},\lambda) \varepsilon^\mu ({\bf k},\lambda)=-1 .
\ee
Summation over polarisations of gauge bosons is done using the formula
\be
\label{summation-bosons}
\sum_{\lambda=1}^2 \varepsilon^\mu({\bf k},\lambda) \varepsilon^\nu ({\bf k},\lambda) = -g^{\mu\nu}.
\ee

Let us add that if the lines represent fermions or scalars of nonAbelian theory, they carry also colour indices and respective propagators include the factor $\delta^{ab}$.

\newpage
\textcolor[rgb]{1.00,1.00,1.00}{.}

\begin{table}[!h]
\caption{\label{table-sqed-vertex} The vertex functions of the ${\cal N}=1$ SUSY QED.}
\centering
\begin{tabular}{cc m{4cm} m{1cm} m{7cm}}
\hline \hline
&& \hspace{1cm} Vertex &&  \hspace{2.5cm} Counterpart 
\\
\hline \hline
\\
        && \centering \includegraphics[scale=.4]{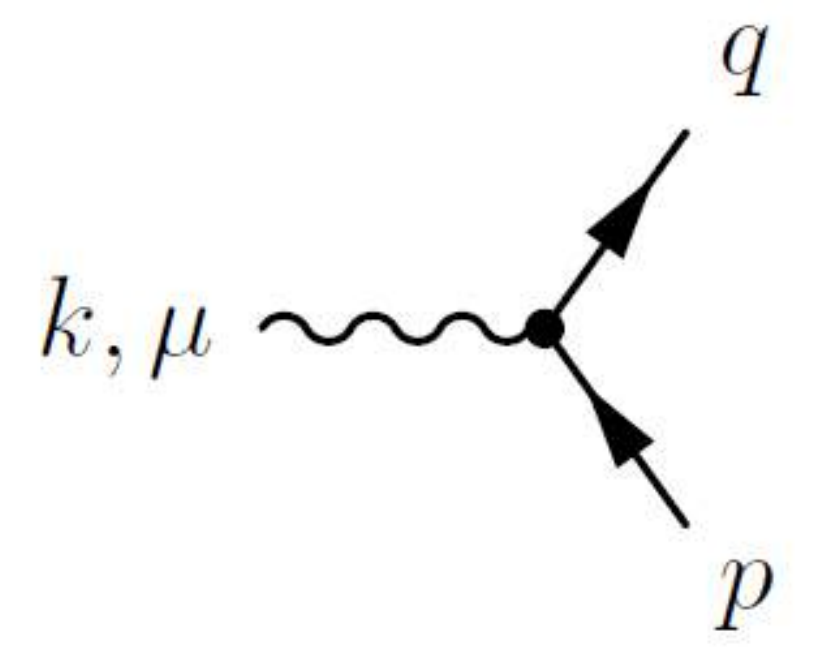}&&
        \ba =-ie\gamma_\mu \nn \ea 
	\\ \\
        && \centering \includegraphics[scale=.4]{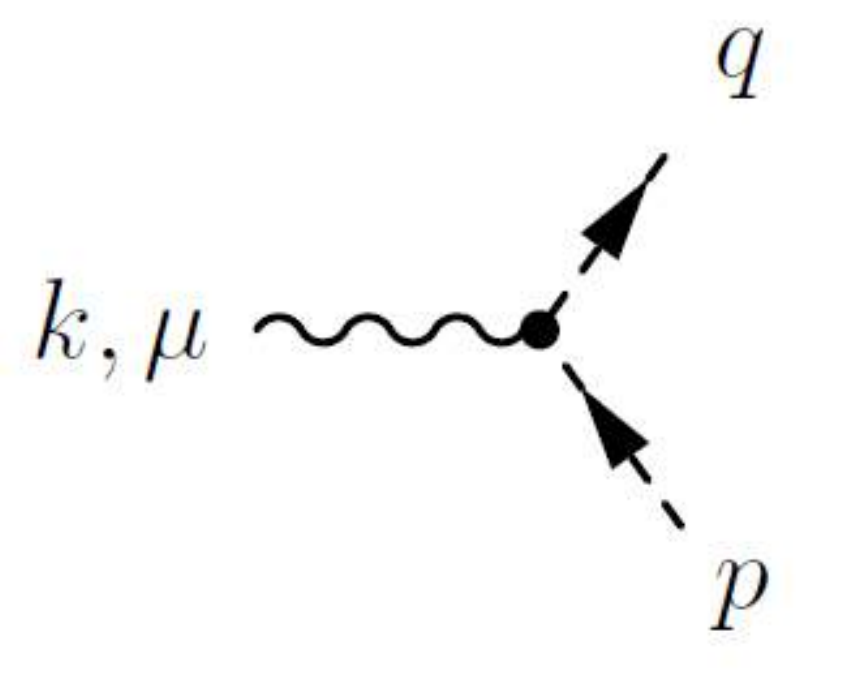}&&
        \ba =-ie(p+q)_\mu \nn \ea
        \\  \\
        && \centering \includegraphics[scale=.4]{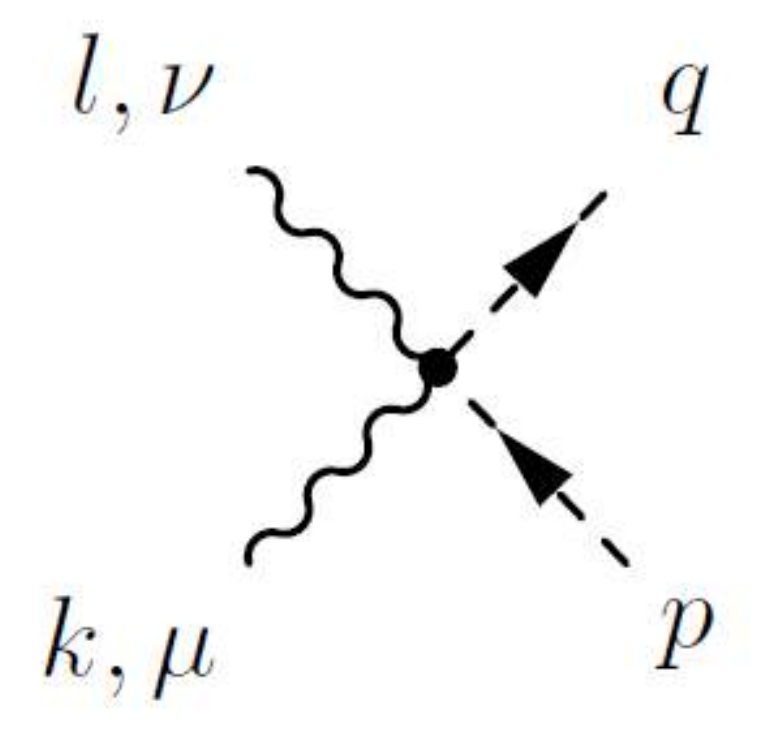}&&
        \ba  =2ie^2 g_{\mu\nu} \nn \ea
	 \\ \\
        && \centering \includegraphics[scale=.4]{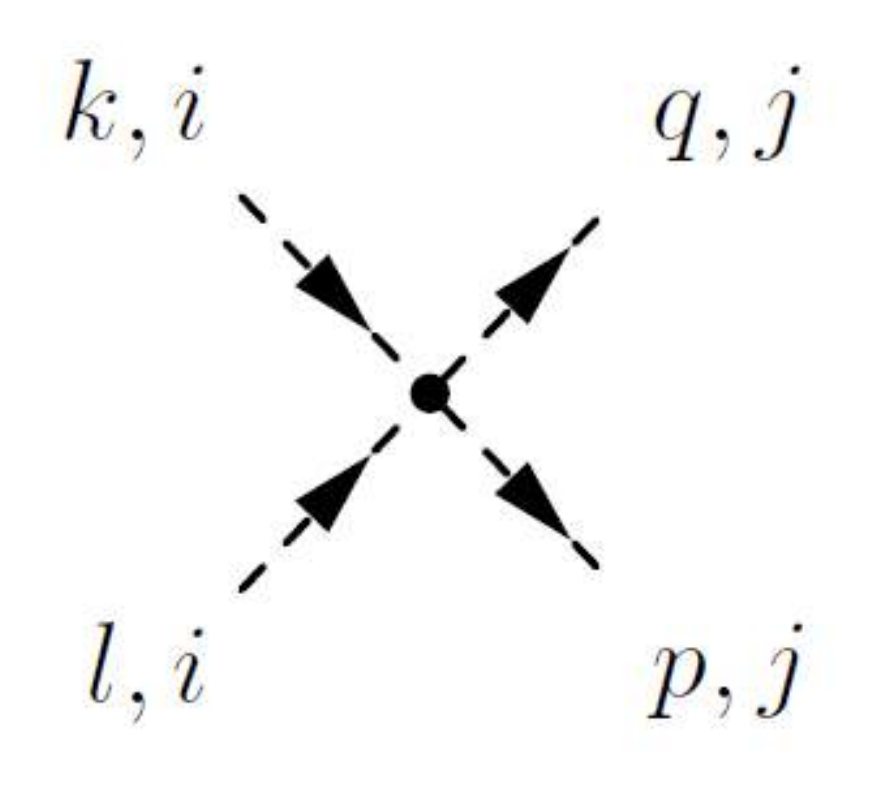}&&
         \begin{eqnarray*} = \left\{ \begin{array}{ll}
           2ie^2 & \textrm{dla $i=j$}\\
           -ie^2 & \textrm{dla $i \neq j$}    
           \end{array} \right. \end{eqnarray*}
          \\ \\
        && \centering \includegraphics[scale=.4]{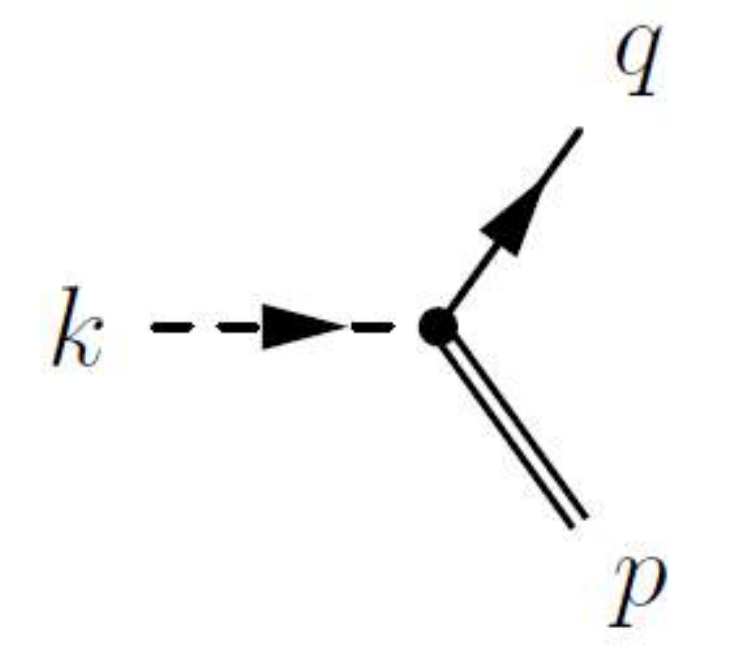}&&
         \begin{eqnarray*} =\left\{ \begin{array}{ll}
          -ie\sqrt{2}P_R & \textrm{dla $\tilde{e}=\varphi_1$,}\\
          +ie\sqrt{2}P_L & \textrm{dla $\tilde{e}=\varphi_2^\dag$.}
           \end{array} \right. \end{eqnarray*} 
	    \\ \\
        && \centering \includegraphics[scale=.4]{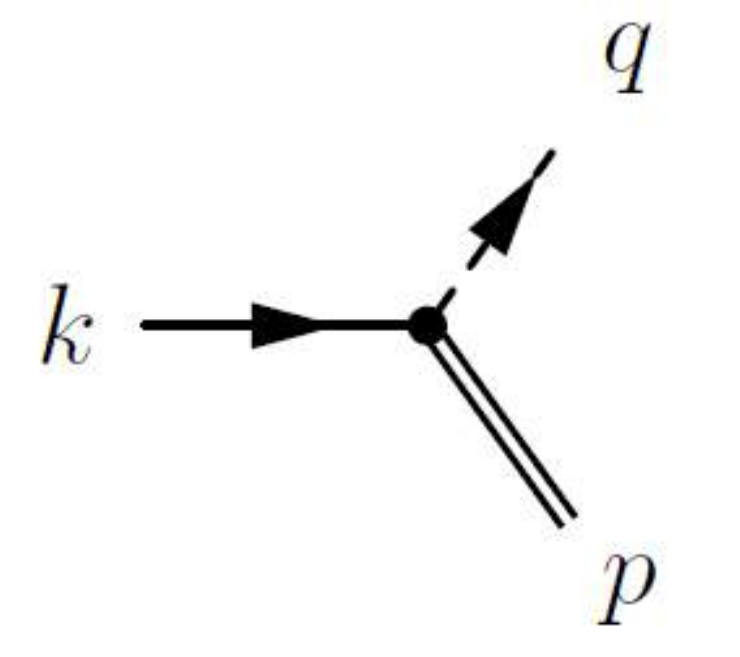}&&
        \begin{eqnarray*} =\left\{ \begin{array}{ll}
          -ie\sqrt{2}P_L & \textrm{dla $\tilde{e}=\varphi_1^\dag$,}\\
          +ie\sqrt{2}P_R & \textrm{dla $\tilde{e}=\varphi_2$.}\\
          \end{array} \right. \end{eqnarray*} \\
            \hline \hline
\end{tabular}
\end{table}

\newpage
\textcolor[rgb]{1.00,1.00,1.00}{.}

\begin{table}[!h]
\caption{\label{table-sym-vertex}The vertex functions of the ${\cal N}=4$ super Yang-Mills theory.}
\centering
\begin{tabular}{cc m{4cm} m{1cm} m{9cm}}
\hline \hline
&& \hspace{1cm} Vertex &&  \hspace{2.5cm} Counterpart  
\\
\hline \hline
\\ 
        & & \centering \includegraphics[scale=.4]{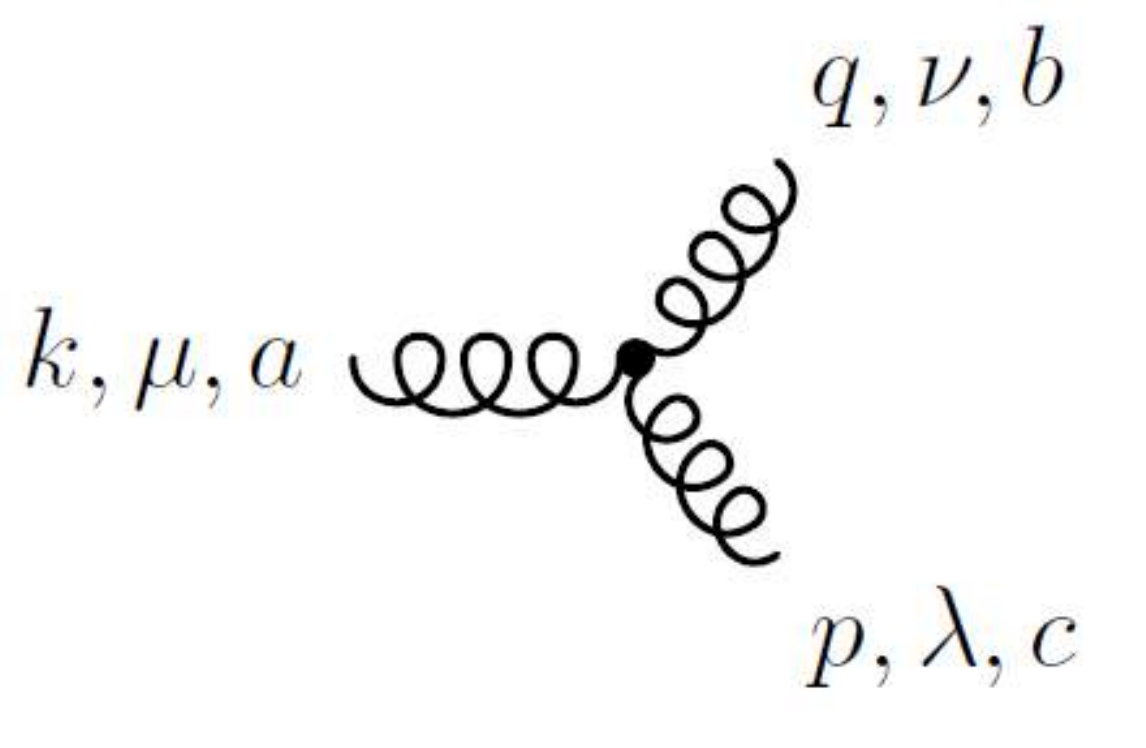}&&
        \ba &=& -gf^{abc} \big[g^{\mu\nu}(k-q)^\lambda + g^{\nu\lambda}(q-p)^\mu \nn \\
	 && \qquad\qquad+g^{\lambda\mu}(p-k)^\nu \big] \nn \ea
        \\ \\
        & & \centering \includegraphics[scale=.4]{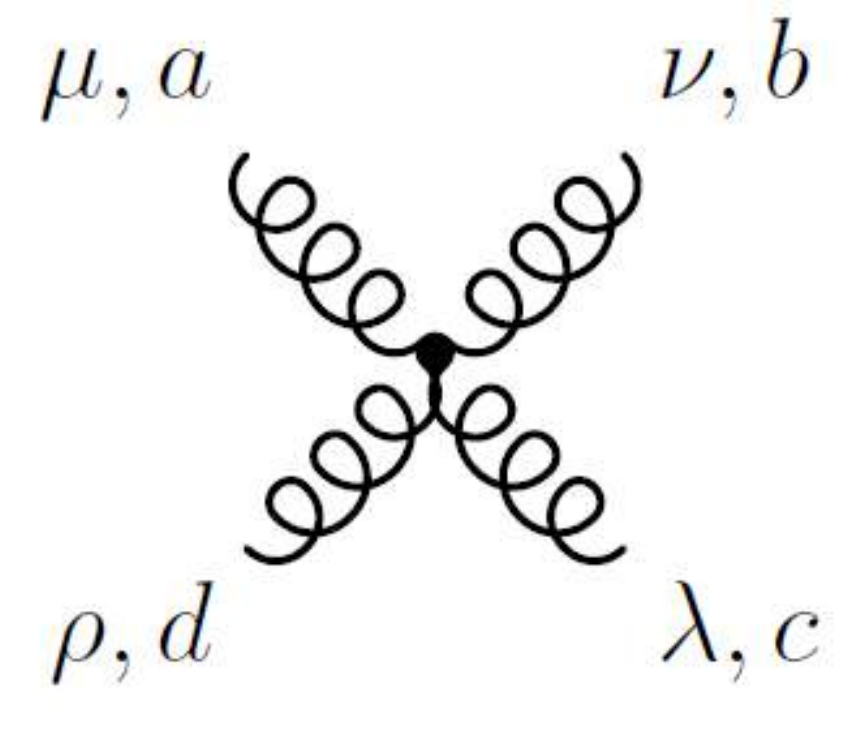} &&
         \ba &=-ig^2& \big[\; f^{abe}f^{cde}(g^{\mu\lambda}g^{\nu\rho}-g^{\mu\rho}g^{\nu\lambda}) \nn \\
         && +f^{ace}f^{bde}(g^{\mu\nu}g^{\lambda\rho}-g^{\mu\rho}g^{\nu\lambda}) \nn \\
         && +f^{ade}f^{cbe}(g^{\mu\lambda}g^{\nu\rho}-g^{\mu\nu}g^{\rho\lambda})\;\big] \nn \ea
         \\ \\
        & & \centering \includegraphics[scale=.4]{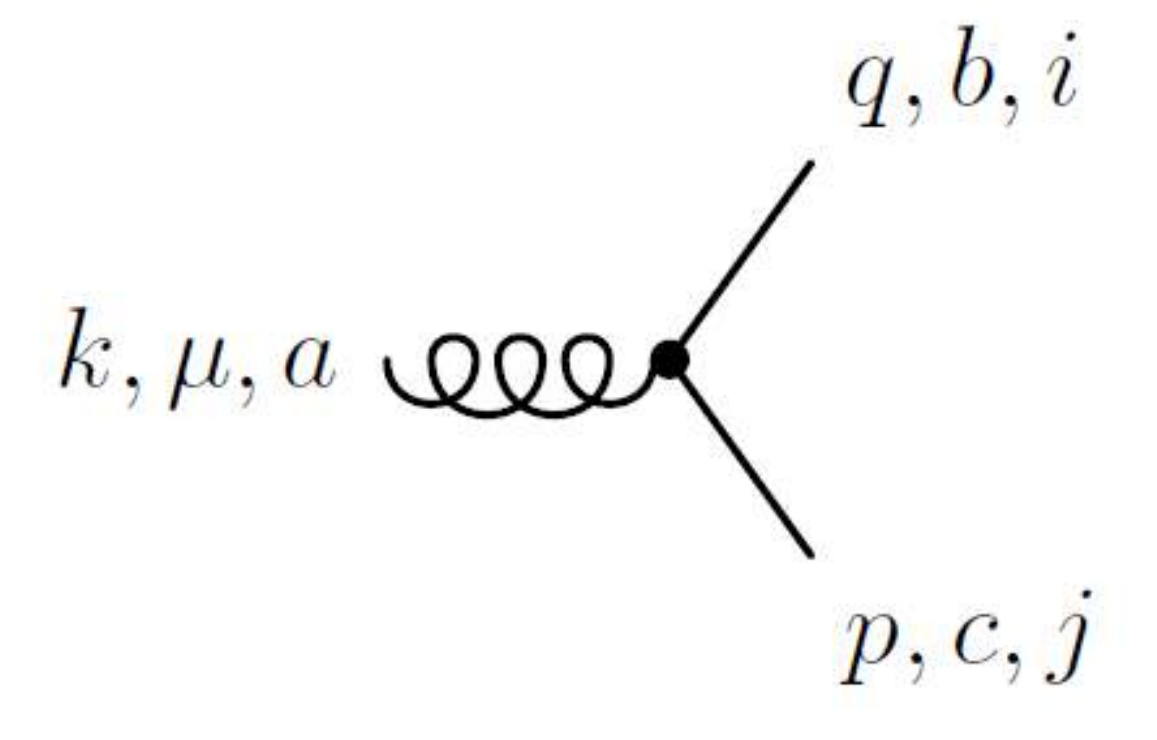} &&
        \ba &=& gf^{abc}\delta^{ij}\gamma^\mu \nn \ea 
        \\ \\
        & & \centering \includegraphics[scale=.4]{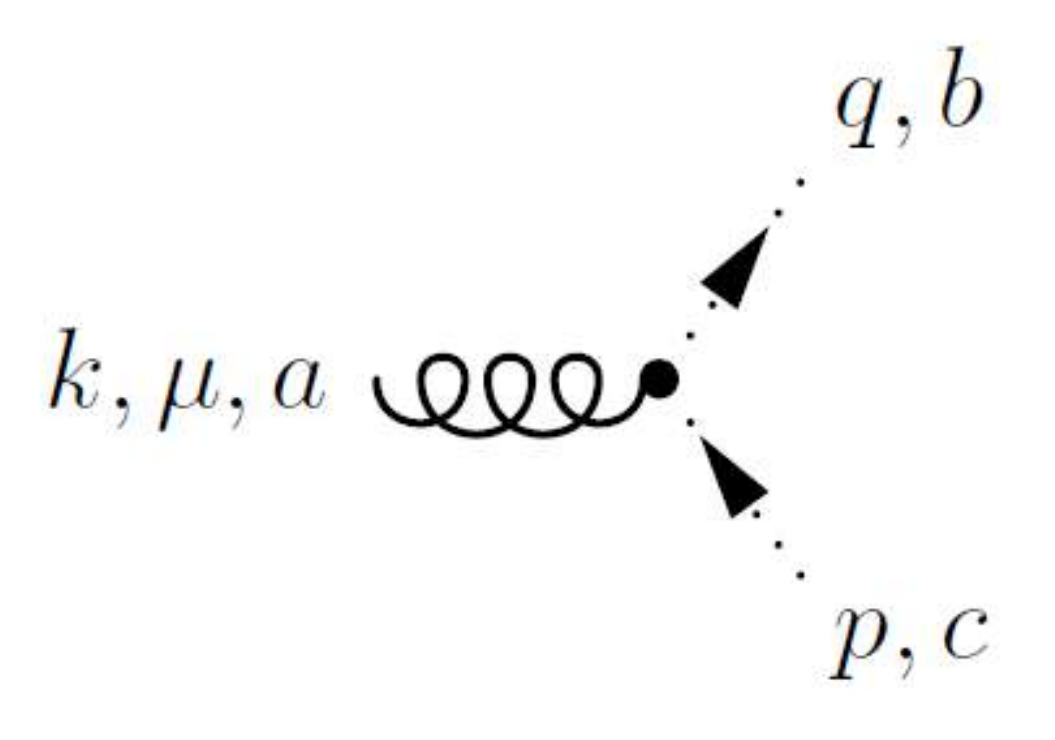} &&
        \ba &=& gf^{abc}q_\mu \nn \ea
         \\ \\
        & & \centering \includegraphics[scale=.4]{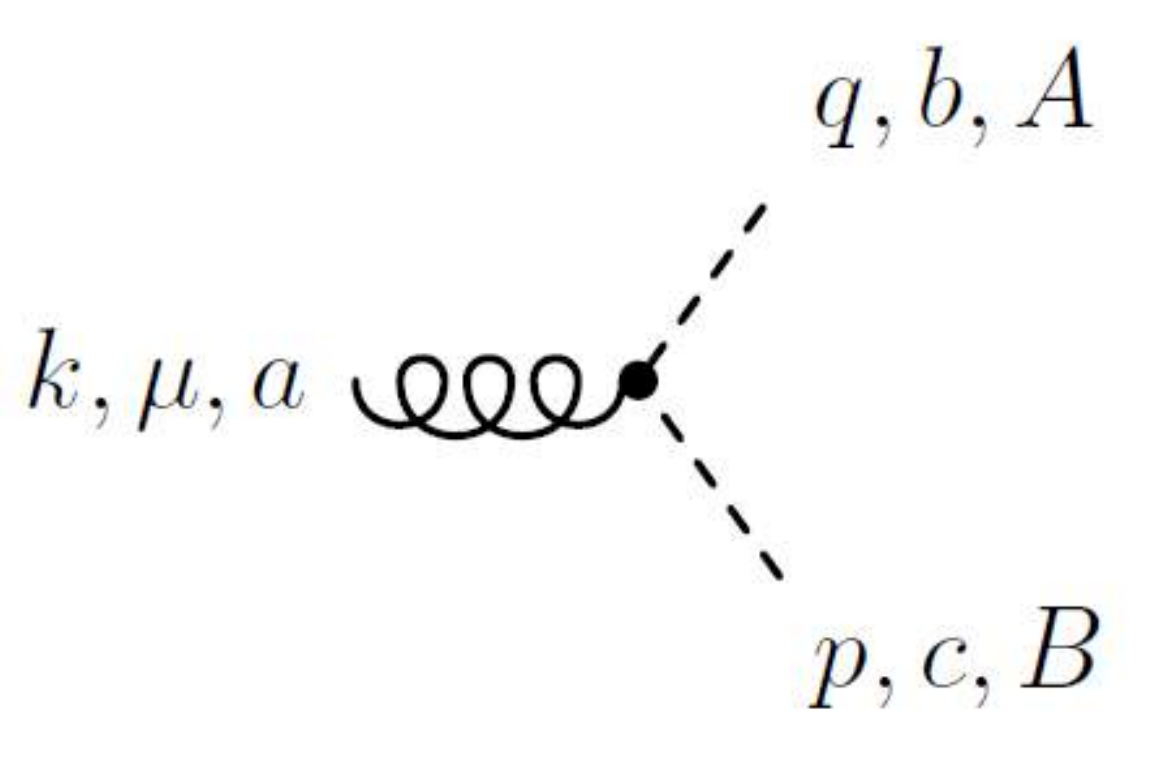} &&
        \ba &=& gf^{abc}\delta^{AB} (p+q)_\mu \nn \ea
        \\ 
\hline \hline
\end{tabular}
\end{table}

\newpage
\textcolor[rgb]{1.00,1.00,1.00}{.}

\begin{table}[!h]
\centering
\begin{tabular}{cc m{4cm} m{1cm} m{9cm}}
\hline \hline
&& \hspace{1cm} Vertex &&  \hspace{2.5cm} Counterpart  
\\
\hline \hline
\\ 
       & & \centering \includegraphics[scale=.4]{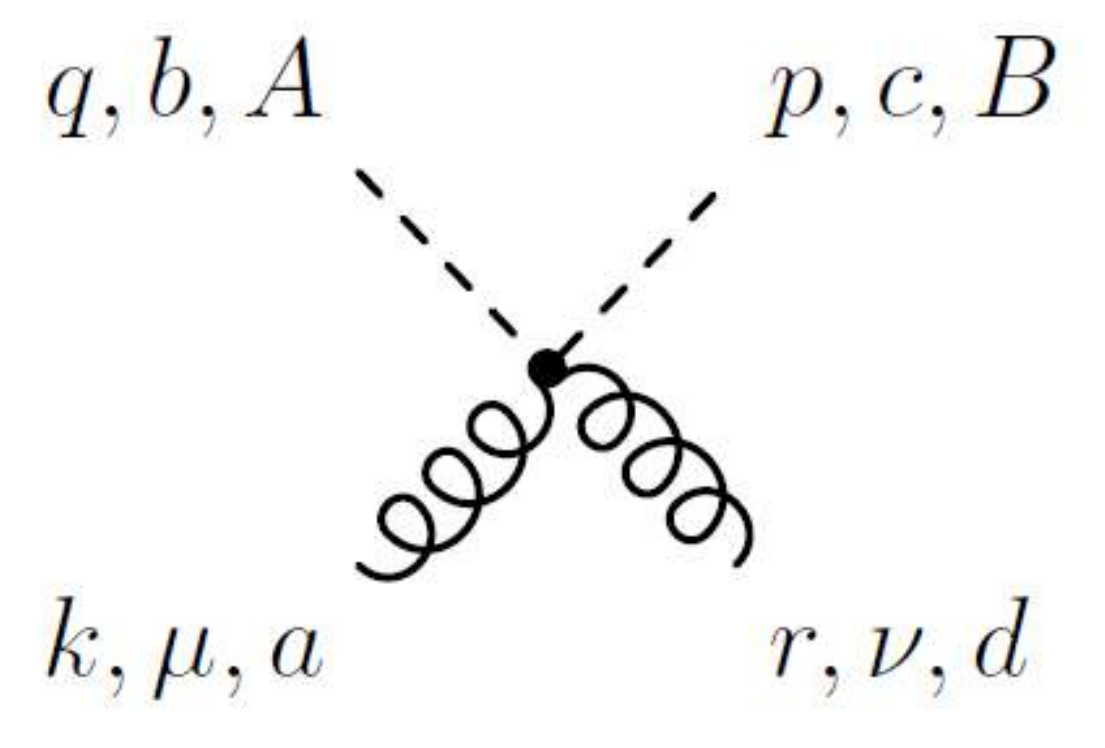} &&
        \ba &=& 2ig^2 g^{\mu\nu} f^{abe}f^{cde}\delta^{AB} \nn \ea
	 \\ \\
        & & \centering \includegraphics[scale=.4]{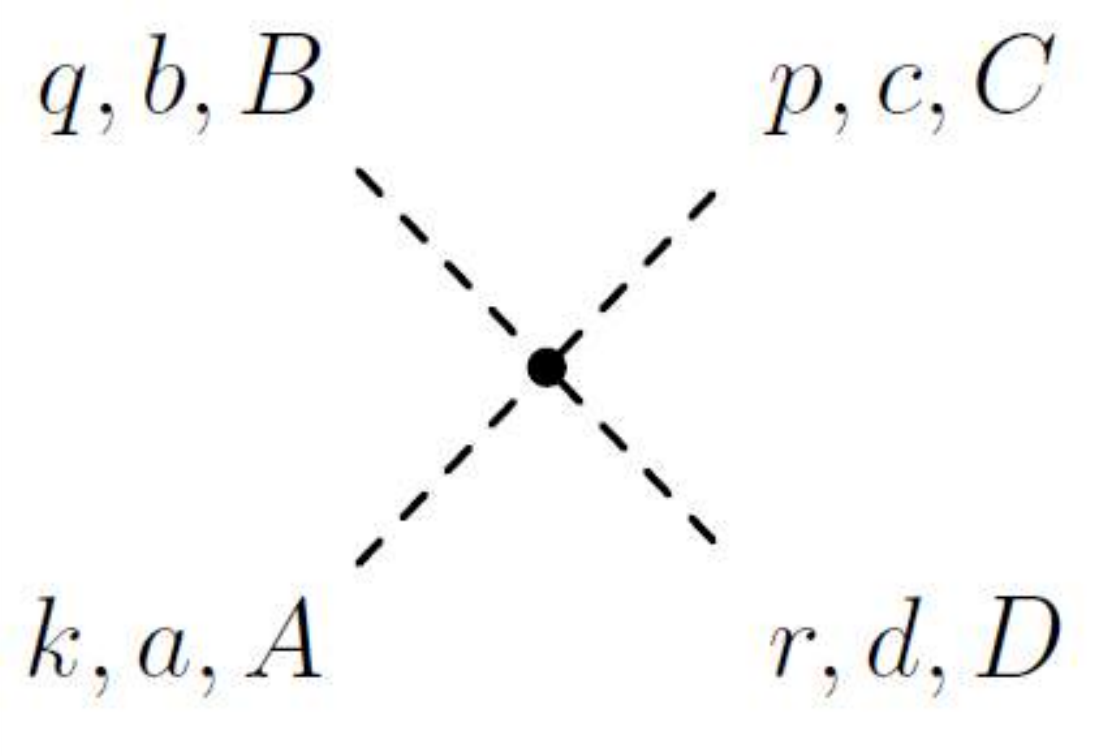}&&\ba
        &=-ig^2&\big[\;f^{abe}f^{cde}(\delta^{AC}\delta^{BD}-\delta^{AD}\delta^{BC})\nn \\
        &&+f^{ace}f^{bde}(\delta^{AB}\delta^{CD}-\delta^{AD}\delta^{BC})\nn \\
        &&+f^{ade}f^{cbe}(\delta^{AD}\delta^{CB}-\delta^{AB}\delta^{DC})\;\big]\nn \ea
         \\ \\
        & & \centering \includegraphics[scale=.4]{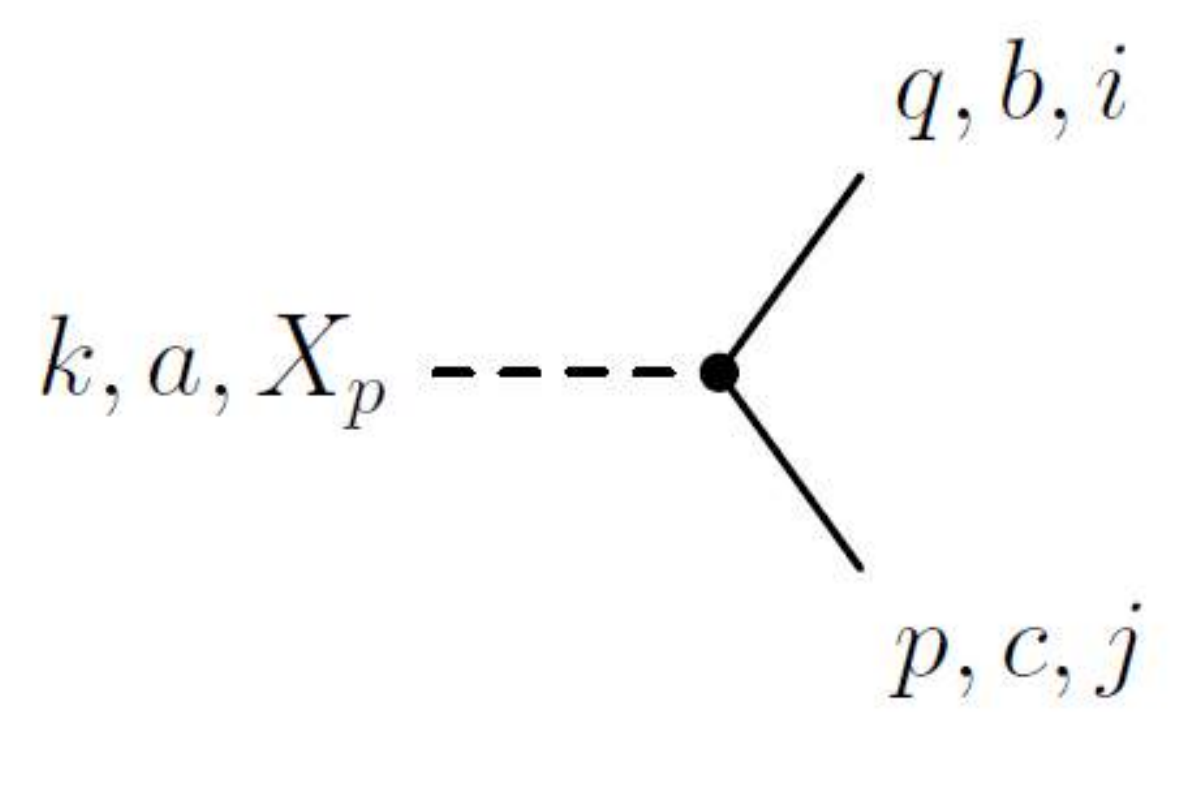} &&
        \ba &=& -ig f^{abc} \alpha^p_{ij} \nn \ea
         \\ \\
        & & \centering \includegraphics[scale=.4]{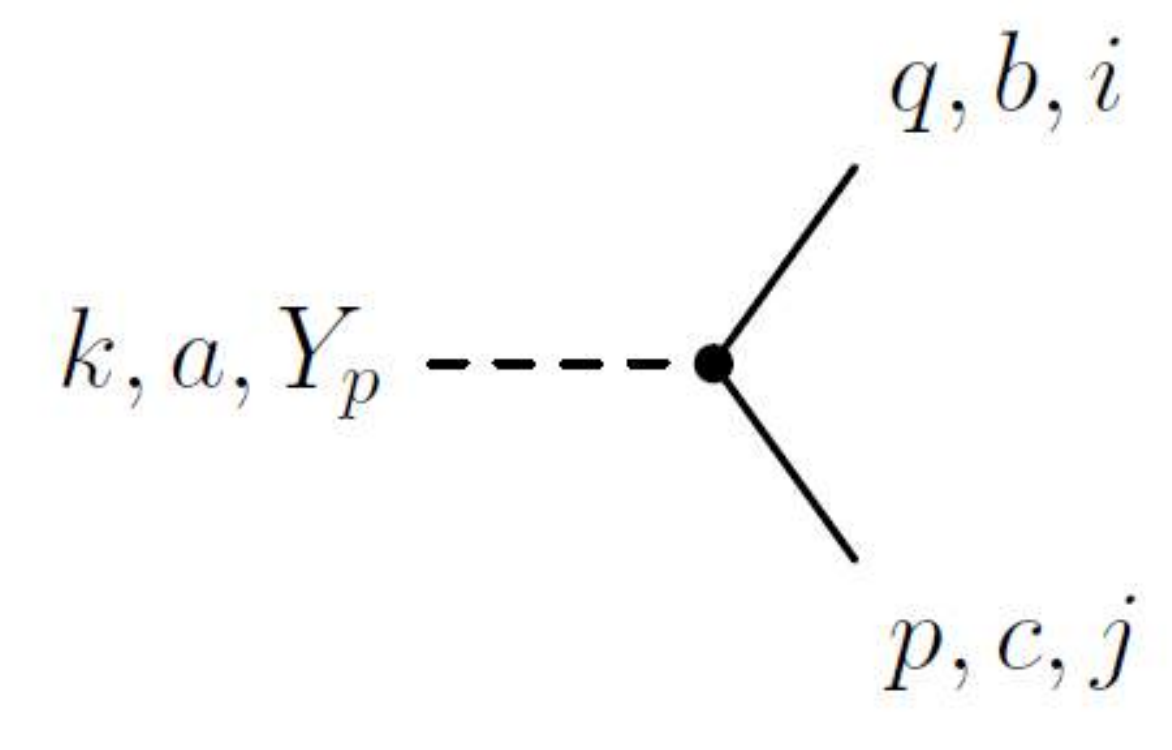} &&
        \ba &=& gf^{abc} \beta^p_{ij}\gamma_5 \nn \ea
        \\  
\hline \hline
\end{tabular}
\end{table}

Since all fields of the ${\cal N}=4$ super Yang-Mills are, except the ghosts, real, there are no arrows orienting the lines. However, one should remember that the momentum of every gluon in the three-gluon coupling is assumed to enter the vertex.  In the case of the gluon coupling to scalars, the momentum of one scalar enters the vertex and the momentum of the other one leaves it.


\newpage
\textcolor[rgb]{1.00,1.00,1.00}{.}



\phantomsection
\addcontentsline{toc}{section}{Bibliography} 
\bibliography{Bibliography}

\bibliographystyle{unsrtnat}    








\todos

\end{document}